# Spatial, Temporal, and Geometric Fusion for Remote Sensing Images

Dissertation

Presented in Partial Fulfillment of the Requirements for the Degree Doctor of Philosophy in the Graduate School of The Ohio State University

By

Hessah Albanwan

Graduate Program in Civil Engineering

The Ohio State University

2022

Dissertation Committee

Dr. Rongjun Qin, Advisor

Dr. Charles Toth

Dr. Alper Yilmaz

# Abstract

Remote sensing (RS) images play an important role in monitoring and surveying the earth's surface at varying spatial scales. Continuous observations from various remote sensing sources over time complement single observations and provide better interpretability and analyzability for various applications. A critical need is to fuse these observations into single or multiple images that are more informative, accurate, complete, and coherent. Studies have intensively investigated spatial-temporal fusion for specific applications, such as pan-sharpening and spatial-temporal fusion of remote sensing images for time-series image analysis. Today, fusion methods have exceeded these typical applications, extending to encompass and accommodate different types of images, modalities, and tasks. Typically, fusion methods are expected to be robust and able to adapt to various types of images (e.g., spectral image, classification maps, and elevation maps) and scene complexities. In this work, we present a few solutions to improve the performance of existing fusion methods by adapting to gridded data with multimodality and considering their type-specific uncertainties. The main contributions of this work include the following:

1) A spatial-temporal filtering method that addresses the spectral heterogeneity of multitemporal images.

2) Three-dimensional (3D) iterative spatiotemporal filtering that enhances the spatiotemporal inconsistencies of classification maps.

3) Adaptive semantic-guided fusion that enhances the accuracy of DSMs, and an analysis of the adaptive fusion approaches comparing them with traditional nonadaptive approaches to show the significant importance of adaptive methods to boost the performance of fusion.

4) A comprehensive analysis of deep learning (DL) stereo matching methods against traditional Census semi-global matching (SGM) to obtain detailed knowledge on the accuracy of the DSMs at the stereo matching level. We analyze the overall performance, robustness, and generalization capability. This analysis is important to identify the limitation of current DSM generation methods.

5) Based on the previous analysis, we develop a novel finetuning strategy to enhance the transferability of DL stereo matching methods, hence, the accuracy of DSMs.

This work reveals the significant importance of using spatial, temporal, and geometric fusion to enhance different kinds of RS applications. It also shows that the fusion problem is case-specific and that it depends on the type of image, scene content, and application. Therefore, a thorough understanding of the nature of the images (i.e., the physical information, uncertainties, etc.), scene content (i.e., land covers, spatial distributions, etc.), application, and limitations of current fusion methods is necessary to be able to accommodate and address these variations and provide optimal fusion solution.

## Dedication

To my mother, my biggest supporter, my role model, and the strongest woman I know, whose hard work and dedication made me the person I am today. This would not have been possible without you.

To my father who always believed in me and called a doctor since I was 14 years old, whose prayers and confidence kept me going through this journey.

To my husband, whose unlimited support, sacrifice, patience, and encouragement kept me going during this challenging time. Thank you for standing by me.

To my daughter, the love of my life, who flipped my world, taught me to multi-task, and made me a better and more efficient person, I hope this work will make you proud and inspire you to achieve great accomplishments.

To my sister and brothers, whose cheerfulness and continuous encouragement always lifted me up.

Thank you for all your love and support. This dissertation would not have happened without you.

**Acknowledgments**

I would like to gratefully acknowledge my advisor Dr.Rongjun Qin who guided me through my master's and Ph.D. journeys from 2016 until 2022. I would like to thank you for all the knowledge, skills, and lessons you taught me throughout the learning process and towards completing my Ph.D. degree. You led me towards great opportunities and engaged me in many projects and collaborations, where I had the chance to publish journal papers, conference papers, and a book chapter. You have always believed in your students and trained them on high standards, you are a true mentor, and I was lucky to be trained under your supervision. Thank you Dr.Qin.

I would like also to thank the committee members for their support and time Dr. Charles Toth and Dr. Alper Yilmaz.

Most importantly, I would like to acknowledge Kuwait university for the sponsorship and opportunity that was given to me to pursue my master's and Ph.D. degrees.

In the end, I would like to thank my family and friends for all their love, encouragement, and prayers during this journey.

# Vita

| | |
|---|---|
| **2013** | B.S., Civil Engineering, Kuwait University |
| **2013- 2015** | Site Engineer, Ministry of Public Works, Kuwait |
| **2015- 2017** | M.S., Civil Engineering, The Ohio State University |
| **2017- Present** | Ph.D. student & candidate, Civil Engineering, The Ohio State University |

## Publications

### Book chapter

- **Hessah Albanwan,** Rongjun Qin (2020). Spatiotemporal Filtering for Fusion in Remote Sensing. In Kwan, C. (Eds.), Recent Advances in Image Restoration with Applications to Real World Problems. IntechOpen

### Journal papers

- Mostafa Elhashash, **Hessah Albanwan**, Rongjun Qin. (2022). A Review of Mobile Mapping Systems: From Sensors to Applications. *Sensors*, *22*(11), 4262.
- Wenxia Gan, **Hessah Albanwan**, Rongjun Qin. Radiometric Normalization of Multi-Temporal Landsat and Sentinel-2 Images Using a Reference MODIS Product through Spatiotemporal Filtering. IEEE Journal of Selected Topics in Applied Earth Observations and Remote Sensing.14 (2021): 4000-4013.
- Yulu Chen, Rongjun Qin, Guixiang Zhang, **Hessah Albanwan** (2021). Spatial Temporal Analysis of Traffic Patterns during the COVID-19 Epidemic by Vehicle Detection Using Planet Remote-Sensing Satellite Images. Remote Sensing, 13(2):208.
- **Hessah Albanwan**, Rongjun Qin, Xiaohu Lu, Mao Li, Desheng Liu, Jean-Michel Guldmann (2020). 3D Iterative spatiotemporal Filtering for Classification of Multi-temporal Satellite Dataset. Photogrammetric Engineering and Remote Sensing. 2020, 86(1): 23-31.
- **Hessah Albanwan,** Rongjun Qin (2018). A Novel Spectrum Enhancement Technique for Multi-temporal, Multi-Spectral Data Using Spatial-temporal Filtering. ISPRS Journal of Photogrammetry and Remote Sensing. 142 (2018) 51-63.

**Conference papers**

- **Hessah Albanwan,** Rongjun Qin (2022). Fine-Tuning Deep Learning Models for Stereo Matching Using Results from Semi-Global Matching. ISPRS. Annals. Photogramm. Remote Sens. Spatial Inf. Sci. (Peer-reviewed)
- **Hessah Albanwan**, Rongjun Qin (2021). Adaptive and Non-adaptive Fusion Algorithms Analysis for Depth Maps Generated using Census and Convolutional Neural networks (MC-CNN). ISPRS. Annals. Photogramm. Remote Sens. Spatial Inf. Sci. (Peer-reviewed)
- **Hessah Albanwan,** Rongjun Qin (2020). Enhancement of depth map by fusion using adaptive and semantic-guided spatiotemporal filtering. ISPRS. Annals. Photogramm. Remote Sens. Spatial Inf. Sci. 3, 227-232. (Peer-reviewed)

**Papers under review and preparation**

- **Hessah Albanwan,** Rongjun Qin (2022). A Comparative Study on Deep-Learning and Semi-Global-Matching for Dense Image Matching of Multi-angle and Multi-date Remote Sensing Stereo Images. Submitted to "The Photogrammetric Record".
- **Hessah Albanwan,** Rongjun Qin (2022). Taxonomy of Image Fusion in Remote sensing: A Comprehensive Review of the Mathematical formulation, Techniques, and Applications.

## Fields of Study

Major Field: Civil Engineering

Studies in:

Topic 1: Remote Sensing Data Fusion

Topic 2: DSM Fusion

Topic 3: Relative Radiometric Normalization

Topic 4: Classification

Topic 5: Stereo Matching

Topic 6: Photogrammetry

## Table of Contents

## List of Tables

## List of Figures

# Chapter 1. Spatial, Temporal, and Geometric Fusion of Remote Sensing Images

This chapter provides the general motivation and background knowledge about this dissertation. First, it covers the main motivations behind this work and the general background on image fusion in remote sensing (RS). Next, this chapter states the research goals and questions, which are important for understanding the potential of spatial, temporal, and geometric image fusion. Finally, this chapter presents the main contributions, dissertation structure, and outline.

## 1.1. Motivation and Background Information

Today, RS images are powerful tools to monitor and survey the surface of the earth continuously. The increased availability of sensors and images has prompted numerous applications to use and process these images. However, the raw RS images often include many issues and errors that develop due to varying acquisition conditions (e.g., sun angle, weather, etc.) and configuration parameters (e.g., viewing angle), which can propagate in the preprocessing algorithms. The most common challenges of RS images can be summarized as follows:

- High spectral variability between multitemporal images causes images covering the same region to have different appearances and spectral values, as illustrated in Figure 1.1 (a).
- Spatiotemporal inconsistency, which is the change that happen over time on the scene due to natural (e.g., disasters or seasons) or human (e.g., urban development projects) transformations, can lead to images with different information, causing misinterpretation. An illustration of the spatiotemporal inconsistency is shown in Figure 1.1 (c).

- Noise and incomplete images due to acquisition conditions like weather and preprocessing algorithms. Figure 1.1 (c) shows an example of noise in the form of a cloud in the spectral image and missing values in the Digital Surface Model (DSM) as indicated by the white pixels showing an incomplete representation of an image.

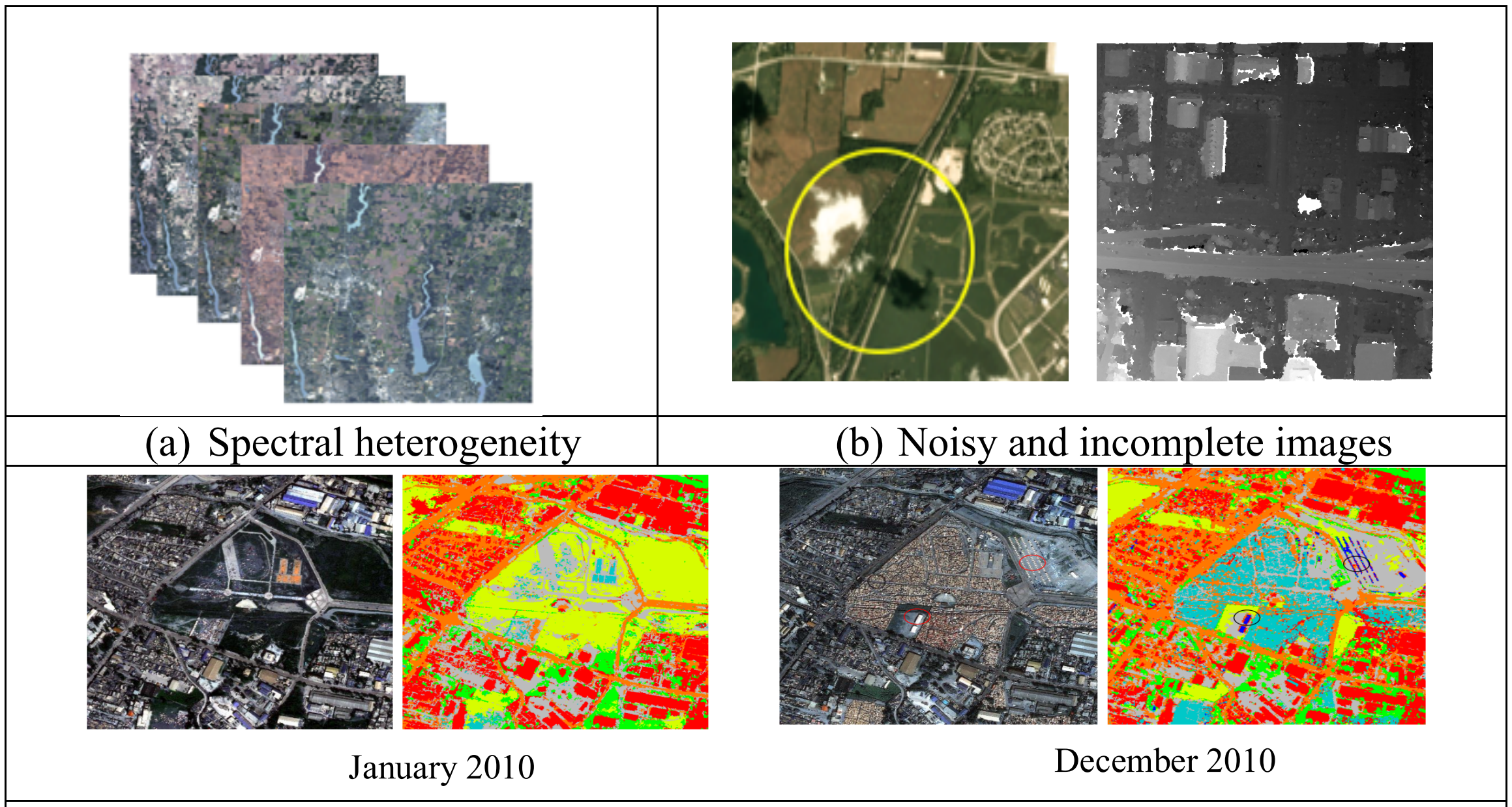

(a) Spectral heterogeneity (b) Noisy and incomplete images

(c) Spatiotemporal inconsistencies

Figure 1.1 Existing challenges of remote sensing images

Image fusion is often proposed to enhance a variety of issues related to RS images by combining information from multiple images into single or multiple images. Fusion allows providing more informative, accurate, complete, and coherent images. The outstanding performance of image fusion has intrigued the RS community to study more aspects of its applications and benefits. Fusion methods have demonstrated a notable progress in numerous RS applications, such as the following:

- Resolution enhancement, such as super-resolution and pan-sharpening (Alparone et al., 2015; Anger et al., 2020; X. Meng et al., 2019; Shen et al., 2008; S. Zhang et al., 2020; H. Zhu et al., 2016, 2018);
- Missing data reconstruction (Fan et al., 2011; J. Kang et al., 2018; Zeng et al., 2013);

- Denoising from acquisition conditions, such as clouds, fog, and shadows (P. Dai et al., 2020; W. Du et al., 2019; K. et al., 2016; C.-H. Lin et al., 2013; F. Meng et al., 2017; Movia et al., 2016; Siravenha et al., 2011);
- Classification map enhancement (Albanwan et al., 2020; Benediktsson et al., 2005; G. Fu et al., 2017; Q. Gao & Lim, 2019; Ji et al., 2020; X. Li et al., 2017; Lucas et al., 2007; Moser et al., 2013; J. Xia et al., 2016);
- Change detection (Z. Li et al., 2017; Luppino et al., 2019; Lv, Liu, Wan, et al., 2018; Lv, Liu, Zhang, et al., 2018; C. Wu et al., 2016; Zerrouki et al., 2022; X. Zhang et al., 2017);
- Relative radiometric normalization (Albanwan & Qin, 2018; Canty et al., 2004; Chavez, 1996; Y. Chen et al., 2018; Y. Du et al., 2002; Nagamani et al., 2021; Syariz et al., 2019; D. Yuan & Elvidge, 1996); and
- Environmental monitoring, such as phenological studies on crop cycles (Erkkilä & Kalliola, 2004; Lunetta et al., 2006; Viana et al., 2019).

The majority of the existing fusion algorithms apply spatiotemporal fusion (sometimes called filtering) on spectral images. Recent literature has also intensively investigated spatiotemporal fusion, primarily covering this concept for specific types of images, such as spectral images, and certain applications, such as pan-sharpening. Because of the standard grid format of RS images, fusion is applicable to process a wide range of image types such as spectral images (e.g., multispectral, hyperspectral, and RGB), topographic/elevation maps, and classification maps. It also enables fusion to explore the spatial and temporal correlations to provide the best predictions based on the most relevant information. The spatial information in the images assumes neighboring pixels are more similar than distant pixels, where the temporal information provided by a set of images provides clues on the most consistent values and outliers. Despite the significance of spatiotemporal fusion, it still lacks robustness when processing spectral images due to their sensitivity to acquisition conditions such as sun angle and weather. Recently, the 3D geometric information (e.g., elevation maps or DSMs) has been included in many of the RS applications, such as 3D change detection (Chaabouni-Chouayakh et al., 2010; MacFaden et al., 2012; Qin et al., 2016a) and that is due to its superiority and robustness to varying acquisition conditions. With height values being less likely to change over time, engaging the geometric information in the fusion process may have promising outcomes to provide stable and acceptable results.

Nevertheless, the accuracy of the input geometric information is critical and can impact the fusion output. With DSMs generated using stereo matching methods, they are most likely to have many issues such as incorrect elevation values, missing points, and incorrect representation boundaries and edges. Fusion concepts can be applied to combine several DSMs and enhance their accuracy, one of the effective methods to boost the performance of the fusion algorithm is through adaptive methods, which takes into consideration similar objects. However, if most DSMs are bad in terms of their accuracy, then the final results will most likely be impacted to produce low accuracy DSM. Therefore, a key solution to enhance the accuracy of the input DSMs is through a comprehensive analysis of the existing DSM generation methods (i.e., stereo matching methods) to understand their capabilities and limitations to be able to conquer these limitations and provide more accurate DSMs.

In this dissertation, we intensively investigate spatial, temporal, and geometric image fusion to enhance different kinds of RS images and applications. We focus on improving existing spatiotemporal fusion methods by engaging 3D geometric information or adaptive algorithms. We explore the underlying relationships between data and combine the grid data with multimodality results (e.g., spectrum, probability maps for classification, and DSM fusion), considering type-specific uncertainties.

## 1.2. Research Scope

This work explores the various ways in which fusion methods can be enhanced through spatial, temporal, and geometric image fusion. It also shows the significant importance of adapting the fusion algorithm based on the unique types of images, uncertainties, and modalities. Therefore, in this dissertation, we apply spatial, temporal, and geometric fusion using different strategies to enhance various types of RS images and applications. We will investigate the potential, capabilities, and limitations of this type of fusion approach. The primary research questions can be summarized as follows:

- What are the most common challenges of RS images?
- What are the existing solutions to address RS image challenges through fusion algorithms?
- What are the capabilities and limitations of existing spatial and temporal image fusion?
- How can we improve existing spatiotemporal fusion methods?
- What type of images and applications can benefit from fusion methods?
- How can 3D geometric information be beneficial to enhancing the performance of spatial and temporal image fusion?
- How can adaptive methods boost the performance of image fusion?
- Is there a standard fusion solution that can be applied to all image types and applications?
- What are the main considerations to consider when applying fusion methods?

## 1.3. Contributions

The performance of RS applications relies on images to be spectrally homogenous, consistent, complete, accurate, and noise-free. Uncontrollable and varying acquisition factors make it impossible to satisfy such conditions, imposing several challenges on RS images and limiting their practical usage in some applications. These challenges can be resolved through fusion methods. Traditional spatiotemporal fusion methods are effective but may be influenced by the varying acquisition conditions that lead to high spectral variability and distortions in images. In some cases, using a supplementary stable and robust type of data as 3D geometric information or advanced fusion methods as adaptive approaches can be beneficial to further enhance the performance of fusion algorithms.

This work explores the various ways in which fusion can be applied and improved in practice. We propose a few fusion solutions that consider and accommodate the different image types, such as multispectral images, DSMs, and classification maps, to improve the final products of RS applications. Primarily, we focus on using spatial, temporal, and geometric information to improve the fused images. We explore the creative approaches as adaptive methods that can boost the performance of fusion algorithms. Figure 1.2 summarizes the main contributions, where the light pink boxes indicate the existing challenge of RS images we attempt to address, and the gray boxes are the proposed fusion algorithms to address this specific challenge. In total we have five projects (that are considered contributions), the first three projects are examples of spatial, temporal, and geometric fusion tailored to address specific problems or challenges in RS images. The fourth project is a follow-up work to evaluate the adaptive fusion methods versus traditional non-adaptive methods. The last two projects are side projects to enhance the accuracy of the input DSMs, where enhancing the accuracy of the inputs can enhance the accuracy of the fusion output. The contributions are described in detail as follows:

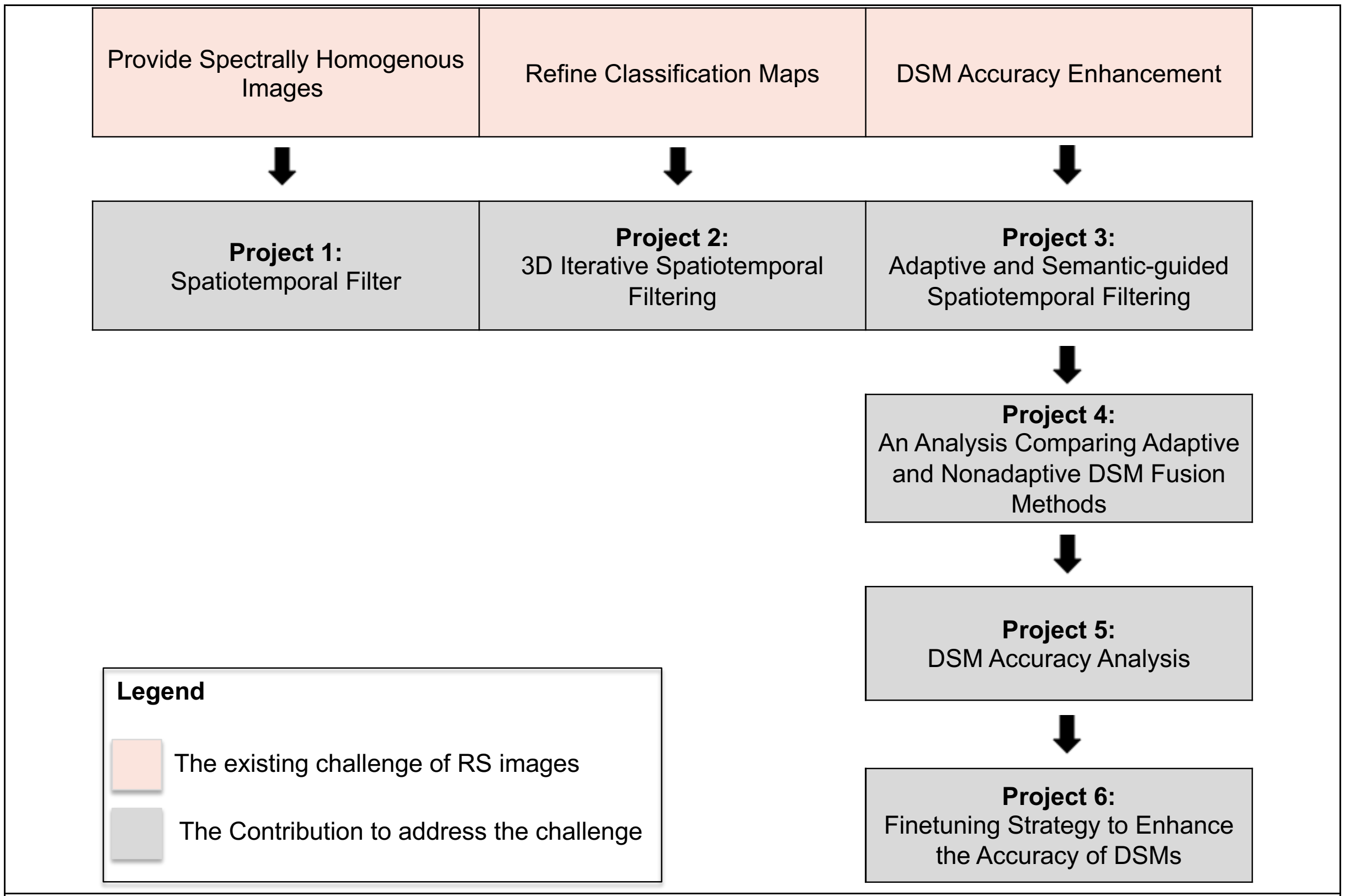

Figure 1.2 Contributions addressing the existing challenges in remote sensing images

***Project 1: Spatiotemporal Filter Providing Spectrally Homogenous, Consistent Multitemporal Images***

Due to the varying acquisition and configuration parameters at the time images are captured, RS images captured at different times, days, and seasons often exhibit noise and lack spectral consistency and heterogeneity. Atmospheric correction may slightly reduce errors like cloud cover; however, it does not solve problems related to non-spectrum distortions. Relative radiometric normalization (RRN) methods reduce the spectral difference between satellite images. Nevertheless, they often depend on noise-free reference images to guide other images.

To address these issues, we propose a simple 3D spatiotemporal filtering method that processes multitemporal images with a spectral difference and generates homogenous and consistent images while maintaining the most important changes and features over time. We experiment on images with medium- (using the Landsat dataset) and high-resolution (using the Planet dataset). We evaluate this method using a supervised classification based on two strategies: 1) classic learning, classifying each image using a sample selected from the current image, and 2) transfer learning, using samples selected from one dataset and applying them to the other dataset. The results indicate that the proposed 3D spatiotemporal filter can enhance the accuracy of classification of the transfer learning approach by ≈5%, ≈15%, and ≈2%. We also perform change detection to evaluate the performance of this method, and the experiment indicates better change detection results using the basic image differencing method.

***Project 2: The 3D Iterative Spatiotemporal Filtering to Enhance the Consistency of Classification Maps***

Land cover and land use maps are essential to monitor and document the components of the earth's surface. However, their accuracy and precision are highly affected by the characteristics and distortion level in the images. Most often, classifying images of the same scene captured at different times leads to inconsistent classification maps because varying acquisition conditions can influence the image appearances. However, 3D geometric features provided by the elevation maps are more stable across time and are invariant to the acquisition conditions, such as weather or sun angle. Therefore, in this work, we propose a spatiotemporal post-classification approach that combines orthophoto with the DSM generated from satellite images to improve the accuracy of the classification maps. First, we generate the initial probability maps using a random forest classifier, then apply the 3D iterative spatiotemporal filtering to each class of probability map. Our experiment indicates consistent improvement in the classification results at between 2% and 6%.

***Project 3: Adaptive Semantic-guided Spatiotemporal Filtering to Enhance the Accuracy of DSMs***

The DSMs are essential components for performing 3D reconstruction. There are two ways to generate DSMs: 1) lidar, which is accurate, highly expensive, and often available to cover the entire surface of the earth, or 2) Multiview stereo (MVS) methods that take a pair of stereo images as input and output disparity triangulated to the DSM. The MVS methods are efficient, low-cost methods; however, the outputted DMSs are often noisy and imprecise, and the DSM accuracy largely depends on the degree of radiometric and geometric consistency between the stereo pair images.

In addition, DSM fusion is often proposed to combine the information from multiple DSMs and gain better accuracy and reliability. However, the fusion process is challenged by the varying uncertainties of the scene components. For example, trees are more affected by seasons, resulting in extreme errors. Similarly, buildings are affected by the visibility and ability of the algorithm to match pixels around edges and boundaries. Thus, they are likely to have high uncertainty. To address this issue, we propose semantic adaptive spatiotemporal filtering that considers different classes during fusion. This method improves the results of the fused DSM.

***Project 4: A Comparison Between Adaptive and Nonadaptive DSM Fusion Methods***

We present the main benefit of using adaptive techniques to boost the performance of the fusion algorithm. The adaptive fusion solutions are well-established approaches in RS and are important to solve problems related to strengthening weak structures, textures, and boundaries of images. This outcome is primarily due to mixed pixels around the edges and obstructions from shadows, trees, viewing angles, and other factors that block the visibility of the complete object and its clear boundaries. This work demonstrates the benefits and advantages of using adaptive versus nonadaptive fusion methods to refine and enhance the DSM accuracy.

***Project 5: An Analysis of the DSMs Accuracy at Stereo Matching Level***

We study the impact of using stereo pair satellite images with variances in locations, acquisition conditions, and configuration parameters (i.e., sun angle difference and intersection angles) on the performance of state-of-the-art deep learning (DL) stereo matching methods that are used to generate the DSMs. We perform a comprehensive review using hundreds of stereo pairs over nine-test sites with two types of DL methods the learning-based presented by MC-CNN and end-to-end learning presented by GCNet, PSMNet, and LEAStereo. We evaluate their overall accuracies, configuration parameter robustness, and generalization capability. A thorough understanding of the performance of DL methods used with satellite images is essential to understanding the gaps and limitations to propose suitable solutions to boost performance.

***Project 6: Fine-tuning Strategy to Enhance the Accuracy of DSMs***

Transferability and generalizability are major challenges for DL stereo matching methods. The instances in the training data limit the performance. Satellite images often cover large areas that vary by location, land cover, spatial object, chrematistics (e.g., shape, size, etc.), and distribution, which may influence the performance of DL methods. Additionally, collecting ground truth data for large-scale areas and the entire globe is impossible due to the expense, leading to a lack of ground truth data that can be used for training. Classical stereo matching approaches, such as Census-SGM, have been proven to work well for most types of data regardless of their source and locations; therefore, we propose a fine-tuning method using SGM's disparity map of the target data. We extract and learn from high-confidence pixels using the energy map derived from SGM's algorithm and the texture-rich regions from the edge map extracted from the left image pair. We experiment on 20 study sites and three well-established DL stereo matching methods, including GCNet, PSMNet, and LEAStereo. Moreover, we demonstrate noticeable improvement in the results numerically.

## 1.4. Overview of the Dissertation Structure

This dissertation is divided into several chapters and sections. Chapter 2 provides the background information, definitions, and existing challenges of RS image fusion. It also describes the concepts of spatial, temporal, and geometric fusion and the existing techniques used to enhance RS applications. Chapters 3 to 7 include the fusion solutions described in the contributions section. In addition, Chapter 3 discusses spatiotemporal filtering to address the spectral heterogeneity and inconsistency of multidate RS images. Chapter 4 addresses the spatiotemporal inconsistency that affects the consistency of the classification maps by providing a multisource image fusion solution that uses spatial, temporal, and geometric information in the fusion process. Next, Chapter 5 proposes a semantic adaptive fusion algorithm to enhance the DSM accuracy. Further, Chapter 6 compares adaptive and nonadaptive DSM fusion algorithms to demonstrate the importance of using adaptive methods in fusion. Chapter 7 evaluates the performance of DL methods used for stereo matching tasks with satellite images and highlights the main gaps and limitations in this analysis. Chapter 8 proposes a fine-tuning solution to address the challenges of DL methods, and generalizability. Finally, this dissertation concludes by describing the major findings and addressing the research question.

.

.

# Chapter 2. Background and Definitions

## 2.1 Remote Sensing Image Definitions and Characteristics

The images comprise two-dimensional (2D) grid data representing physical information related to specific areas on the earth's surface as observed from space. Images comprise pixels distributed in rows and columns, where each pixel withholds a value, such as intensity, elevation, and temperature, defining the image type. Examples of RS images include topographic/elevation maps, classification maps, satellite images, aerial photos, and thermal maps (Figure 2.1).

| 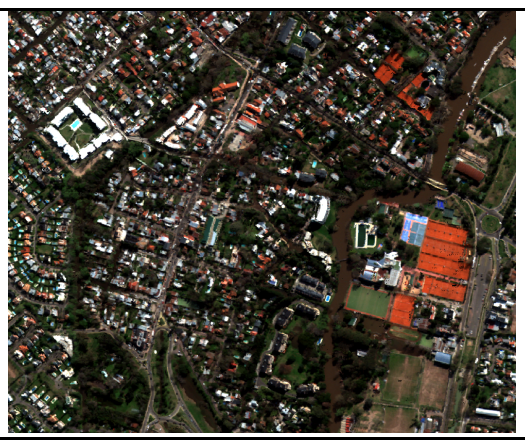 | 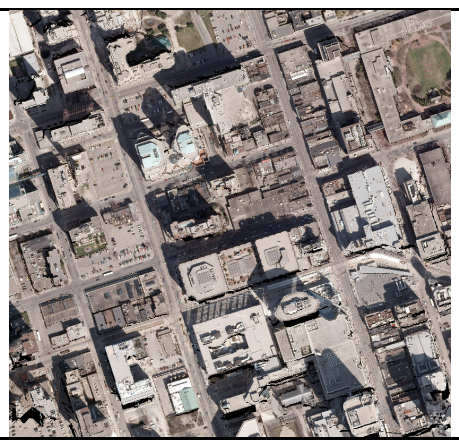 | 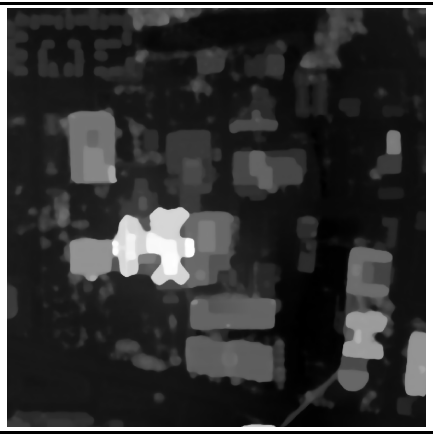 | 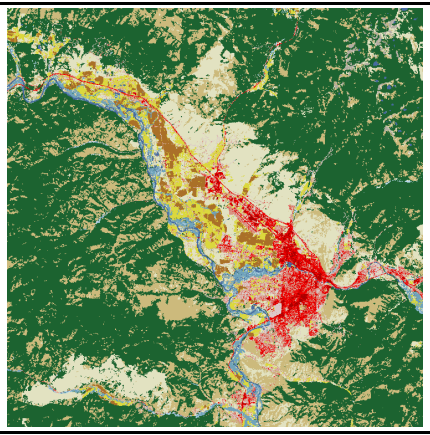 |
|---|---|---|---|
| Satellite image. Argentina IARPA benchmark. | Airborne image. ISPRS Toronto benchmark. | Elevation map. From fused Census-SGM, the 2019 DFC benchmark | Classification map. From NLCD, USA |

Figure 2.1. Image types in remote sensing

Typically, there are two ways to acquire images: directly from sensors or by processing the raw image using specialized algorithms to deliver the final product (e.g., stereo matching to obtain DSMs). In addition, RS images have different characteristics based on several factors related to sensor properties. First, spectral resolution refers to a sensor measuring specific spectrum wavelengths. Images from different sensors may have varying numbers and widths of bands in terms of reading wavelengths. Images can include different number and types of bands, for example, RGB (i.e., red, green, and blue), multispectral, hyperspectral, and others.

Second, the spatial resolution which indicates the extent of pixel coverage on the ground; for example, a 0.3 meters (m) resolution indicates that each pixel covers 30 centimeters (cm) on the ground. Smaller spatial resolutions have more precision and details in the image. However, in many cases, there is a tradeoff between spatial and spectral resolutions, where high spatial resolution images are more likely to have a low spectral resolution (i.e., few bands) because of the signal-to-noise ratio (SNR) and volume of the images. The typical spatial resolution for RS images ranges from low to high and can be as large as 1000 m, like the MODIS satellite sensor, or as small as 0.3 m, like the Worldview-3 sensor.

Third, the temporal resolution, which is the revisiting time, for example, daily, weekly, or monthly, of the satellite to a specific location on the earth's surface. The type of research and application are major factors in determining the appropriate sensor, method, and image type to use.

## 2.2 The Advantages of Image Fusion in Remote Sensing

The RS images are highly complex, and this is due to various reasons, which may include: 1) processing high-volume multimodal data with rich content, and 2) the varying acquisition conditions and configuration parameters that can cause many errors in the raw images, hence propagate through the preprocessing algorithms. Image fusion combines information from multiple images into single or multiple images to provide more informative, accurate, reliable, and complete images (Albanwan & Qin, 2021b; Belgiu & Stein, 2019; Ghassemian, 2016). The concept of fusion is basic and can be applied to process varying types of multimodal data that has a grid structure. It is mainly based on the redundancy of observations, where more information improves the confidence and interpretation of the images.

## 2.3 Existing Challenges of Image Fusion in Remote Sensing

Despite the many benefits of fusion methods in RS applications, some challenges can still limit its performance and degrade the output. These challenges include the existence of noise, spatiotemporal inconsistencies, spectral heterogeneity, propagated algorithmic errors, fusing multimodal images, and computational efficiency. In this section, we discuss these challenges in detail.

### 2.3.1. Noise in Images

Noise in raw acquired RS images is most likely to occur and is often difficult to estimate and quantify (Zhang et al., 2019). The acquisition conditions, such as weather, sun illumination, season, and others, are major sources of noise leading to ambiguous and contaminated images. The acquisition condition is the main cause of noise in the images, which can be in the form of haze, clouds, season, and shadows (Figure 2.2). The noise can provide incorrect information and block some information in the scene. For example, the snow cover and extent of the casted shadow due to the sun angle in Figure 2.2 hides the land cover type and objects in the scene. The noise in the images eventually accumulates and propagates in the fusion algorithm to generate artifacts and low visible images.

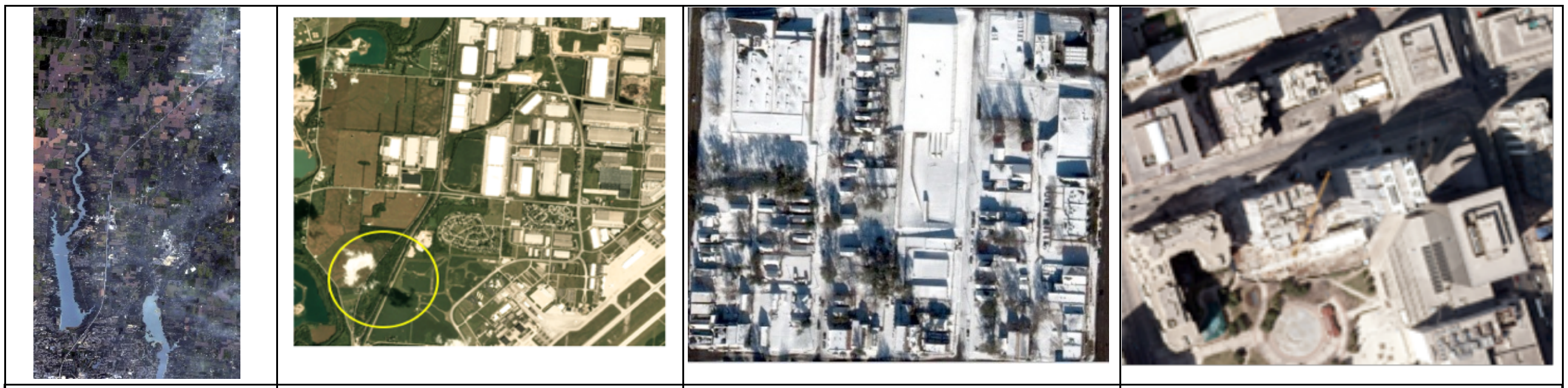

**Haze** | **Cloud cover** | **Snow cover** | **Shadows**

Figure 2.2. Examples of noise in remote sensing images

### 2.3.2. Spatiotemporal Inconsistency Between Images:

Fused images must cover the same geographical area simultaneously, and the scene content must be consistent with the information in all images. Due to natural and human changes on the earth's surface, images can vary greatly over time, leading to severe inconsistencies between images and presenting different content or information. These inconsistencies can result from natural changes, such as season or weather, or human-made transformations, such as urban development projects over time (Figure 2.3). In both cases, this variation can yield different information in the fusion algorithm and cause misinterpretation of the type of temporal change.

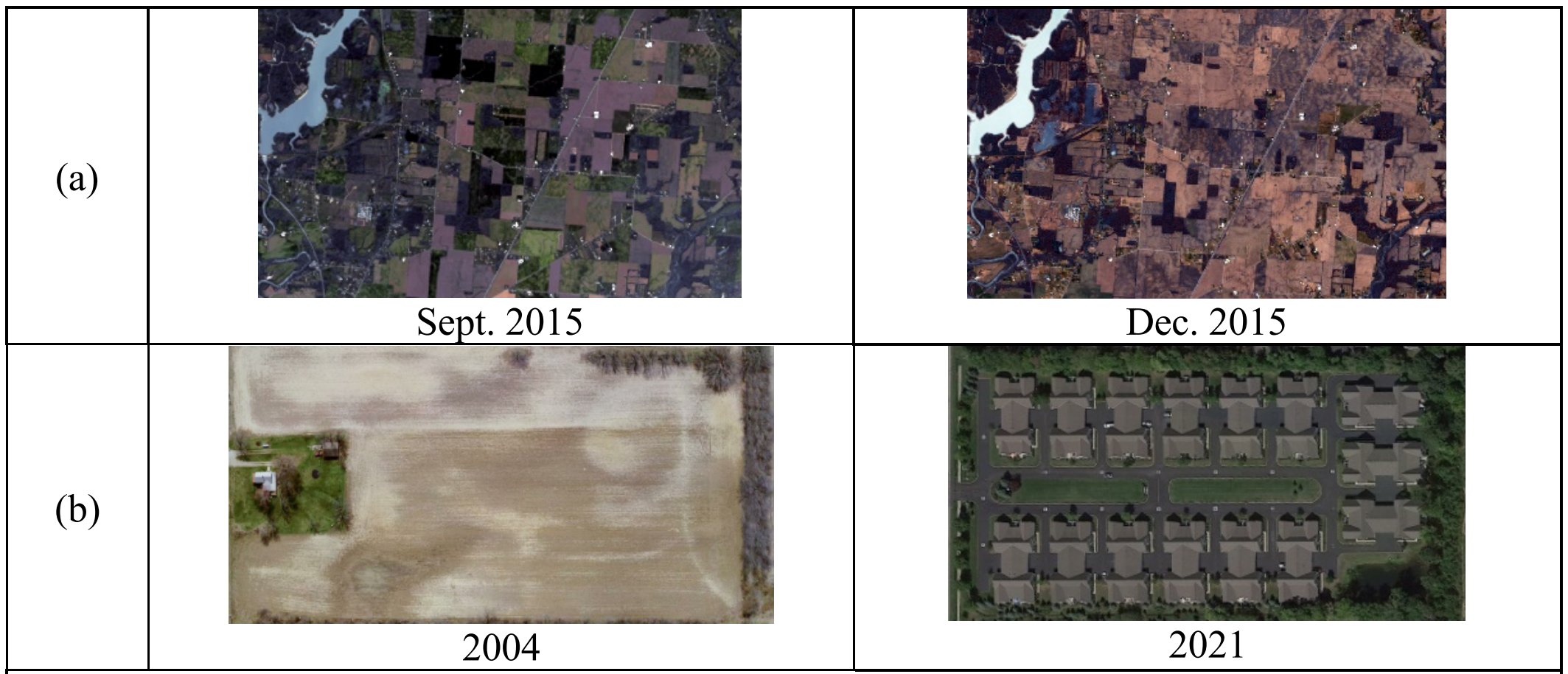

Figure 2.3. Examples of spatiotemporal inconsistencies due to (a) natural causes like crop cycle and (b) human causes like urban development of new buildings.

### 2.3.3. Spectral Heterogeneity:

Acquired images often have high spectral variability because of the constantly changing acquisition conditions (e.g., sun angle, time, season, etc.), resulting in different appearances. In RS, this is often referred to as spectrally heterogeneous images. Spectral heterogeneity refers to the resemblance between radiometric properties and the general image appearance. Figure 2.4 depicts an example of five images of the same location with different radiometric appearances, representing the typical situation of spectrally heterogeneous images due to varying acquisition conditions.

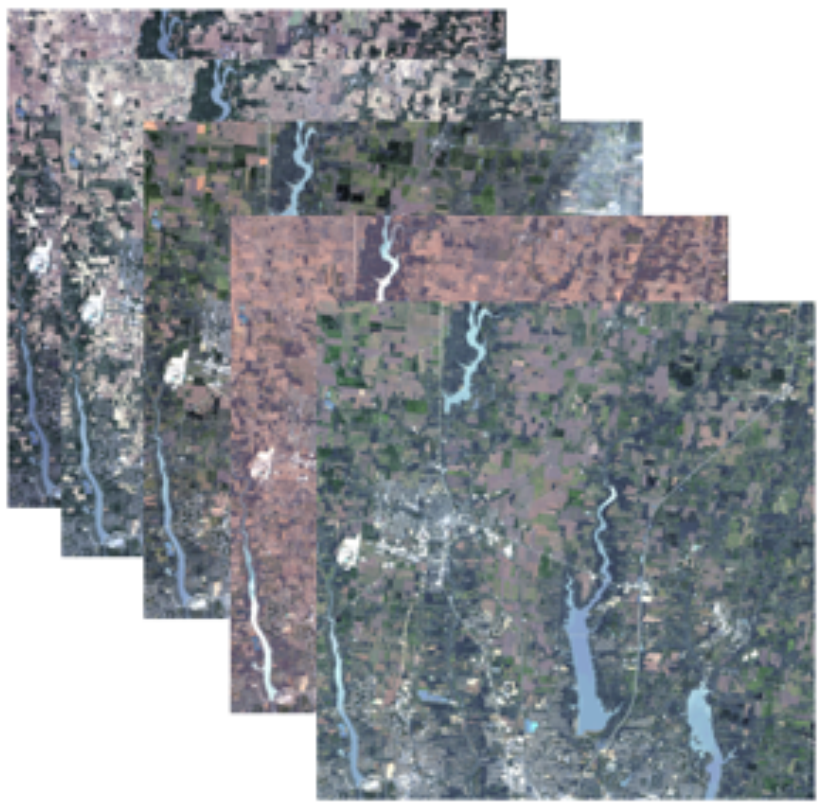

Figure 2.4. Examples of spectrally heterogeneous satellite images due to different acquisition conditions (i.e., time, sun angle, and season).

### 2.3.4. Fusing Multimodal Images:

Multimodal data refer to images of different types and sensors, for example, RGB images and digital elevation models (DSMs) or MODIS and Landsat satellite images. Combining multimodal data requires special considerations to account for unique types of images and their uncertainties. Additionally, considering the numerous existing approaches for RS tasks and applications, the performance and capabilities may vary based on the method. Figure 2.5 shows an example of DSMs derived from disparity maps generated using different stereo matching algorithms like Census-SGM, MC-CNN-SGM, and PSMNet, (Chang & Chen, 2018; Hirschmuller, 2005b; Žbontar & LeCun, 2015). Combining multimodal data requires a complete understanding of the image type, characteristics, source, and uncertainties.

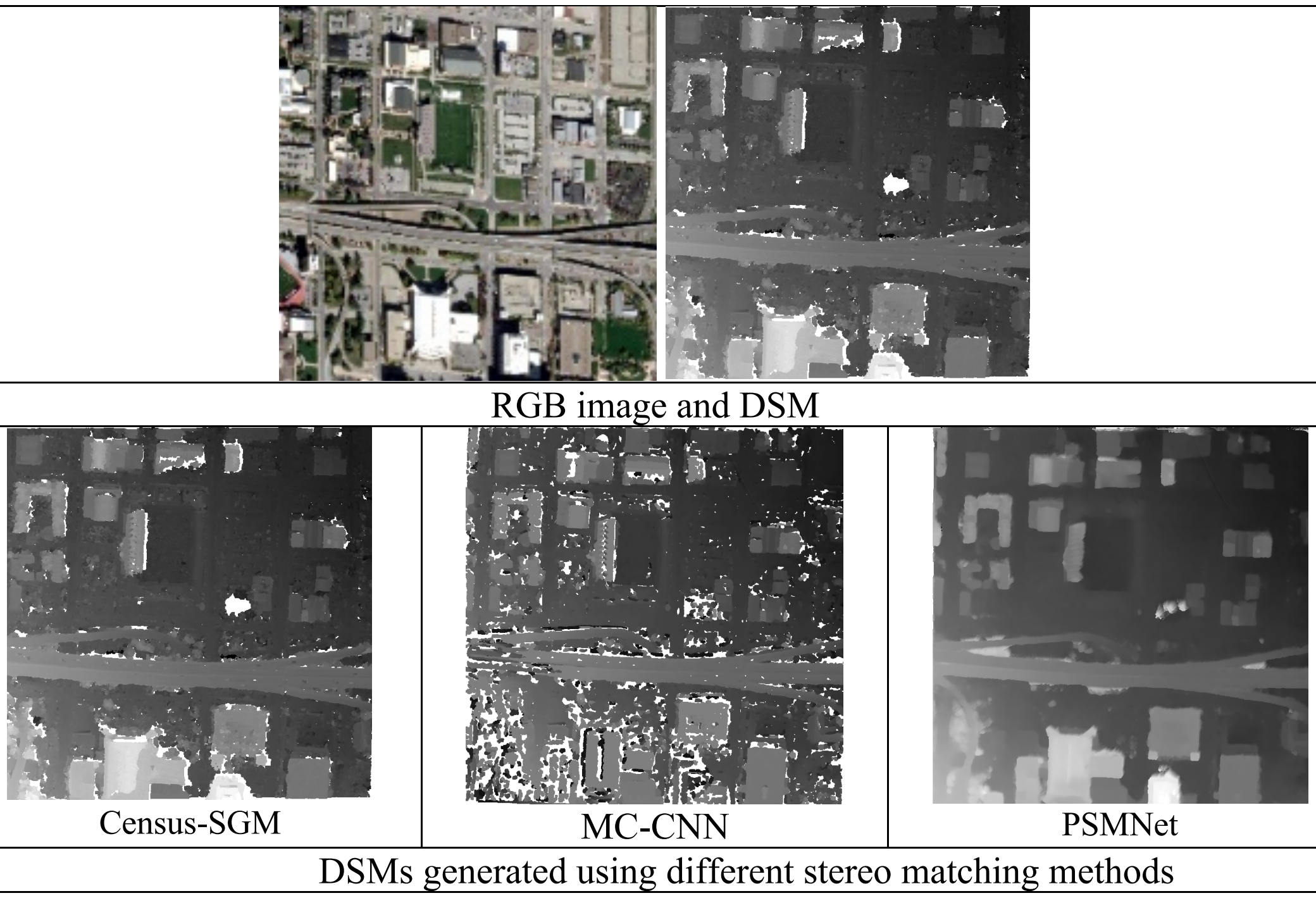

Figure 2.5. Examples of multimodal images.

### 2.3.5. Preprocessing Propagation Errors:

Some RS image types require preprocessing to reach their final product, such as classification or elevation maps, which are generated through classification and stereo matching algorithms. Raw images acquired directly from sensors often contain errors propagating through the preprocessing algorithms, leading to inaccurate, incomplete, and erroneous images. Generating digital elevation models (DSMs) from inconsistent stereo pair images can result in many mismatches and incorrect or missing elevation values (Figure 2.5(a)). Similarly, images with high spectral variability can lead to similar scene content being classified into different classes, resulting in imprecise and incorrect output images (Figure 2.5(b)). Fusing inconsistent noisy images consequently affects the fusion output.

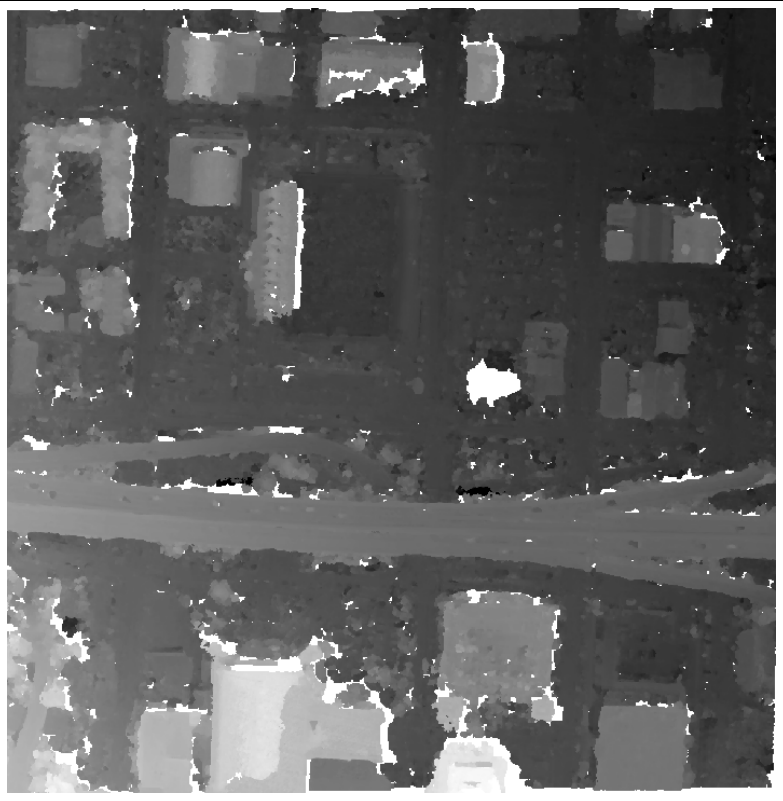

(a) DSM generated using Census-SGM shows noise (black pixels) and missing values (white pixels).

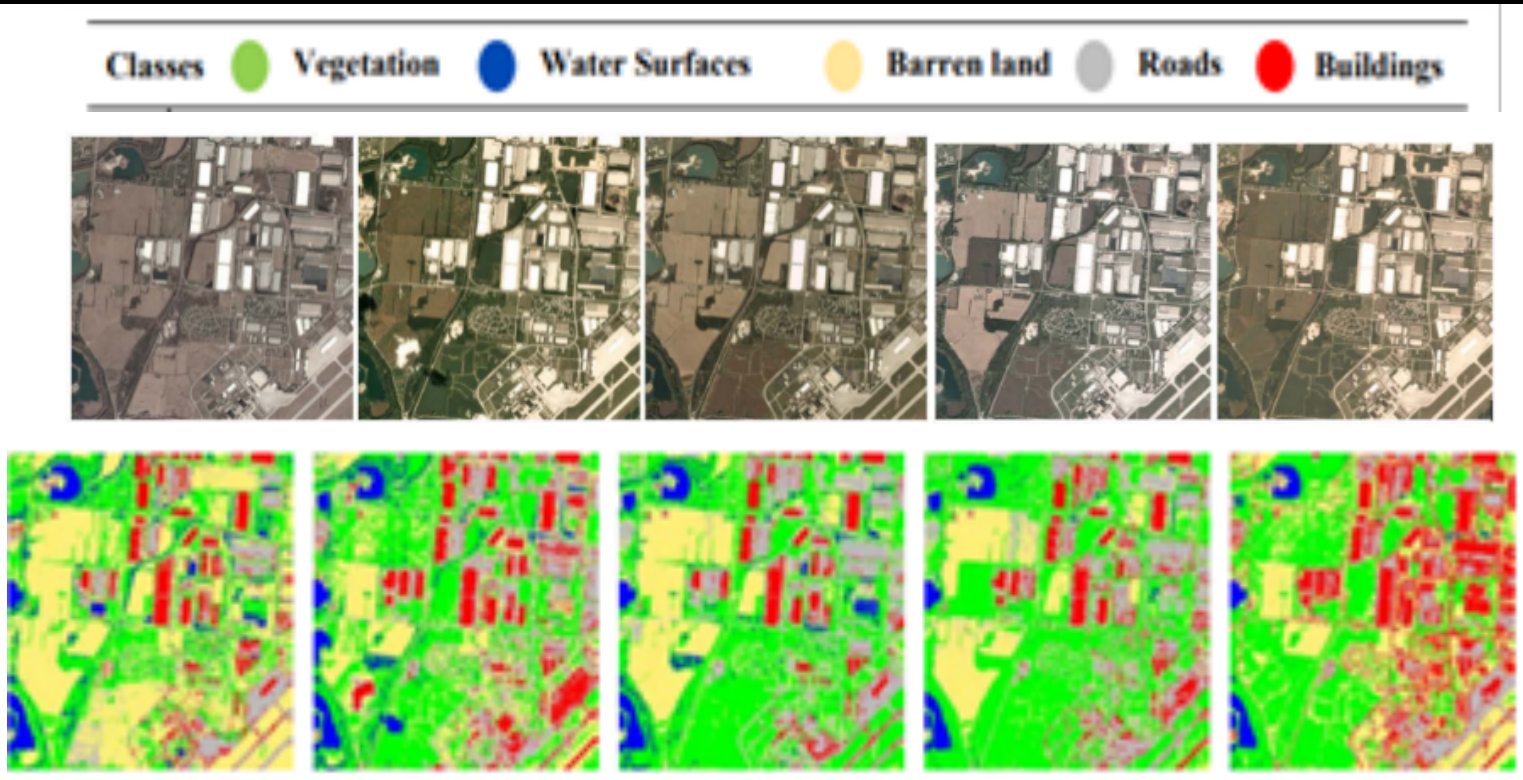

(b) Noise and Inconsistencies between the classification maps generated using a support vector machine (SVM).

Figure 2.6. Examples for preprocessing and algorithmic noise and inconsistencies from (a) stereo matching and (b) classification

### 2.3.6. Computational Efficiency:

Processing large-scale images with dense content in per-pixel operations can compromise the cost efficiency and processing time of the fusion algorithm, reducing its significant value in practice. This compromise introduces a challenge for processing voluminous RS data with minimum cost.

## 2.4 Fusion Taxonomies, Approaches, and Applications in Remote Sensing

Fusion has been extensively used for decades to enhance images and a wide range of RS applications. Numerous fusion approaches exist to handle different types of images, modalities, and applications, which has prompted the RS community to provide several taxonomies to classify the fusion problem based on varying perspectives. This section discusses some existing taxonomies and approaches and the most common applications that apply fusion, highlighting and describing spatial, temporal, and geometric fusion.

### 2.4.1. Image Fusion General Taxonomies in Remote Sensing

Over the years, several taxonomies have been proposed to classify fusion based on different aspects. These aspects include 1) the source of the input images, that is, similar or different sensors (Pohl & Van Genderen, 1998; Stein, 2005; J. Zhang, 2010), 2) the fusion operation level, which could be the pixel, feature, or decision level (Albanwan & Qin, 2020; Ghassemian, 2016; Schmitt & Zhu, 2016), 3) the method type, such as unmixing-based, Bayesian-based, or learning-based (X. Zhu et al., 2018), and 4) the fusion strategy (i.e., the type of fused information, which can be one type or a combination of spatial, spectral, temporal, and geometric information).

### 2.4.2. Concept of Spatial, Temporal, and Geometric Fusion

The grid structure of images allows exploring them in high dimensions using spatial, temporal, and geometric information. Spatial information is important for exploring spatial contextual correlations, where nearby pixels are assumed to be more similar to distant pixels (Figure 2.7- second column), allowing the investigation of the similarity locally using a small window. However, this may not be sufficient to determine inconsistencies or recover large areas.

In contrast, temporal information is more useful for comparing several images of the same geographical region and can assist in finding outliers and inconsistencies by comparing the images (Figure 2.7-third column). Nevertheless, images are sometimes noisy and spectrally variable across time due to varying acquisition conditions, which may affect the performance of the spatiotemporal fusion algorithm. Including supplementary 3D data, such as geometric information, can provide more stability to the fusion algorithm due to its robustness to acquisition conditions, such as the sun angle and weather. Combining these pieces of information in image fusion can provide a more robust and efficient algorithm and may provide better images.

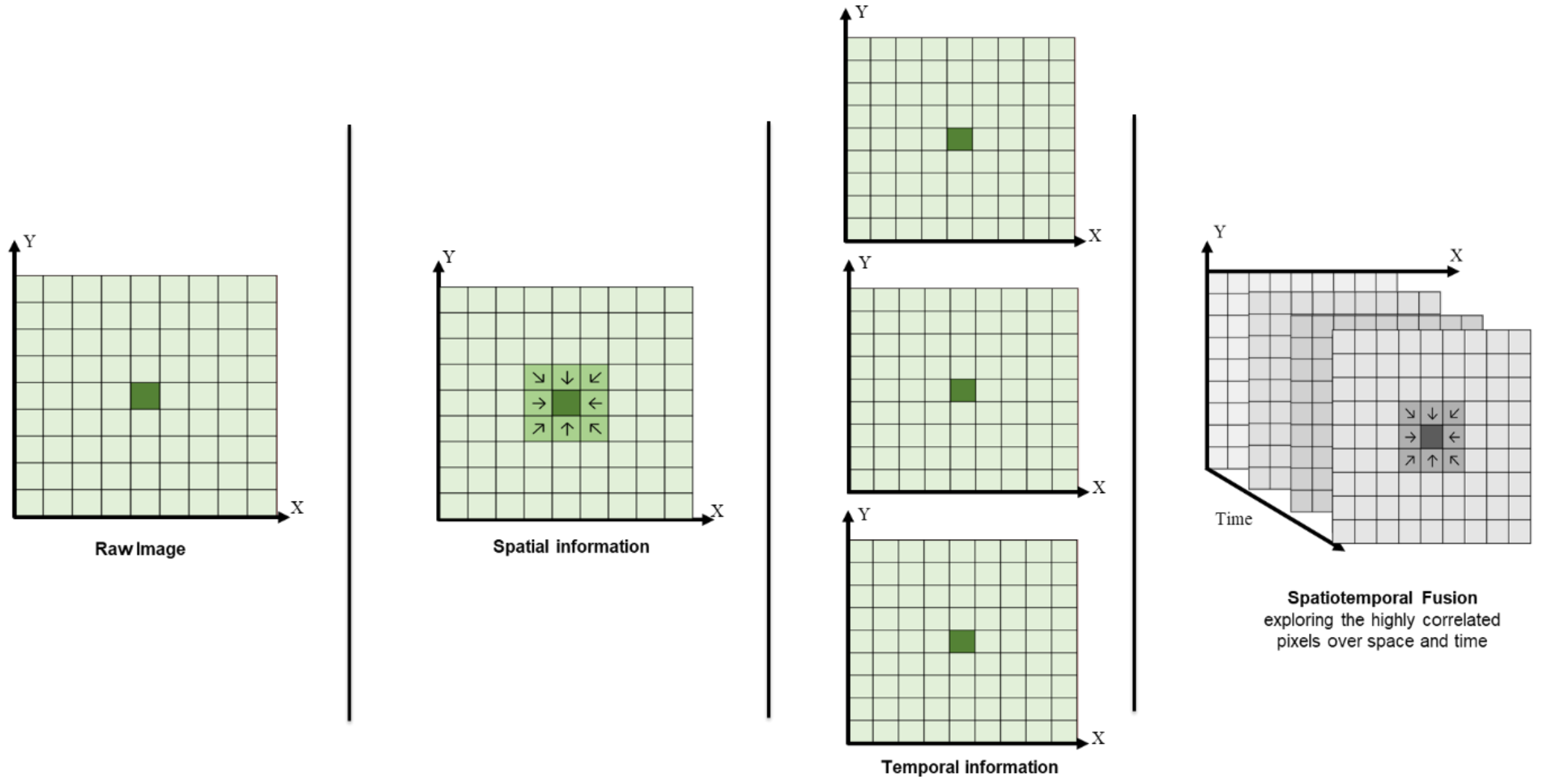

Figure 2.7. The concept of spatial, temporal, and geometric image fusion

### 2.4.3. Strategies and Techniques of Image Fusion in Remote Sensing

Various fusion approaches are used for RS applications, ranging from simple approaches, such as taking the average and median, to complex approaches, such as regression models and optimization approaches. Generally, no specific approach is superior or ideal for all image types. The fusion task is usually considered a case-by-case situation and can vary based on the available resources (i.e., number of inputs, computational power, etc.), image type, and application type.

Typically, the fusion process follows two major stages: removing and recovering contaminated pixels. Both stages are crucial for boosting the performance of fusion algorithms. The image errors and distortions can be of any type, such as noise, spatiotemporal inconsistencies, spectral heterogeneity, and others. Distortion removal requires identifying error locations and positions in the images, which can be performed manually, automatically using algorithms, or via a direct update of the pixel value. Manually masking contaminated pixels is very time-consuming and may be subject to human errors. Most RS algorithms prefer using automated methods to identify distortions through clues, rule-based algorithms, or comparative analysis of multiple images. The clues and rule-based methods must include prior information about the distortions, such as the pixel value, shape, and size. In contrast, comparative analysis tends to compare normal and abnormal events across images using such methods as differencing, standard deviation, averaging, and the median. However, this may require at least two images to operate in some cases. In certain cases, such as in image filtering, all image pixels are considered contaminated; thus, they are directly refined. However, recovery aims to fill and predict the contaminated pixels, which is often performed using spatiotemporal inferences. These methods include interpolation, filtering, probabilistic, machine learning, regression, adaptive, iterative optimization, and DL.

#### 2.4.4. Fusion Applications in Remote Sensing

Fusion has been used for a tremendous number of applications. These applications include the following:

- Spatial resolution enhancement, such as pan-sharpening and super-resolution (Alparone et al., 2015; Anger et al., 2020; X. Meng et al., 2019; Shen et al., 2008; S. Zhang et al., 2020; H. Zhu et al., 2016, 2018),
- Missing pixel reconstruction (Fan et al., 2011; J. Kang et al., 2018; Zeng et al., 2013),
- Denoising acquisition conditions effects, such as clouds, fog, and shadows (P. Dai et al., 2020; W. Du et al., 2019; K. et al., 2016; C.-H. Lin et al., 2013; F. Meng et al., 2017; Movia et al., 2016; Siravenha et al., 2011),
- Classification map enhancement (Albanwan et al., 2020; Benediktsson et al., 2005; G. Fu et al., 2017; Q. Gao & Lim, 2019; Ji et al., 2020; X. Li et al., 2017; Lucas et al., 2007; Moser et al., 2013; J. Xia et al., 2016),

- Change detection (Z. Li et al., 2017; Luppino et al., 2019; Lv, Liu, Wan, et al., 2018; Lv, Liu, Zhang, et al., 2018; C. Wu et al., 2016; Zerrouki et al., 2022; X. Zhang et al., 2017),
- RRN (Albanwan & Qin, 2018; Canty et al., 2004; Chavez, 1996; Y. Chen et al., 2018; Y. Du et al., 2002; Nagamani et al., 2021; Syariz et al., 2019; D. Yuan & Elvidge, 1996), and
- Environmental monitoring, such as phenological studies on crop cycles (Erkkilä & Kalliola, 2004; Lunetta et al., 2006; Viana et al., 2019).

# Chapter 3. Spatiotemporal Filter to Provide Spectrally Homogenous and Consistent Multitemporal Satellite Images

This chapter is based on the paper called "Novel Spectrum Enhancement Technique for Multi-temporal, Multi-Spectral Data Using Spatial-temporal Filtering" that was published in the "ISPRS Journal of Photogrammetry and Remote Sensing" by (Albanwan & Qin, 2018).

## 3.1. Abstract:

Time-sequence remote sensing images are usually captured under varying acquisition conditions due to atmospheric differences, lighting conditions, humidity, etc. Comparing the spectral values of the well-registered images taken at different times is a complicated issue due to the non-linear spectrum distortion caused by these effects. Atmospheric correction can eliminate part of the errors, while precisely removing it requires many other in-situ data such as the weather condition, optical aerosol depth, etc. We propose an algorithm that performs spatial-temporal inferences that correct the spectral values through a data-driven approach - we developed a simple 3D spatiotemporal filtering method that uses the time-sequence imagery themselves to homogenize the spectral property of similar objects while being heterogeneous to objects with significant differences. We have performed extensive experiments using medium-resolution Landsat datasets and high-resolution Planet imagery, by evaluating the classification results from both classic machine learning (sample selected from the current image) and transfer learning (samples selected from one dataset and applied to the other dataset). The experiment results show that the proposed 3D spatiotemporal filter can improve the accuracy of classification using transfer learning by ≈5%, ≈15%, and ≈2%. We have also demonstrated that the enhanced time-sequence image offers much better change detection outputs using just a simple image differencing method. The improved results in typical remote sensing tasks indicate our proposed method being effective for time-sequence data preprocessing.

## 3.2. Introduction

### 3.2.1. Background

Satellite remote sensing images are a great source of information to study the land, water, atmosphere, and natural phenomena. Time series analysis of satellite images acquired over time is an important field of study in remote sensing since multi-temporal data can be used to extract the plant phenology, and track human activities and dynamics of the urban/natural systems. However, one of the most critical issues of analyzing multi-temporal satellite images is the requirement of radiometric/spectral consistence for correlating, comparing, and processing the data. Very often, the acquired images vary in their appearances, due to different lighting conditions such as the sunlight intensity and direction, atmospheric scattering and absorption, and metrological conditions such as the existence of clouds, snow, and rain. In addition, the changing condition of satellite sensors due to aging and operating environments will also cause spectral distortion and imaging qualities. Therefore, one of the consistent endeavors is to minimize such effects and homogenize the spectral reflectance of similar objects on the ground (Lu et al., 2004; Paolini et al., 2006), which is subsequently beneficial to boosting the performance and success of relevant applications that utilize time-series remote sensing data, e.g. change detection, spatial-temporal analysis, and land-cover and land-use change mapping (LCLUC).

Image classification is regarded as an effective way of evaluating the radiometric/spectral consistency of remote sensing images in many of the existing works (Paolini et al., 2006), and the accuracy of the classification tells to which extent the images can be used for automated interpretation. A classification framework examines the spectrum of an image through:

(1) Intra-class similarity: The spectrum of a class should be similar, such that the class can be uniquely characterized.

(2) Inter-class dissimilarity: The spectrums of two different classes should be different enough to be identified as different classes of objects.

Theoretically, earth surfaces with different materials should reflect different spectral curves. Practically, such intra-class similarity and inter-class dissimilarity are often saturated by factors such as resolution, image noise, and atmospheric absorption. An effective and actively investigated approach to resolving such ambiguities is to introduce spatial-spectral classifications, where the spatial distribution of spectrum value around a pixel was taken into account (Bernabe et al., 2014; Bernard et al., 2012; Fauvel et al., 2012, 2013; M. Li et al., 2014). However, when time-series data are being processed, a more important question about these two factors is, whether the same level of intra-class similarity and inter-class similarity can be obtained consistently through time? It is very often observed that the classification results vary greatly between two temporal images of the same location, even with a similar set of training data. Such discrepancy is reflected by the fact that: (1) the pixel values of the same ground unit are different; (2) the difference of neighboring pixels of the same ground unit is different. Temporal images taken under different acquisition conditions are highly disparate, the resulting pixel values/spectrums of which are too complex to be modeled through simple linear/quadratic radiometric correction methods.

In multi-temporal data analysis, image classification is usually applied through the temporal datasets, either independently or concurrently to derive the spatial-temporal dynamics of the land classes. It is usually time-consuming to generate training samples for each dataset for classification, and one approach to avoid time-consuming training sample collection process, termed transfer learning, is to apply classifiers trained from one dataset to other datasets (Dai et al., 2009). This approach usually requires a process of performing feature space transformation to fit the reference classifier to the target image (Arnold et al., 2007), and this is due to the dramatic non-linear differences between image spectrums/pixel values of the same class of objects. Such feature space transformation, usually apply a parametric model with fixed degree of freedom (Duan et al., 2012; Pan et al., 2008; Shao et al., 2015), which might be hard to model the complex image radiometric/spectrum discrepancies brought by the image acquisition conditions.

In this paper, we develop a simple but effective non-parametric approach to directly correct the relative radiometric discrepancies of multi-temporal, multi-spectral remote sensing images. The approach is termed a 3D spatiotemporal filter, which simultaneously incorporates temporal and spatial aspects of time series multispectral data into an edge-aware filter: it homogenizes similar spectrums while keeping dissimilar spectrums disparate. The unique characteristic of this approach is that it neither requires any prior information nor assumes a fixed parametric model for data correction. The enhanced (corrected) multi-temporal dataset will allow the classifier trained from one dataset, to be directly used in another dataset. The remainder of our work is organized as follows: Section 3.2.2 reviews the relevant methods of relative radiometric correction. Section 3.3 introduces the proposed 3D spatiotemporal filter to enhance the radiometric qualities of multi-temporal datasets, and then introduces our evaluation methods based on classic machine learning and transfer-learning-based classification, with an additional change detection experiment. It also includes a comparative study with other state-of-the-art relative radiometric normalization approaches. The experimental results and the discussion are presented in Section 3.4, with the conclusions drawn in Section 3.5 discussing the pros and cons of the proposed method and future improvements.

### 3.2.2. Related Works and the Proposed Method

Radiometric correction methods were normally presented as a part of the preprocessing steps for tasks such as image classification and change detection. While being critical to the classification and change detection results, their impact has rarely been comprehensively discussed (Paolini et al., 2006; Vanonckelen et al., 2013). Our proposed method aims to enhance the spectral quality of the multi-temporal for spatial-temporal image analysis, specifically for classification and change detection. In this section, we will review the existing efforts in radiometric correction and pinpoint the existing challenges of classification and change detection associated with it.

- *Radiometric Correction Using Time Series Analysis*

The radiometric consistency of multi-temporal images is often affected by metrological conditions, illumination differences, and satellite sensor conditions. Traditional remote sensing atmospheric correction techniques using radiative transfer models have demonstrated success in reflectance recovery (Berk et al., 1999), while utilizing such methods accurately requires much in-situ information such as AOD (aerosol optical depth) (Schroeder et al., 2006), camera looking angle (Chander et al., 2009; S. Lin et al., 2004), sensor calibration parameters (Chander et al., 2009), temperature, humidity, etc. The recovery of surface reflectance is important when deriving the actual physical property of the earth's surfaces (Gordon & Wang, 1994; Kaufman et al., 2001; Tucker & Sellers, 1986), while not necessary for applications such as image classification and change detection. Particularly, the required in-situ information might not be always available which increases the workload and expenses. Therefore, it is important to find simple and effective ways to correct non-linear spectral reflectance prior to performing classification and change detection.

Radiometric correction for multi-temporal images can be either absolute or relative. The absolute radiometric correction refers to the recovery of the actual surface reflectance (atmospheric correction), normally achieved through the process of atmospheric correction, which is complicated and requires additional in situ measurements for accurate correction (Moran et al., 1992; Slater et al., 1987; Vermote & Kaufman, 1995). The relative radiometric correction and normalization (RRN) use one image as a reference to adjust and relate the radiometric properties of the rest of the images (Q. Xu et al., 2012). The relative methods (relative correction or normalization) are preferred as it does not require additional observations except for the images themselves. This normally refers to the correction of a multiplicative and additive intensity change, and the correction parameters can be either estimated through a few reference pixels or all pixels and patches (El Hajj et al., 2008; Lo & Yang, 2000). In the process of estimating the correction parameters (scale and offset, relevant to multiplicative and additive intensity change), blunder pixels, such as clouds or significantly changed areas need be eliminated (Paolini et al., 2006). However, the radiometric difference between temporal images is sometimes too complex to be simply modeled as linear or quadratic functions. For instance, the widely used Dark Object Subtraction (DOS) algorithm (Chavez, 1988) is a relative correction method that was developed particularly for haze reduction, while this method assumed the presence of the haze over the entire image, and potentially leads to "over-correction" – areas not affected by the haze will lose its fidelity. One of the most relevant works is the spatiotemporal filter used in our previous work (Qin et al., 2016b), where the height information is incorporated in the temporal domain to reduce the noises of the building classification map. However, this work did not incorporate the spectrum component in the temporal direction. Moreover, the selection of a single reference image for relative correction methods can be challenging. Several conditions govern the choice of the reference image such as having minimum spectral changes over time (e.g. minimum vegetation) and ensuring the regions be flat to minimize shadows or any blockage in the scene, in addition to having clear and atmospheric effect-free images (Biday & Bhosle, 2010). Since a clear noise-free image as a reference is very difficult to obtain, it would be more beneficial and flexible to use the image pixels adaptively in the dataset for correction.

- *Supervised Image Classification and Transfer Learning*

Image classification is a heavily investigated, yet still very active research topic. It directly yields land-cover and land-use maps for local and regional mapping. A variety of methods exist to perform supervised classification (Foody & Mathur, 2004; Xiong et al., 2010) on images where spectral and spatial properties of the pixels are decisive (Bernabe et al., 2014; Fauvel et al., 2013; M. Li et al., 2014; L. Wang et al., 2017), in some situations auxiliary information such as height is introduced to further enhance the outcomes (Qin et al., 2016b). Classification using machine learning techniques often involves extracting features and attributes (labels) and training a classifier using the available known information to be applied to the unlabeled testing data.

Very often classifiers trained from one temporal dataset can be hardly applied to another dataset, mainly due to the fact that the features are not invariant to the complex discrepancies of the temporal images. Transfer learning was introduced primarily to address this problem in classification (J. Gao et al., 2015; Yang et al., 2009; Y. Zhu et al., 2011), where the core research issue is to find a transformation in the feature space based on the differences of the data and spectrum distribution, such that the transformed features of the new dataset can be adaptive to the domain where training dataset. Several domain adaptation (DA) algorithms have been developed over years to tackle the problem of inconsistently transferred data in transfer learning (Paul et al., 2016; Shi et al., 2010; Yang et al., 2009). A classical approach towards solving the DA problem attempts to find an invariant feature space to project both the source and target data – as will be mentioned next. Dimensionality reduction-based methods used in DA project the data into a new space by minimizing the marginal distribution across domains (Pan et al., 2008, 2011). Since the dimensionality reduction techniques only maintain partial information during the projection to a lower-dimensional space, they must be modified and incorporated with other statistical methods to be used in transfer teaching (Pan et al., 2008). For instance, Pan et al. (2008) modified the Maximum Mean Discrepancy (MMD) used in transfer learning, and further extended their work to create feature extraction method known as Transfer Component Analysis (TCA) (Pan et al., 2011). Another form of feature transformation is correlation; Yeh et al. (2014) developed a new SVM associated with Canonical Correlation Analysis (CCA) to project different feature space distributions into a consistent space to enhance DA. However, in the case where there is a lack of labels (training samples between datasets), the CCA may have a higher potential for negative transfers (Liu et al., 2017). (Cortes & Mohri, 2011) suggested using regression-based algorithms for DA, where the main idea is to minimize the distances between source and target distributions using kernel functions. However, regression-based models might be susceptible to over-fitting (Mansour & Schain, 2014). Many of the existing methods are parametric models which are known for their complexity and demanding for in-situ data. Unlike parametric models, the nonparametric models (Ghahramani, 2015; Lopez-Paz et al., 2013; Wood & Teh, 2009) are known for their better performances, higher flexibility, and robustness (few or no prior information required). Nonparametric models can work with complicated data distributions and provide better generalization and performance when working in a higher-dimensional space.

- *Change Detection*

Remote sensing change detection is a highly relevant topic of spatial-temporal processing, the aim of which is to detect the differences of multi-temporal or bi-temporal remote sensing images for monitoring and change mapping (Justice et al., 2015). One of the most fundamental and basic processing for change detection is image differencing, where the spectrums or parametrically transformed data are compared under the metric of vector norms of Euclidean distances (Lu et al., 2004; Singh, 1989), where consistency of image spectrums through time is critical. Very often the change detection tends to overestimate the "changes", where "false positives" are the dominant error sources. To achieve good change detection accuracy, the fundamental issues go back to the relative radiometric correction (as introduced in Section 3.2), where the increased consistency of spectrum data of multi-temporal images is expected to greatly reduce the false positives.

- *Our Proposed Method and Rationale*

Having the spectrum values consistent through time for multi-temporal data is a key to improving the classification and change detection algorithms. The discussions above identify the following three issues: (1) Due to the complex variation of acquisition conditions, a parametric model to correlate spectrum values of the same object in the multi-temporal dataset may not be sufficient to render good correction results. (2) The solution of using transfer-learning for classification share the same issue: the feature-space transformation between different dataset remains parametric. (3) The relative correction results may be sensitive to the selected reference image.

Inspired by the 2D edge-aware filtering (e.g. Bilateral filer) (He et al., 2013; Tomasi & Manduchi, 1998) which homogenizes similar areas while being sharp at the object boundary, we develop a novel 3D spatiotemporal filter that incorporates both the spatial and temporal domain to increase the spatiotemporal consistencies of the multi-temporal data. In addition to considering the object boundaries, we consider the significant spectrum differences as "temporal boundaries", such that the algorithm can produce smoothing effects in the temporal direction to reduce the spectrum variation, and at the same time keep necessary temporal changes.

In this paper, we proposed a solution that covers the trade-offs resulted from preprocessing large-format remote sensing satellite images (e.g. full frame Landsat data) with the following properties: (1) it is a non-parametric, non-iterative, and relative method that does not assume fixed functional relationship between the temporal spectrums; rather, it incorporates an adaptively weighted joint-Gaussian filter that locally refines the spectrum values based on the temporal data itself, allowing efficient implementation and processing. (2) Reference image in the RRN will not be necessary since the proposed method enhances the time-series data concurrently using the entire multi-temporal dataset, such that the consistency is formed globally. (3) Finally, we expect the transferability of the trained data between different datasets to improve by the modification and correction of the spectrum values.

### 3.3. Methodology

Figure 3.1 presents a general workflow of our proposed study. The proposed 3D spatiotemporal filter is applied after conditioned by a few preprocessing steps to remove the initial misalignment of the data. The proposed method will be validated through the accuracy evaluation of typical remote sensing tasks of classification and change detection, which will be described in Section 3.3.

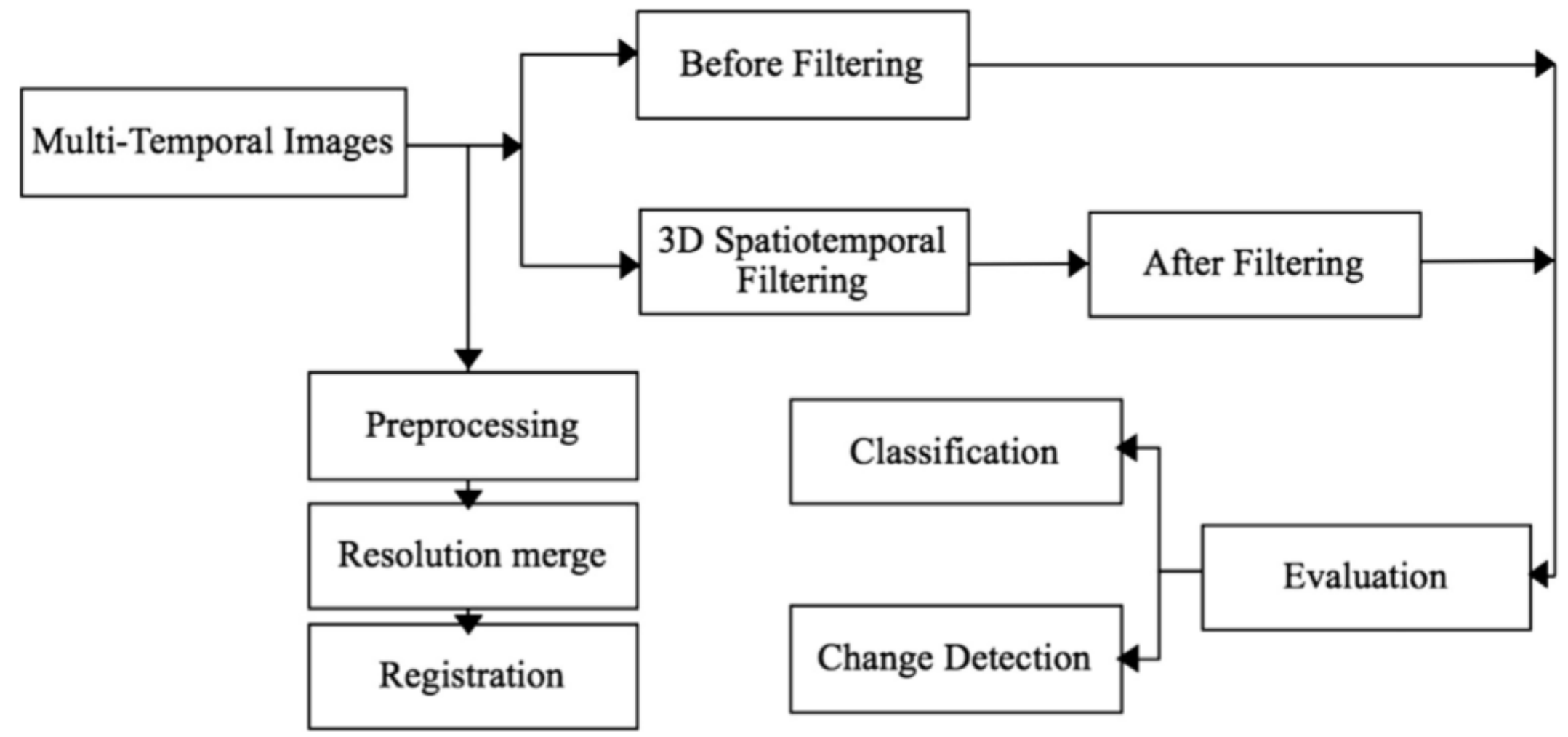

Figure 3.1. The general pipeline for the spatiotemporal filtering method

### 3.3.1. Preprocessing Steps

We apply a few typical preprocessing steps for the multi-spectral and multi-temporal images. The first step is to fuse the multispectral and panchromatic images. Pan-sharpening was performed using the Subtractive resolution merge (SRM) method provided by ERDAS Imagine, 2016 (Ashraf et al., 2013a, 2013b; Witharana et al., 2013). In our case, we use Landsat data, thus the pan-sharpen processing generates 8 bands of multispectral images with 15-meter spatial resolution. The second step performs accurate co-registration between multi-temporal images. Although these datasets have been roughly geo-referenced, it still requires precise refinement to facilitate per-pixel operation. We applied a least-square co-registration method that estimates the optimal affine transformation based on the similarity of the sum of squared errors (Mattes et al., 2003; Styner et al., 2000; Thevenaz & Unser, 2000).

### 3.3.2. The 3D Spatiotemporal Filtering

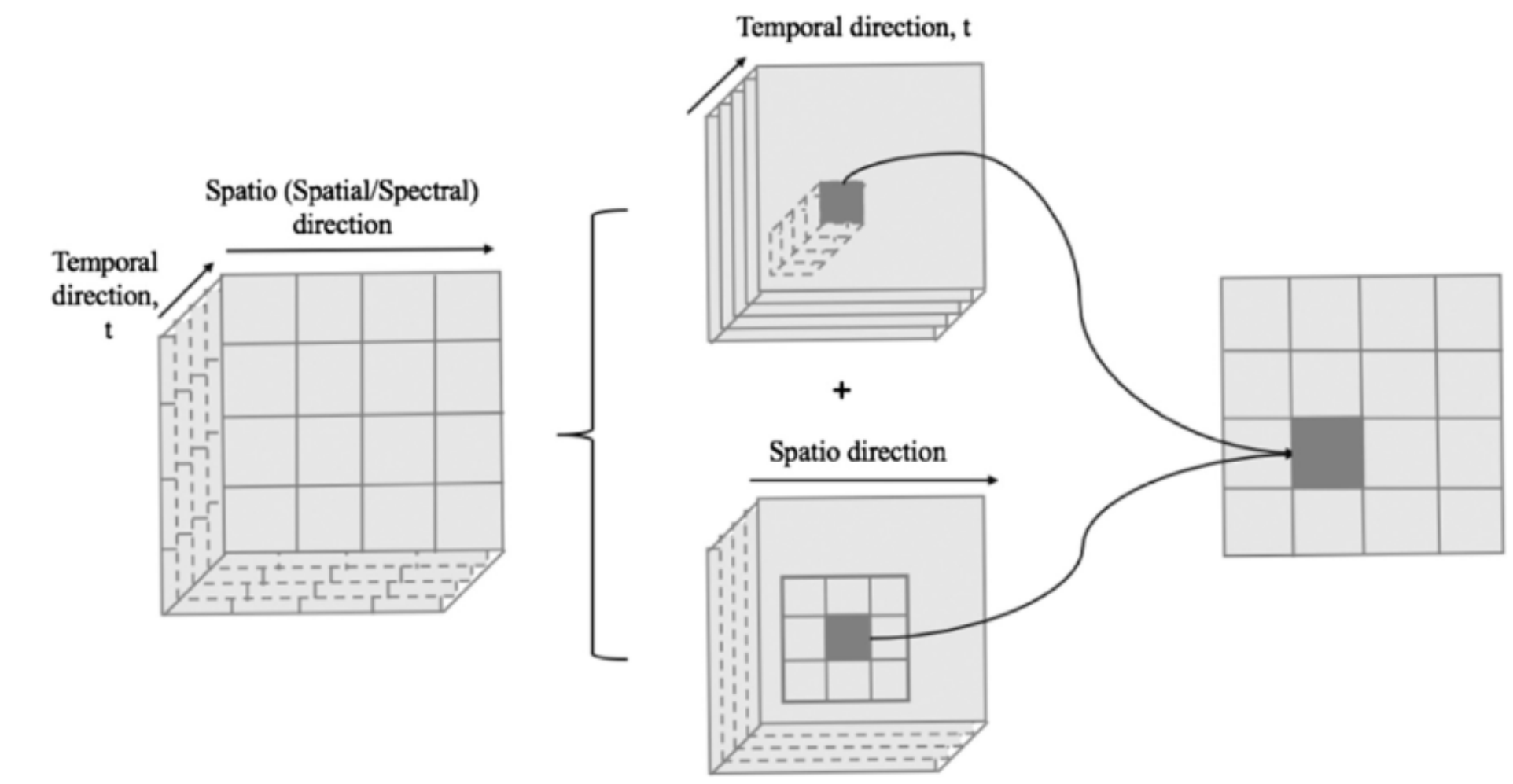

Figure 3.2. An overview of our 3D spatiotemporal filter.

- *A Generic Spatiotemporal Filter for Multispectral and Multitemporal Dataset*

Our proposed 3D spatiotemporal filter is derived from the general weighted average filter:

$$\bar{I}_i = \int_{\Omega} w_{j,i} . I_j . dj \tag{3.1}$$

Where $\bar{I}$ and I are the filtered data and original data; $w_{j,i}$ is the weight of each data point $j$ being aggregated to $i$ over a space Ω (aggregated space), with $i, j \in \Omega$ being the data point indices. For instance, low-pass filtering (e.g. Gaussian filtering) can be represented as for each pixel $i$ of the filtered image $\bar{I}$, and its value $\bar{I}_i$ is determined by original pixel values of $j$ (being $I_j$) aggregated over a 2D space $\mathbb{R}^2$ (pixel coordinate system), weighted by the geometric proximity of the pixel positions $j_{\vec{x}}$ to the central pixel position $i_{\vec{x}}$ within a Gaussian kernel ($\sigma_{\vec{x}}$ being the bandwidth responsible for the amount of filtering in the spatial direction):

$$w_{j,i} = \exp\left(-\frac{|j_{\vec{x}} - i_{\vec{x}}|^2}{\sigma_{\vec{x}}}\right) \tag{3.2}$$

Where $\vec{\boldsymbol{x}}$ is a vector of x and y components, such that:

$$-\frac{|j_{\vec{x}}-i_{\vec{x}}|^2}{\sigma_{\vec{x}}} = -\left(\frac{|j_{\mathrm{x}}-i_{\mathrm{x}}|^2}{\sigma_{\vec{x}}} + \frac{|j_{\mathrm{y}}-i_{\mathrm{y}}|^2}{\sigma_{\vec{x}}}\right) \quad (3.3)$$

Multispectral and multitemporal data are a stack of images as an image cube (Figure 1), with multiple bands in each image. Our proposed spatiotemporal filter naturally considers the aggregation over a four-dimensional domain consisting of spatial ($\mathbb{R}^2$), spectral ($\mathbb{R}$) and temporal ($\mathbb{R}$) dimension for each pixel in the image cube ($\Omega = \mathbb{R}^4 = [\mathbb{R}^2, \mathbb{R}, \mathbb{R}]$). The weight for each pixel can be then written in the following form:

$$w_{\mathrm{j},i} = \exp\left(-\frac{|j_{\vec{x}}-i_{\vec{x}}|^2}{\sigma_{\vec{x}}} - \frac{|j_t-i_t|^2}{\sigma_T} - \frac{|j_b-i_b|^2}{\sigma_B} - \frac{|I_j-I_i|^2}{\sigma_I}\right), j, i \in \Omega \quad (3.4)$$

Where $j_{\vec{x}}, j_t, j_b$ refers to $j$'s spatial, temporal, and spectral components in the aggregated space (the same applies to $i_{\vec{x}}$, $i_t$, $i_b$). The weight computation is grounded on an intuitive rationale: pixels that are closer to the centric pixel have a larger weight, where being "close" is measured through similarity in different domains. As formulated in Equation (4), $\exp\left(-\frac{|j_{\vec{x}}-i_{\vec{x}}|^2}{\sigma_{\vec{x}}}\right)$, $\exp\left(-\frac{|j_t-i_t|^2}{\sigma_T}\right)$, $\exp\left(-\frac{|j_b-i_b|^2}{\sigma_T}\right)$, $\exp\left(-\frac{|I_j-I_i|^2}{\sigma_I}\right)$ respectively refer to the similarity measurement in the spatial, temporal, spectral, and band values, meaning that a pixel being close to a centric pixel in terms of its position in the 2D image grid ($|j_{\vec{x}} - i_{\vec{x}}|^2$), acquisition time ($|j_t - i_t|^2$), wavelength ($|j_b - i_b|^2$), and pixel values ($|I_j - I_i|^2$), which provides a generic weight computation for multispectral and multi-temporal image filtering.

- *3D Spatiotemporal Filter*

Remote sensing images normally have relatively lower temporal frequencies (compared to video sequences), e.g. Landsat 8 and Planet satellite with 16 days and weekly revisiting time. It is preferable to consider filtering process aggregates information over the entire temporal direction to gain maximal contributions, leading to $\sigma_T = \infty$.

The multispectral images contain a number of bands with discrete wavelengths, therefore the band wavelength similarity $|j_b - i_b|^2$ should be considered as a discrete function (of band order out of m bands, b ∈ $[\,1, 2, \ldots, m]$). Moreover, aggregation through different bands might bring cross-band inference and we naturally prefer different bands being independent, we thus only consider the aggregation over the same bands, by replacing the wavelength similarity $\exp\left(-\frac{|j_b - i_b|^2}{\sigma_B}\right)$ with a Dirac delta function $\delta(j_b - i_b)$:

$$\delta(j_b - i_b) = \begin{cases} 1, & if\ j_b = i_b \\ 0, & otherwise \end{cases} \qquad (3.5)$$

Where $j_b$ and $i_b$ being the band number. Therefore, the weight of the generic weighted average filter can be reduced as:

$$w_{\mathrm{j},i} = \frac{1}{||w||} \exp\left(-\frac{\left|j_{\vec{x}} - i_{\vec{x}}\right|^2}{\sigma_{\vec{x}}} - \frac{\left|I_j - I_i\right|^2}{\sigma_I}\right) \delta(j_b - i_b) \qquad (3.6)$$

Where $||w||$ is used to normalize the weight such that $\sum w_{\mathrm{j},i} = 1$. By inserting this weight computation (equation (3.6)) to the general weighted filter formulation (equation (3.1)), and replacing the integration with a discrete sum, we obtain:

$$\bar{\mathrm{I}}_{i_{\vec{x}}, i_t, i_b} = \sum_{j_{\vec{x}}} \sum_{j_t} \exp\left(-\frac{\left|j_{\vec{x}} - i_{\vec{x}}\right|^2}{\sigma_{\vec{x}}} - \frac{\left|I_{j_{\vec{x}}, j_t, j_b} - I_{i_{\vec{x}}, i_t, i_b}\right|^2}{\sigma_I}\right) \cdot I_i \qquad (3.7)$$

Where we separate the indices of *j* and *i* to their sub-component. Since $\left|I_{(j_{\vec{x}}, j_t, j_b)} - I_{(i_{\vec{x}}, i_t, i_b)}\right|^2$ represents the pixel value difference in a 3D space (spatial $j_{\vec{x}}$ and temporal $j_t$ ($j_b = i_b$)), which assumes a single bandwidth over both the spatial and temporal domain, to separate these we, therefore, consider an approximation using the following formulation:

$$\exp\left(\frac{\left|I_{(j_{\vec{x}}, j_t, j_b)} - I_{(i_{\vec{x}}, i_t, i_b)}\right|^2}{\sigma_I}\right) \approx \exp\left(\frac{\left|I_{(j_{\vec{x}}, i_t, i_b)} - I_{(i_{\vec{x}}, i_t, i_b)}\right|^2}{\sigma_s} + \frac{\left|I_{(i_{\vec{x}}, j_t, i_b)} - I_{(i_{\vec{x}}, i_t, i_b)}\right|^2}{\sigma_{T_s}}\right) \qquad (3.8)$$

Where $\left|I_{(j_{\vec{x}}, i_t, i_b)} - I_{(i_{\vec{x}}, i_t, i_b)}\right|^2$ computes the similarity of pixel values in the spatial direction (with $j_t = i_t$) and $\left|I_{(i_{\vec{x}}, j_t, i_b)} - I_{(i_{\vec{x}}, i_t, i_b)}\right|^2$ computes the similarity of pixel values in the temporal direction ($j_{\vec{x}} = i_{\vec{x}}$). $\sigma_s$ and $\sigma_{T_s}$ are the spectral and temporal bandwidths respectively. This formulation provides the control of the bandwidth in both the spatial and temporal domain when computing pixel similarities. Incorporating equation (3.8) to (3.7), we obtain our proposed 3D spatial-temporal filter:

$$\bar{\mathrm{I}}_{i_{\vec{x}},i_t,i_b} = \sum_{j_{\vec{x}}} \sum_{j_t} \exp\left(-\frac{|j_{\vec{x}}-i_{\vec{x}}|^2}{\sigma_{\vec{x}}} - \frac{\left|I_{(j_{\vec{x}},i_t,i_b)} - I_{(i_{\vec{x}},i_t,i_b)}\right|^2}{\sigma_s} - \frac{\left|I_{(i_{\vec{x}},j_t,i_b)} - I_{(i_{\vec{x}},i_t,i_b)}\right|^2}{\sigma_{T_s}}\right) \cdot I_i \tag{3.9}$$

By further organizing the terms, we obtain:

$$\bar{\mathrm{I}}_{i_{\vec{x}},i_t,i_b} = \left[\sum_{j_{\vec{x}}} \exp\left(-\frac{|j_{\vec{x}}-i_{\vec{x}}|^2}{\sigma_{\vec{x}}} - \frac{\left|I_{(j_{\vec{x}},i_t,i_b)} - I_{(i_{\vec{x}},i_t,i_b)}\right|^2}{\sigma_s}\right) \sum_{i_t} \exp\left(-\frac{\left|I_{(i_{\vec{x}},j_t,i_b)} - I_{(i_{\vec{x}},i_t,i_b)}\right|^2}{\sigma_{T_s}}\right)\right] \cdot I_i \tag{3.10}$$

The output of the filtering process is still a multispectral and multi-temporal dataset. The first component $\sum_{j_{\vec{x}}} \exp\left(-\frac{|j_{\vec{x}}-i_{\vec{x}}|^2}{\sigma_{\vec{x}}} - \frac{\left|I_{(j_{\vec{x}},i_t,i_b)} - I_{(i_{\vec{x}},i_t,i_b)}\right|^2}{\sigma_s}\right)$ in equation (3.10) is essentially a typical bilateral filter. With the remaining part being the additional term for restricting the temporal consistency, the proposed 3D spatiotemporal filter is supposed to perform edge-aware filtering in both spatial and temporal domains, smoothing inconsistencies (noise) while maintaining significant inconsistencies (edges and temporal changes).

### 3.3.3. Results and Discussion

We evaluate the data enhancement capability of the proposed method by performing the classification and change detection applications on the original and enhanced multi-temporal data and compare their results. As our intention is to test the effectiveness of the proposed 3D spatiotemporal filter, we only apply the most basic but widely used classification and change detection method: being 1) pixel-based classification using Support Vector Machine (SVM) with & without transfer learning (Hussain et al., 2013; Pan & Yang, 2010); 2) image differencing change detection method (Hussain et al., 2013; Radke et al., 2005) with a thresholding method suggested by Otsu (1979), which utilizes the histogram of gray-level images to find the optimum (maximized) separability between the classes within the gray levels. Specifically, the evaluation of classification application considers 1) a simple spectral-value based classification using SVM and 2) transfer-learning based classification, where the 3D spatiotemporal filter is regarded as an explicit domain adoption transformation between datasets, with a classifier trained completely from a separate image. The change detection considers a simple but often widely used method: Image-differencing based classification for the multi-temporal dataset. To validate our method being effective under different scenarios, we consider experiments with both medium-high resolution datasets (Landsat, Experiment I) and high-resolution datasets (Planet images, Experiment II), with sub-regions covering scenarios of both urban and suburban areas (in the area of Columbus, Ohio).

### 3.4. Experimental Results and Discussion

#### 3.4.1. Data Description

Medium-resolution images usually cover large geographical areas, where the spectral complexity differs based on the region. For example, urban areas have large spectral variability due to various materials and radiance reflectance from the scene (used in infrastructures, buildings, roads, etc.), where each has its own spectral response, unlike suburban areas with most of the content being barren lands and vegetation (see Figure 3.3), thus with less spectral variability. Therefore, in Experiment I, we present two datasets varying in their spectral complexity covering urban and suburban areas taken from Landsat 8 (medium-resolution images) with dimensions cropped at 2000 × 2000 pixel sub-regions, with each dataset consisting of five images within two years across different seasons. Experiment II contains high-resolution images from Planet satellite, a satellite constellation capable of capturing images (spatial resolution of approx. 3.0 meters) at a weekly/daily frequency. A 1534 × 1534 pixel sub-region containing both natural and manmade objects is cropped for this experiment. Such high-resolution data normally show significant variance in reflectance, particularly considering the fact that the images are captured across different and distinct seasons (Table 3.1. shows the acquisition time, and we will use an image in the table to present thereafter instead of referring to the date).

In the classification tasks, we consider six typical land-cover classes (shown in Table 3.2), covering most of the land-cover types considering the resolution of the datasets. It can be noticed that the temporal images in each dataset vary greatly in their radiometric and spectral appearances (Figure 3.3 shows the varying appearances using the same band composition and scaling), highly non-linear even for the same class (discussed in 3.2.2). Both the environmental condition (season, humidity, lighting, optical depth of the atmosphere) and acquisition condition (incidence angle, sensor sensitivity) contribute significantly to this variation. For example, the river and lake in image 5 of experiment I (in both datasets) have some locations (see Figure 3.5. (b)) where part of the water surfaces appearing bright may be caused by direct sunlight reflection. Similar issues can be found in impervious manmade objects such as buildings and roads. Such anomalies may directly lead to failures of classic algorithms of classification and change detection.

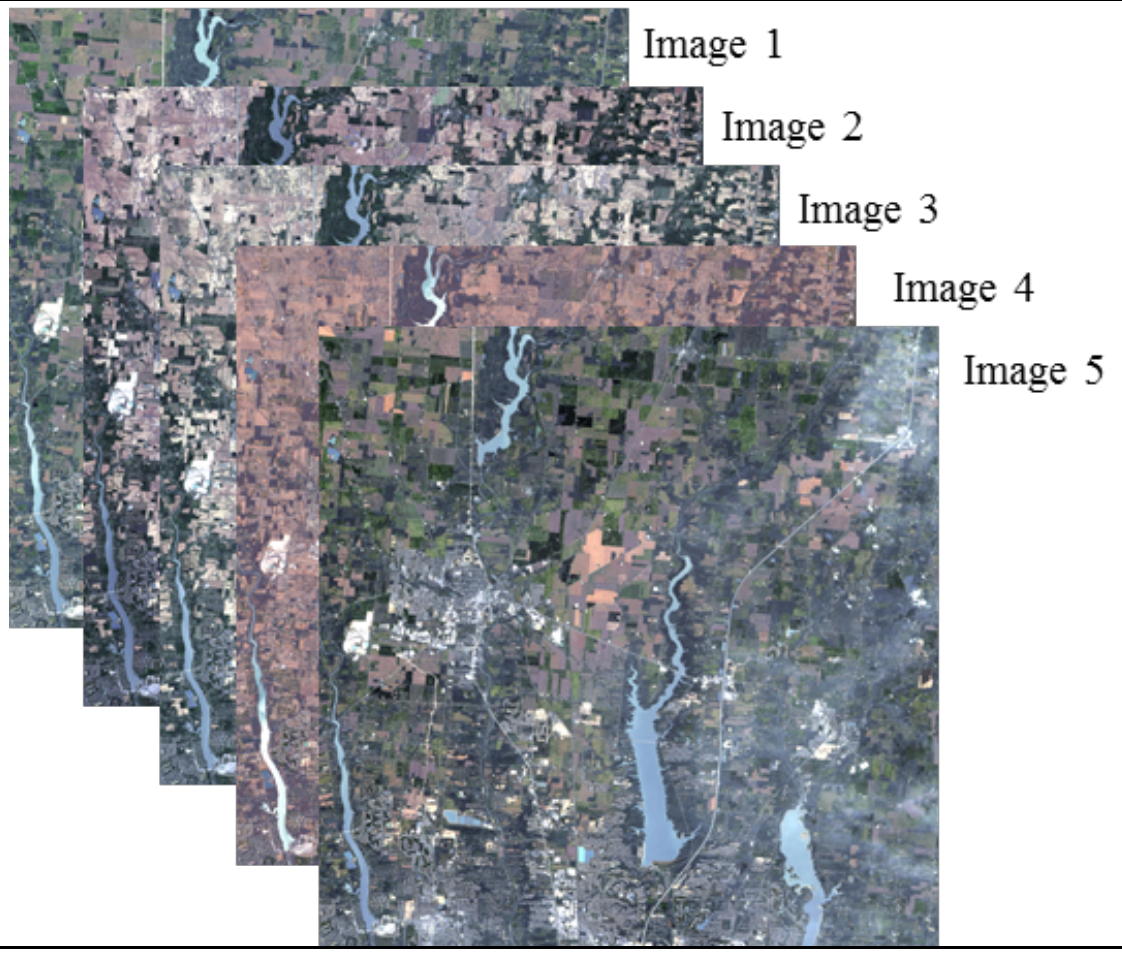

Figure 3.3. A sample of input images (from experiment I suburban dataset) with heterogeneous radiometric properties.

Table 3.1. Input Images Details

| Experiment | Resolution | Date | Season |
|---|---|---|---|
| I (suburban) | 15-30 m | Image 1- 5/27/16 | Spring |
| | | Image 2- 6/12/16 | Spring |
| | | Image 3- 9/16/16 | Summer |
| | | Image 4- 2/21/16 | Winter |
| | | Image 5- 9/14/15 | Summer |
| I (urban) | 15-30 m | Image 1- 5/27/16 | Spring |
| | | Image 2- 6/12/16 | Spring |
| | | Image 3- 9/16/16 | Summer |
| | | Image 4- 2/21/16 | Winter |
| | | Image 5- 9/14/15 | Summer |
| II | 3 m | Image 1- 2/26/17 | Spring |
| | | Image 2- 9/2/16 | Spring |
| | | Image 3-10/14/16 | Summer |
| | | Image 4- 10/5/16 | Winter |
| | | Image 5- 8/28/15 | Summer |

Table 3.2. Classes and Their Descriptions

| Classes | | Description | Miscellaneous |
|---|---|---|---|
| 1 | Water Surfaces | Lakes, rivers, ponds | For Landsat and Planet data |
| 2 | Barren Land | Sand, rock, leafless trees | For Landsat and Planet data |
| 3 | Vegetation | Trees, grass | For Landsat and Planet data |
| 4 | Impervious Surfaces | Roads, buildings, infrastructures | For Landsat data |
| 5 | Roads | Highways, minor roads | For Planet data |
| 6 | Buildings | Houses, commercial, and industrial buildings | For Planet data |

### 3.4.2. Experimental Results

The spatial and spectral bandwidth parameters $\sigma_x$ and $\sigma_r$ were fixed at 7 and 50 (for an 8-bit scaled image, this parameter would become 50/255 = 0.19 for [0,1] scaled image) as their influence has already been studied in previous works (Tomasi and Manduchi, 1998; (X. Kang et al., 2014); the window size 5×5 (pixels) in all the datasets as empirical values (offering good trade-offs on denoising and maintain boundary sharpness as normally used in bilateral filters (Patanavijit, 2015). The temporal parameter $\sigma_{Ts}$ has been tested on each dataset with values ranging from 0 to 0.9 with increments of 0.1 to obtain the optimal performance (for each band data we normalized the value within [0, 1], $\sigma_{Ts} = 0$ essentially refers to a 2D bilateral filter and $\sigma_{Ts} = 1$ to a simple averaging in the temporal direction). The performance of the filter is evaluated by applying classification (with and without transfer learning) to provide visual and numerical assessments. We adopt the widely used SVM with radial basis function RBF as the kernel function as the classifier and the labels used are manually identified for each class. After densely labeling all the images, we used only 30 randomly selected labeled pixels per class as the training samples (Table. 3.2), while using the rest for test and accuracy assessment. To test the capability of transfer learning, for each dataset (five temporal images in our example), we choose a training image out of the five which has the most land-cover classes and samples in this image, which the classifier for transfer learning experiment will be trained from.

- *Filtering Results*

The proposed 3D spatiotemporal filter shows its ability to maintain heterogeneity in the temporal direction, accounting for different spectrum variations. Figure 3.4. shows the histogram of images processed by 3D spatiotemporal filter with different temporal bandwidths ($\sigma_{Ts}$). It can be seen that the variations among the temporal dataset reduce as $\sigma_{Ts}$ increase, and the histogram of each temporal image looks almost identical when $\sigma_{Ts} = 0.9$. However, the respective spectral properties for each image in the dataset are preserved when $\sigma_{Ts} <= 0.3$. We identify that the parameter $\sigma_{Ts}$ within the range of [0.1, 0.3] may be able to eliminate partial noises while maintaining the necessary information in each temporal image. To elaborate, Figure 3.5. Shows a section of two images from experiment I (image 1 and 5 with different dates) before and after the filter with the temporal parameter $\sigma_{Ts}$ =0.2. Before filtering, the two images show distinct appearances and similar features (e.g. water surfaces and vegetation location) have different colors/intensities due to radiometric effects (see the first column in Figure 3.5). Nonetheless, after filtering, some regions like the water surface (marked in dark-blue dashed line – the river in Figure 3.5) experience a large change in intensity leading to reduced heterogeneity between similar regions. Meanwhile, other regions being unique due to seasonal differences (e.g. grass vs. barren land) can preserve their uniqueness (see solid yellow polygons in Figure 3.5) through the proposed 3D spatial-temporal filter.

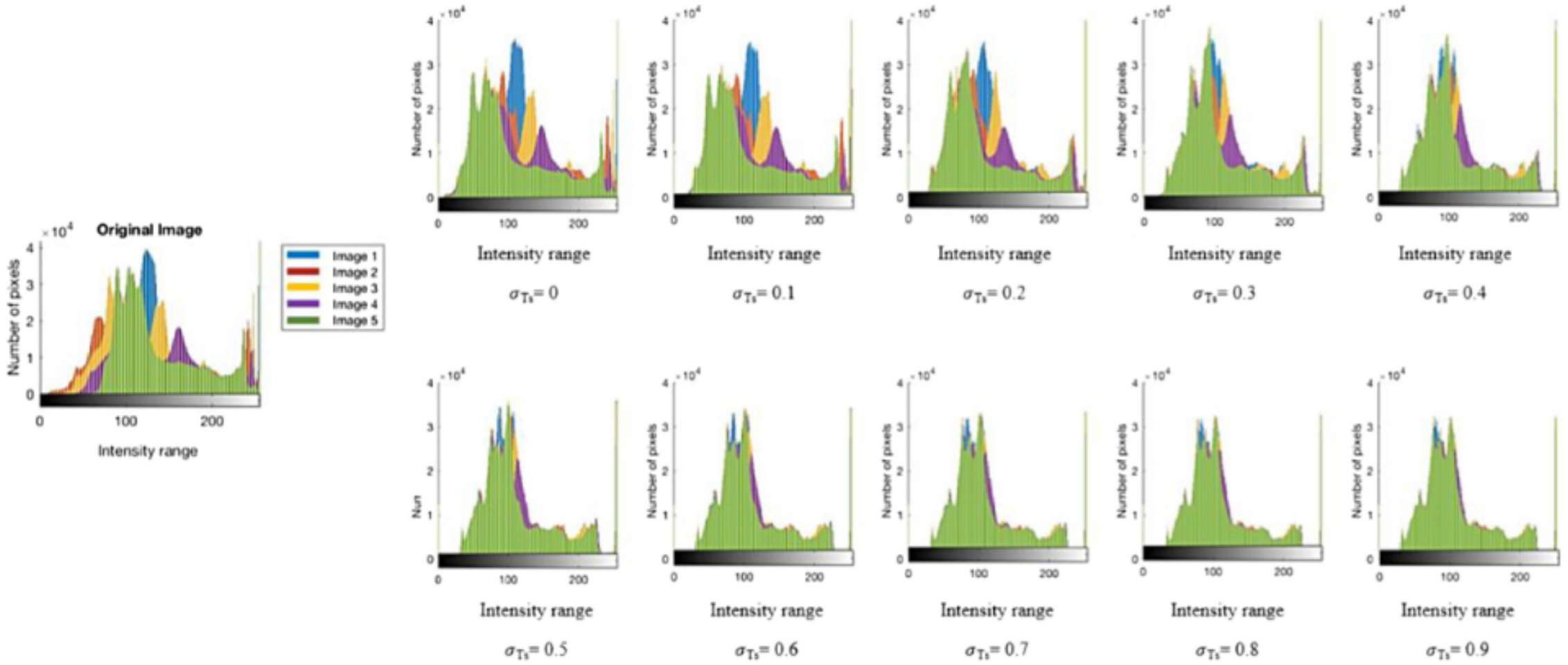

Figure 3.4. The spectrum histograms for the multitemporal images before and after the filter.

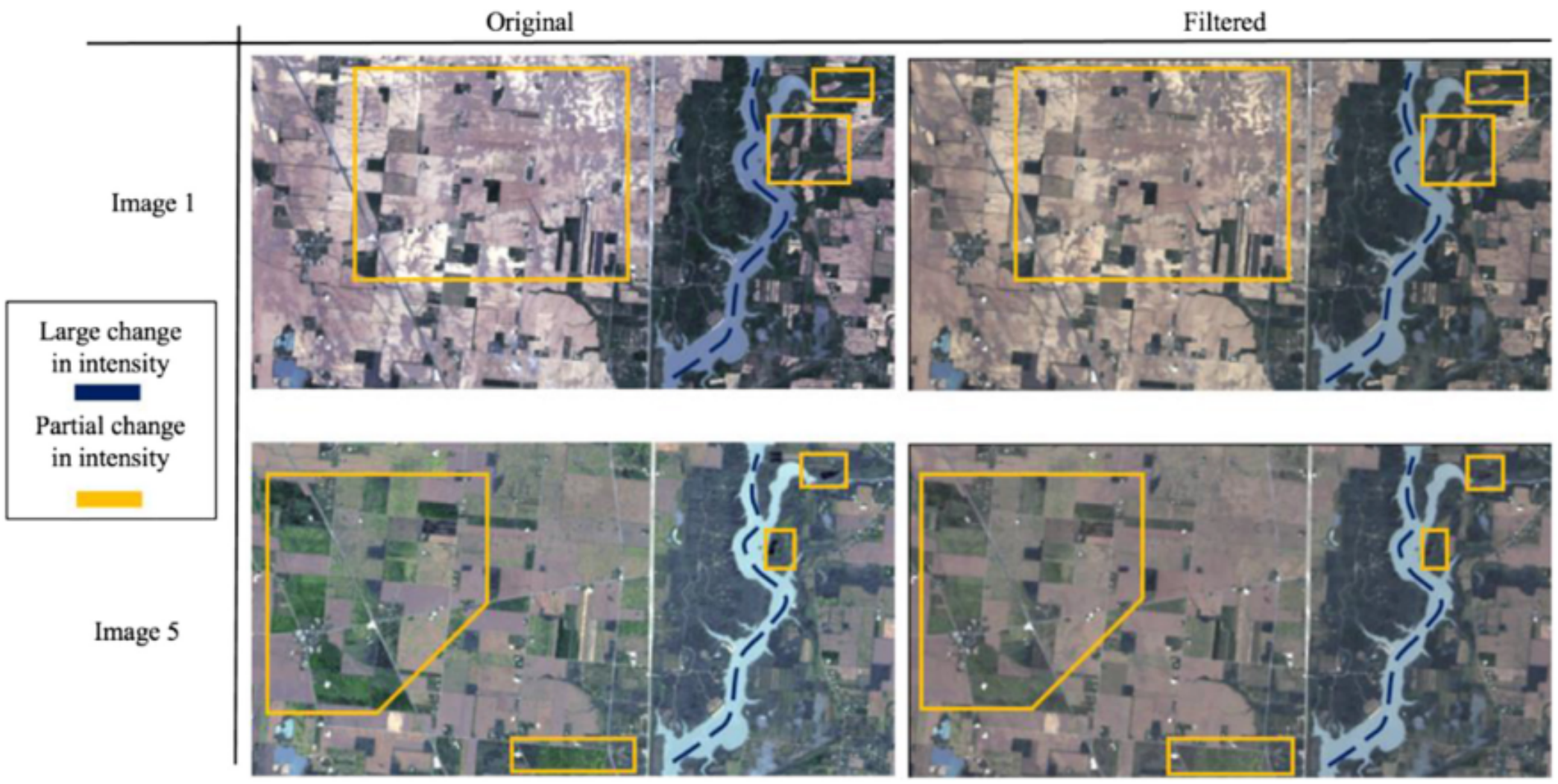

Figure 3.5. Sample of the filtered images from the medium resolution of the suburban areas showing significant improvements.

### 3.4.3. Evaluation and Classification

- *Experiment I*

We applied the classification (both classic and transfer learning) on both the original and filtered data (with different $\sigma_{Ts}$) with the same training samples, and Figure 3.6 shows the overall accuracy of the classification when transfer learning is applied. The dashed lines are the classification accuracy of the original dataset using transfer learning (since this is not processed by a 3D spatiotemporal filter, it is invariant of $\sigma_{Ts}$), and the solid curves present the accuracy of the filtered data (at varying $\sigma_{Ts}$). We observe that the classification accuracies of the filtered image are improved (compared to the original image) in general at various $\sigma_{Ts}$ values. This conclusion has been consistently observed when $\sigma_{Ts} \leqslant 0.3$ for most of the images in these two datasets. In this experiment image 3 for the "Suburban" dataset (Figure 3.6(a)) and image 5 for the "Urban" dataset (Figure 3.6(b)) are used as the training images (where the classifier is trained), therefore there are no corresponding curves for them. These images were chosen as references for training due to their variable spectrums, for instance, images taken in summer allow a variety of elements in each class to be captured under minimum atmospheric effects. In general, we find the curves achieve the peak accuracy at $\sigma_{Ts}$= 0.2 or $\sigma_{Ts}$= 0.3. As $\sigma_{Ts}$ increase, the overall accuracies start to drop. It is an expected trend as the larger the temporal bandwidth, the more "blurring" effects it has in the temporal direction, which eventually turns the filtered image into another image (the "averaged image" of the temporal dataset). Note when $\sigma_{Ts}$= 0, our proposed 3D spatiotemporal filter is equivalent to a typical bilateral filter, which shows slightly higher accuracy while being lower than the results of filtered data at $\sigma_{Ts}$= 0.2 or $\sigma_{Ts}$= 0.3. This implies that incorporating the temporal information can indeed enhance the results to a notable level, and the rate of improvement varies dramatically across different images (compared to the transfer learning results for the original dataset). However, there are two factors that affect the filter and the data used in training the classifier, (1) if an image is significantly different (in terms of spectral appearance) from other images in the dataset, and (2) the season in which the image is captured. To demonstrate, image 4 (from the first dataset in experiment I) was taken in winter and is the most variant image in the dataset, thus, the spectral appearance and the availability of the training samples are limited and highly affect the classification accuracy, which explains the different behavior of the green line in Figure 3.6(a) Moreover, it can be seen that the "urban" dataset has higher improvement, which might be as the result of the higher resolution of the imagery, where larger spectral variability across different temporal imagery can be homogenized.

The third and fourth row of Table 3.3 shows the classification accuracy of the original and filtered data at $\sigma_{Ts}$ =0.2 to numerically compare the overall classification accuracy before and after the filtering. It clearly shows that the improvement of the transfer learning-based classification can reach as high as 15% for the "suburban" dataset (image 2) and 27% for the "urban" dataset (image 2) in overall accuracy, on average 5.9% for the "suburban" dataset and 19.3% for the "urban" dataset over all the images. However, we also observe that the classical classification (training samples are selected from the image to be classified) has gained improvements in some of the images, primarily due to the de-nosing effect of the proposed filter, while it does bring down the classification accuracy in some of the images, this may indicate that the parameter $\sigma_{Ts}$ = 0.2 may not be always optimal such that it brings in spectrums of other land-cover classes.

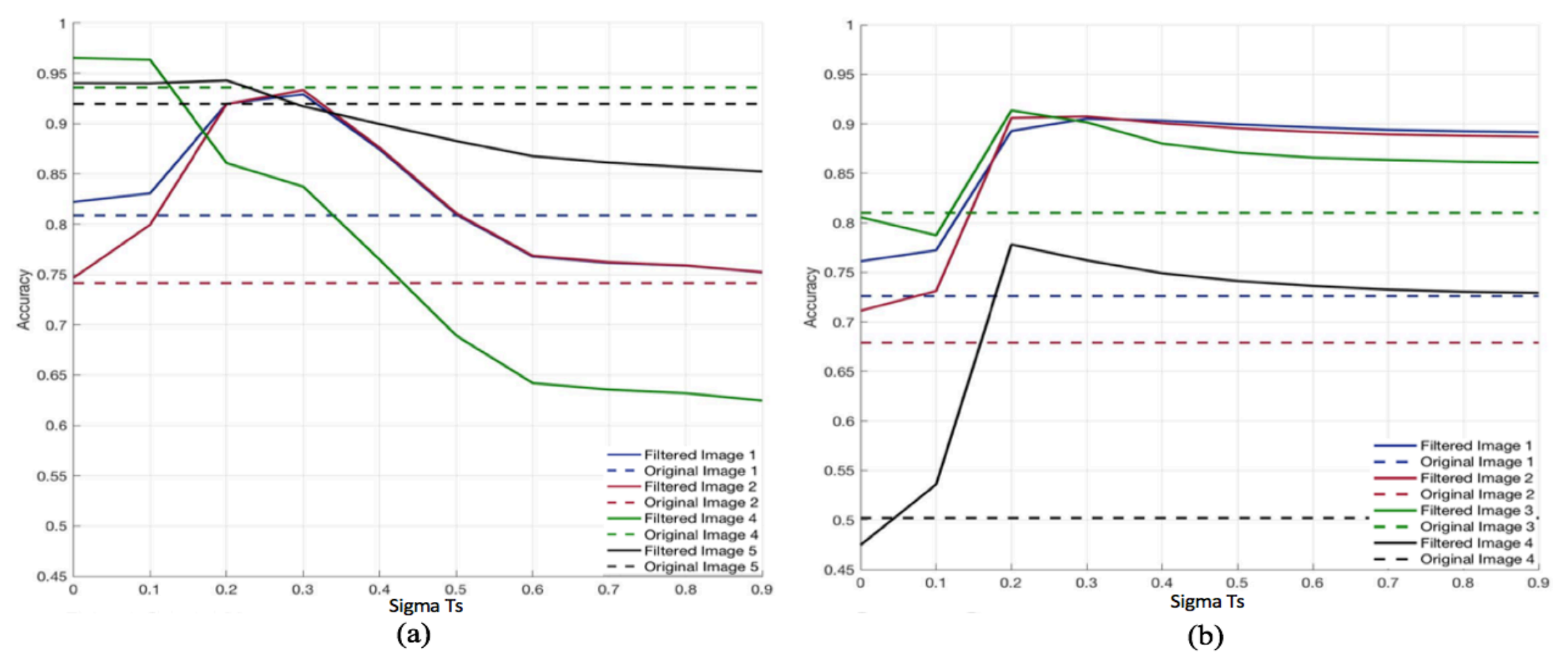

Figure 3.6. Classification accuracy comparison between original and filtered data (Experiment I). (a) Suburban dataset (b) Urban dataset. Details are explained in texts.

We found that the regions such as barren lands, narrow roads, rivers, etc. that were originally classified incorrectly, were recovered with more complete results; this is for instance demonstrated in Figure 3.7, where the road across the center red-box region is recovered correctly, as well as the surrounding impervious surfaces, and similarly, in the lower red-box, the shaded area (by the cloud) incorrectly identified as water surface are recovered to the correct impervious class. This is because the proposed 3D spatiotemporal filter can homogenize similar spectrums across the temporal dataset, and the reduced variation of the similar object in the temporal direction will allow the classifier trained from one image more successfully applied to the other.

Table 3.3. Accuracy assessment with $\sigma_{Ts}$ =0.2 for classic and transfer learning classification. The bold numbers indicate higher accuracy, while the gray highlight represents the reference image used for transfer learning.

| | Classic classification | | | | | Transfer learning | | | | |
|---|---|---|---|---|---|---|---|---|---|---|
| Image | 1 | 2 | 3 | 4 | 5 | 1 | 2 | 3 | 4 | 5 |
| Exp. I - Suburban | | | | | | | | | | |
| Without filter | **97.11** | **97.13** | 93.30 | 90.59 | **97.05** | 80.88 | 74.14 | 93.30 | **93.59** | 91.97 |
| With filter | 96.04 | 96.72 | **93.95** | **91.53** | 96.76 | **91.99** | **91.96** | **93.95** | 86.08 | **94.30** |
| Exp. I - Urban | | | | | | | | | | |
| Without filter | **92.62** | 93.39 | **92.69** | **90.80** | **93.75** | 72.61 | 67.91 | 81.00 | 50.21 | **93.75** |
| With filter | 92.16 | **94.03** | 91.72 | 85.75 | 93.10 | **89.29** | **90.60** | **91.35** | **77.82** | 93.10 |
| Exp. II | | | | | | | | | | |
| Without filter | **68.12** | 71.64 | 70.08 | **67.17** | **75.29** | 66.48 | 74.14 | 68.35 | **67.17** | 73.30 |
| With filter | 66.17 | **74.56** | **70.34** | 65.10 | 73.86 | **66.75** | **76.20** | **72.06** | 65.10 | **78.54** |

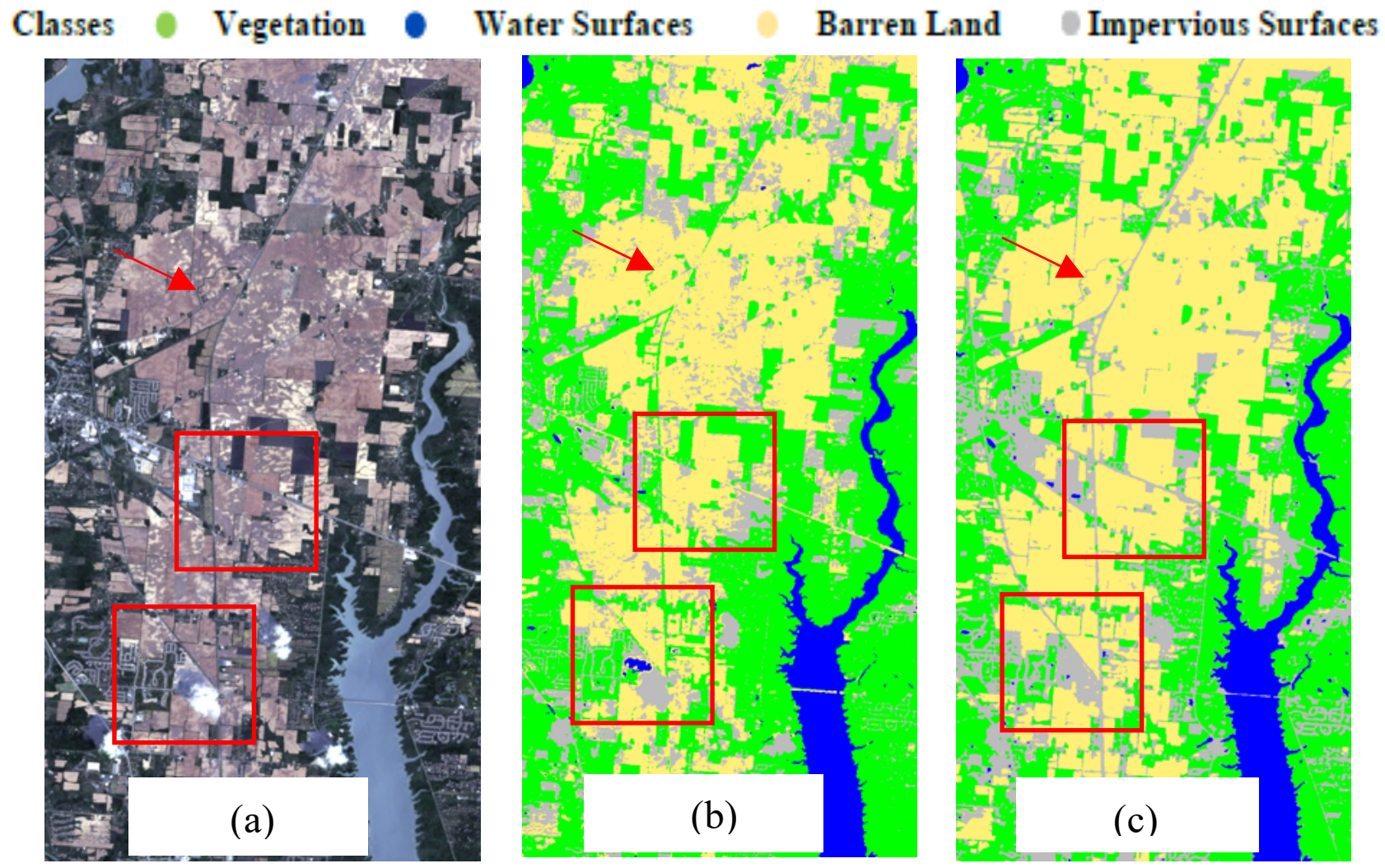

Figure 3.7. Transfer learning-based classification output – a subregion of the urban dataset of Experiment I: a comparison of the original and spatiotemporally filtered image. Note: The red squared boxes represent clouds and barren land covered by shadows corrected over the filter. The red arrow indicates the appearance of a feature (road). (a) Original image; (b) original image classification results, and (c) filtered image classification result at $\sigma_{Ts}$=0.2.

- *Experiment II*

The Planet constellation high-resolution images are captured by a series of satellites instead of a single sensor, and this accounts for another dimension of variation for images taken from different satellites in this constellation. This experiment demonstrates that the proposed method has the potential to correct such variations, therefore, in this section, we will assess the performance of the spatiotemporal filter under multiple classification scenarios (classical and transfer learning). Figure 8. shows the classification accuracy graph for the spatiotemporally filtered images under several $\sigma_{Ts}$ parameters for the transfer learning approach; the peak points (optimum values) were reached when $\sigma_{Ts}$ is 0.2 or 0.3, which is the same range where image spectral properties are moderately preserved (as noted previously in section 3.2.1.). Similar to experiment I, the range of accuracy improvement of classification on the optimum filtered results (dashed and solid lines respectively in Figure 8.) over the original data is between 1%- 5.24%, where image 4 is the reference image. The peak points (optimum accuracies) also indicate higher accuracies than the regular bilateral filter ($\sigma_{Ts}$ =0) (see Figure 8.). The classification accuracies (of the filtered data) at $\sigma_{Ts}$= 0.2 is indicated in the last row in Table 3., where each filtered image shows an increase in its accuracy value resulting an enhancement in the overall accuracy by 1.84%≈ 2%. We also observe that large values of $\sigma_{Ts}$ (>0.3) drive the images to have less discrepancies and more resemblance, and this leads their spectral distribution maps to converge gradually as $\sigma_{Ts}$ increase as shown in Figure 4. As a result, this will saturate the original appearances of the image dramatically leading to the change of classified labels, thus causing the drop in classification accuracy after $\sigma_{Ts}$>0.2-0.3.

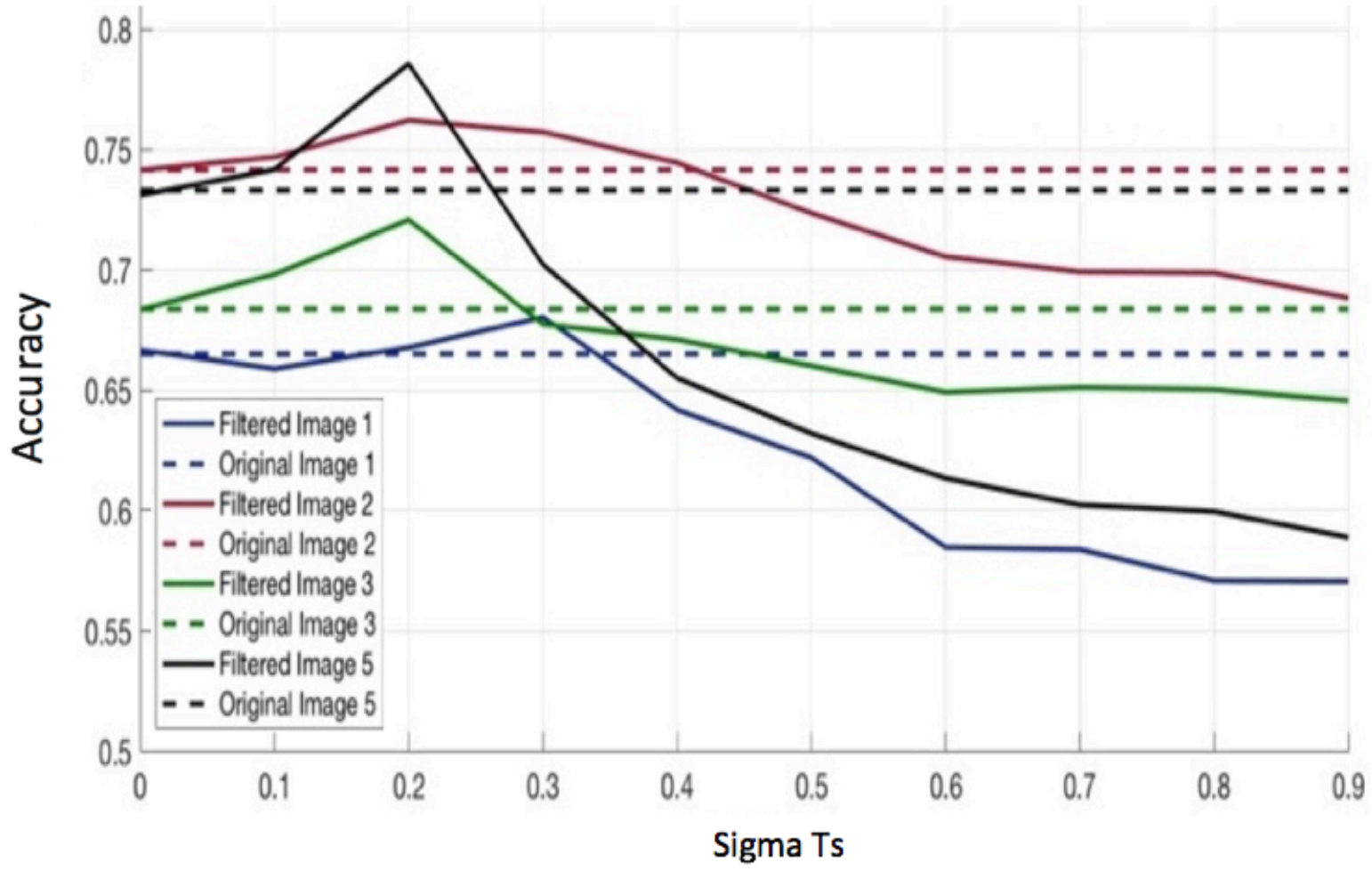

Figure 3.8. Classification Accuracy comparison between original and filtered data (Experiment II).

We observe that the spatiotemporal filter can enhance the classification results of high-resolution images. For instance, water surfaces with a similar color to vegetation (due to radiometric variations) are often misinterpreted (see Figure 3.9(b), black dashed region). The proposed 3D spatiotemporal filter enables to transfer of distinctive spectral information of the water surfaces from other images; thus the classifier can correctly identify the water surfaces with the filtered image (see Figure 3.9(c)). Moreover, buildings and roads often have similar intensities and are more likely to be misclassified (as shown in the black box in Figure 3.9), similarly, the proposed filter can enhance the distinctiveness of the spectrum and thus be able to restore part of the information of such regions.

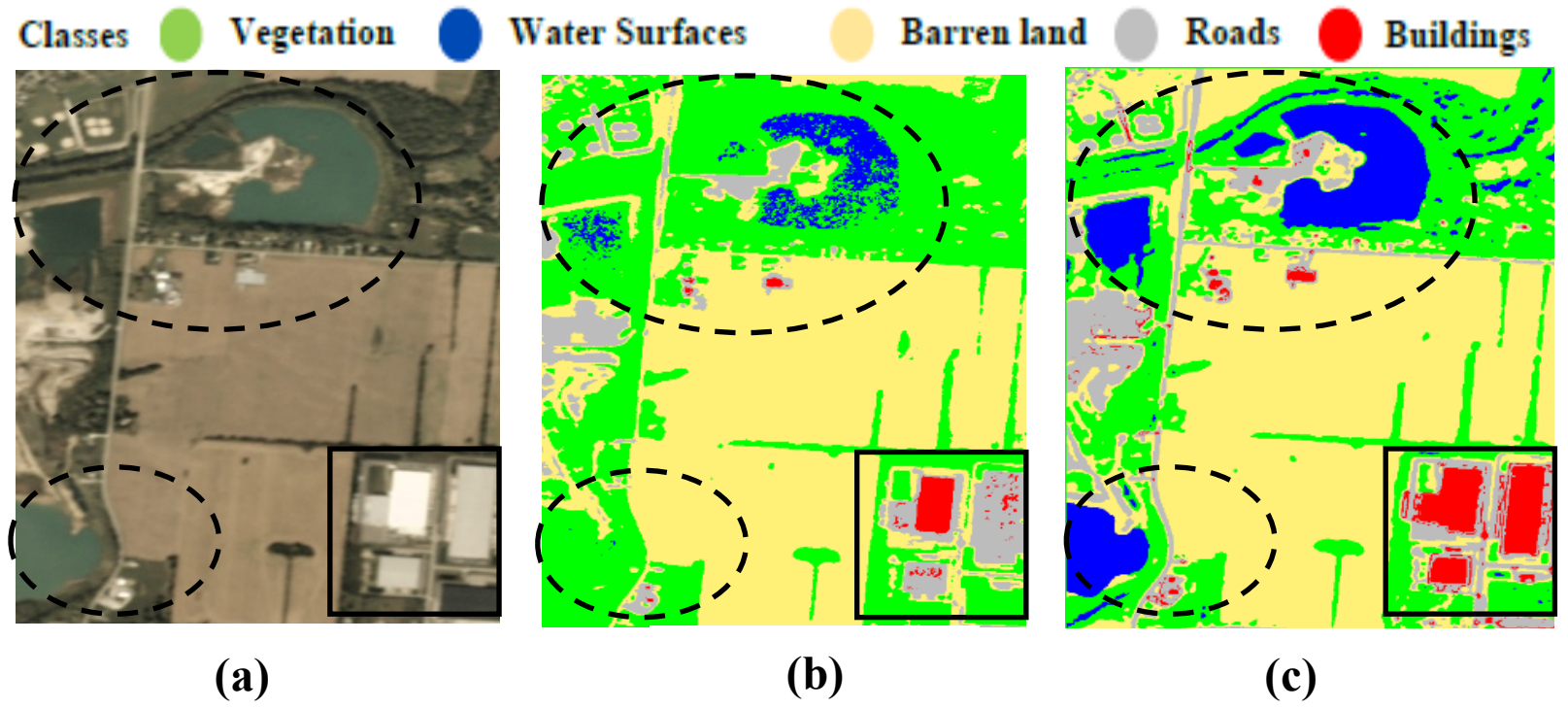

Figure 3.9. Classification enhancement by the filter (a sub-region of the Planet dataset).Note: the solid black rectangle shows the roads and buildings correction, while the black dashed ellipses show water surfaces correction. (a) Original image, (b) Original image classification results, and (c) Filtered image classification results at $\sigma_{Ts}$=0.2.

- *Evaluation Using Other Relative Radiometric Normalization (RRN) Methods*

To evaluate the results, we compared our proposed method with five other state-of-the-art methods in relative radiometric normalization: (1) Dark object subtraction (DOS), (2) Histogram matching (HM), (3) Regression, (4) Mean-standard deviation (5) IRMAD (Iteratively weighted multivariate alteration detection) (Canty & Nielsen, 2008; D. Yuan & Elvidge, 1996). These methods are parametric, and all rely on a single reference image for adjusting their radiometric properties. The comparative results in both classic classification and transfer learning are shown in Figure 3.10. As expected, the results of classic classification before and after the RRN methods do not show significant differences and enhancement, as the relative correction is intended to enhance the radiometric consistencies among temporal images. Therefore, these methods should be better compared under the transferring learning context (Figure 3.10(b)). Our experiment in the transferring learning scenario (Figure 3.10 (b)) shows that our proposed spatiotemporal filter (displayed in the orange bar) outperforms all the other tested methods (a minimal 10% higher in accuracy), and in the best case (image 2), yields an accuracy of approximately 20% higher than others. The classification accuracy of our method is consistently higher than the non-filtered image, while some of the tested RRN methods show an even decrease in accuracy.

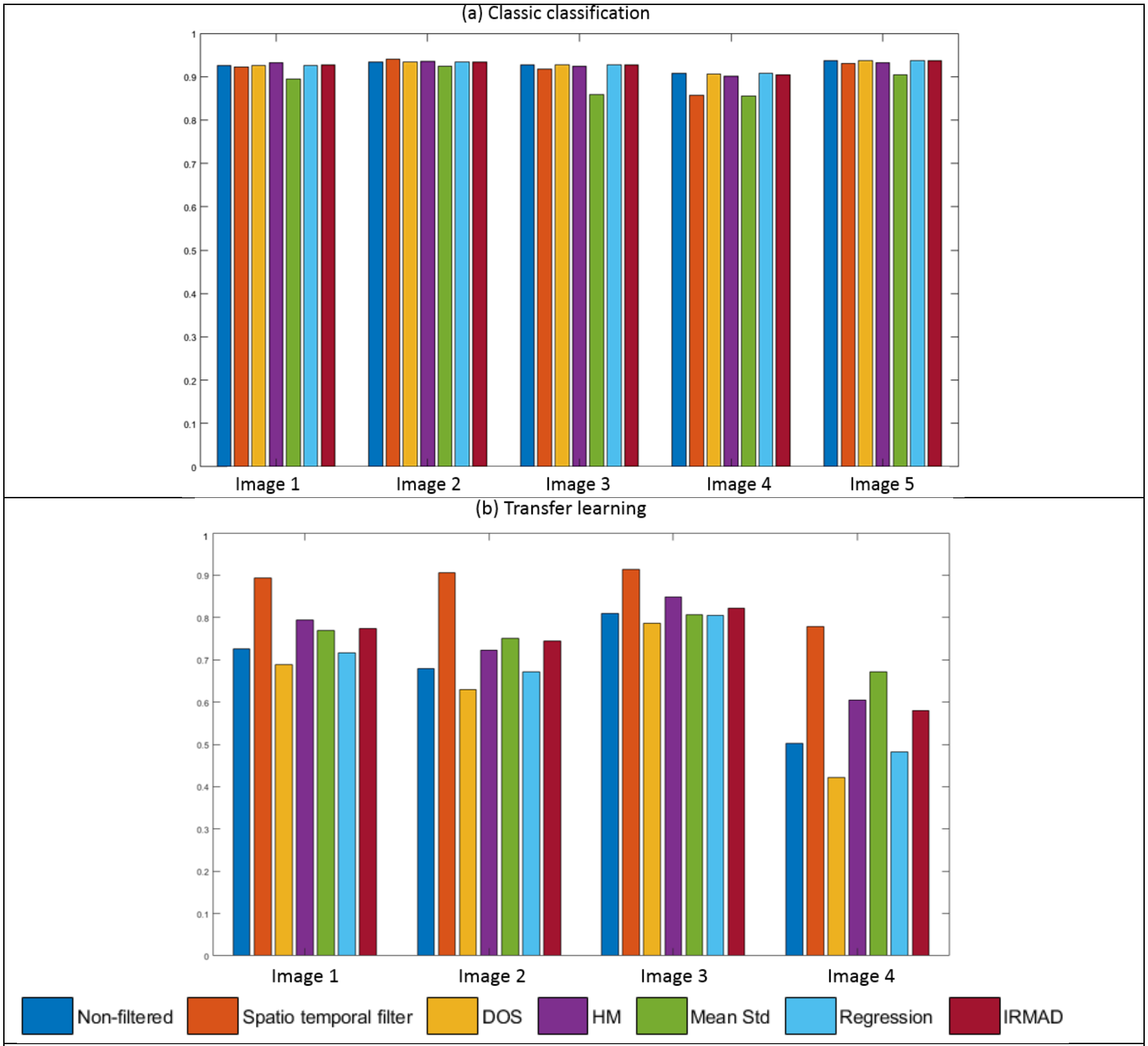

Figure 3.10. A comparison between different RRN methods using the urban dataset (from Experiment I). (a) Classic classification. (b). Transfer learning. Note in figure (b) we have four images since image 5 was used as a reference to train the classifier.

### 3.4.4. Evaluation Using Change Detection

We conduct a simple change detection analysis experiment based on image differencing to understand the capability of the proposed 3D spatial-temporal filter, and this experiment can be further extended using advanced change detection algorithms. The spectrum difference of bi-temporal images is computed, followed by a quantitative evaluation of changes against a manually labeled ground truth change mask. The experiment is performed on two bi-temporal pairs, one pair from the original (unfiltered) image dataset and the other from the filtered dataset, to compare the enhancement capability of the 3D spatial-temporal filter on the change detection application. The change mask from the difference is generated using a threshold, which is optimally computed using Otsu method (1979) (A histogram-based method). In Figure 3.11, we present the results of a sample of change detection experiment, using image 5/image 1 from the suburban dataset in experiment I; it can be seen that the change mask from the filtered images has less noises. These noises are dominated by false positives, for instance, the red box in Figure 3.11 shows the water surface remains through the bi-temporal images, while the original change detection result identified this region as changed areas (Figure 3.11(a)), and that of the filtered dataset can correctly identify this area as being non-change, leading to much fewer false positives (see Figure 3.11(b)). We use the typical statistical metrics including recall, precession, and F1-score, to quantitatively evaluate the change masks:

$$recall = \frac{Tp}{Tp+FN} \qquad (3.11)$$

$$precision = \frac{Tp}{Tp+FP} \qquad (3.12)$$

$$\mathrm{F1-score} = \frac{2\,recall*precision}{recall+precision} \qquad (3.13)$$

Where, the TP (true positives), TN (true negatives), FN (false negatives) and FN (false positives) are normalized according to the actual number of positives and negatives to eliminate sampling biases. Table 3.4 shows the quantitative results using bi-temporal pairs image 5/image 1 and image 5/image 4 from experiment I suburban dataset. The actual change in these bi-temporal data is relatively insignificant (see the ground truth in Figure 3.11), since the time span of the dataset is relatively short (maximum a year, see Table 3.1). The change detection result of the original dataset detected many positive changes, leading to a slightly higher recall value in the original than filtered (see the first column in Table 3.4). However, the low precision measurement of the results from the original dataset indicates that many of these detected changes are false positives, while the results of the filtered dataset show a much larger position of the detected changes are correct (true positives), giving rise to 30% and 6% improvement in terms of precision, at the cost of only 1-2% drop in the recall. This leads to a maximal 14% of F-1 score improvement in our experiment.

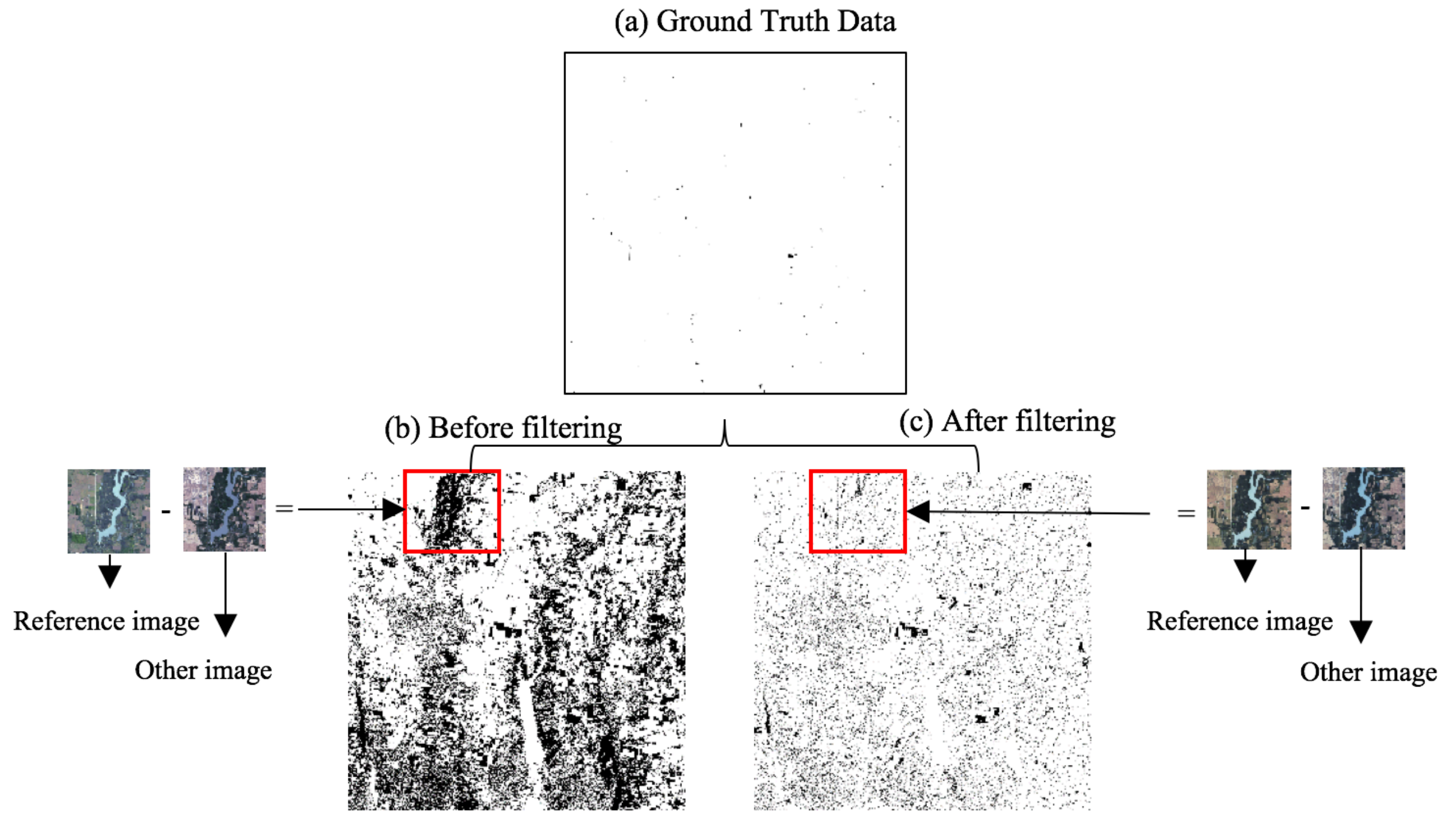

Figure 3.11. Change detection results of Experiment I - suburban dataset, image 5/image 1, before and after filtering.Note: the red box shows the change detection section, where (a) ground truth data (b) before filtering, and (c) after filtering.

Table 3.4. Quantitative evaluation of change detection between images 5 and 4 from the experiment I suburban dataset with temporal bandwidth $\sigma_{Ts}$=0.2 using recall, precession, and F1-score. The bold numbers represent higher accuracies.

| $\Delta I = \lvert I_{reference} - I_i \rvert$ | Image type | Recall | Precession | F1-score |
|---|---|---|---|---|
| $I_5$-$I_1$ | Original (Unfiltered) | **56.37%** | 42.42% | 0.4841 |
| | Filtered | 55.76% | **72.06%** | **0.6287** |
| $I_5$-$I_4$ | Original (Unfiltered) | **68.86%** | 79.78% | 0.7392 |
| | Filtered | 66.04% | **85.64%** | **0.7457** |

## 3.5. Conclusion

In this paper, we propose a novel 3D spatial-temporal filter to enhance the spectrum quality of the time sequence satellite images to facilitate multi-temporal image processing. We demonstrated that the proposed method can enhance the land-cover classification and change detection applications on the multi-temporal dataset to a notable level. This proposed 3D spatial-temporal filter is a derivation of the general weighted filter and is developed as an edge-aware filter in both spatial and temporal directions. The proposed method surpasses other existing relative radiometric correction techniques used in the multi-temporal image sequence processing by being (1) a non-parametric correction method that does not require additional in-situ data or model fitting procedures. (2) A reference-free technique that does not require a single image to correct the rest of the images. (3) The proposed method is efficient and easy to implement and process since it is derived from a local filter.

Our experiments on both the low-resolution Landsat data (in experiment I) and high-resolution images (Planet lab) (in experiment II) have shown that the proposed 3D spatial-temporal filter is capable of improving classification accuracy using transfer learning, in which we have obtained a maximal 27% (overall accuracy) improvement in our classification and 14% (F-1 score) for change detection in our experiments. The proposed method only contains three tunable parameters: $\sigma_x, \sigma_r$ and $\sigma_{T_s}$, being spatial, spectral and temporal bandwidth. $\sigma_x, \sigma_r$ is insensitive as those in typical bilateral filters, we also demonstrated through our experiments that the temporal bandwidth $\sigma_{T_s}$ is rather stable, where the optimal accuracies of classification (using transfer learning) were obtained when $\sigma_{T_s}$ = 0.2-0.3, this is consistently observed to be the optimal value in our experiments. A larger value of $\sigma_{T_s}$ (e.g. ≥ 0.4), may bring potential drops in the accuracy values due to the large "blurry" effect in the temporal direction, this is expected as the filter tends to incorporate different spectrums from other images that eventually diffuse the original spectrums.

During our experiments, we also observed that although a $\sigma_{T_s}$ = 0.2-0.3 is normally optimal, a single $\sigma_{T_s}$ might not be general enough to cover the nature of heterogeneity of the scene, e.g. vegetation may be more tolerant towards spectrum differences due to seasonal differences, and man-made objects may require a small $\sigma_{T_s}$ as a large radiometric change may indicate the change of the object. Moreover, such heterogeneity can also be considered for images taken from different seasons and different time spans. Therefore, in our future work, we will attempt the incorporation of semantic contexts to our proposed spatiotemporal filter, such that the three bandwidths can be adaptively adjusted for different pixels, we also intend to further test the fixed values of the optimal $\sigma_{T_s}$ (bandwidth). Future works also include accelerating the proposed method to reduce the computational expenses using GPU (graphics Processing Unit) parallel implementation or approximate solutions. Furthermore, the inclusion of more external information (e.g. height) in the filtering process might achieve further improvements.

# Chapter 4. The 3D Iterative Spatiotemporal Filtering to Enhance the Consistency of the Classification Maps

This chapter is based on the paper "3D Iterative spatiotemporal Filtering for Classification of Multi-temporal Satellite Dataset" that is published in the "Photogrammetric Engineering & Remote Sensing" by (Albanwan et al., 2020).

## 4.1. Abstract

The current practice in Land-cover and land-use change analysis heavily relies on the individually classified maps of the multi-temporal dataset. Due to varying acquisition conditions (e.g. illumination, sensors, and seasonal differences), the yielded classification maps are often inconsistent over time for robust statistical analysis. 3D geometric features are shown to be stable to assess differences across the temporal dataset, therefore, in this paper, we investigate the use of multi-temporal orthophoto and digital surface model (DSM) derived from satellite data for spatial-temporal classification. Our approach consists of two major steps: (1) generating per-class probability distribution maps using the Random Forest classifier with limited training samples, and (2) spatiotemporal inferences using an iterative 3D spatiotemporal filter operating on per-class probability maps. Our experiments results have demonstrated that the proposed methods can consistently improve the individual classification results by 2% ~ 6%, thus can be an important post-classification refinement approach.

## 4.2. Introduction

In land-cover land-use change mapping, obtaining consistent data and classification results are critical to improving the representation and analysis for remote sensing applications, as the varying acquisition conditions, such as meteorological conditions, viewing angles, sun illumination, sensor characteristics, etc., largely affect the radiometric values of the similar objects through different datasets. In recent years, state-of-the-art satellite imaging platforms including IKONOS (1999, decommissioned), Geoeye-1 (2008), and Worldview-series (2008 and onwards) are able to collect very-high-resolution (VHR) multispectral images, with a spatial resolution that can reach as high as 0.3 meters. The availability of such VHR satellite imagery, increases the capability to interpret and monitor the land-surfaces, urban/natural dynamics in much higher levels of detail: VHR imageries can reduce the mixed-pixel effects that often exist in low-resolution images and allow information extraction to be carried out on an individual object basis, such as shapes, sizes, and patterns of buildings, trees, roads, etc. Although the advantage of higher resolution brings human operators more information for decision-making, the algorithmic development for automatic interpretation of VHR images do not offer equivalent advancements. The major challenge for such algorithms is to overcome low intra-class (within-class) similarity and inter-class (between-classes) separability (Salehi et al., 2011). The spectral ambiguities of VHR images increase with scene complexities (Qin & Fang, 2014). For instance, in dense urban regions where many different object classes exist such as buildings, pavements, ground, roads, etc. (Kotthaus et al., 2014), it is possible that object classes may reflect similar spectral reflectance (e.g. buildings/roads and ground/plazas are concretes), which consequently leads to misclassifications for an algorithm that merely considers spectral information ( Moran, 2010). Efforts that utilize the shape information of the objects have achieved considerable improvement (Ghamisi et al., 2015; Qin, 2015), but these efforts often suffer from unreliable segmentations and inconsistent spectral information. Therefore, there is a critical need in developing consistent and robust methods for land-cover classification and change detections for VHR satellite data.

Existing studies have shown that integrating the spectral and 3D geometric information (e.g. height, depth, etc.) reduces the uncertainty of classification and change detection for objects with similar spectral characteristics (Chaabouni-Chouayakh et al., 2010; MacFaden et al., 2012; Qin et al., 2016a). Unlike image information where appearances vary over time due to illumination and acquisition conditions, height is relatively robust, which provides more valuable information for multi-temporal dataset comparison. With the growing number of optical satellite sensors in operation, the possibilities for satellites to view an area from multiple angles have dramatically increased. 3D information such as digital surface models (DSM) generated from such multi-view or stereo-view satellite images present useful data sources for mapping and monitoring. As compared to typical LiDAR-based DSM data, the satellite-derived DSMs are more affordable and can capture images of inaccessible regions, but it presents a higher level of noise. Thus, to better utilize such DSM data, the uncertainties of the data and the derived results (i.e. classification) need to be well accounted for.

In this paper, we introduce an iterative 3D spatiotemporal filter that applies probability inference to refine land-cover classification of VHR multi-temporal data. This paper further extends a past work (Albanwan & Qin, 2018) that has shown that a single-step spatiotemporal filter has the capability to improve the spectrum consistency across the multi-temporal dataset. The proposed iterative 3D spatiotemporal filter further enhances such capability for multi-modal time-series data (i.e. spatial, spectrum and height) for classification problems, which aims to achieve higher classification accuracy for multi-temporal datasets through spatial-temporal inferences than independent temporal data classification. The proposed method is applied to per-class probability maps and can adaptively reduce the heterogeneity of the probability maps, leading to more consistent classification results. The rest of this work is organized as follows:

Section 4.3 briefly introduces relevant works and the rationale of our proposed work. Section 4.4 presents the proposed iterative 3D spatiotemporal filter. The experimental results and evaluation are shown in Section 4.5. Section 4.6 concludes the paper by discussing the effectiveness, limitations of our work, and potential future works for improvements.

### 4.3. Related Works and Rationale

The proposed work aims to enhance the classification accuracy through inferring information across the multi-temporal dataset, thus is very relevant to both classification problems and multi-temporal data processing in a land-user and land cover change (LULCC) context. Therefore, this section briefly summarizes both these two relevant topics.

#### 4.3.1. Related Works

- *Classification of VHR Satellite Images*

VHR image classification is a widely studied research topic in remote sensing (Fauvel et al., 2013). Traditional classification is categorized into pixel- or object-based methods. In most cases, the object-based methods perform reasonably better on VHR images due to the availability of the spatial information that allows the use of patterns and local regions, which are often hard to obtain through pixel-based approaches (G. Chen et al., 2012; Keyport et al., 2018; Weih & Riggan, 2010). The major challenge in classifying multi-temporal VHR datasets is obtaining and maintaining consistency in the dataset through the entire process, which can be difficult due to the varied acquisition conditions. Object-based methods often require segmentation of the image to extract spatial information (Qin, 2015). However, improper segmentation parameters can lead to under- or over-segmentation, which impacts the classification accuracy. Incorporating 3D geometric information (i.e. height) can provide more accurate, robust, and stable solutions since the height of objects is insensitive to spectral variations (MacFaden et al., 2012; Minh & Hien, 2011; Qin et al., 2016a). Including height in the classification involves fusing multi-source data (i.e. spectral and height information) (X. Huang et al., 2011; Kim, 2016; Salehi et al., 2011), which can be performed at either the pixel, feature, or decision level (J. Zhang, 2010). One of the challenges in including height in classification is its source and quality. For instance, DSMs are either generated directly from sensors like LiDAR or indirectly using modern photogrammetric techniques (e.g. stereo matching). Although the direct approach is more accurate, the indirect approach is preferred in many current studies due to its higher availability and lower cost (Salehi et al., 2011). However, stereo-matching algorithms used to generate DSM depend on the pair of images and their acquisition conditions and thus may result in noise, artifacts, and uncertainties in the DSM (Minh & Hien, 2011). Statistical approaches are often used to process multisource data to fuse, filter, and determine the class membership of pixels using probabilities (Moser et al., 2013; Qin, 2015). However, they may be complex depending on the complexity of the scene (W. Y. Yan et al., 2015), since they may create high-dimensional feature space and may be responsible for producing noise and outliers (Fu, 2011).

- *Spatiotemporal Inferences of Land-Use Land-Cover Change (LULCC) Maps*

LULCC products are highly dependent on the quality of land cover classification maps. The inconsistencies and uncertainties in the classification maps due to the varying data spectrums can negatively influence applications that directly rely on them, such as time-series analysis, change detection, object recognition, and modeling. In such a context, improving single-date classification on one hand may require ad-hoc techniques specifically tailored for factors influencing the images, and on the other hand, might not yield optimally consistent classification maps over time. In this regard, spatiotemporal inference approaches applied to the images or classification results can be particularly effective, as they simultaneously homogenize information in both spatial and temporal directions (Floberg & Holden, 2013). Current spatiotemporal inferences models are categorized into local, non-local, and global models. The local and non-local methods are more efficient in terms of their computational complexity since they operate over small pixel-neighborhoods and local regions (Floberg & Holden, 2013). For instance, (Cheng et al., 2017) proposed a spatial and temporal non-local filter-based fusion model (STNLFFM), to predict the land-cover class for every pixel based on the spectral values through a weighted sum of the neighboring pixels in the dataset. Albanwan and Qin (2018) proposed a local spatiotemporal bilateral filter as a preprocessing step for multi-temporal images to enhance their radiometric characteristics and consistency for classification. Global approaches for spatiotemporal inference are mainly based on probabilistic graphical models and Markov Random Field (MRF), which models each pixel as a statistical variable where local smoothness and global consistency serve as objectives when performing maximal likelihood estimation (W. Gu et al., 2017; Kasetkasem & Varshney, 2002; D. Liu & Cai, 2012). Such inference algorithms help to generate more consistent change maps and prevent noises caused by inaccuracy classification results. The solution for such inference algorithms usually uses fixed spatial weight parameters which might potentially lead to over-smoothing for the change maps (Gu et al., 2017). Gu et al. (2017) proposed a linear weighting scheme, where the spatial weights are estimated adaptively for every pixel based on its status changed, unchanged, or uncertain change. Although global methods used in change detection (such as MRF) provide high-quality results, they involve per-pixel processing which increases the computational complexity exponentially as the number of pixels and time-series data increase.

### 4.3.2. Proposed Method and Rationale

The goal of this work is to achieve an efficient spatiotemporal inference model that allows accurate LULCC mapping. Inspired by Albanwan and Qin (2018), where a 3D spatiotemporal filter is developed for the classification of images using spectral, spatial, and temporal knowledge, we propose here a simple but effective extension leading to a 3D iterative spatiotemporal filter. Using global methods for per-pixel classification is a time-consuming process, (Krähenbühl and Koltun, 2012) suggest using iterative approaches through local filters to approximate global inferences, with the benefit of reducing computational costs. Therefore, we here apply an iterative approach that filters the probability maps along with the corresponding DSMs in an iterative fashion to achieve optimal classification results. Figure 4.1. illustrates our rationale for using spatiotemporal filtering: given three per-class probability maps generated from three temporal datasets, the highest probability normally determines the class of the pixel (represented as ci for the ith class). In the first and third images, the yellow pixel is classified as the second class c2 and the second image is classified as cn. Knowing that if the feature of the central pixel in the second image is similar to that of the first and the third, this central pixel in the second image can be a misclassification. Probability inferences in the temporal direction might help correct such errors by providing higher confidence in the correct class. Our proposed iterative solution is expected to perform such probability inferences in a global sense, to refine and improve the consistency of the classification maps and their accuracy.

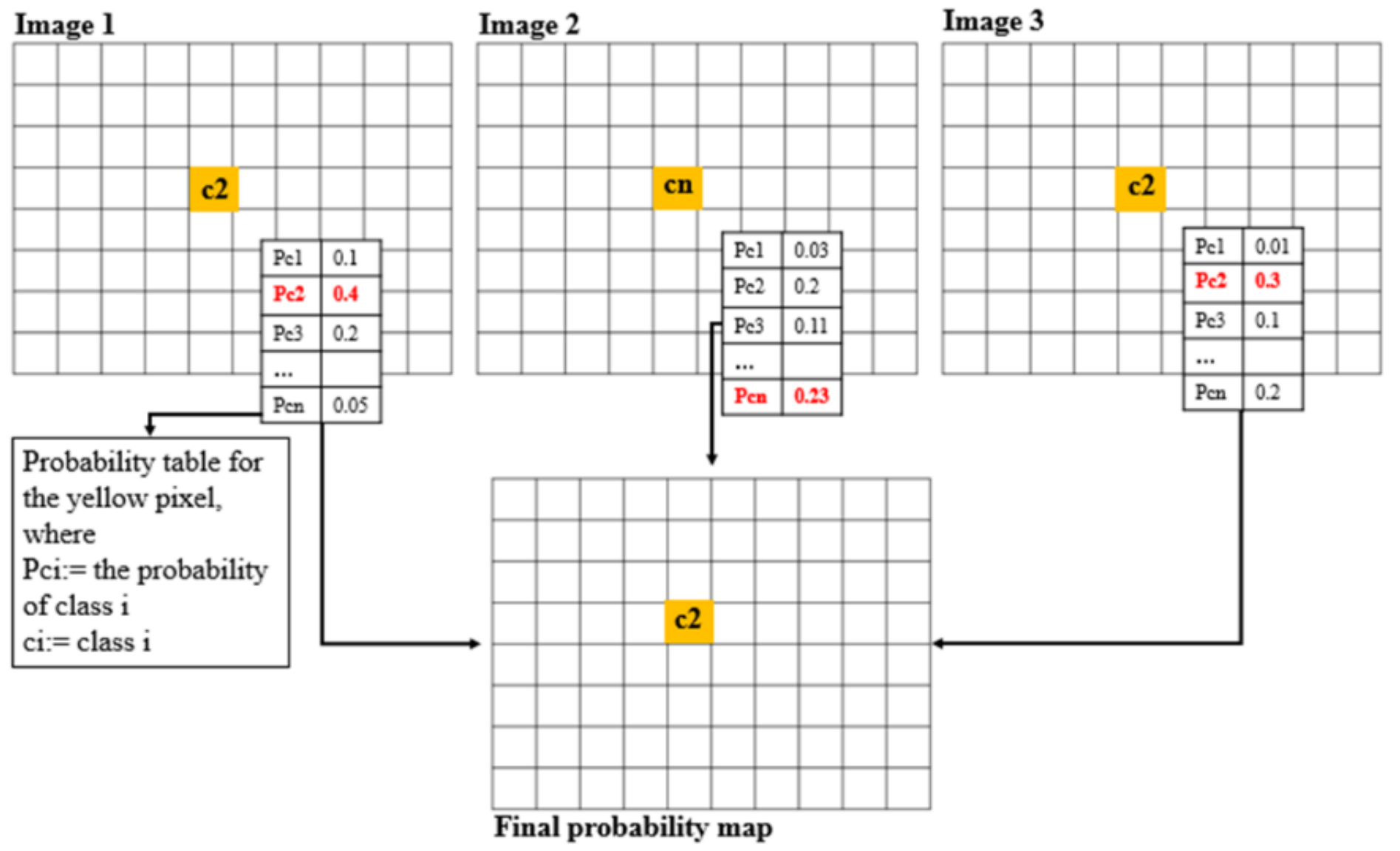

Figure 4.1. The concept of the proposed spatiotemporal inference method

## 4.4. Methodology

Our workflow consists of three main steps (Figure 4.2): 1) data preprocessing of multi-temporal stereo satellite images, 2) object-based classification, and 3) 3D iterative filtering for spatiotemporal inference. In this section, we provide an overview of the first two steps with emphasized details on step 3) where our main technical contribution lies in. The preprocessing step mainly consists of Digital Surface Models (DSM) and Orthophoto generation and registration using stereo satellite images. Once the DSM and orthophotos are registered, we take advantage of the DSM and orthophoto differences to infer training labels from one temporal dataset to others, and then utilize these labels to train random forest classifiers individually to generate initial and per-class probability maps. These probability maps are then fed into our proposed 3D iterative spatiotemporal filter, where probabilities maps are globally inferred (concept as per described in Figure 4.1) as the final probability map. We assess the classification results by estimating the overall accuracy of each iteration.

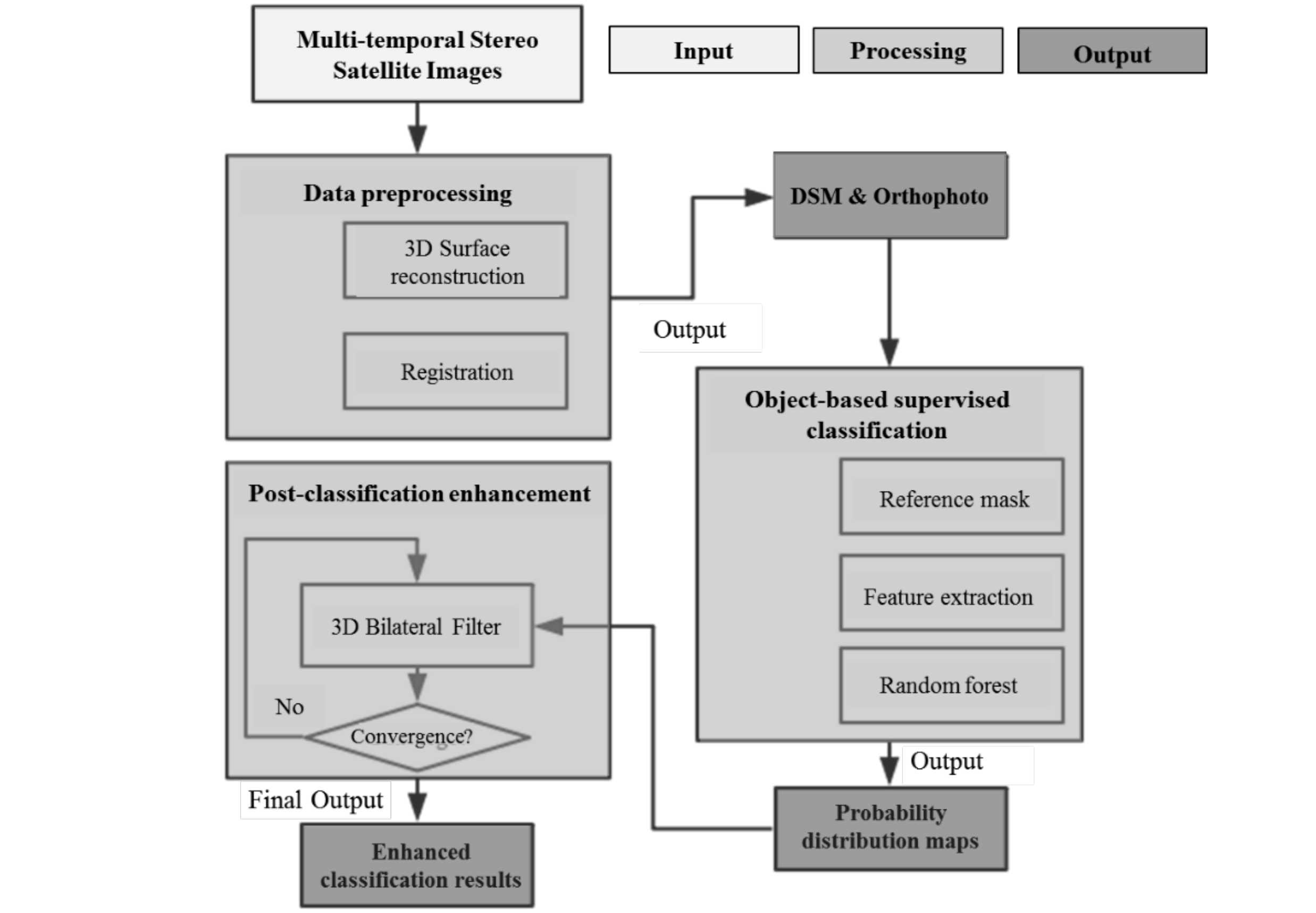

Figure 4.2. General workflow for the proposed method.

### 4.4.1. Data Preprocessing Steps

The preprocessing stage involves the generation of the DSM and orthophoto datasets, precise geometric alignment of the data, and normalization to nDSM. We generate high-quality DSMs and orthophotos of the related satellite image pairs using RSP (RPC Stereo Processor) (Qin, 2016), which adopts a hierarchical semi-global matching method (Hirschmuller, 2008a). Due to the systematic satellite positioning error, the generated DSMs and orthophotos may not be well aligned. Co-registration of a DSM requires estimating the transformation parameters (rotation and translation) between the reference and target DSMs as described in Gruen and Akca (2005). Since for the satellite overview dataset, the rotation differences are often regarded as neglectable (Waser et al., 2008), we can then only compute the shift parameters in three directions x, y, and z. Using a simple approximation to the least-squares surface matching (Gruen & Akca, 2005), we estimate the shift parameters that minimize the sum squared Euclidean distances. We eliminate the systematic error by discarding the potential noise and outliers due to the DSM generation process by applying a threshold (in our experiment we used 6 meters with an empirical estimation of the height uncertainties), which regards points with errors bigger than the threshold as outliers and discards them for error computation. Finally, we generate nDSM using the morphological top-hat reconstruction strategy as explained in Qin and Fang (2014).

#### 4.4.2. Object-based Classification and Probability Distribution Map Generation

- *Training Label Generation Through Empirical Inferences*

Since generating masks for every single image is tedious and time-consuming, we simplify this process of propagating training samples for the entire dataset through an intuitive and automated mask propagation approach, with empirically-set thresholds to ease the extraction of ground truth/ reference map. Given a labeled image (i.e. the reference map); we can obtain the training samples of the rest of the temporal dataset through inferences from unchanged labels. Our automated method utilizes preliminary features such as nDSM and normalized difference vegetation index (NDVI), as change indicators and determines if a pixel in the labeled images should be labeled as the same class in the data of a different date. The nDSM offers higher robustness towards spectral variations that result due to different seasonality between the multi-temporal data. We apply an intuitive threshold-based change detection on nDSM and NDVI with pre-defined thresholds determined empirically to mask out objects in each category/class. The seven classes used in our work are 'Building', 'Temporary Lodge', 'Long-Term Lodge', 'Tree', 'Grass', 'Ground', and 'Road.

- *Feature Extraction, Classification, and Probability Map Generation*

The per-class probability (or sometimes called confidence) maps are important to determine the class membership of each data point, this is provided by most of the statistical classifiers: for example, SVM weight the distances of the output to the support boundaries, and random forest computes the probability by counting the votes from all the decision trees forming the forest. In our experiment, we use the Random Forest (RF) classifier and adopt an object-based classification approach, where mean-shift is used as the segmentation approach due to its wide use in relevant applications. The spectral feature is computed using PCA (principal component analysis) on the spectrum bands. We have also considered spatial features extracted from the DSMs and Orthophotos as part of the feature vector, which includes: dual morphological profiles (DMP), and morphological top hat reconstruction (MTHR) (Qin, 2015; L. Zhang et al., 2006). Details of these features are shown in Table 4.1.

Table 4.1. Extracted features for classification, the light boxes are types of features and the dark boxes provide additional details

| PCA | DMP & MTHR of DSM | Adaptive structural elements (SEs) |
|---|---|---|
| All columns from PCA of the spectrum (which includes only the RGB bands) of segment S (Jolliffe, 2005; Wold et. al., 1987). | The radius sequence of disk shape used in morphological profile reconstruction [24 146 260] pixels.<br>DMP of the first component from PCA in the first feature.<br>DMP for the color inverse with the first component from PCA in the first feature. | Used to construct DMP and address the problem of multiscale information of segments (Qian Zhang et al. 2015) |

As discussed, the random-forest classifier is an ensemble classifier that decides the labels based on the number of votes per class among the decision trees, and the normalized voting numbers (rationing the total number of votes) are then used as the probability/chance that a segment or pixel belongs to a specific class; 500 decision trees were used in our experiment, and their per-class probability maps yield in total N (number of classes) probability maps for each temporal data set.

### 4.4.3. Post-Classification Enhancement by a 3D Iterative Spatiotemporal Filter

- *The General Algorithm of the 3D Spatiotemporal Filter*

The proposed filter is based on a 3D spatiotemporal bilateral filter (Albanwan and Qin 2018), the general form of which is represented as:

$$Pnew(x_i, y_i, t_m) = \frac{1}{N * T} \sum \sum_{j \in N, n \in T} W_{3D}(x_j, y_j, t_n) * P(x_j, y_j, t_n) \quad (4.1)$$

Where $N$ denotes the spatial neighborhood centered on the point $(x_i, y_i)$ and $T$ refers to the number of observations (dates) through time; $P(x, y, tn)$ is the probability of pixel at location $(x, y)$ in the probability map taken on the date $t_n$. $W_{3D}$ are adaptive weights that consider spectral, spatial, and temporal differences of the datasets. Since in our context, the DSM data is available and are regarded robust to illumination changes, we hereby take the nDSM differences as means of weighting the temporal coherences for probability inference. Thus, the weight $W_{3D}$ can be decomposed into three components:

$$W_{3D} = W_{Spatial} * W_{Spectral} * W_{nDSM} \quad (4.2)$$

where

$$W_{spatial}(x_j, y_j, t_n) = e^{-\frac{\|x_i - x_j\|^2 + \|y_i - y_j\|^2}{2\sigma_s^2}} \quad (4.3)$$

$$W_{spectral}\left(x_j, y_j\right) = e^{-\left(\frac{\|I(x_i, y_i) - I(x_j, y_j)\|^2}{2\sigma_r^2}\right)} \quad (4.4)$$

The $\sigma_s$, $\sigma_r$, and $\sigma_h$ parameters are the spatial, spectral, and elevation/temporal bandwidths of the weighted filter; $I(x_i, y_i)$ refers to the values of the color image (transformed color space), and nDSM refers to the normalized height values. In the spatial and spectral domains, the pixels are weighted based on their closeness and spectral similarities to the central pixel in a defined window, where larger weights are assigned to spectrally similar and spatially proximate pixels (and vice versa). The value of $\sigma_s$ is empirically determined based on the size of the window; a large value of $\sigma_s$ can lead to an overly smoothed image with blurry edges. $\sigma_r$ serves as the spectral bandwidth of a typical bilateral filter to allow edge-aware filtering (Tomasi and Manduchi, 1998). The CIELAB color is used as the transformed color space for computing the spectral differences, where three bands (near-infrared, red, and green) are used for the color transformation (Joblove & Greenberg, 1978; Tomasi & Manduchi, 1998). The height variation is class-dependent, for example, the ground class normally has smaller variation thus it should have a smaller bandwidth to panelize larger weight for incorporating changed area in the filtering processing. Also, tree classes normally have a higher uncertainty and thus a larger bandwidth is needed to average out noises. Therefore, the relevant bandwidth $\sigma_h$ here is estimated per class to achieve the optimal performance. The height range of each class is approximated by the labeled pixels in each of the training labels; for a certain class $c$, the elevation bandwidth $\sigma_h$ and the relevant weight is computed using the following equations:

$$\sigma_h = \left(\frac{70\%}{2}\right) * [\,range\ of\ nDSM\ of\ class\ c] \tag{4.5}$$

$$W_{nDSM}\left(x_j, y_j, t_n\right) = e^{-\frac{\|nDSM(x_i,y_i,t_m) - nDSM(x_j,y_j,t_n)\|^2}{2\sigma_h^2}} \tag{4.6}$$

The "70%" value comes from the consideration that the height measurements from the stereo-matching introduce noises. We, therefore, take the one sigma rule to determine $\sigma_h$, and by assuming the height difference variation as a zero-centered Gaussian distribution, computes the one-sigma interval as 68.27%$\approx$ 70% of the nDSM. These all together give the final form of the 3D weight as computed in Equation (4.7).

$$W_{3D}\left(x_j, y_j, t_n\right) = e^{-\left(\frac{\|x_i - x_j\|^2 + \|y_i - y_j\|^2}{2\sigma_s^2} + \frac{\|I(x_i,y_i) - I(x_j,y_j)\|^2}{2\sigma_r^2} + \frac{\|nDSM(x_i,y_i,t_m) - nDSM(x_j,y_j,t_n)\|^2}{2\sigma_h^2}\right)} \tag{4.7}$$

Considering that both the object's heights and the stereo-matching errors introduce noises, we take the one sigma rule to determine the bandwidth of the height component $\sigma_h$. Thus, we eliminate this noise for every class by only considering the height values that fall in the range ($\mu_{nDMS} \mp 1 * \ \sigma_{nDMS}$) of the mean nDSM ($\mu_{nDMS}$) and the standard deviation ($\sigma_{nDMS}$), which accounts for 68.27%$\approx$ 70% of the entire nDSM. The rationale of this approach is that since height uncertainties are correlated with different classes: for example, the uncertainty of the tree class in terms of the geometry is generally higher than the ground. Therefore, higher bandwidth is expected, which can be estimated from the nDSM variation of that class.

$$W_{3D}\left(x_j, y_j, t_n\right) = e^{-\left(\frac{\|x_i - x_j\|^2 + \|y_i - y_j\|^2}{2\sigma_s^2} + \frac{\|I(x_i, y_i) - I(x_j, y_j)\|^2}{2\sigma_r^2} + \frac{\|nDSM(x_i, y_i, t_m) - nDSM(x_j, y_j, t_n)\|^2}{2\sigma_h^2}\right)} \quad (4.8)$$

This 3D weighted filter performs local spatiotemporal filtering in a limited receptive field for each pixel. To achieve optimal probability maps, we further globalize the inference algorithm through iterative application to probability maps. The iterative filtering is global by means of gradually propagating information from locally processed cells to wider receptive fields. We use the probability maps generated from the random forest classifier (RF) (as described in section 4.4.2) as the initial probability maps (see Figure 4.3) for inference. For each date, there will be in total *n* (number of classes) probability maps, and the spatiotemporal inference is performed on probabilities of the same classes. In the filtering process, the total weight ($W_{3D}$) for each pixel is constant through all iterations since it is a function of the spectral, spatial, and height information of the image. The probability maps are updated in each iteration:

$$P_c^k(x_i, y_i, t_n) = \frac{1}{N*T} \sum \sum_{j \in N, n \in T} W_{3D}(x_j, y_j, t_n) * P_c^{k-1}(x_j, y_j, tn) \quad (4.9)$$

Where $P_c^k(x_i, y_i, t_n)$ is the estimated probability of pixel $(x_i, y_i)$ at date $t_n$, class $c$, and at the $k^{th}$ iteration, derived from the probability $P_c^{k-1}(x_j, y_j, t_n)$ at the *(k-1)*$^{th}$ iteration. We compute this iterative process until the differences between $P_c^{k-1}$ and $P_c^k$ smaller than a threshold, which in our work, is set as 5% of relative changes (as shown in Equation 4.10):

$$convergence\ criterion = \frac{P_c^k(x_i, y_i,\ t_n) - \ P_c^{k-1}(x_i, y_i,\ t_n)}{P_c^k(x_i, y_i,\ t_n)} * 100\% < 5\% \quad (4.10)$$

We take the processed probability map and determine the class label for every pixel as the one with the highest probability. Using the ground truth data, we can evaluate the accuracy of the resulting classification maps to understand how the proposed iterative spatiotemporal filter improves the results. Algorithm 1 presents the pseudo-code of the algorithm.

---

Algorithm 1: Pseudo-code of the proposed 3D iterative spatiotemporal filter

Input: Initial probability maps $P_c^0\ (x_i, y_i,\ t_n\ )$, ortho-photos I, and band widths ($\sigma_s$,$\sigma_r$)
Output: Final probability maps $P_c^f\ (x_i, y_i,\ t_n\ )$

*For every category/class c do*

*While not converge do*

*For every pixel (x,y) in the window w do*

$W_{spatial}$ → $\exp(\frac{\|x_i - x_j\|^2 + \|y_i - y_j\|^2}{2\sigma_s^2})$

$W_{spectral}$ → $\exp(\frac{\|I(x_i, y_i) - I(x_j, y_j)\|^2}{2\sigma_r^2})$

*Compute* $\sigma_h$ *for class c*

$W_{nDSM}$ → $\exp\ (\frac{\|nDSM(x_i, y_i,\ t_m) - nDSM(x_j, y_j,\ t_n)\|^2}{2\sigma_h^2})$

W3D=$W_{Spatial} *\ W_{Spectral} * W_{nDSM}$

*Update the probability distribution map*
$$P_c^k(x_i, y_i,\ t_n) = \frac{1}{N*T}\sum\sum_{j\in N, n\in T} W_{3D}(x_j, y_j,\ t_n) * P_c^{k-1}(x_j, y_j,\ tn)$$

End For

*Check convergence*
$$\frac{P_c^k(x_i, y_i,\ t_n) - P_c^{k-1}(x_i, y_i,\ t_n)}{P_c^k(x_i, y_i,\ t_n)} * 100\% \ \rightarrow \begin{cases} \leq 5\% & Stop \\ > 5\% & Continue \end{cases}$$

*End While*

*End For*

*Compute overall accuracy*

---

### 4.5. Experiments and Results:

#### 4.5.1. Data Description

The dataset used in our experiment is a multi-temporal dataset in Port-au Prince, Haiti, through the 2010 earthquake, where a catastrophic earthquake with a magnitude of 7.0 $M_w$ caused a large number of fatalities and extreme damages to the area, forcing Haitians to migrate into temporary (tents) and long-term lodging (i.e. shelters last for a longer time for accommodations). Therefore, our classification work particularly includes changes in buildings with different functionalities (i.e. long-term, temporary lodgings, and normal built-up areas). The satellite dataset contains seven on-track stereo pairs (data collected on the same day with the satellite on the same track) from 2007 to 2014 and one incidental image pair in 2015 over Port-au-Prince, Haiti, with their acquisition date and details of the data shown in Table 4.2.

Table 4.2. An overview of the satellite stereo pair images used in this experiment

| Satellite | Date | Resolution (Panchromatic) | Resolution (Multispectral) |
|---|---|---|---|
| IKONOS | 03/22/2007 (on-track) | 0.82m | 3.28m |
| Geoeye-1 | 01/16/2010 (on-track) | 0.41m | 1.84m |
| IKONOS | 06/6/2010 (on-track) | 0.82m | 3.28m |
| IKONOS | 12/21/2010 (on-track) | 0.82m | 3.28m |
| Geoeye-1 | 03/8/2012 (on-track) | 0.41m | 1.84m |
| Geoeye-1 | 09/11/2013 (on-track) | 0.41m | 1.84m |
| Worldview | 07/24/2014 (on-track) | 0.46m | 1.84m |
| Geoeye-1 | 06/23/2015 & 07/01/2015 | 0.41m | 1.84m |

We select three test regions with a size of 1×1 $km^2$ (i.e., 2001×2001 pixels for images resampled at a 0.5 GSD (Ground Sampling Distance)) of various urban scenes (see Figure 4.3 for details). Test region 1 is an open area around the airport in which refugees and the government built up their temporary and long-term lodgings spontaneously. We consider seven classes in the classification: buildings, ground, trees, roads, grass, temporary lodging, and long-term lodging (see Figure 4.3). The three test regions vary in terms of their urban forms: Test region 1 indicates an open area surrounded by dense buildings, test regions 2 are specifically dense residential areas with narrow streets, and test region 3 is a school zone with moderate and larger buildings amongst dense and small buildings. The selection of the three test regions is based on our visual inspection of representative regions across the dataset, and apparent surface changes were observed as well in these regions which will be quantified in detail in subsequent sections.

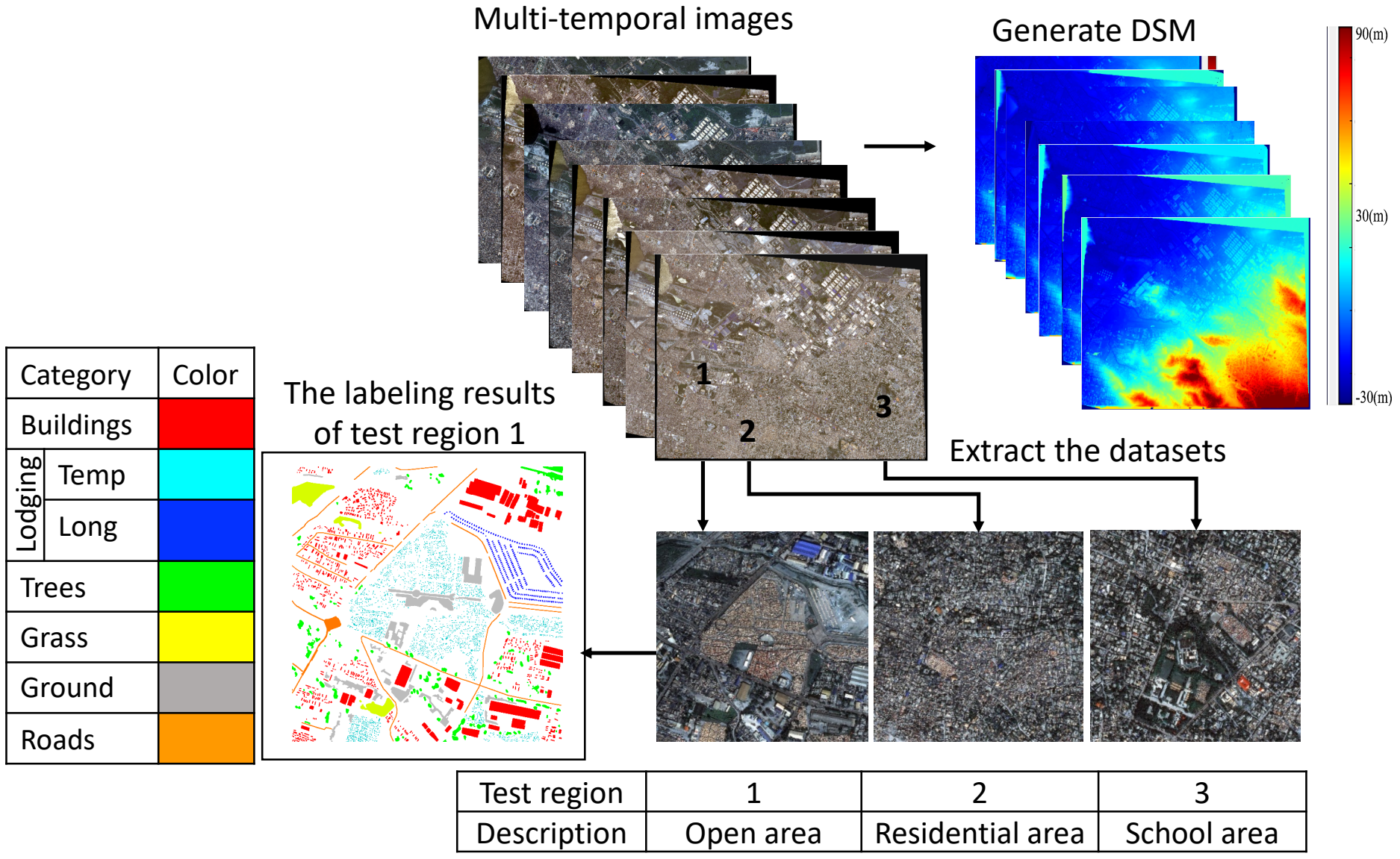

Figure 4.3. Test datasets, where we show multi-temporal orthophotos and DSMs sequences, as well as the used reference labels for training and testing (for the first test region in the time epoch). Note: Temp=temporary, Long=long-term.

### 4.5.2. Data Preprocessing: DSM and Orthophoto Generation

In this preprocessing step, we generate the orthophoto and DSMs for the entire area of Port-au-Prince, Haiti, using the RSP software and the stereo satellite imageries; the DSM generation is based on the semi-global matching (SGM) algorithm. Then, the DSMs are geometrically aligned and nDSMs are generated, as described in section 4.4.2 of the methodology. Figure 4.3 shows an example of the generated orthophoto and DSMs for the dates 2007 and 2010 (right after the earthquake) dates. During this step, all datasets with varying spatial resolutions are sampled to the highest cell size of the entire datasets (0.5 m).

### 4.5.3. Classification and Probability Map Generation

The initial object-based classification is generated using methods as described in sections 4.4.2. We keep the per-class probability map for further spatiotemporal inference. This initial classification yields a relatively satisfactory classification map to start with, with the overall accuracy (OA) shown in Table 3. In general, test region 2 (dense residential regions) yields the lowest OA (ca. 80%), primarily due to the complex rooftops and potentially the higher uncertainties of the DSM.

Table 4.3. Overall classification accuracies for all the three datasets

| Date | Accuracy | | |
|---|---|---|---|
| | Dataset 1 | Dataset 2 | Dataset 3 |
| 2007/03 | 91.04% | 83.47% | 91.12% |
| 2010/01 | 93.21% | 81.50% | 93.06% |
| 2010/06 | 91.93% | 83.52% | 88.82% |
| 2010/12 | 89.08% | 80.81% | 88.58% |
| 2012/03 | 92.19% | 81.43% | 91.44% |
| 2013/09 | 90.40% | 81.03% | 94.99% |
| 2014/07 | 95.11% | 82.19% | 90.39% |
| 2015/07 | 92.74% | 83.22% | 94.61% |

### 4.5.4. Experimental Details of the Proposed 3D Iterative Spatiotemporal Filter

- *Parameters*

As per described in section 4.3, the 3D iterative spatiotemporal filter has spatial, spectral, and temporal components (height difference through the temporal dataset). Based on the described rationale for determining these bandwidth parameters, we apply the following parameters based on our dataset.

1) Bandwidth for the spectral and spatial domains

Because of the size of each test image (over 2000×2000 pixels), the window size is set to 5×5 pixels to balance the efficiency and accuracy of the results. The bandwidth $\sigma_s$ for each dimension of the spatial domain in the filter, the window is set to 3. In addition, for 8-bit images (ranging from 0-255) the bandwidth of the spectral domain $\sigma_r$ is set empirically to 5, and this is shown in the original bilateral filter (Tomasi and Manduchi, 1998) to have good leverage on smoothing and edge-preserving.

2) Bandwidth for temporal (height) range domain $\sigma_h$

The elevation bandwidth is determined separately for each class as specified in Equation (4.8). Table 4.4 shows the bandwidth $\sigma_h$ for each class, calculated using statistics in the test region, and this is to be done for each test region as they have their own training labels. We can see that $\sigma_h$ of buildings and trees are much higher than the ground and grass classes. This is reasonable because the building and tree classes, in general, have larger height variations, and for the same amount of height difference, the ground class should have smaller bandwidth to act sensitive to large changes, while the trees should have moderate tolerance to height differences when considering weighted summing probability maps.

Table 4.4. Temporal (height) domain bandwidth σ_h for each class on test region 1

| $\sigma_h$ value for each class | | | |
|---|---|---|---|
| Buildings | 5.98 (m) | Trees | 4.45 (m) |
| Temporary Lodging | 0.64 (m) | Grass | 0.56 (m) |
| Long-term Lodging | 1.50 (m) | Roads | 1.08 (m) |
| Ground | 0.53 (m) | | |

### 4.5.5. Visual and Statistical Analysis of the Proposed Iterative Spatiotemporal Filter

- *Visual Analysis*

Our method enhances classification by utilizing the per-class probabilities of time-series data via spatiotemporal inference. This is reasonable as unchanged objects can be used to mutually enhance the fidelity of the classification probability maps. We compare the initial and enhanced classification results for all the datasets in Figure 4.5. For instance, the mislabeled objects in the initial classification map of the dense areas in test regions 2 and 3 are corrected after being processed by the proposed spatiotemporal inference method. As shown in regions within the black circle, open grounds primarily dominated by temporary lodgings are misclassified to ground or grass classes in the initial classification map, and the proposed spatiotemporal inference method shows that these misclassifications can be effectively corrected and demonstrated classification maps visually consistent with the images (quantifiable accuracy improvement shown under the statistical analysis section).

| | (a) Subsection in test region 2 (2013/09) | (b) Subsection in test region 3 (2010/01) |
|---|---|---|
| Original image | 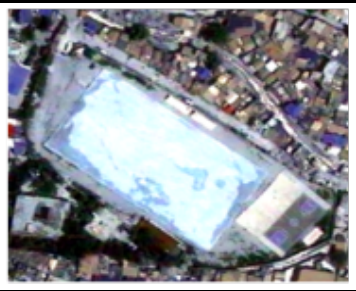 | 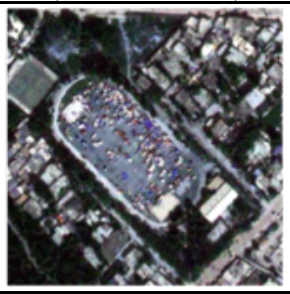 |
| Initial classification map | 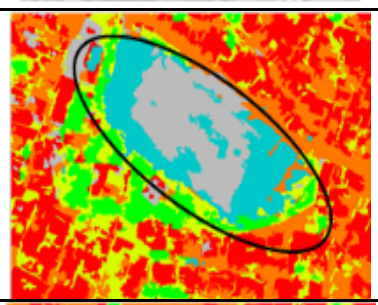 | 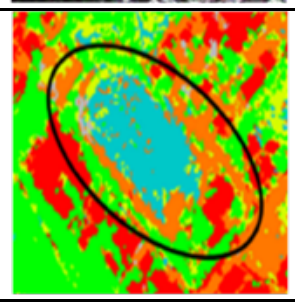 |
| Final classification map after 3D iterative spatiotemporal filtering | 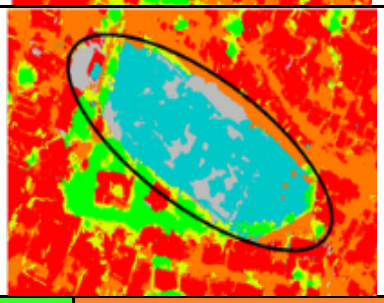 | 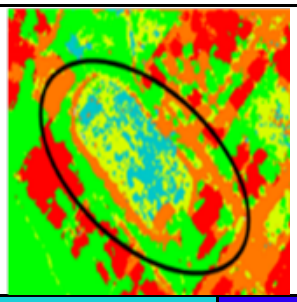 |

| Building | Ground | Tree | road | Grass | Temp | Long |
|---|---|---|---|---|---|---|

Figure 4.4. Two sample patches in test regions 2 and 3 illustrate the enhancement of classification after the proposed filter.

To demonstrate how the iterative process improves the classification results, we select a small patch from the date of 2014/07 in test region 1 (Figure 4.6) to analyze their classification map in each iteration of the process. The patch shows two buildings with close adjacency and its initial classification map (Figure 4.6(b)) shows the enlarged view where the buildings are partly misclassified as long-term lodgings. As can be seen from the figures, the iterative process gradually recovers the classification map whereas in the $11^{th}$ iteration (convergence), the final result (Figure 4.6(g)) shows a much more complete building segment

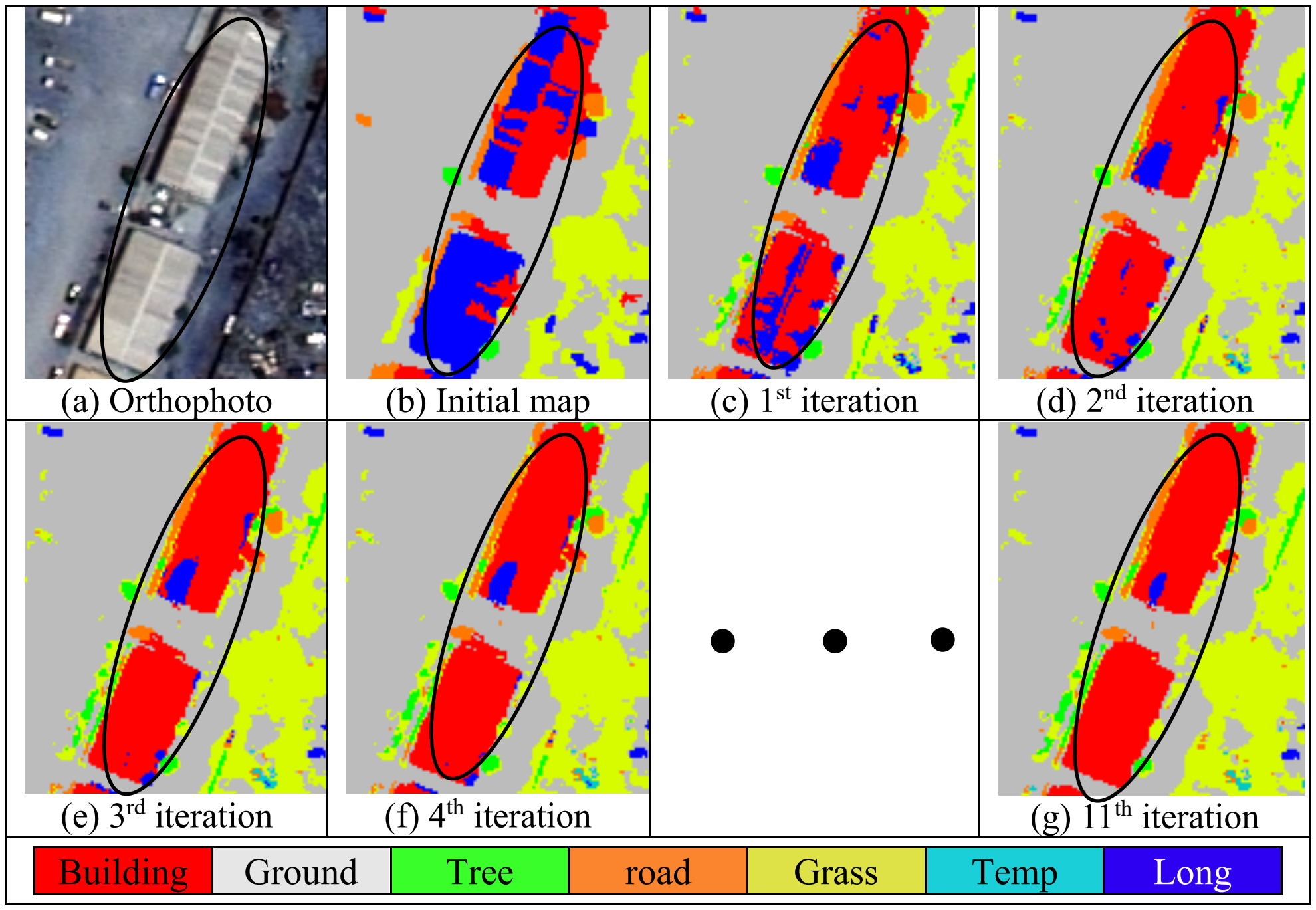

Figure 4.5. Detailed classification result for each iteration of enhancement for a patch in test region 1.

- *Statistical Analysis*

Table 4.5 shows the classification accuracies of each image in each test region before and after their probability maps being processed by the proposed method, where "Before" refers to the overall accuracy of initial classification results, and "After" indicates the enhanced accuracy through our proposed method, and "Δ" presents the accuracy change in the enhanced results over the initial ones. The average increase in the accuracy of the three test regions is 4.2363%, 5.285%, and 4.723%, respectively.

Table 4.5. Overall accuracy before and after probability enhancement for each dataset (Note: the bold numbers indicate the maximum increase in the accuracy)

| Date | Dataset 1 | | | Dataset 2 | | | Dataset 3 | | |
|---|---|---|---|---|---|---|---|---|---|
| | Before | After | Inc | Before | After | Inc | Before | After | Inc |
| 2007 | 91.04% | 95.21% | +4.17% | 83.47% | 88.14% | +4.67% | 91.12% | 95.85% | +4.73% |
| 2010/1 | 93.21% | 96.45% | +3.24% | 81.50% | 85.67% | +4.17% | 93.06% | 96.82% | +3.76% |
| 2010/6 | 91.93% | 96.26% | +4.33% | 83.52% | 89.79% | **+6.27%** | 88.82% | 94.87% | **+6.05%** |
| 2010/12 | 89.08% | 95.57% | **+6.49%** | 80.81% | 87.59% | **+6.78%** | 88.58% | 94.86% | **+6.28%** |
| 2012/3 | 92.19% | 95.92% | +3.73% | 81.43% | 86.92% | +5.49% | 91.44% | 97.08% | +5.64% |
| 2013/9 | 90.40% | 96.56% | **+6.16%** | 81.03% | 87.29% | **+6.26%** | 94.99% | 97.54% | +2.55% |
| 2014/7 | 95.11% | 97.27% | +2.16% | 82.19% | 88.90% | +6.17% | 90.39% | 96.58% | +6.19% |
| 2015 | 92.74% | 96.35% | +3.61% | 83.22% | 85.69% | +2.47% | 94.61% | 97.19% | +2.58% |
| Average | 92.09 | 96.20 | +4.24 | 82.15 | 87.50 | +5.29 | 91.63 | 96.35 | +4.72 |

In Table 4.6, we took a small patch of pixels in test region 2 to show probability changes with the number of iterations. The first patch is a part of a building area, and the second patch shows trees. In both patches, we notice that the probability of the correct class increases with iterations (see the arrows), and those of the remaining classes decrease.

Table 4.6. Mean probabilities (%) for each class changed through iterations using a sample patch of buildings and trees in test region 2.

Building patch

| Iter/Class | Buildings | Grass | Ground | Road | Temp | Tree |
|---|---|---|---|---|---|---|
| 1 | 73.16% | 1.68% | 17.69% | 11.53% | 0.92% | 0.24% |
| 2 | **74.95%↑** | 1.85% | 16.36% | 10.65% | 1.09% | 0.25% |
| 3 | **75.45%↑** | 1.96% | 15.87% | 10.28% | 1.19% | 0.25% |
| 4 | **75.53%↑** | 2.05% | 15.70% | 10.11% | 1.25% | 0.26% |
| 5 | 75.53% | 2.14% | 15.70% | 10.11% | 1.30% | 0.27% |
| 6 | 75.53% | 2.14% | 15.70% | 10.11% | 1.35% | 0.27% |
| 7 | 75.53% | 2.14% | 15.70% | 10.11% | 1.35% | 0.28% |

(b) Trees patch

| Iter/Class | Buildings | Grass | Ground | Road | Temp | Tree |
|---|---|---|---|---|---|---|
| 1 | 0.12% | 24.66% | 0.23% | 0.01% | 0.01% | 80.69% |
| 2 | 0.10% | 23.76% | 0.22% | 0.01% | 0.01% | **82.97%↑** |
| 3 | 0.09% | 23.18% | 0.22% | 0.01% | 0.01% | **84.03%↑** |
| 4 | 0.08% | 22.75% | 0.22% | 0.01% | 0.01% | **84.58%↑** |
| 5 | 0.08% | 22.41% | 0.22% | 0.01% | 0.01% | **84.86%↑** |
| 6 | 0.08% | 22.41% | 0.22% | 0.01% | 0.01% | **84.95%↑** |
| 7 | 0.08% | 22.41% | 0.22% | 0.01% | 0.01% | 84.91% |

Figure 4.6 presents the overall accuracy for the entire iterations and three experiments. We can that the first iteration plays a major role in the accuracy enhancement. In all three experiments, the first iteration increased about 4~6% of the OA and the rest of the iterations (until convergence) contribute approximately 1~2% of the accuracy enhancement. On average, we note that all experiments converge around the fifth iteration.

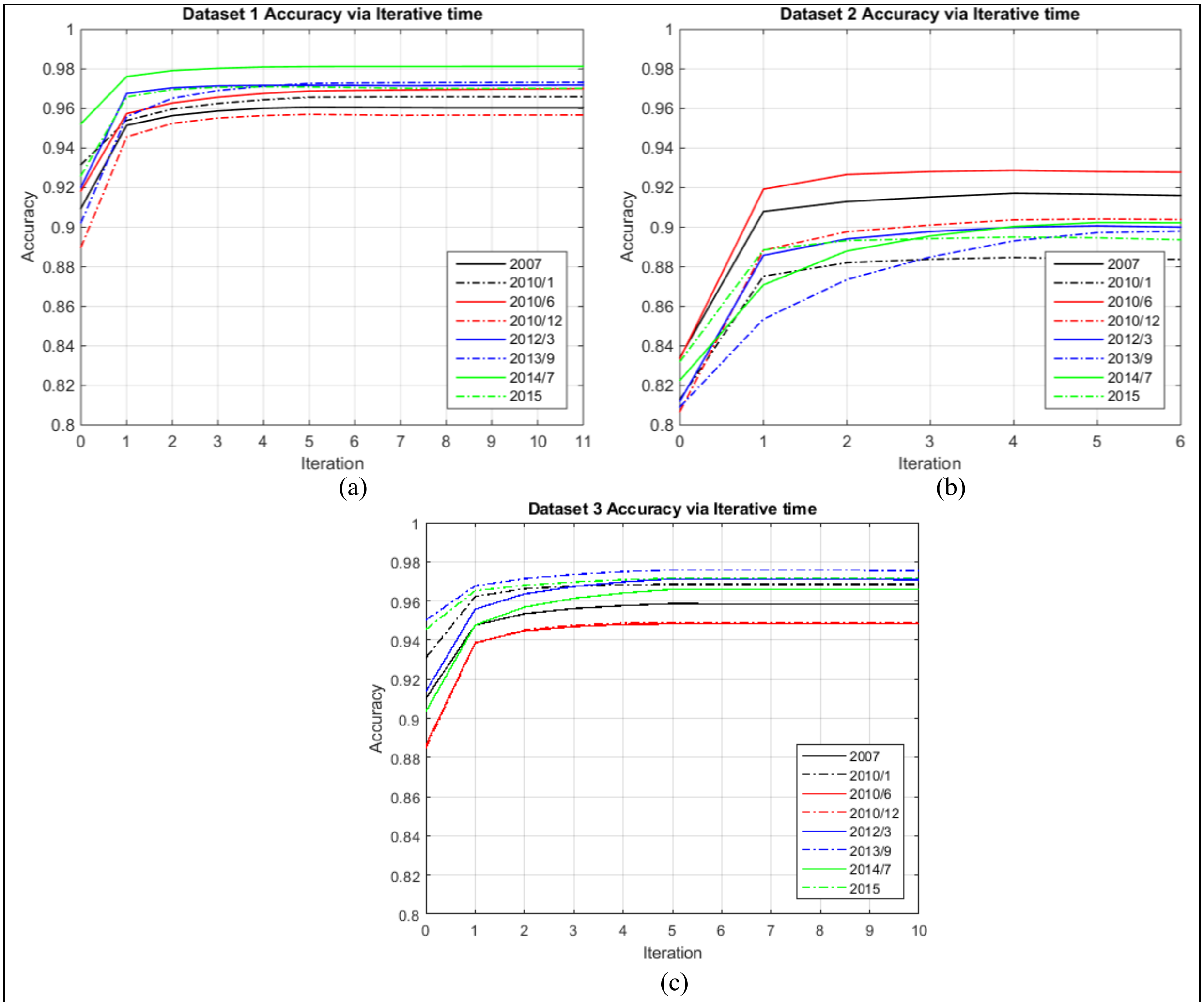

Figure 4.6. Accuracy-iteration curve of all datasets

## 4.6. Conclusion and Future Work

We have proposed a spatiotemporal inference method that improves the classification accuracy of multi-view and multi-temporal VHR satellite images. Knowing that temporal classification maps often suffer from inconsistencies in the results due to highly variable features that change over time, we propose using 3D geometric features (i.e. DSM generated from multi-view images) for better stability. We applied our proposed spatiotemporal inference method to process VHR imageries and DSMs to enhance the classification accuracy of each temporal dataset. Our proposed method approximates the global inference model through iterative processing of 3D bilateral filters utilizing spatial, spectral, and geometric information. Our method achieved a 2% to 6% increase in the overall accuracy for all experiments with varying scene complexities. Notable improvements were observed in certain scenarios, for example, for the open (sparse) areas in the first and third datasets, the accuracy increased from ≈89% to ≈97%, whereas for dense regions such as in test region 2, the accuracy increased from ≈80% to ≈90%. Since the proposed method is a post-processing method operating on probability maps, it is able to work on probability maps generated by different types of classifiers as long as the probability maps can be derived. In addition, the proposed method can be further generalized to less restrictive datasets by considering the temporal component to be the image themselves. For instance, it can be also applied to moderate and low-resolution satellite imageries and can be used to process video sequences or different types of time-series raster data to enhance the image data qualities.

# Chapter 5. Adaptive Semantic-guided Spatiotemporal Filtering to Enhance the Accuracy of Digital Surface Models (DSMs)

This chapter is based on the paper called "Enhancement of Depth Map by Fusion Using Adaptive and Semantic-Guided Spatiotemporal Filtering" that is published in the "ISPRS Annals of the Photogrammetry, Remote Sensing and Spatial Information Sciences, Volume V-3-2020, 2020 XXIV ISPRS Congress (2020 edition)" by (Albanwan & Qin, 2020).

## 5.1. Abstract

Digital surface models (DSMs) are essential components to perform the 3D model reconstruction. There are two ways to generate DSMs either 1) from LiDAR which is accurate highly expensive and often available to cover the entire surface of the earth or 2) by using multi-view stereo (MVS) methods that take a pair of stereo images as input and output disparity that is triangulated to the DSM. The MVS methods are efficient low-cost methods, however, the outputted DMSs are often noisy and imprecise, where the accuracy of the DSMs largely depends on the accuracy and consistency of the stereo pair images. DSM fusion is often proposed to combine the information from multiple DSMs and gain better accuracy and reliability. However, the fusion process itself is challenged by the varying uncertainties of the scene components, for example, trees are more impacted by seasons and can result in extreme errors, similarly, buildings are impacted by the visibility and ability of the algorithm to match pixels around edges and boundaries, thus, they are likely to have high uncertainty. To address this issue, we propose semantic adaptive spatiotemporal filtering that takes into account different classes during fusion. our method shows an improvement in the results of the fused DSM.

## 5.2. Introduction

### 5.2.1. Background

Over the last few decades, a large number of Very High Resolution (VHR) Satellites are established to provide sub-meter resolution imageries, with frequent re-visiting times during the year to allow extracting comprehensive 3D geometrical information about the scene. Algorithms such as Multi-view stereo (MVS) highly depend on the spatial and temporal resolutions of sensors to facilitate generating reliable 3D reconstructed models. However, dense stereo image matching algorithms often rely on the image pairs and the nature of the captured surface. Images captured by satellite sensors are prone to spectral inconsistencies and distortions, which may affect the dense stereo matching algorithm and produce incorrect or missing height information, and therefore, degrade the accuracy of the depth map. In particular, MVS algorithms are very sensitive to temporal inconsistencies between the images, thus, they cannot be used directly to obtain 3D models and generate height information like the Digital surface model (DSM). The acquisition conditions and measurement errors such as distance to sensor, the lighting conditions, occlusions in the scene (e.g. Tree obstructing a building), and not enough overlap between the images can also complicate unique feature matching (Qin, 2019). Object properties and their pattern in the scene can also increase the uncertainty of the generated height, such as thin structures, texture-less surfaces, featureless areas, and repeated patterns or structures directly affecting stereo matching. All these errors can cause the height data to be temporally inconsistent and lead to holes, noise, missing data, blurry artifacts, and fuzzy edges and boundaries, hence, incomplete and unreliable representation of the 3D information.

To enhance depth maps (i.e. height maps) generated from MVS algorithms, researchers suggest fusing several depth maps to utilize the redundant information in the temporal data. A common approach to fusing depth maps is simple median filtering (Kuschk, 2013; Matyunin et al., 2011; Ozcanli et al., 2015; Qin, 2017). The median filter is preferred in many works related to depth refinement and processing due to its simplicity, and ability to eliminate outliers while preserving the details. Other methods to process fusion include global approaches such as Markov Random Field (MRF) and total variation (TV) which optimizes the solutions (Zhu et al., 2010; Liu et al., 2015 (Kuschk & d'Angelo, 2013; Lasang et al., 2016); Kuschk, et al., 2017). However, they are mostly used to fuse depth maps resulting from RGB-D images from Kinect or video scenes, and despite the effectiveness of these algorithms, there are still some limitations that strict their usage. One limitation is the necessity to acquire a noise-free and clear set of images to improve stereo matching and the corresponding depth map, and unlike Kinect and video scenes where many images are captured indoors within a few milliseconds with consistent acquisition and lighting conditions, satellite images are more exposed to noise and outliers due to atmospheric conditions, seasonal variations, sun and satellite angles, image time, occlusions, etc. (Qin, 2019). Another issue is that current fusion algorithms are not adaptive to the scene objects. Urban features made of concrete and asphalt-like buildings and roads are time-invariant, which means they experience a low rate of change in the temporal depth map. Vegetation, on the other hand, is time-variant, where they tend to change frequently during the year due to atmospheric conditions and seasonal changes. The time variance is not the only factor that should be considered while designing the fusion algorithm; the characteristics of the object should also be considered. For example, narrow objects like roads and ground tend to be blocked by shadows, or trees, which increase their height uncertainty. The discrepancies in the height of objects in the depth maps produce a nonlinear type of change, which cannot be resolved directly using simple filters or fusion techniques that process all objects in the scene similarly. Therefore, in our work, we emphasize the importance of analysing the type of class to develop an adaptive spatiotemporal fusion that processes each pixel based on its class stability.

### 5.2.2. Related Works and Rationale

Generating accurate depth maps using very high-resolution multi-view satellite images to improve the 3D reconstruction model has been an ongoing research topic in the past few years. Multi-view Stereo (MVS) methods are widely used to extract elevations from multiple satellite views due to their efficiency and lower cost in comparison to direct methods like LiDAR. However, the depth map generated from MVS may include noise, outliers, and missing data due to the temporal and spectral inconsistencies between the images pair used to generate the depth map. In a dense image matching the number of matched points between the image pair can play a major role in depth accuracy, if no or few points can be matched between the image pair due to other factors such as object surface properties (e.g. smooth or texture-less surface, repeated patterns, etc.), it can produce inaccurate depth map.

One solution to recover the depth map generated from MVS algorithms is multi-view depth fusion, which has been explored by researchers in two contexts either global or local approaches. The global approaches mainly involve optimization and energy function to minimize the losses and sometimes include smoothing and regularization terms as additional constraints. Markov Random Field (MRF) is one example that has been widely used in depth fusion. For instance, (Jiejie Zhu et al., 2010; J. Liu et al., 2015) both used spatiotemporal MRF to fuse depth maps by achieving temporal coherence. Weighted total variation (TV) and total generalized variation (TGV) methods are also popular approaches to the global fusion of depths (Kuschk & d'Angelo, 2013; Lasang et al., 2016; Kuschk, et al., 2017). Neural networks are also effective fusion techniques that have been mostly used to fuse depth extracted from video scenes. Although global methods are useful and can provide accurate depth results, they are mostly applied to Kinect RGB-D sensors or video scenes, which in comparison to satellite images have an optimal indoor environment to capture numerous numbers of frames in a few milliseconds, therefore, providing lower distortions and better dense image matching. Additionally, background and foreground objects in the frames can be separated easily because of the fixed acquisition and lighting conditions, unlike satellite images where all objects are interrelated and difficult to extract directly. For depths generated from satellite images, local approaches are more popular, where fusion is performed mostly using local filtering. The most common fusion technique is the median filter because of its stability and robustness to outliers (Kuschk, 2013; Matyunin et al., 2011; Ozcanli et al., 2015; Qin, 2017). Other algorithms such as (Reinartz et al., 2005) used average filtering to fuse DSMs obtained from stereo techniques using SPOT-5 and radar data obtained from SRTM, but because average filtering techniques smooth high-frequency data it tends to discard high levels of details and generate outliers. Recent methods include median clustering which has been proposed in (Facciolo et al., 2017; Rumpler et al., 2013), where clustering is considered an effective method to assess the temporal homogeneity of height data by measuring the inter- and intra-class similarity and dissimilarity. Local strategies are efficient in terms of time and robust to small outliers but are not able to solve the problems of extensive nonlinear noise and object boundaries.

Moreover, the current filtering methods used in fusion often ignore objects class and process the image using fixed parameters. Since objects in the scene and nature have different responses and reflection properties in the captured satellite image, it is important to assess the elevation uncertainty for each class and incorporate it into the fusion algorithm.

### 5.2.3. Proposed Method and Rationale

The spatiotemporal analysis is an effective way to solve problems related to data consistency, noise, missing data, outliers, etc. Using redundant data, we can fuse the depth maps to result in a reliable and accurate single depth map. Fusing depth maps using an adaptive spatiotemporal algorithm is an ongoing research topic that very few researchers have investigated this area of research (Qin, 2017a). Therefore, in our work, we aim to investigate and analyse the role of class to improve the multi-depth fusion algorithm that is adaptive to scene elements, efficient, robust, and able to recover the gaps mentioned in the literature.

This paper is organized in the following order, in Section 5.2, we will mention the methodology and analysis, where we will discuss the dataset used, the pre-processing methods, and the analysis that supports our proposed work. Section 5.3 includes the experimental results, with reasonable explanations and validations. Finally, the conclusion and future works are in Section 5.4.

## 5.3. Methodology and Analysis

### 5.3.1. Data Description and Preprocessing

In our experiment, we chose 3 different datasets with varying complexities to examine and fuse; dataset I, is the area of a commercial building, dataset II is a more open area with natural objects such as vegetation and a water surface, and dataset III is a condensed housing area (for more details see Figure 5.1).

We follow the same pre-processing method for all datasets, wherein for each dataset we use VHR image pairs from the Worldview 3 satellite to generate the corresponding multispectral orthophoto and the temporal DSMs. We use RSP (RPC Stereo Processor) software developed by (Qin, 2016) that performs hierarchical semi-global matching (SGM) algorithm (Hirschmuller, 2008a) to generate and register the Orthophoto and the DSMs. We then generate the mask for each class using the 8-band multispectral orthophoto. We categorize the classes into trees, grass, buildings, and a combined category for ground/road for all datasets. We use indices such as the Normalized difference vegetation index (NDVI) along with normalized DSM (nDSM) generated using top har reconstruction (Qin & Fang, 2014) to extract the masks. For instance, the NDVI helps to determine the locations of trees and grass, and with the appropriate nDSM we can find the position of trees based on their heights, and determine the location of buildings, ground, and roads accordingly. For more details on the pre-processing steps, see the diagram in Figure 5.2.

| Dataset | 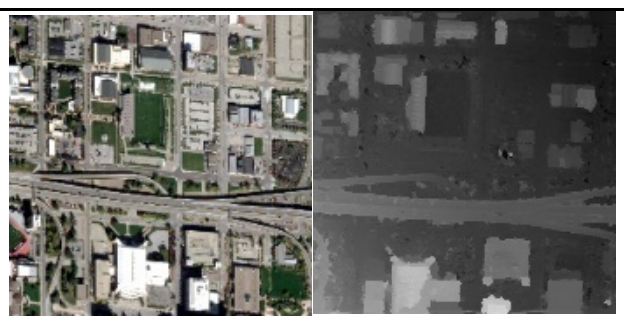 I | 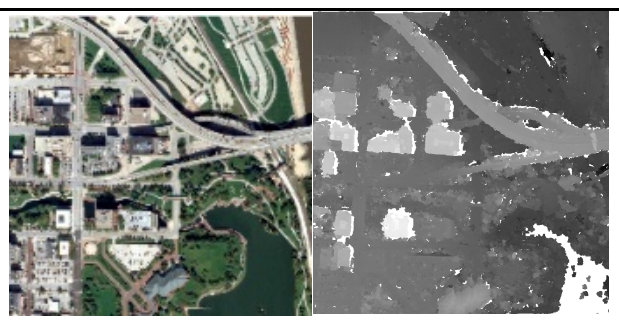 II | 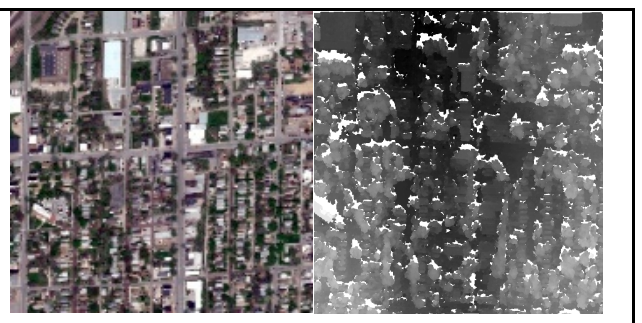 III |
|---|---|---|---|
| Number of stereo pair images | 400 | 51 | 51 |
| DSM generated from good pairs | 219 | 37 | 26 |
| DSM dimensions | 1089 × 1113 | 1080 × 1109 | 1043 × 949 |

Figure 5.1. Details on the dataset.

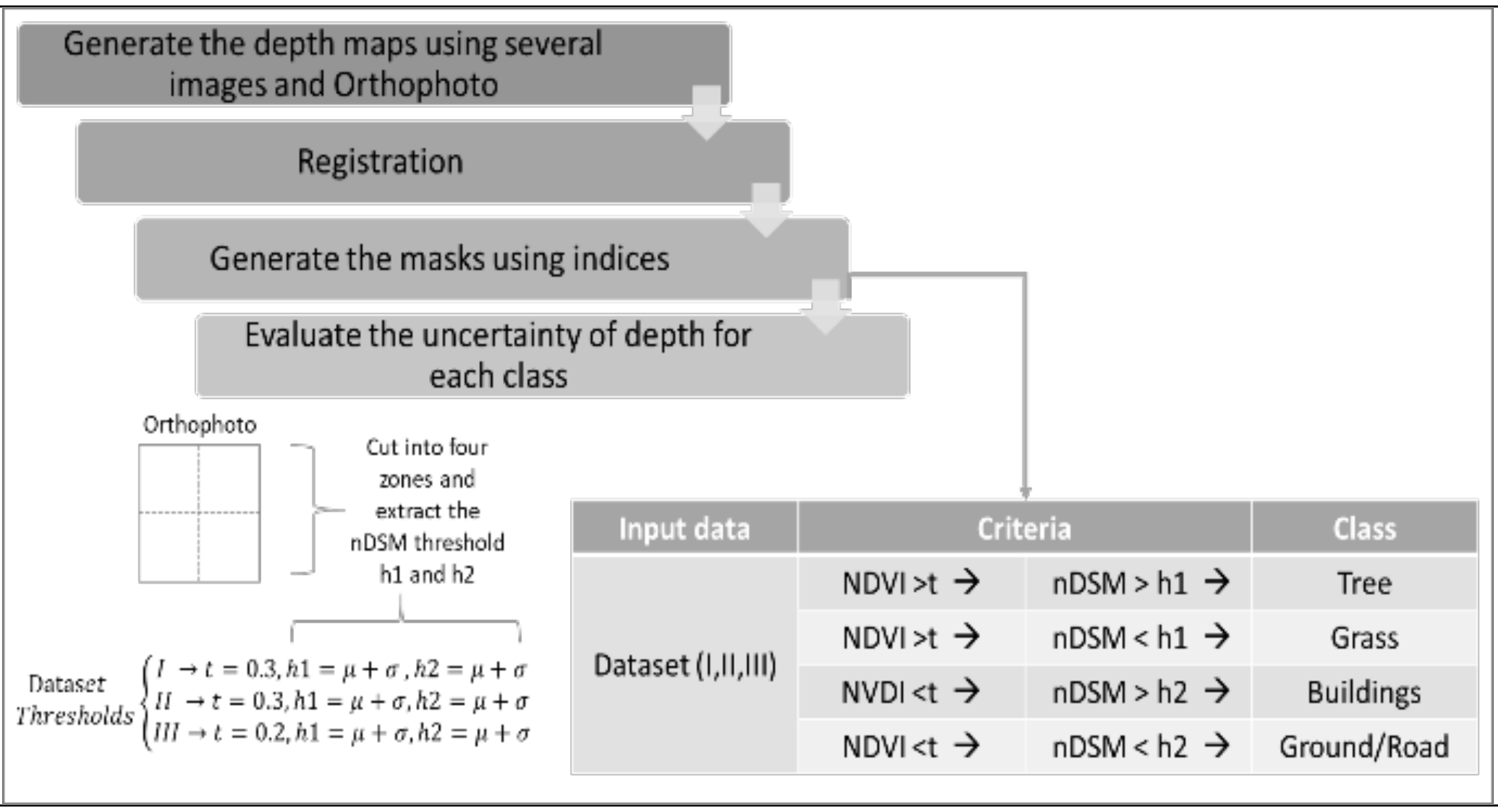

Figure 5.2. Preprocessing steps. Note: h1, h2, and t are empirically determined.

### 5.3.2. Data Analysis

In our data analysis, we aim to investigate the stability and uncertainty of the height of objects and the way it varies according to the class. We categorize the objects in the scene into either:

1) Time-invariant (e.g. buildings, ground/roads.) or time-variant (e.g. vegetation) objects.
2) Based on its surface properties (e.g. smooth texture-less, flat, etc.).

To assess the height of classes, we perform per-pixel processing by taking the standard deviation in the temporal DSMs of the DSMs for each class. We then summarize the analysis results by viewing the histogram of standard deviations to be able to see the distribution of every class and its average standard deviation (see Figure 5.3). From Figure 5.3, we can see that each class has different standard deviation distribution. Most classes follow a normal distribution, which can give us a clue about which type of probability to use in the weight measurement of our adaptive spatiotemporal fusion. The average standard deviation in Table 5.1 tells us the average uncertainty of height for each class. We can notice from Table 5.1 that for all three datasets vegetation (including trees and grass) has higher uncertainty than other classes, which means that height in the temporal DSM varies more than other objects. The buildings class comes in second to have higher height uncertainty. Among all classes, ground/ road appears to have better and higher elevation stability, whereas in all datasets it had the lowest values of uncertainty.

Table 5.1. The standard deviation of the uncertainty per class (meters)

| Class | Dataset I | Dataset II | Dataset III |
|---|---|---|---|
| Building | 4.0968 | 8.8098 | 7.4692 |
| Ground/ road | 3.8197 | 8.5997 | 7.3950 |
| Tree | **4.1097** | **8.9717** | **8.8059** |
| Grass | **4.6147** | **9.0144** | **9.6021** |

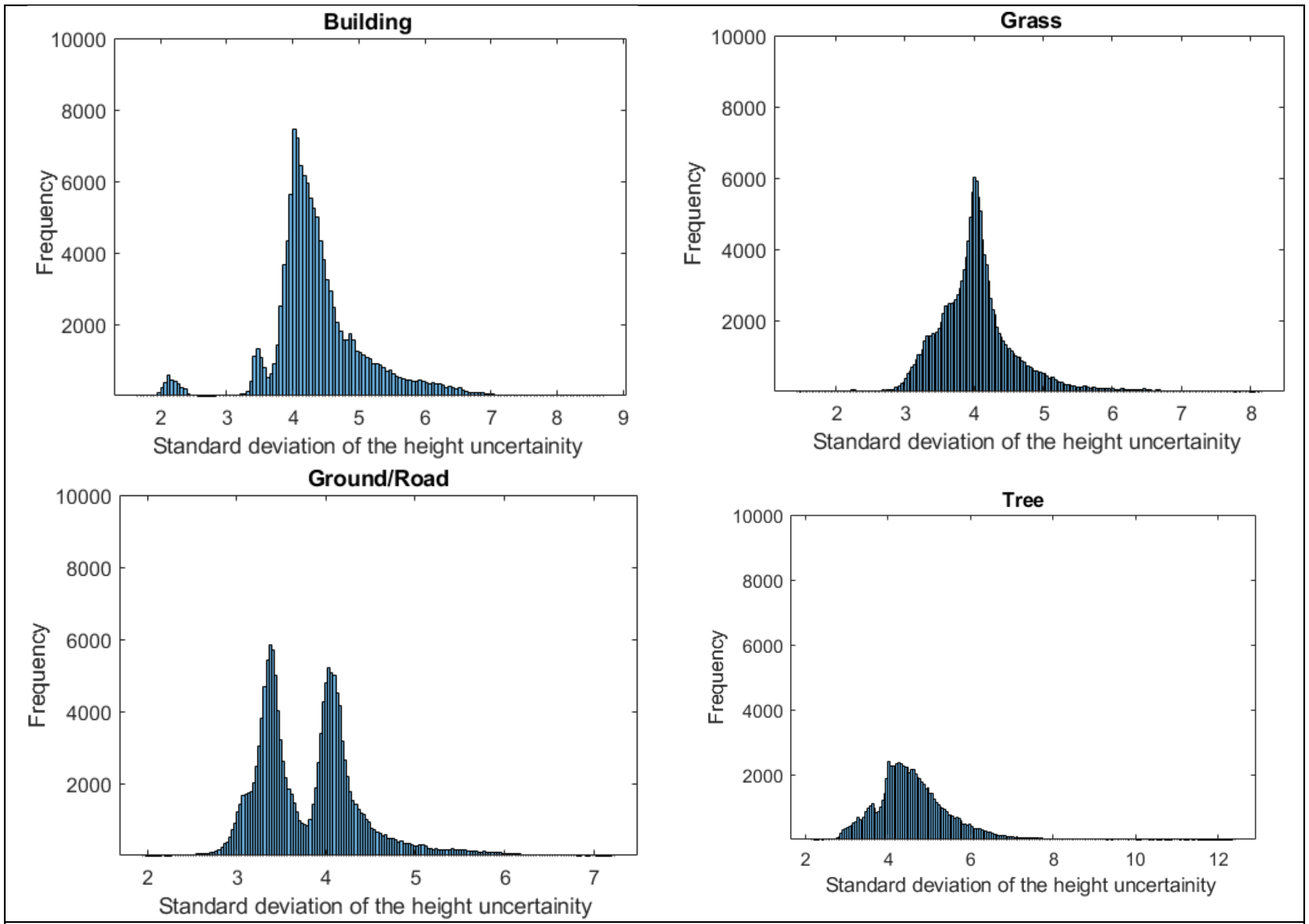

Figure 5.3. The distribution of the standard deviations of temporal heights in dataset I. The x-axis shows the standard deviation.

### 5.3.3. The Proposed Spatiotemporal Fusion Algorithm

Our fusion method is based on the spatiotemporal filter; the generic formula for the fusion process is as follows:

$$DSM_f(i,j) = \frac{1}{W_T} * \sum_{i=1}^{Width} \sum_{j=1}^{Height} W_r * W_s * W_h * h(i,j,t)_{med} \tag{5.1}$$

Where the $DSM_f$ is the fused pixel at position (i, j), the $h_{med}$ is the median height measure from all input DSMs, we use median filtering as an initial estimate because of its robustness to noise and ability to maintain boundaries and sharp edges. The computed weights are the spectral $W_r$, spatial $W_s$, and temporal geometric weights $W_h$. The spatial and spectral weights are used to measure similarities from the neighboring pixels. For the spectral weight, we use the RGB image provided by the orthophoto to further regularize and smooth the fused DSM. We compute the spatial and spectral weights ($W_r, W_s$) as follows:

$$W_r(i,j) = exp^{\frac{-||(I(i,j)-I(k,l)||^2}{2\,\sigma_r^2}} \tag{5.2}$$

$$W_s(i,j) = exp^{\frac{-((i-l)^2+(i-k)^2)}{2\,\sigma_s^2}} \tag{5.3}$$

Where I is the orthophoto RGB images, (i,j) is the current central pixel under process, (l,k) are the position of neighboring pixels in the window. The spectral and spatial bandwidths that determine the degree of filtering are $\sigma_r$ and $\sigma_s$ respectively. For the computation of the geometric height weight, we use Euclidean distance to measure the difference between the median value and the elevation at date t.

$$W_h(i,j) = exp^{\frac{-||hmed-h(i,j,t)||^2}{2\,\sigma_h^2}} \tag{5.4}$$

Where $\sigma_h$ is the height bandwidth, and it is determined empirically for each type of class. We use the masks that are created in the preprocessing steps as an indication of the location of each class as follows:

$$\sigma_h = \begin{cases} \sigma_{Building} \rightarrow if\ pixel\ (i,j)\ is\ building \\ \sigma_{Ground/road} \rightarrow if\ pixel\ (i,j)\ is\ ground/road \\ \sigma_{tree} \rightarrow if\ pixel\ (i,j)\ is\ tree \\ \sigma_{grass} \rightarrow if\ pixel\ (i,j)\ is\ grass \\ \sigma_{water} \rightarrow if\ pixel\ (i,j)\ is\ water \end{cases} \tag{5.5}$$

## 5.4. Experimental Results

The results of our proposed spatiotemporal fusion are shown and discussed in this section. We will specify the parameters used and the outcomes for the three datasets. We validate our results per pixel, and against other existing methods.

### 5.4.1 Parameters

The two main parameters that we require in this work are the window size and bandwidths (spectral, spatial, and height), the choice of these parameters is made empirically. The window size is set and fixed to a moderate value of 7. Similarly, the spatial and spectral bandwidths ($\sigma_r$, $\sigma_s$) are set to (11, 50) (Tomasi and Manduchi, 1998). Validating the adaptive spatiotemporal fusion concept requires examining the bandwidth under different scenarios. For example, for objects with high uncertainty such as the time-variant objects like vegetation, we might need high $\sigma_h$. For flat, narrow, and featureless objects like ground and road, we might also need high $\sigma_h$ . We also want to compare our adaptive approach to the fixed value of $\sigma_h$. We also chose the height bandwidth ($\sigma_h$) empirically, but in a manner that allows validating the assumption and the analysis. Therefore, we chose $\sigma_h$ based on several criteria indicated in Table 5.2.

Table 5.2. The category list for the height bandwidth $\sigma_h$

| $\boldsymbol{\sigma_h}$ | **Notes** |
|---|---|
| High for time-invariant objects | Give more emphasis to urban structures as buildings and ground |
| High for time-variant and flat texture-less objects | Give more emphasis to vegetation and ground/road |
| High for flat texture-less objects | Give more emphasis to urban ground/road |
| Fixed for all classes | |

### 5.4.2 Results and Discussion

The fused DSM for all datasets is shown in Figure 5.4, where we can see that most of the distortions such as noise, missing datasets, and holes are filled and taken into consideration. We can notice that the adaptive spatiotemporal fusion algorithm produces good results visually, in comparison to other methods. For instance, if we compared it to the simple median filter, we can see that median filter fusion results can produce overly smoothed outcomes, in addition to blurring some details. The adaptive median, on the other hand, is better in terms of capturing the details of the buildings since they use an adaptive window for buildings, but smaller details in the ground are also overly smoothed. The C-median clustering can generate fuzzy and partially noisy results as in dataset II and III.

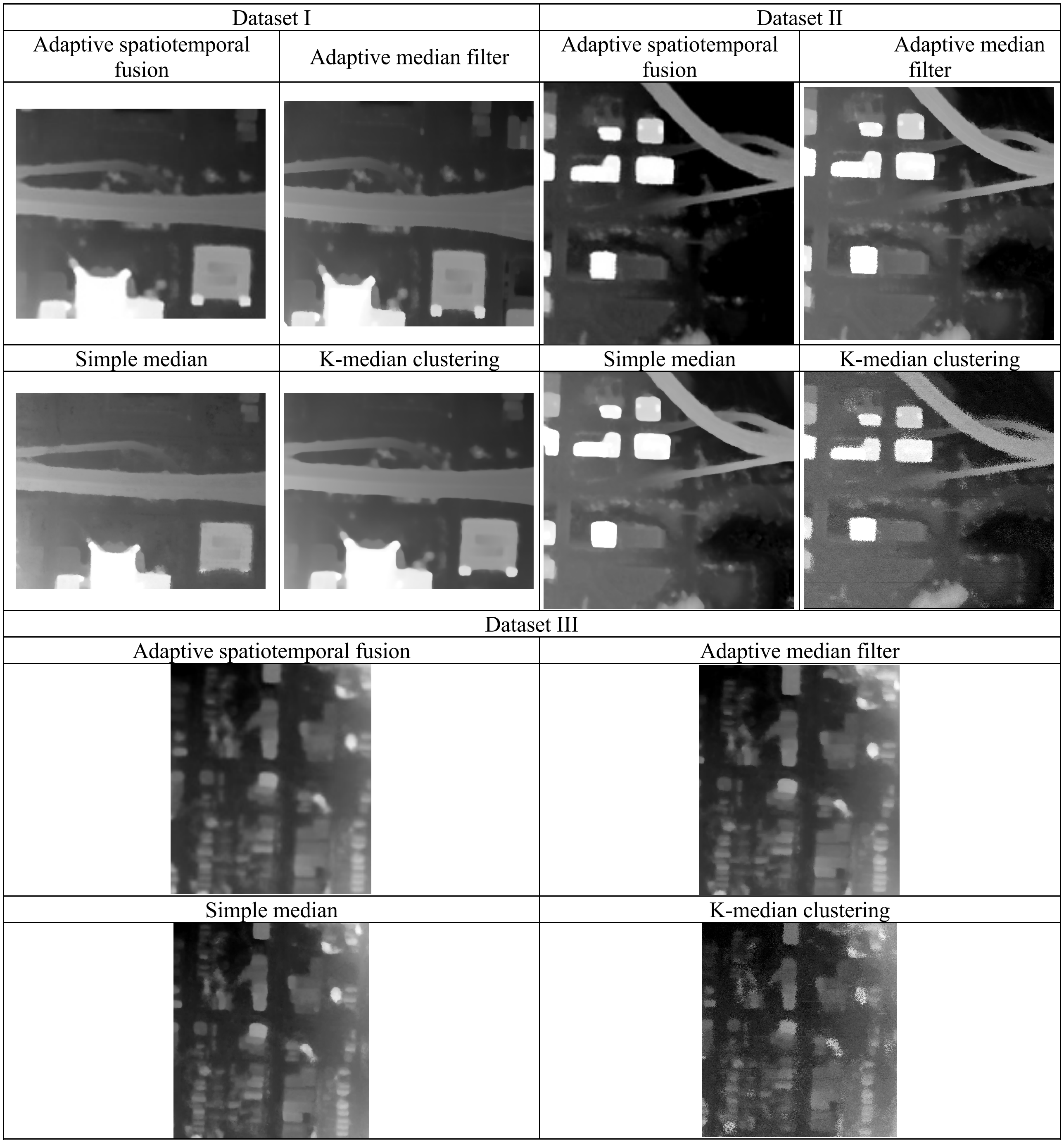

Figure 5.4. The DSMs fused results from all datasets using the adaptive spatiotemporal fusion, the adaptive median filter, the simple median filter, and C-median clustering.

We evaluate the results of the adaptive spatiotemporal fusion statistically by showing a comparison between different values of $\sigma_h$ for all classes, and a comparison between other existing approaches. We compared the results of the proposed adaptive spatiotemporal fusion to the ground truth height data (from LiDAR) and measured the accuracy of each dataset to a 6-meter level of difference. The results are shown in Table 5.3.

Table 5.3. Accuracy assessment of all datasets.

| **Dataset I** | | | | | | | | |
|---|---|---|---|---|---|---|---|---|
| **sigma h** | | | | **Overall Accuracy (OA %)** | **Accuracy per class (%)** | | | |
| **Tree** | **Grass** | **Building** | **Ground** | | **Tree** | **Grass** | **Buildings** | **Ground/Road** |
| 11 | 11 | 17 | 13 | 95.400 | 96.038 | 94.325 | 92.727 | 95.683 |
| 13 | 13 | 15 | 12 | 95.407 | 96.083 | 94.366 | 92.745 | 95.674 |
| 15 | 18 | 15 | 14 | 95.481 | 96.168 | 94.369 | 92.817 | 95.760 |
| 15 | 15 | 13 | 14 | 95.493 | 96.168 | 94.369 | 92.781 | 95.782 |
| 23 | 27 | 20 | 23 | 95.597 | 96.140 | 94.369 | 92.907 | 95.929 |
| 30 | 33 | 23 | 27 | 95.615 | 96.111 | 94.369 | 92.933 | 95.956 |
| 18 | 18 | 15 | 17 | 95.542 | 96.140 | 94.369 | 92.880 | 95.846 |
| 23 | 28 | 33 | 30 | 95.589 | 96.140 | 94.369 | 92.898 | 95.925 |
| 15 | 15 | 15 | 15 | 95.504 | 96.173 | 94.369 | 92.817 | 95.795 |

| **Dataset II** | | | | | | | | |
|---|---|---|---|---|---|---|---|---|
| **sigma h** | | | | **Overall Accuracy (OA %)** | **Accuracy per class (%)** | | | |
| **Tree** | **Grass** | **Building** | **Ground** | | **Tree** | **Grass** | **Buildings** | **Ground/Road** |
| 13 | 13 | 17 | 12 | 99.035 | 99.925 | 99.525 | 96.212 | 99.221 |
| 23 | 28 | 33 | 30 | 99.002 | 99.911 | 99.686 | 96.443 | 99.418 |
| 13 | 13 | 17 | 17 | 98.970 | 99.930 | 99.290 | 95.870 | 99.430 |
| 15 | 15 | 11 | 13 | 99.092 | 99.965 | 99.669 | 95.564 | 99.428 |
| 18 | 18 | 15 | 17 | 99.054 | 99.960 | 99.675 | 95.701 | 99.464 |
| 37 | 37 | 35 | 35 | 98.993 | 99.938 | 99.726 | 96.250 | 99.440 |
| 18 | 18 | 17 | 23 | 98.960 | 99.960 | 99.510 | 95.830 | 99.470 |
| 17 | 33 | 35 | 27 | 98.989 | 99.831 | 99.741 | 96.502 | 99.338 |
| 15 | 15 | 15 | 15 | 99.067 | 99.960 | 99.588 | 95.786 | 99.438 |

| **Dataset III** | | | | | | | | |
|---|---|---|---|---|---|---|---|---|
| **sigma h** | | | | **Overall Accuracy (OA %)** | **Accuracy per class (%)** | | | |
| **Tree** | **Grass** | **Building** | **Ground** | | **Tree** | **Grass** | **Buildings** | **Ground/Road** |
| 11 | 11 | 17 | 13 | 99.991 | 99.956 | 99.895 | 99.987 | 99.997 |
| 13 | 13 | 15 | 11 | 99.993 | 99.985 | 99.956 | 99.987 | 99.996 |
| 23 | 28 | 33 | 30 | 99.994 | 99.989 | 99.974 | 99.987 | 99.995 |
| 15 | 15 | 13 | 14 | 99.994 | 99.989 | 99.956 | 99.985 | 99.996 |
| 18 | 18 | 15 | 17 | 99.995 | 99.989 | 99.965 | 99.987 | 99.997 |
| 21 | 27 | 20 | 23 | 99.996 | 99.989 | 99.974 | 99.990 | 99.997 |
| 15 | 18 | 15 | 21 | 99.991 | 99.948 | 99.878 | 99.987 | 99.998 |
| 30 | 30 | 25 | 35 | 99.9962 | 99.993 | 99.983 | 99.990 | 99.998 |
| 15 | 15 | 15 | 15 | 99.994 | 99.989 | 99.956 | 99.985 | 99.996 |

From Table 5.3, we note that the highest overall accuracies (95.625%, 99.093%, and 99.996%) for all datasets are located in the second category (explained in Table 5.2), where greater values of weight $W_h$ are given to objects with high elevation uncertainty. We also note that the adaptive approach provides slightly better results (0.02% higher) than the fixed bandwidth parameter (see last rows in table 3). The second part of the table explains how different classes correspond to different sigma values. We also show this in Figure 5.5. For instance, we can notice from Figure 5.7 that trees and grass achieve the highest accuracies at large values of $\sigma_h$ (30 and 35), while buildings require lower $\sigma_h$ to achieve high accuracy at $\sigma_h$ = 20 to 25. In all three datasets, the grass always requires a larger $\sigma_h$, whereas urban objects such as buildings and ground can achieve high accuracies at moderate values of $\sigma_h \approx$ = 20 to 25. This can lead us to conclude that the height of time-variant objects is less certain and requires larger compensation using higher $\sigma_h$. We can also conclude that fixed bandwidth parameters do not achieve optimal results; this can be seen in Figures 5, 6, and 7, and confirmed by the overall accuracy in table 3 at a fixed value of $\sigma_h$=15. We can use the information in these figures, to determine the optimal sigma values for each class and use it to get the optimal fused depth map. However, we can see that the optimal height bandwidth patterns between the classes differ with the dataset depending on the complexity and objects in the scene. For instance, for areas with few trees and many large commercial buildings as in dataset I, trees and grass required the least $\sigma_h$ of values 13 and 18 respectively, while dominant objects like buildings and ground required larger $\sigma_h$ of values 23 and 27 respectively. On the contrary, datasets II and III had trees and grass as dominant objects, thus, they require larger $\sigma_h$ than the other classes.

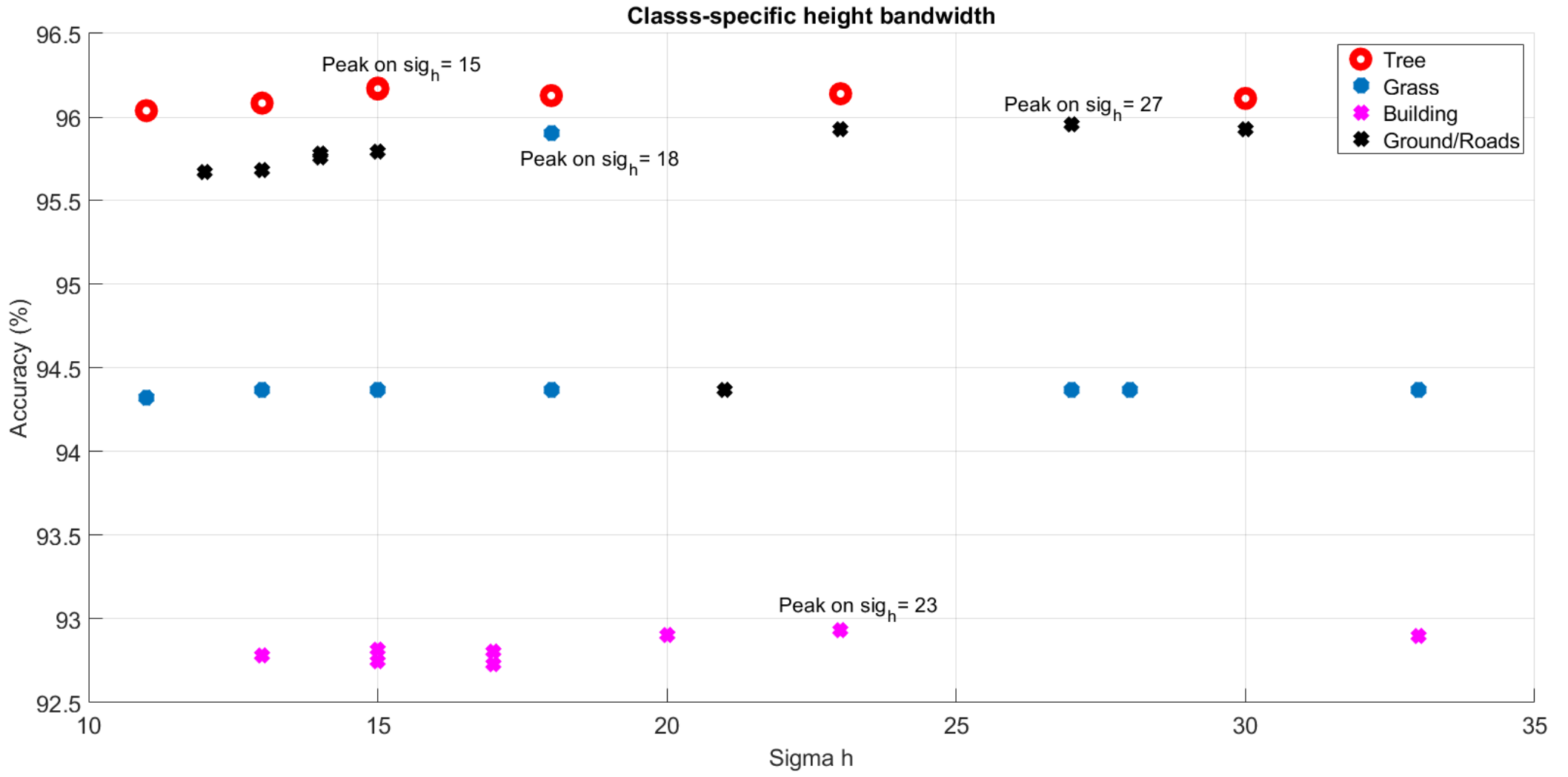

(a) The Accuracy of each class in dataset I according to the height bandwidth.

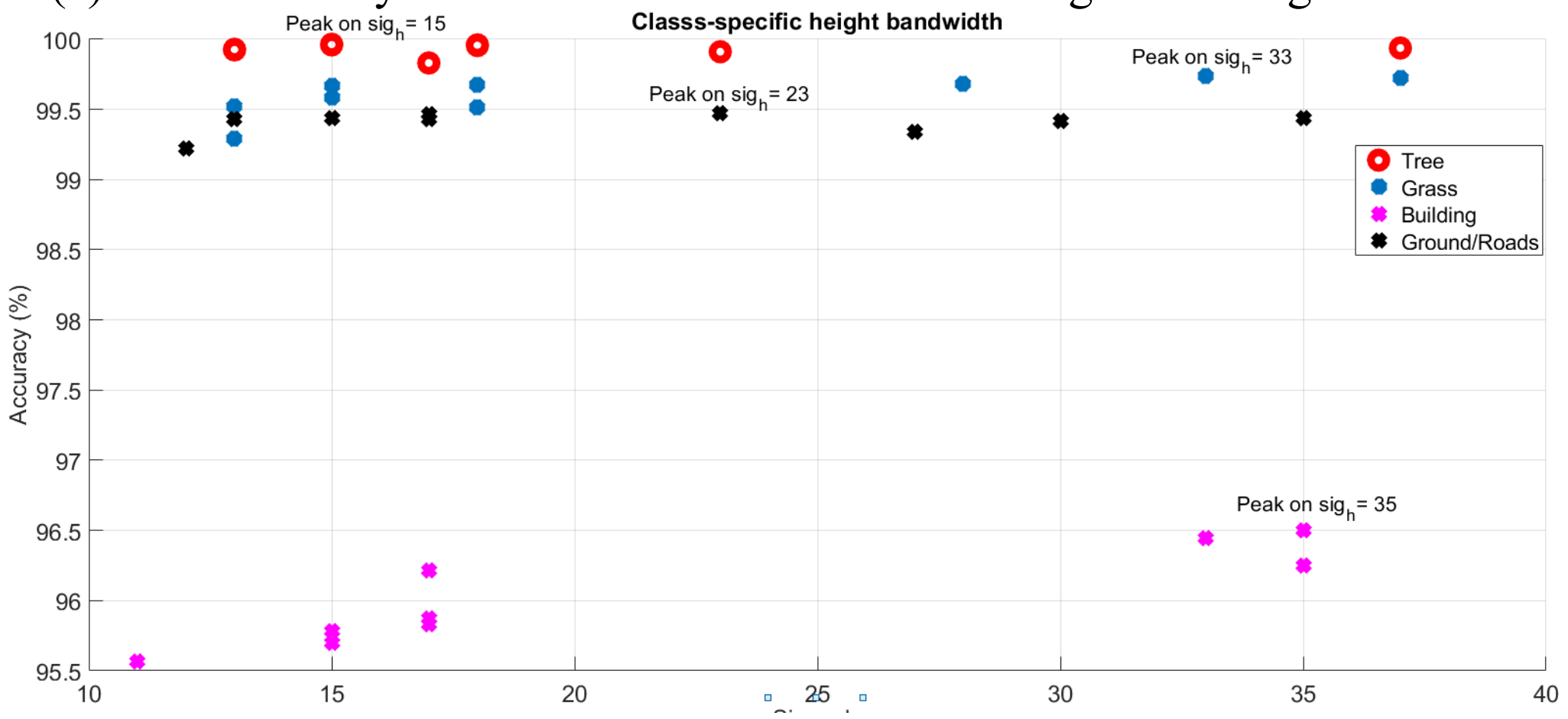

(b) The Accuracy of each class in dataset II according to the height bandwidth.

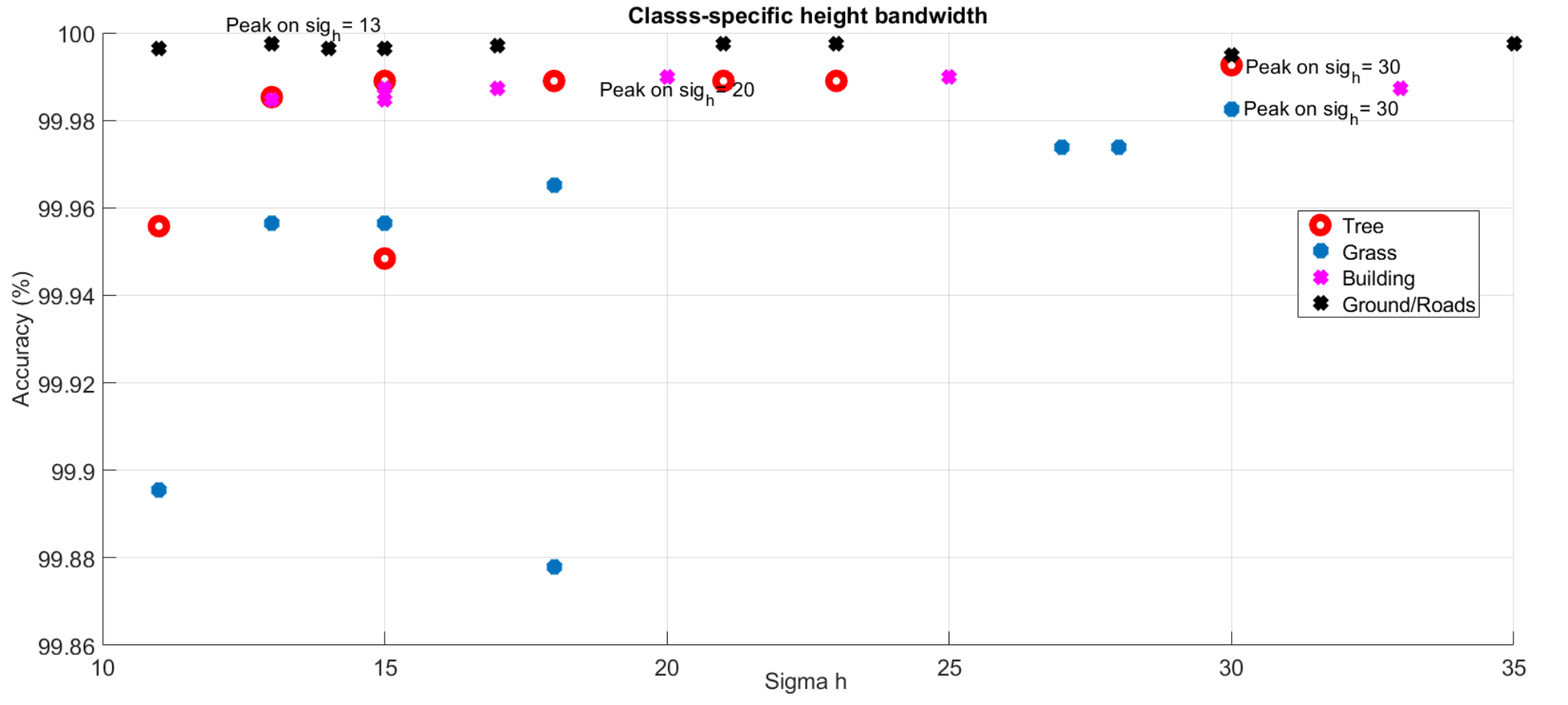

(c) The Accuracy of each class in dataset III according to the height bandwidth.

Figure 5.5. The accuracy analysis for each specific class after the proposed adaptive semantic DSM fusion algorithm.

We also use existing methods such as simple median filter, adaptive median filter by (Qin et al., 2017), and K-median clustering by (Facciolo et al., 2017) to evaluate our method (see Table 5.4). We find that our adaptive method provides marginally higher accuracy with an increased range between almost 0.01-2%. Similarly, the accuracy of the majority of classes has higher accuracy in the adaptive case than the other methods with fixed bandwidths.

Table 5.4. Accuracy comparison between the adaptive spatiotemporal fusion and other fusion methods.

| **Dataset I** | **Overall accuracy (OA) (%)** | **Accuracy per class (%)** | | | |
|---|---|---|---|---|---|
| | | **Tree** | **Grass** | **Buildings** | **Ground** |
| Adaptive median | 95.469 | 95.937 | 94.371 | 92.933 | 95.763 |
| Simple median | 95.447 | 95.949 | 94.369 | 92.960 | 95.709 |
| K-median clustering | 94.695 | 92.599 | 94.305 | 92.835 | 95.028 |
| Spatiotemporal fusion | 95.615 | 96.111 | 94.369 | 92.933 | 95.956 |

| **Dataset II** | **Overall accuracy (OA) (%)** | **Accuracy per class (%)** | | | |
|---|---|---|---|---|---|
| | | **Tree** | **Grass** | **Buildings** | **Ground** |
| Adaptive median | 98.749 | 99.916 | 99.176 | 95.634 | 98.967 |
| Simple median | 98.643 | 99.228 | 98.644 | 89.097 | 94.894 |
| K-median clustering | 95.960 | 96.802 | 97.026 | 94.484 | 95.687 |
| Spatiotemporal fusion | 99.092 | 99.965 | 99.669 | 95.564 | 99.428 |

| **Dataset III** | **Overall accuracy (OA) (%)** | **Accuracy per class (%)** | | | |
|---|---|---|---|---|---|
| | | **Tree** | **Grass** | **Buildings** | **Ground** |
| Adaptive median | 99.680 | 99.322 | 98.675 | 99.169 | 99.871 |
| Simple median | 99.987 | 93.210 | 84.260 | 98.715 | 99.562 |
| K-median clustering | 98.196 | 93.129 | 87.249 | 99.090 | 98.633 |
| Spatiotemporal fusion | 99.996 | 99.993 | 99.983 | 99.990 | 99.997 |

### 5.5. Conclusion and Future Works

In our work, we show that the adaptive spatiotemporal fusion technique can provide a better solution for objects with a high level of elevation uncertainty. The overall accuracy in all three datasets showed that optimal results could be achieved using a class-adaptive approach rather than the fixed parameter. Our analysis also shows that for classes with a high level of uncertainty like vegetation, more emphasis should be given by adjusting their height bandwidths to larger values. We also compare our results to existing work and found that it achieved slightly better overall accuracy ranging from 0.01 to 2%. In the next step, we will extend this work to determine the value of the height bandwidth automatically based on the scene information-using machine learning (ML) methods. We also would like to obtain the label image more efficiently using better nonparametric classification methods with indices such as NDVI, Morphology index, etc. to extract varying objects, their class, and the corresponding classification map.

# Chapter 6. A Comparison Between Adaptive and Nonadaptive Digital Surface Models (DSMs) Fusion Methods

This chapter is based on the work "Adaptive and Non-adaptive Fusion Algorithms Analysis for Depth Maps Generated Using Census and Convolutional Neural networks (MC-CNN)" that is published in the "The International Archives of the Photogrammetry, Remote Sensing and Spatial Information Sciences, Volume XLIII-B2-2021 XXIV ISPRS Congress (2021 edition" by (Albanwan & Qin, 2021a).

## 6.1. Abstract:

The digital surface models (DSM) fusion algorithms are one of the ongoing challenging problems to enhance the 3D models, especially for complex regions with variable radiometric and geometric distortions like satellite datasets. DSM generation using Multiview stereo analysis (MVS) is the most common cost-efficient approach to recover elevations. Algorithms like Census-semi global matching (SGM) and Convolutional Neural Networks (MC-CNN) have been successfully implemented to generate the disparity and recover DSMs; however, their performances are limited when matching stereo pair images with ill-posed regions, low texture, dense texture, occluded, or noisy, which can yield missing or incorrect elevation values, in additions to fuzzy boundaries. DSM fusion algorithms have proven to tackle such problems, but their performance may vary based on the input and the type of fusion which can be classified into adaptive and non-adaptive. In this paper, we evaluate the performance of the adaptive and nonadaptive fusion methods using median filter, adaptive median filter, K-median clustering fusion, weighted average fusion, and adaptive spatiotemporal fusion for DSM generated using Census and MC-CNN. We perform our evaluation on nine testing regions using stereo pair images from the Worldview-3 satellite to generate DSMs using Census and MC-CNN. Our results show that adaptive fusion algorithms are more accurate than non-adaptive algorithms in predicting elevations due to their ability to learn from temporal and contextual information. Our results also show that MC-CNN produces better fusion results with a lower overall average RMSE than Census.

## 6.2. Introduction

The accuracy of the digital surface model (DSM) generated from satellite images has always been a crucial element in most remote sensing and photogrammetry applications. DSM generated using Multiview stereo (MVS) algorithms is very common due to its high efficiency and low cost, but it is limited performance due to its sensitivity to radiometric and geometric distortions in the stereo images, which lead to noise, occlusions, missing elevation values, etc. in the DSM. One of the most promising techniques that raised significant attention to improving the accuracy of the DSM is fusion (Albanwan & Qin, 2020; Cigla et al., 2017; Papasaika et al., 2011). Fusion is the process of combining multi-temporal DSMs into a single accurate DSM; it takes advantage of the redundant temporal information to compensate for incorrect representations or missing elevation points (Albanwan and Qin, 2020; Cigla et al., 2017; Papasaika et al., 2008). DSM fusion algorithms can be categorized into 1) adaptive and 2) nonadaptive; the prior approach learns from the context, shape, and type of objects in the scene, in addition to the temporal information between DSMs, whereas non-adaptive approaches simply learn and predict elevation from the temporal information (Cigla et al., 2017; N. Wang & Gong, 2019). One of the oldest non-adaptive algorithms to perform fusion is median filtering, it is known to be robust to outliers and preserve the boundaries of the objects (Kuschk, 2013; Ozcanli et al., 2015). Many fusion algorithms have upgraded simple median filter to a more robust approach by including the concept of adaptivity to scene objects, for instance, Qin, (2017), has proposed adaptive median filtering where he incorporates a flexible window that is formed on the shape of the object and applies median filtering on each object instead of using a fixed-sized window. Such adaptive methods are able to retain the boundaries and shapes of objects in the scene (e.g., buildings, roads, etc.). Other studies have shown that the uncertainty of the DSM can be correlated with the class cover type (Albanwan & Qin, 2020), for instance, trees and grass changes based on the acquisition date and season, which can adversely influence the performance of MVS algorithm, and reduces the DMS uncertainty, whereas structures like buildings and roads are less changeable over time and most of the times have lower uncertainty. This led to the development of class-oriented fusion algorithms, as example, (Albanwan and Qin, 2020) developed adaptive spatiotemporal fusion to impose different bandwidths based on the class of objects. On the other hand, other methods used the concept of k-median clustering to locate the cluster with the most consistent elevation points to reduce the number of outliers (Facciolo et al., 2017), whereas others used a pair ranking scheme based on a scoring technique to evaluate and sort stereo pairs and merge the pairs with the best scores (Facciolo et al., 2017; Qin, 2019; Qin et al., 2020).

There are many factors influencing the fusion outcomes including the type of input and the fusion algorithm. Traditional MVS algorithms generate DSMs using local window-based approaches, which extract and match similar feature correspondences to compute the disparity and then transform it into a DSM. Census-Semi global matching (SGM) proposed by (Hirschmuller, 2005b) is one of the simplest and most cost-efficient algorithms to generate DSMs; it is one of the widely used methods from 2005 until today. Although it is considered invariant to radiometric changes, it is still sensitive to illumination changes and window size, therefore many have proposed adaptive windows to capture the size and shape of objects instead of rigid windows (Han et al., 2020; Hirschmuller & Scharstein, 2009; Loghman & Kim, 2013). Nowadays, deep learning algorithms have captured a great interest in the area of dense image matching and elevation generation, where they are intended to enhance dense image matching by better understanding and learning from the scene components (Chang & Chen, 2018; Hamid et al., 2020; Žbontar & LeCun, 2015). (Žbontar & LeCun, 2015) are the first to introduce matching cost Convolutional neural networks (MC-CNN) for stereo analysis and disparity and DSM generation; their target was to enhance the matching cost and produce faster and better similarity matching results. Although MC-CNN is able to capture the shapes of objects well, it still requires a lot of post-processing and filtering. Many deep learning MVS algorithms are inspired by MC-CNN, which are developed to further enhance predictability, cost matching, and time efficiency. One of the drawbacks to deep learning algorithms is their limited performance due to the training process; they require a great amount of time and data along with the ground truth data for training. Additionally, the generalization of the training is a very critical matter, the trained model should be well reflected on the testing dataset or any other dataset regardless of the domain difference, otherwise, the network must be retrained.

In this work, we aim to evaluate the fusion of adaptive and non-adaptive algorithms for DSM generated using Census and MC-CNN, we use these algorithms as they have been widely used in the area of MVS algorithms. Understanding such work can help to close the holes in current algorithms of fusion and provide insights to improving their performance and output, in addition, to help the user understand how each data works under different fusion algorithms.

The chapter is organized as follows: Section 6.3 includes the data description, pre-processing steps like DSM generation, and the fusion algorithms used, Section 6.4. Include the discussion and analysis of the results, and finally, we present our conclusion in the last section.

## 6.3. Methodology

### 6.3.1. Dataset Description

In Our work, we use three different datasets from Omaha (OMA), Jacksonville (JAX), and Argentina (ARG) conducted from a very high-resolution satellite Worldview-3. Each dataset includes hundreds of multispectral image pairs with 0.3 meters spatial resolution. Every dataset includes three testing regions with varied spatial complexity and density, as some locations may be urban or suburban areas with either dense small houses or sparse large buildings, or a mix of both as can be seen in Table 6.1. The total number of DSMs generated ranges from 91 – 482. The dataset images were captured within almost a one-year time span from September 2014-November 2015, October 2014- February 2016, and January 2015-January 2016 for OMA, JAX, and ARG respectively. For evaluation, we use a reference ground truth DSM generated from the LiDAR dataset.

Table 6.1. Dataset information

| Dataset ID | 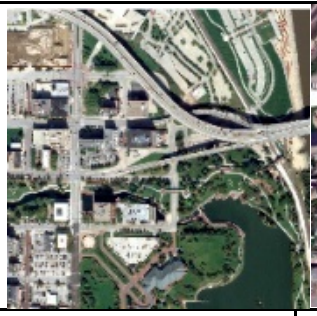 | 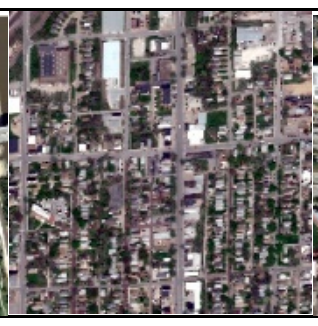 | 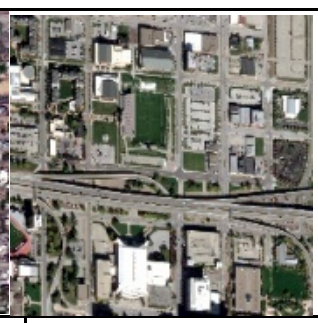 | 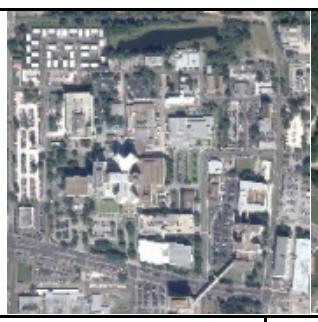 | 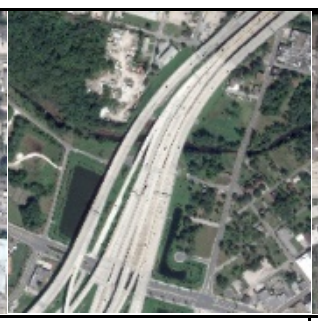 | 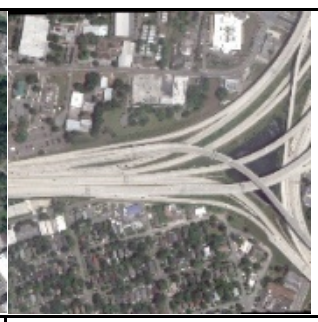 |
|---|---|---|---|---|---|---|
| | OMA I | OMA I | OMA II | JAX III | JAX II | JAX III |
| Dimensions (WxH) | 1168 x 1165 | 1168 x 1165 | 1206 x1202 | 1163x1164 | 1122x1152 | 1133x1164 |
| Number of DSMs | 460 | 460 | 293 | 138 | 91 | 247 |
| Region properties: Urban or Suburban, dense or sparse | Urban sparse | Urban sparse | Suburban sparse | Urban +sparse | Suburban sparse | Urban dense |
| Dataset ID | 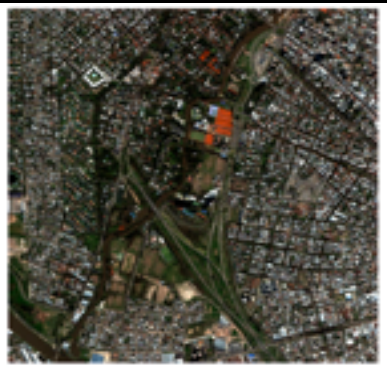 ARG I | | 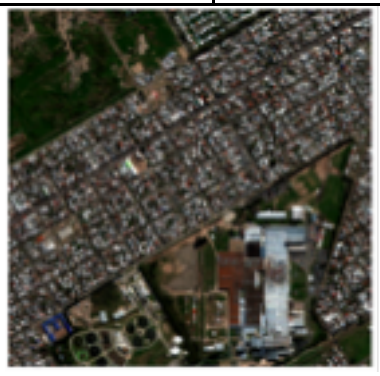 ARG II | | 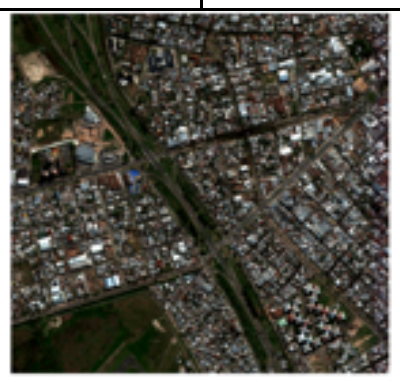 ARG III | |
| Dimensions (WxH) | 2565x2561 | | 4199x4256 | | 4062x4208 | |
| Number of DSMs | 479 | | 472 | | 482 | |
| Region properties: Urban or Suburban, dense or sparse | Urban dense | | Urban dense | | Urban dense | |

### 6.3.2. Data Preprocessing

Our pre-processing steps can be summarized as the following: 1) Image pair selection based on specific criteria, 2) geo-registration to ensure alignment, and finally 3) Disparity and DSM generation using census and MC-CNN cost metrics followed by semi-global matching for optimization (Hirschmuller, 2005).

- *Image Pair Selection*

Since we have about 20 images for every dataset, we can obtain hundreds of DSMs for the stereo pairs, however, in practice, only a few numbers of DSMs can be available. Therefore, we choose the best 20 pairs to generate elevation and perform the fusion. The selection of the stereo pairs is performed using a scoring or ranking scheme that sorts the images based on the metadata information from the pair of images, we include geometrical information such as the intersection angles, sun angle difference, and the number of days in which the images were acquired. We rank the stereo pairs based on their scores that are computed from the optimal values of sun angle difference and intersection angles as mentioned in (Qin, 2019).

- *Ortho-Ready Image, Geo-Registration, and Image Rectification*

Each image is transformed into an ortho-ready image and registered to assure alignment of images using a reference image and using RPC Stereo Processor RSP software (Qin, 2016). We also rectify the images to produce epipolar images which we use in the following steps for disparity and DSM generation using RSP software.

- *Disparity and Digital Surface Models (DSM) Generation*

Initially, we generate the disparity from the rectified epipolar images, where Census (Hirschmuller, 2005) and MC-CNN (Zbontar and LeCun, 2015) are used as the cost matching metrics to determine the horizontal displacement between feature correspondences, which leads to the disparity images. Census requires a predefined window to perform string bit/binary transformation, where any pixel lARGer than the central pixel takes a value of 1 and 0 otherwise. This transformation process is followed by a hamming distance to measure the score and compute the disparity based on the minimum score. On the other hand, we follow the CNN architect as in (Zbontar and LeCun, 2015), where we first train the MC-CNN using a satellite dataset, then we extract image patches with a size of 9x9. The image patches are then fed into a single layer of MC-CNNs each with size 5x5 and 32 kernels, followed by a couple of series of fully connected layers each with 200 neurons, which are then concatenated into a single layer of 400 neurons and passed to several layers with 300 neurons until the last layer which classified each pixel into a match or no match. The DSM is finally generated using Semi-global matching as proposed by (Hirschmuller, 2005) and implemented in RSP software.

### 6.3.3. DSM Fusion and Evaluation

For DSM integration, we use different fusion algorithms that vary between adaptive and nonadaptive approaches including:

1) Non-adaptive fusion algorithms: median filter, K-median clustering fusion (Facciolo et al., 2017), and weighted average fusion (Papasaika et al., 2008)
2) Adaptive fusion algorithms: adaptive median fusion (Qin, 2017) and adaptive spatiotemporal fusion (Albanwan and Qin, 2020)

The median filter simply takes the median of the temporal DSMs at any pixel position and produces median DSMs. The adaptive median filter as suggested by (Qin, 2017) applies the same concept except using an adaptive window that captures the shape of the object and applies median filtering on individual objects. K-median clustering in fusion is proposed by (Facciolo et al., 2017) to merge several pre-ranked stereo pairs into a single DSM by clustering the input temporal elevations and picking the cluster with minimum cost to fuse its elevation points. Weighted average fusion on the other hand is a broader algorithm (Papasaika et al., 2008); first, it computes the residual map between two DSM images, then use these residuals to get weights and multiply by the elevations as follows:

$$WAF = \frac{\sum_{i=1}^{t} Wi * DSMi}{\sum_{i}^{t} Wi} \quad (6.1)$$

Where F is the fused DSM, $w_i$ is the weight computed from the residual maps, and $t$ is the number of temporal DSMs.

Finally, we perform adaptive spatiotemporal fusion by applying different bandwidths for different classes, where highly complex and seasonally variable classes like trees, grass, and water take higher bandwidths, while static rigid objects like buildings and roads take smaller bandwidths. In Adaptive spatiotemporal fusion, we first compute the median and generate masks as suggested by (Albanwan and Qin, 2020), then difference the spatial and temporal DSMs from the median to calculate the weight and impose different bandwidths for each type of class, the algorithm is as follows:

$$DSM_{fused}(x,y) = \frac{1}{W_T} * \sum_{i=1}^{Width} \sum_{j=1}^{Height} W_r * W_s * W_h * h(x,y,t)_{med} \tag{6.2}$$

Where $DSM_{fused}$ (x,y) is the fused pixel, (x, y) is the position of pixels in the depth map, $h_{med}$ is the median height from the temporal DSMs, $W_r$ is the spectral weight, $W_s$ is the spatial weight, $W_h$ is the temporal height weight, and $W_T$ is total weight. The $Wr$ and $Ws$ compute the range and spatial weights from the orthophoto as the bilateral filter, as follows:

$$W_r(x,y) = exp^{\frac{-||(I(x,y)-I(k,l)||^2}{2\,\sigma_r^2}} \tag{6.3}$$

$$W_s(x,y) = exp^{\frac{-((x-k)^2+(y-l)^2)}{2\,\sigma_s^2}} \tag{6.4}$$

Where I is the orthophoto image, (x, y, l, k) are the position of current and neighboring in the window, $\sigma_r$ is the range bandwidth, and $\sigma_s$ is the spatial bandwidth. The adaptivity for classes is computed using the following equation:

$$W_h(x,y) = exp^{\frac{-||hmed-h(i,j,t)||^2}{2\,\sigma_h^2}} \tag{6.5}$$

Where $\sigma_h$= the height bandwidth relative to each class. We perform the fusion on the 20 best pairs, and evaluate the fusion output using root mean squared error (RMSE) and the LiDAR data as the ground truth (GT) as follows:

$$RMSE(F_DSM) = \sqrt{\frac{\sum_{i,j=1}^{N,M}(F_DSM(x,y) - GT(x,y)^2}{N \times M}} \tag{6.6}$$

Where *F_DSM* is the fused DSM, (x, y) are the pixel location of the depth point in the image, and N, M are the dimensions of the fused image.

## 6.4. Results and Analysis

### 6.4.1. Parameter Selection for Fusion Methods

For the adaptive median filtering, we use the default setting from (Qin, 2017), and for the k-median clustering, we use elevation point from the temporal and spatial domains, where the window size is set to 5 and the threshold to determine the cost and when to stop clustering is set to less than 10. For the weighted average filter, we used the median fused DSM as a reference to compute the residuals map and compute the corresponding weights. Finally, for the adaptive spatiotemporal fusion we used a window size of 5, we obtain 4 to 5 classes for each dataset including buildings, roads and ground, trees, grass, and water. The spectral and spatial bandwidths are chosen empirically and set to $\sigma_r = 30$ and $\sigma_s = 5$; the elevation bandwidths for the classes are set to $\sigma Buildings = 3$, $\sigma Road_Ground = 3$, $\sigma Trees = 7$, $\sigma_{Grasss} = 7$, and $\sigma_{Water}_surface = 7$.

### 6.4.2. DSM Analysis

The first step in the analysis is DSM accuracy inspection, which is necessary for the evaluation and comparison between the before and after fusion results. Figure 6.1. Represents a sample of the initial DSM generated by Census and MC-CNN from two OMA testing regions (OMA I and OMA III). We can notice that regardless of the method, the generated DSM is always associated with problems such as noise, outliers, missing elevation points, and fuzzy representation of object shapes, which can raise due to mismatching in the dense image matching process. These issues can vary DSMs and may be addressed in some of the generation algorithms, for instance, Census produces DSMs that are smoother, fuller, and well distributed as can be seen in Figure 6.1(a) and highlighted by the blue boxes. On the other hand, MC-CNN includes more missing elevation points which can be obvious from the black spots in the right image of Figure 6.1. Nevertheless, MC-CNN captures the edges and boundaries of buildings better than Census, which can be indicated in the red and yellow boxes around the buildings in Figure 6.1(b).

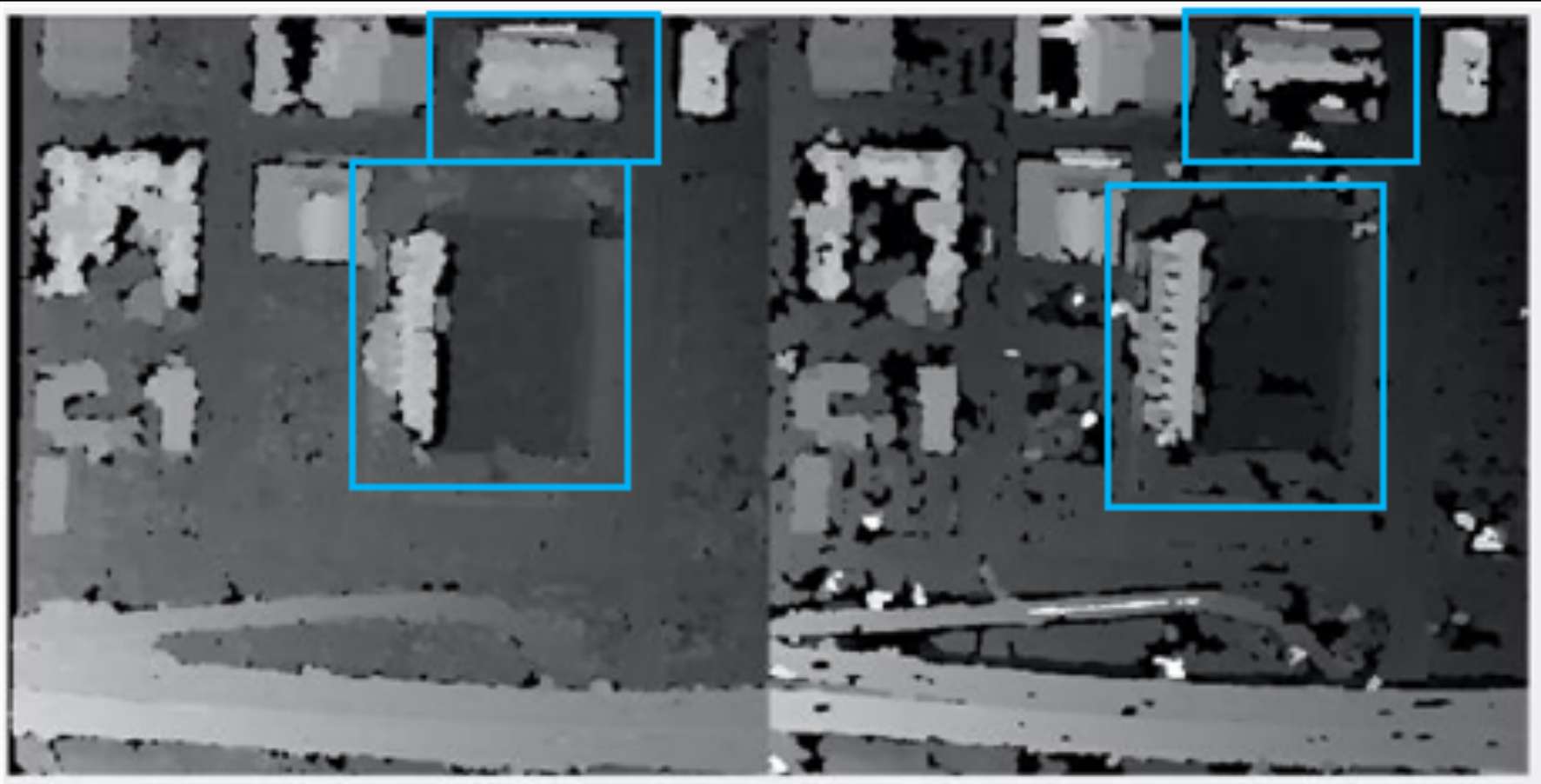

(a) DSM from OMA I, showing Census (on the left) and MC-CNN (on the right)

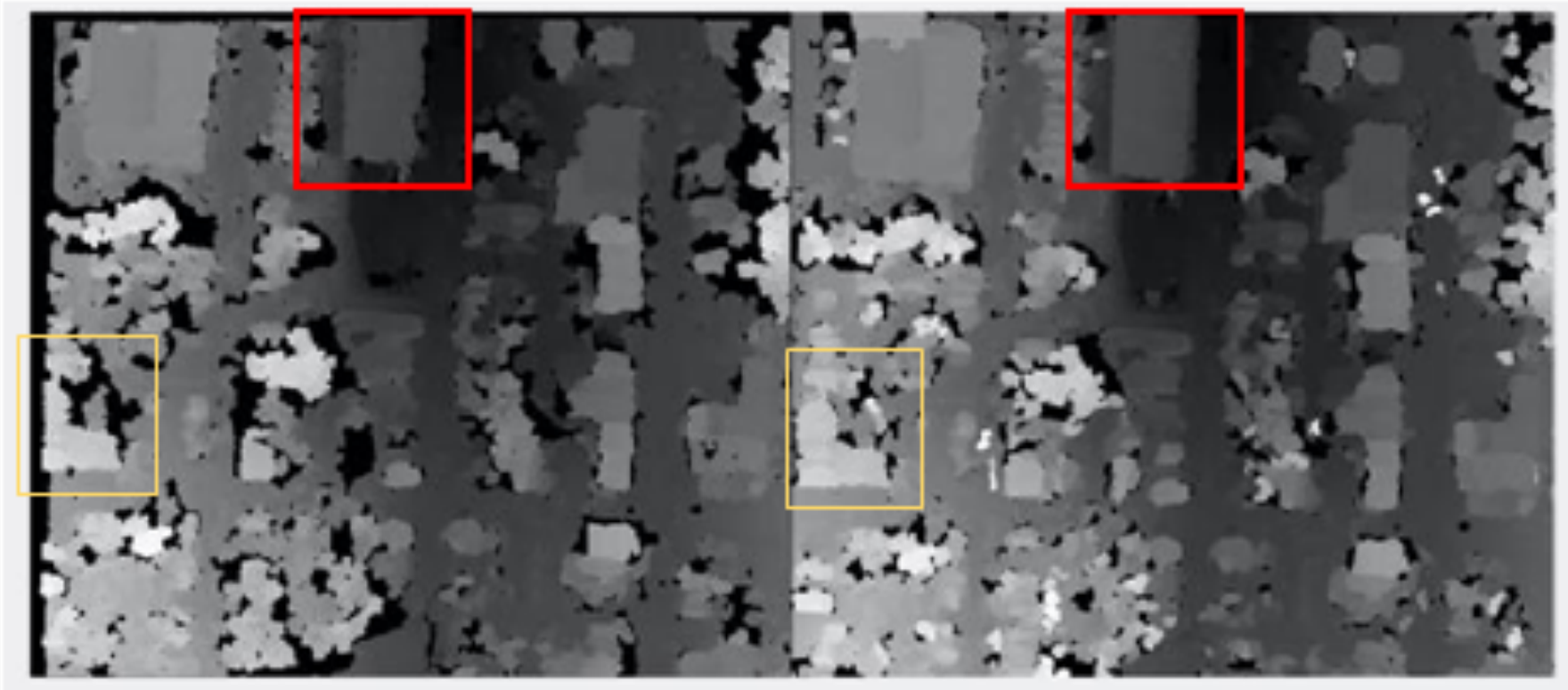

(b) DSM from OMA III, showing Census (on the left) and MC-CNN (on the right)

Figure 6.1. The initial DSMs generated by Census-SGM and MC-CNN-SGM

### 6.4.3. Statistical Analysis

We provide a comprehensive analysis and comparison of the overall performance of the adaptive and nonadaptive fusion algorithms, in addition to the performance of fusion algorithms for the non-deep learning and deep learning (i.e., Census and MC-CNN) DSM generation algorithms. We present a visual and statistical evaluation of the results as presented in Figure 6.2. and Table 6.2. From the bar representation in Figure 6.2, we can notice that the adaptive methods such as the adaptive median fusion and adaptive spatiotemporal fusion always produce less uncertainties in comparison to the other fusion algorithms, which can be indicated in the orange and purple bars in Figure 6.2. They also had the lowest average RMSE of 5 meters as can be seen in Table 6.2. Their performance is consistent over different datasets regardless of the DSM generation algorithm, which is indicated by having less varying RMSE values (See Figure 6.2 and Table 6.2). On the other hand, we can notice that fusion methods like weighted average fusion and k-median clustering fusion have the highest ranges of RMSE as indicated by the green and yellow bars in Figure 6.2, their RMSE also ranges between 5 to 35 meters.

We can also notice during our analysis that MC-CNN performs better than Census in terms of robustness to outliers, like in the case of median filter, k-means filtering, adaptive spatiotemporal fusion, and weighted average fusion for dataset ARG I, MC-CNN was able to suppress these invalid elevation errors that were not resolved in Census. This can lead to concluding that fusing Census DSMs is not robust to outliers as in MC-CNN. Table 6.2 also shows that the error range for MC-CNN is about 4-12 meters, whereas for Census the range is between 6 to 31 meters.

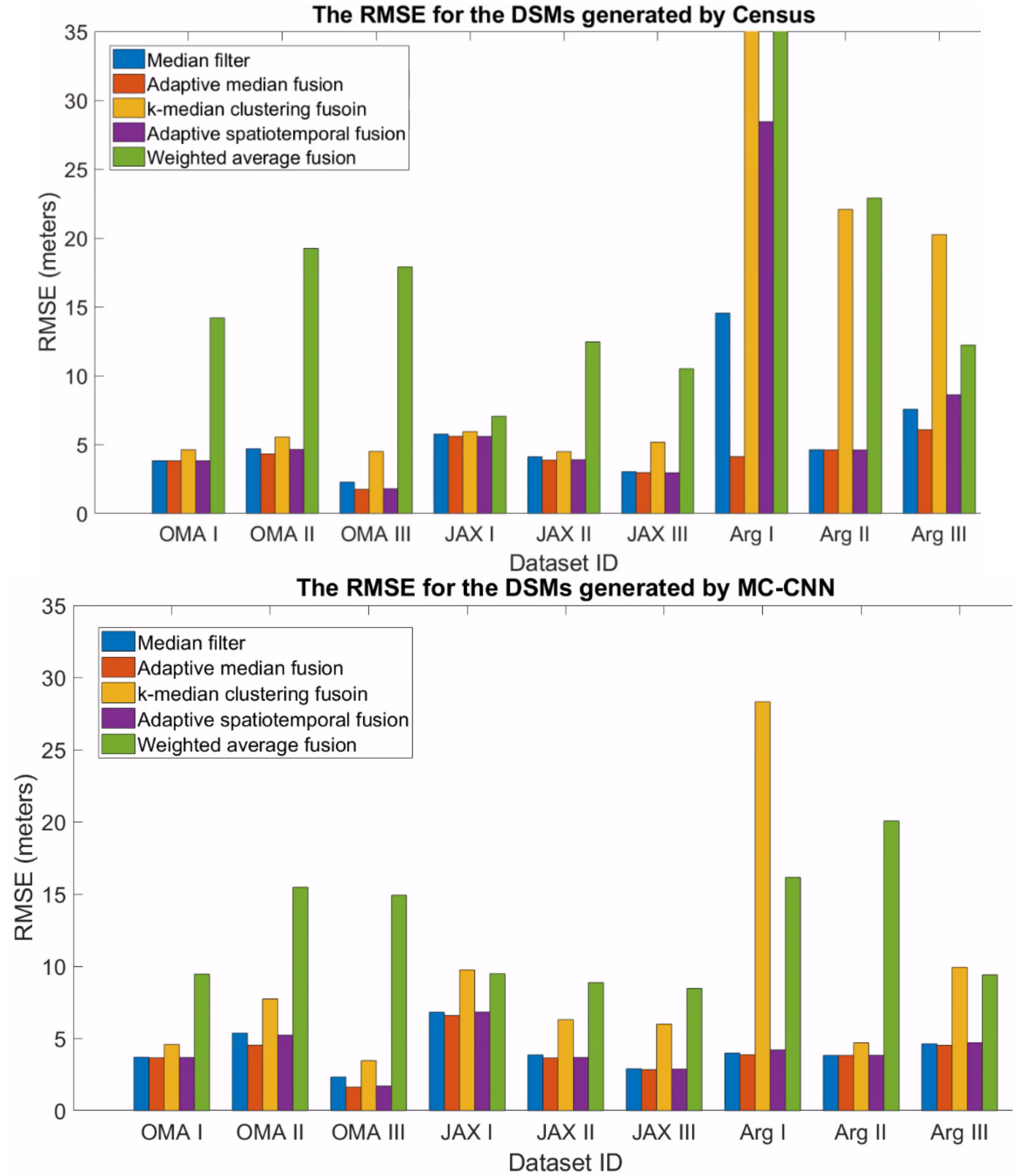

Figure 6.2. A comparison between fusion algorithms for DSMs generated from Census and MC-CNN.

Table 6.2. A comparison between the RMSE of the fused DSMs using different adaptive and non-adaptive fusion algorithms.

| OMA | | Median filter | Adaptive median filter | K-median filter | Adaptive spatio-temporal fusion | Weighted average fusion |
|---|---|---|---|---|---|---|
| I | Census | 6.39 | 6.38 | 13.96 | 6.36 | 18.29 |
| | MC-CNN | 7.85 | 7.68 | 7.79 | 7.68 | 11.84 |
| II | Census | 6.47 | 4.33 | 7.82 | 4.65 | 36.33 |
| | MC- CNN | 7.21 | 4.53 | 8.53 | 5.30 | 33.57 |
| III | Census | 2.36 | 1.76 | 4.83 | 1.79 | 18.36 |
| | MC- CNN | 2.41 | 1.64 | 3.81 | 1.70 | 15.28 |
| **JAX** | | **Median filter** | **Adaptive median filter** | **K-median filter** | **Adaptive spatio-temporal fusion** | **Weighted average fusion** |
| I | Census | 5.76 | 5.60 | 5.95 | 5.60 | 7.06 |
| | MC- CNN | 6.81 | 6.59 | 9.86 | 6.82 | 9.92 |
| II | Census | 4.11 | 3.87 | 5.27 | 3.90 | 14.92 |
| | MC- CNN | 3.86 | 3.64 | 22.88 | 3.68 | 9.68 |
| III | Census | 3.03 | 2.97 | 5.57 | 2.96 | 17.19 |
| | MC- CNN | 2.89 | 2.84 | 7.53 | 2.87 | 8.51 |
| **ARG** | | **Median filter** | **Adaptive median filter** | **K-median filter** | **Adaptive spatio-temporal fusion** | **Weighted average fusion** |
| I | Census | 14.55 | 4.12 | 74.65 | 28.45 | 37.85 |
| | MC- CNN | 3.98 | 3.87 | 28.31 | 4.20 | 16.15 |
| II | Census | 4.63 | 4.61 | 22.08 | 4.61 | 22.91 |
| | MC- CNN | 3.82 | 3.82 | 4.70 | 3.83 | 20.07 |
| III | Census | 7.57 | 6.07 | 20.25 | 8.62 | 12.22 |
| | MC- CNN | 4.63 | 4.52 | 9.93 | 4.70 | 9.40 |

Note: Bold numbers indicate the lowest RMSE for all fusion methods in each testing region and algorithm.

### 6.4.4. Visual Analysis

In general, we can notice that fusion has solved many problems related to missing or incorrect elevation points (See Figure 6.3.). Moreover, the visual illustration in Figure 6.3. confirms our findings in Section 6.4.3. that adaptive methods produce the best fused results not only statistically but also visually, this can be seen from the second and last rows in Figure 6.3, where they produced smooth and sharp DSMs, we can also see that the buildings are well captured and most of the noise in the original DSMs have been reduced. Other methods like k-median clustering fusion and weighted average fusion are less robust to noise and outliers in the DSM, which is evident from the third and fourth images in Figure 6.3.

Additionally, we can see that fusion using Census is better in terms of generating smoothed fused DSM for all fusion methods (Seen Figure 6.2.), especially in the case of k-median clustering fusion and weighted average fusion, whereas fusion of MC-CNN DSM can produce noisy results in these methods. Nevertheless, unlike Census, MC-CNN fused DSMs seem better in edge and boundary preservation in all fusion methods, as can be noticed in Figure 3 where buildings are sharp and better outlined.

| Method | Census | MC-CNN |
|---|---|---|
| **Median filter** | 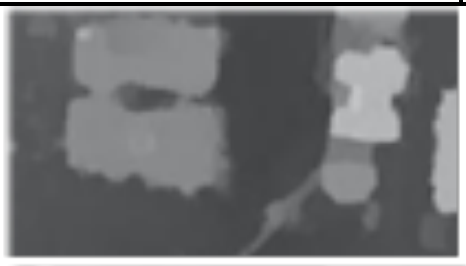 | 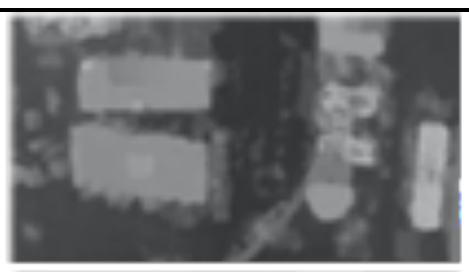 |
| **Adaptive median filter** | 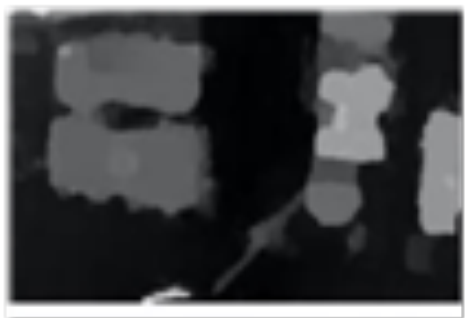 | 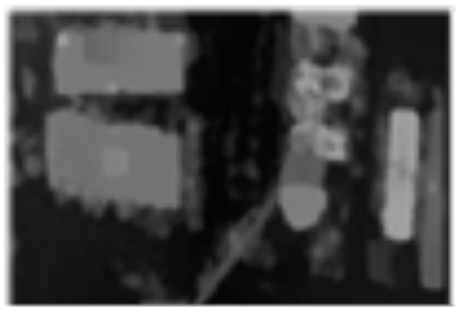 |
| **K-median clustering fusion** | 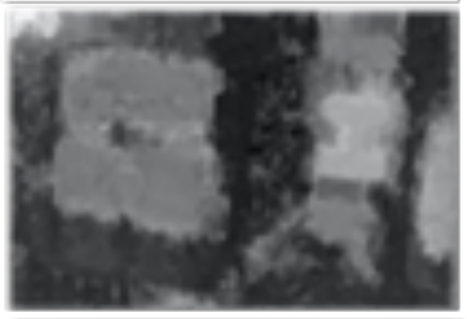 | 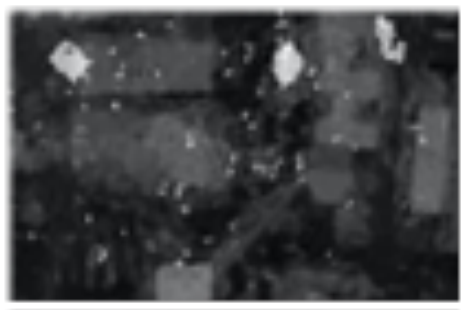 |
| **Weighted average fusion** | 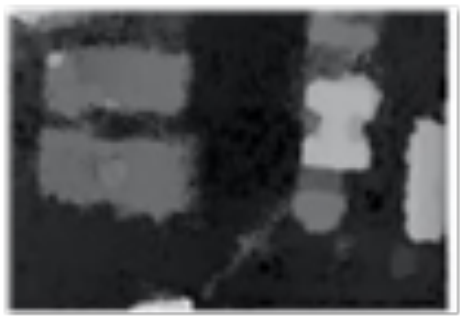 | 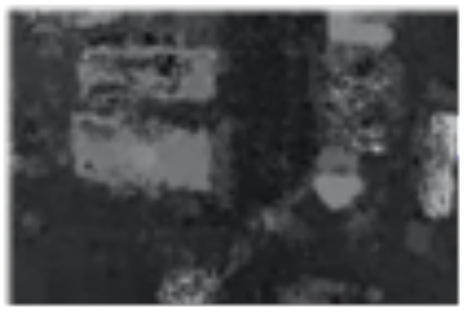 |
| **Adaptive spatiotemporal fusion** | 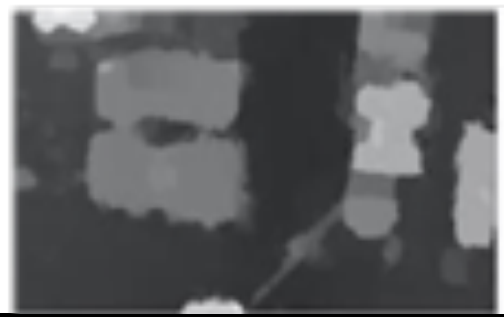 | 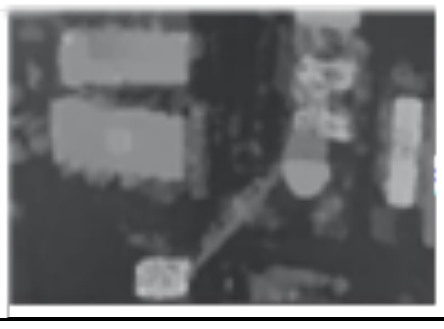 |

Figure 6.3. A subsection of fused DSMs from Census-SGM and MC-CNN-SGM from JAX II dataset using different fusion algorithms.

### 6.5. Conclusion

To conclude, our work has shown that in general median filter produces fairly good results with low computational cost. However, adaptive fusion methods such as adaptive median fusion and adaptive spatiotemporal fusion always produce the best results due to their robustness towards outliers and flexibility to learn and predict elevation from homogeneous objects and a consistent set of neighboring pixels. The generalized non-adaptive fusion method did not perform as well due to the lack of contextual and temporal information and correlation in the elevation prediction and fusion process. We also evaluate the usage of DSMs generated using different dense image matching algorithms in the fusion process. We found that in general, MC-CNN performs better than Census for most fusion algorithms, due to the architecture and mechanism of MC-CNN and since it can learn similarities from highly complex features and produces a more detailed DSM. Therefore, combining its results can help achieve better fusion outcomes. Census on the other hand may generate less accurate DSMs than MC-CNN but leverages accuracy and computational extensive algorithms. Overall, Census can generate satisfactory results and is appropriate for when there are limited resources and expertise to avoid training and deep learning algorithms. Although MC-CNN has shown its ability to preserve the shape of objects, they still require post-processing and refinement to reduce the noise and outliers in the resultant fused DSMs. Such work can further be extended to improve fusion methods by combining DSMs from both MC-CNN and Census to take the advantage of both methods, however, the distribution of data must be taken into consideration, since different methods generate different DSMs.

# Chapter 7. An Analysis of the Digital Surface Model's (DSMs) Accuracy at Stereo Matching Level

This chapter is based on the paper called "A Comparative Study on Deep-Learning Methods for Dense Image Matching of Multi-angle and Multi-date Remote Sensing Stereo Images" that is submitted in the "The photogrammetric record" by (Albanwan & Qin, 2021a).

## 7.1. Abstract:

Deep learning (DL) stereo matching methods have recently shown their superior performance on several Computer Vision benchmarks such as Middlebury and KITTI and have gained great attention on their use in remote sensing satellite datasets as well. However, most of these existing studies for remote sensing images conclude the assessments based only on a few or single stereo images, which lack a systematic evaluation on how robust these typical DL methods are on satellite stereo images with varying radiometric and geometric configurations. This paper provides an evaluation on four of the most used DL stereo matching methods through hundreds of multi-date satellite stereo pairs with varying geometric configurations over multiple sites, against the traditional, but well-practiced Census-based SGM (Semi-global matching), to comprehensively understand the accuracy, robustness, and generalization capabilities of DL stereo matching methods and their practical potential. These DL methods include a learning-based cost metric through convolutional neural networks (MC-CNN) followed by SGM, and three end-to-end (E2E) learning models using Geometry and Context Network (GCNet), Pyramid Stereo Matching Network (PSMNet), and LEAStereo. We choose a few of the most well-established DL methods that have a high impact and readily available codes as representatives in our study. Our experiments show that the E2E algorithms are able to achieve the upper limit of geometric accuracies, while they might not generalize well for unseen data. The learning-based cost metric and the Census-based SGM algorithms are rather robust and are able to consistently achieve acceptable results. All DL algorithms are robust to geometric configurations of stereo pairs and are less sensitive in comparison to the Census-based SGM, while learning-based cost metric, due to the nature of learning similarity as a much simpler task, can generalize on satellite images when trained on different datasets (airborne and ground view).

### 7.2. Introduction

Stereo dense image matching (DIM) has been an active area of study over the years, as it offers a cost-effective and efficient approach to generating digital surface models (DSM) for applications such as 3D modeling, forestry mapping, and change detection (Furukawa & Hernández, 2015; Navarro et al., 2018; Nebiker et al., 2014; Tian et al., 2014; H. Wang et al., 2017). This is especially relevant when these applications are fueled by satellite-based 3D reconstructions due to their wide data coverage and consistent data collection over time. However, it has been noted (Brown et al., 2003; Qin, 2019; Seitz et al., 2006) that the accuracy of the DSM not only depends on a specific DIM algorithm, but also on various factors of stereo pairs including: 1) sensor and image characteristics (e.g., spatial and radiometric resolutions), 2) acquisition conditions including the atmosphere and position and orientation of the sun, camera, and objects, 3) scene structure and texture across different geographical regions, for instance, different land patterns including urban, suburban, or forest areas, water surfaces, and parallaxes formed by stereo pairs, etc. Although there have been studies evaluating the achievable quality of these satellite-based reconstructions using different DIM algorithms, most of them only take one or a few pairs for quality assessment, thus the resulting conclusions are often not comprehensive to cover data with various stereo configurations appearing in practice.

A typical stereo DIM algorithm performs disparity estimation on rectified stereo images (i.e., epipolar images), which generally follows several key steps, including cost matching, aggregation, disparity optimization, and filtering (Scharstein et al., 2001). Cost matching is the main step for measuring feature similarities for disparity computation. For the last decade, Census has been known as the classical cost matching metric for stereo DIM problems and has been studied intensively due to its robustness to radiometric differences, at the same time it can be computed efficiently (B. Chen et al., 2019a; Ma et al., 2013; Y. Xia et al., 2018; Zabih & Woodfill, 1994). On the other hand, recently developed deep learning (DL) algorithms have shown a promising performance that is assumed to surpass classical cost metrics like Census (B. Chen et al., 2019a; Hamid et al., 2020). There are mainly two ways that DL algorithms are implemented in stereo DIM, for example, 1) Learning-based cost metric and 2) end-to-end (E2E) learning. The learning-based cost metric learns similarity from image patches given examples of similar and dissimilar patches, such methods were first introduced by (Žbontar & LeCun, 2015) through convolutional neural networks (MC-CNN) as learnable feature extractors. Like Census, such patch-based cost metrics do not handle texture-less regions well, and often require multistage optimization after cost computation, followed by cost aggregation for regularization and smoothness of the disparity map (Bobick & Intille, 1999; Hirschmuller, 2005a, 2008b; Kolmogorov & Zabih, 2001). There are plenty of cost aggregation methods ranging from local, semi-global, and global approaches. A well-known example is the semi-global matching (SGM) algorithm (Hirschmuller, 2005a), which is known to effectively leverage both accuracy and speed well, thus nowadays it is used as a standard cost aggregation approach. However, even with cost aggregation, studies found that MC-CNN still suffers from poor performance in ill-posed regions with occlusion, high-reflective surfaces, lack of texture, and repetitive patterns (B.

Chen et al., 2019a; Ma et al., 2013; Y. Xia et al., 2018; Zabih & Woodfill, 1994) demanding further postprocessing and refinement. Alternatively, E2E learning methods directly generate disparity from stereo pair rectified images without additional optimization steps. They have become a popular line of stereo DIM algorithms because they can directly predict highly accurate disparity maps through learning from geometry and context (e.g., cues such as shading, illumination, objects, etc.) rather than low-level features (Chang & Chen, 2018; X. Cheng et al., 2020; X. Gu et al., 2020; Kendall et al., 2017; H. Xu & Zhang, 2020; F. Zhang, Prisacariu, et al., 2019). Examples of this type of work are Geometry and Context Network (GCNet), pyramid stereo matching networks (PSMNet), and LEAStereo, which due to their high accessibility and performances (i.e., winning the best rank in the KITTI 2012 and 2015 leaderboards (Chang & Chen, 2018; X. Cheng et al., 2020; Kendall et al., 2017), have become popular in the field.

However, DL algorithms often suffer from generalization problems, and at the same time, face challenges to process large-volume and large-format remote sensing images, limiting their practical values (Pang et al., 2018; Song et al., 2021; F. Zhang, Qi, et al., 2019). The degree of generalization may vary with both DL models and tasks. For instance, studies indicate that most E2E approaches enjoy deep feature representations, which however, may encode scene-specific information (Pang et al., 2018; Song et al., 2021; F. Zhang, Qi, et al., 2019), thus may poorly perform on unseen scenes or data (Song et al., 2021). Most of the existing DL stereo DIM algorithms are trained using Computer Vision (CV) benchmark datasets such as KITTI or Middlebury, etc. (Geiger et al., 2012; Hamid et al., 2020), which mainly consist of images of everyday scenes. These images are distinctively different from satellite images in terms of scene content, view perspectives, and object granularity, thus DL models trained from these CV datasets may not perform well when applied to satellite datasets. However, acquiring a large number of satellite datasets for training is a challenge: first, the cost of ground-truth data is usually high and requires extensive data preprocessing when converted to ground-truth disparity images; second, the scene content of the earth's surface is extremely diverse across the entire Globe, thus it is difficult to create a reprehensive dataset by only using data of a few regions, although these regions can be as large as an entire city. A few remote sensing benchmarks such as the IARPA (The Intelligence Advanced Research Projects Activity) Multiview stereo 3D mapping challenge (Bosch et al., 2016) and the 2019 Institute of Electrical and Electronics Engineers (IEEE) Geoscience and Remote Sensing Society (GRSS) Data Fusion Contest (DFC) (Le Saux, 2019) provide the ground-truth dataset in the form of LiDAR point clouds and raster DSMs, which must be post-processed and converted to the ground-truth disparity for training. Unfortunately, this step may generate undesired errors in the training data due to geometric errors in the orientation parameters, inconsistencies in the temporal and spatial resolutions, or other operations such as triangulation, projection, and interpolation (Bosch et al., 2016; Cournet et al., 2020; Patil et al., 2019; T. Wu et al., 2021).

In addition, a significantly under-evaluated criterion for DIM algorithms on satellite images is their robustness to varying stereo configurations such as sun illuminations, intersection angles, different sensors, etc. A robust DIM algorithm pertaining to these factors is extremely important because this will allow us to make full use of the already limited satellite images (as compared to everyday images). Ultimately, we want these algorithms to be agnostic and less selective to stereo configurations of data. It has been reported that the result of a typical stereo algorithm, i.e., Census with SGM, directly correlates with geometrical acquisition factors such as sun angle difference and intersection angle (Qin, 2019). Unfortunately, the same analysis has not been covered for DL algorithms. Since DL methods (especially E2E ones) can learn context, it is of interest to understand their performance under varying stereo configurations.

In this paper, we aim to comprehensively explore the limitations and strengths of recent stereo DIM algorithms for satellite datasets. We consider representative DIM algorithms of three main categories: 1) traditional approaches (e.g., Census cost metric with SGM), 2) deep learning-based cost metrics (e.g., MC-CNN cost matching metric with SGM), and 3) three E2E learning methods (e.g., GCNet, PSMNet, and LEAStereo). Since "Census+SGM" has been well studied and widely used in satellite stereo-photogrammetry (Qin, 2019), it serves as a baseline method in this study. Deep learning-based cost metrics are regarded as a simpler task than E2E learning for DIM because it learns patch-level similarity as a binary classification problem (similar or not similar). The very one algorithm in this category is MC-CNN, which applies a Siamese network to learn the similarities (Žbontar & LeCun, 2015). Despite the very many E2E methods in the CV community (Laga, 2019), we choose three State-of-the-art (SOTA) methods that are frequently used by the community, well performed in the leader board (Chang & Chen, 2018; X. Cheng et al., 2020; Kendall et al., 2017), and have well-organized codes available. To achieve a comprehensive evaluation and analysis, we use nine satellite datasets from different locations, and each dataset contains ~100-500 stereo pairs with their respective ground truth LiDAR data. We train and test the models on the same and different datasets, and analyze the results to understand their performance, robustness, and generalization. To be more specific, this paper presents three contributions:

1) We comprehensively evaluate five stereo DIM algorithms (including four DL approaches) on satellite stereo images using hundreds of pairs from nine test-sites, to inform the community of the performance of such DIM algorithms under varying configurations.

2) We analyze the accuracy of the evaluated DL methods and study their robustness against stereo configurations of data that were reported to be critical for the resulting accuracy of DSMs for traditional methods (Qin, 2019).
3) We study the generalization capability (or transferability) of these DL stereo DIM methods trained on and applied to datasets across different geographical regions and resolution/sensors (including satellite, airborne and ground-view images).

The remainder of the paper is organized as follows: Section 7.3 introduces relevant work including a brief review of stereo DIM algorithms and existing comparative studies. Section 7.4 describes the datasets, the stereo DIM algorithms, and the overall workflow of our analysis. Section 7.5 presents the results, evaluation, and discussion. Finally, the conclusion, limitations, and potential future directions are discussed in Section 7.6.

## 7.3. Related Works

### 7.3.1. Stereo DIM Algorithms

There has been a tremendous development in stereo DIM algorithms over the years, they are broadly classified into traditional and DL methods (Zhou et al., 2020). Traditional methods are the very early algorithms with basic cost matching metrics as the sum of absolute differences (SAD), normalized cross-correlation (NCC), mutual information (MI), and Census transformation (Brown et al., 2003; Seitz et al., 2006). With the development of DL methods, the cost-matching pipeline was replaced by convolutional neural networks (CNN) (Žbontar & LeCun, 2015). DL-based algorithms have attracted great attention due to their superior performances in benchmark testing (Geiger et al., 2012). Depending on the task of learning, DL stereo methods can be further categorized into learning-based cost metrics (Žbontar & LeCun, 2015, 2016) and E2E learning (Chang & Chen, 2018; X. Cheng et al., 2020; Kendall et al., 2017;

H. Xu & Zhang, 2020; F. Zhang, Prisacariu, et al., 2019). Learning-based cost metric was first introduced by (Žbontar & LeCun, 2015) to learn similarities from image patches. Both traditional and learning-based algorithms process low-level features as intensity or gradient patches to indicate similarity. As a result, their performance is limited to repetitive patterns and texture-less regions. Thus, post-processing like cost aggregation, optimization, and refinement based on these metrics is necessary to enhance the disparity map (Hirschmuller, 2005a, 2008b; Scharstein et al., 2001). DL methods rapidly evolved to E2E learning algorithms, where their main contribution is to replace the classical multistage optimization with a trainable network to directly predict the disparity from stereo images (Hamid et al., 2020; Laga, 2019). The underlying concept is that these neural networks can directly capture more global features, hence, they may better perform (Chang & Chen, 2018; X. Cheng et al., 2020; Kendall et al., 2017). In addition to these intensively studied methods, there are a few methods that perform context learning for part of the traditional pipeline but do not fully fall into either of these DL categories, for example, SGM-Net (Seki & Pollefeys, 2017) learns the per-pixel smoothness penalty and GA-Net (F. Zhang, Prisacariu, et al., 2019) learns networks to guide the cost-aggregation process.

### 7.3.2. Existing Comparative Studies of Stereo Dense Matching Algorithms

Most of the existing review papers on stereo DIM algorithms take upon single to few stereo pairs for evaluation (Hamid et al., 2020; Laga, 2019; Xia et al., 2020; Zhou et al., 2020), which may be insufficient to provide an accurate and conclusive evaluation. There are a few but limited studies concerning the use of more pairs to study DIM algorithms: they indicate that the accuracy of the DSM is significantly correlated with the radiometric and geometric characteristics and the configurations of the stereo pairs (Facciolo et al., 2017; Qin, 2019; L. Yan et al., 2016; X. Zhou & Boulanger, 2012). For instance, Facciolo et al., (2017) found that selecting stereo pairs based on specific heuristics as minimum seasonal differences improves DIM and reduces the uncertainties in the DSM. In (Qin, 2019), the author observed a direct relation between the geometric configurations at the time of acquisition as the sun angle difference and intersection angle (base-height ratio) and the accuracy of DSMs. In addition, the spatial distribution of objects in space (e.g., buildings density, sizes, and distances) and land cover types (e.g., trees, roads, water surfaces, etc.) are highly diverse across different test-sites (Chi et al., 2016), hence, may impact the performance of stereo DIM algorithms. With satellite images being rich in content and acquisition configurations, it is necessary to analyze the performance of stereo DIM algorithms on a large number of datasets covering a variety of regions, complexities, and configurations to understand their practical values in various applications.

### 7.3.3. Deep Learning Training Models and Generalization

Despite the superiority of DL algorithms in DIM, the generalization issue remains a major challenge. Learning across different domains such as data collected from different sensors, data of different locations, data with different spatial resolutions, etc., often leads to a deep drop in the accuracy of DIM because of their inability to predict disparity from unseen data (Pang et al., 2018; Song et al., 2021). One possible solution is to encapsulate a large number of training datasets covering all scenarios and instances that a network may encounter (Najafabadi et al., 2015). However, in practice obtaining a large training dataset with their ground truth (such as LiDAR) is costly and often unavailable for large-scale areas (Chi et al., 2016), it also requires manual or post-processing to convert to the ground-truth disparity which may produce systematic errors (Cournet et al., 2020; Patil et al., 2019; Song et al., 2021). Although some benchmark datasets are available (Bosch et al., 2016; Le Saux, 2019; Rottensteiner et al., 2012), existing evaluations are mostly performed on a dataset-by-dataset basis. Moreover, these datasets, although seem to be large in data volume, mostly present typical urban scenes of a few major cities and remain very sparse when considering the generalization in terms of scene contexts around the globe, sensor types, and resolutions.

## 7.4. Methodology

In this section, we describe our approach to comprehensively evaluate DL-based algorithms in stereo DIM. First, we introduce the datasets in our analysis, then, we briefly present the selected stereo DIM algorithms including: 1) Census+SGM, 2) MC-CNN+SGM, and 3) E2E methods using GCNet, PSMNet, and LEAStereo. Lastly, we describe our validation and evaluation schemes for these algorithms. In particular, MC-CNN has two architectures called MC-CNN-fst and MC-CNN-acc (Žbontar & LeCun, 2016); the former uses a product of two feature vectors to decide the similarity and the latter uses fully connected layers following the concatenated feature vectors. Although the latter was reported to be slightly more accurate, we use the former version (MC-CNN-fst) in our evaluation, given its good leverage of performance and spend for large-sized satellite images. Without specifically stated, MC-CNN mentioned hereafter refers to the MC-CNN-fst version.

### 7.4.1. Dataset Description

In this work, we have two types of data that are used for training and evaluation purposes. All our data are collected from publicly available benchmarks. First, we will describe our evaluation dataset and then we will describe the training datasets.

The evaluation is based on satellite images from IARPA (Bosch et al., 2016) and the 2019 DFC (track 3) (Le Saux, 2019) benchmarks. They provide stereo images from the WorldView-3 satellite sensor with a spatial resolution of 0.3 meters. In addition, they provide the airborne LiDAR which we convert to the ground-truth DSM and use for evaluation. The IARPA benchmark provides 50 overlapping images for a 100 $km^2$ area near San Fernando, Argentina (ARG) collected between January 2015 and January 2016. The 2019 DFC benchmark provides 16 to 39 overlapping images for 100 $km^2$ of Omaha, NE, USA (OMA) and Jacksonville, FL, USA (JAX) collected from September 2014 to November 2015. We select three sub-regions as test-sites from OMA, JAX, and ARG datasets. Our selection covers test-sites with a variety of densities, complexities, land types, and covers. The 16-50 stereo images provided for all test-sites, yield approximately 6,278 stereo pair images. After omitting stereo pairs that fail during feature matching (Kuschk et al., 2014) due to large radiometric variations or extremely small and large intersection angles, the final total stereo pairs is 2,861. The stereo pair images have a wide range of geometric configurations. The sun angle difference range between 0 and 50 degrees and the intersection angle range between 0 and 67 degrees. Figure 7.1(a) provides detailed information about the evaluation dataset and test-sites.

The training dataset is used to train the DL algorithms. We include a variety of datasets from satellite, airborne, and ground-view sensors. For training with a satellite dataset, we use another set from the 2019 DFC benchmark known as track 2. The dataset provides 4,293 rectified stereo pair images of size 1024 × 1024 with their ground truth disparities derived from LiDAR. We also use datasets that are distinctively different from satellite images including ground-view images from ***KITTI*** dataset (Geiger et al., 2012) and airborne dataset from Toronto, Canada ***ISPRS*** (International Society for Photogrammetry and Remote Sensing) benchmark (Rottensteiner et al., 2012). The airborne dataset contains 13 images from an aerial block, captured using the UltraCam-D camera covering an area of 1.45 $km^2$ at a ground sampling distance (GSD) of 15 (cm). Each image is at a size of 11500 × 7500 pixels, and an intersection angle between neighboring images ranges from 15 to 30 degrees. The provided images have a varying overlap, thus in total, we have 8 stereo pair images. The **ISPRS** airborne dataset also provides LiDAR point clouds from Optech's airborne laser scanner (ALS), which we converted to ground-truth disparity and DSM for training and testing. The Toronto dataset includes a mixture of small to high-rise buildings, as well as other classes such as roads, trees, grass, etc. For more details refer to Figure 7.1(b).

| OMA I | OMA II | OMA III |
|---|---|---|
| 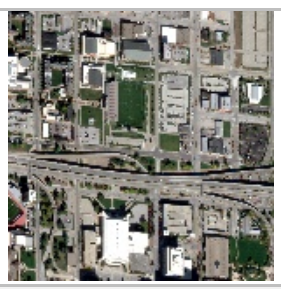 | 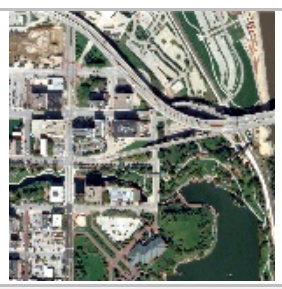 | 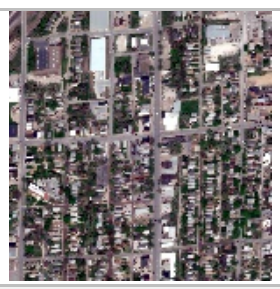 |
| Number of stereo pairs: 361<br>Image dimensions: 1168x1165<br>Sun angle difference range: 0°-27.7°<br>Intersection angle range: 1.53°-46.36°<br>Ground truth data: LiDAR | Number of stereo pairs: 293<br>Image dimensions: 1206x1202<br>Sun angle difference range: 0°-25.1°<br>Intersection angle range: 1.25°-45.89°<br>Ground truth data: LiDAR | Number of stereo pairs: 239<br>Image dimensions: 1187x1190<br>Sun angle difference range: 0°-25.3°<br>Intersection angle range: 1.43°-46.99°<br>Ground truth data: LiDAR |
| JAX I | JAX II | JAX III |
| 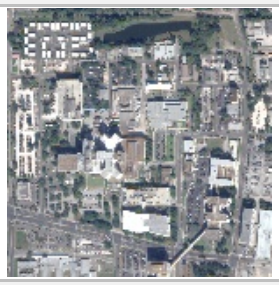 | 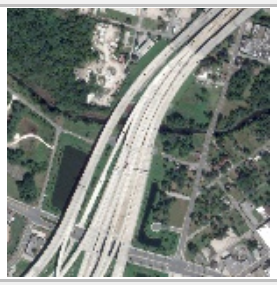 | 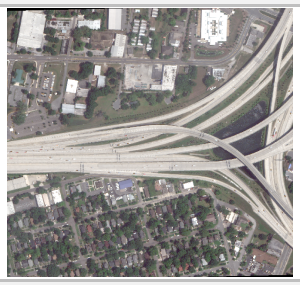 |
| Number of stereo pairs: 138<br>Image dimensions: 1163x1164<br>Sun angle difference range: 0.10°-50.90°<br>Intersection angle range: 2.64°-48.51°<br>Ground truth data: LiDAR | Number of stereo pairs: 91<br>Image dimensions: 1122x1152<br>Sun angle difference range: 0°-50.60°<br>Intersection angle range: 2.64°-48.51°<br>Ground truth data: LiDAR | Number of stereo pairs: 247<br>Image dimensions: 1133x1164<br>Sun angle difference range: 0°-55.30°<br>Intersection angle range: 2.64°-53.22°<br>Ground truth data: LiDAR |
| ARG I | ARG II | ARG III |
| 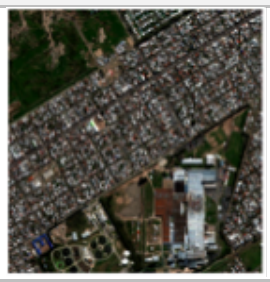 | 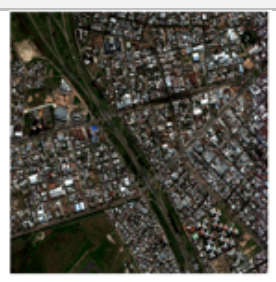 | 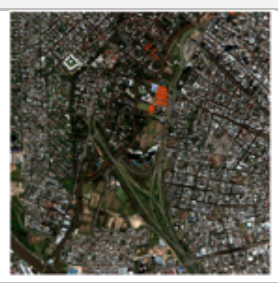 |
| Number of stereo pairs: 503<br>Image dimensions: 2565x2561<br>Sun angle difference range: 0.04°-49.02°<br>Intersection angle range: 1.42°-66.20°<br>Ground truth data: LiDAR | Number of stereo pairs: 475<br>Image dimensions: 4199x4256<br>Sun angle difference range: 0.02°-49.04°<br>Intersection angle range: 3.26°-54.55°<br>Ground truth data: LiDAR | Number of stereo pairs: 472<br>Image dimensions: 4062x4208<br>Sun angle difference range: 0.02°-49.04°<br>Intersection angle range: 1.44°-60.90°<br>Ground truth data: LiDAR |

(a) Satellite datasets stereo pair images information for OMA, JAX, and ARG test-sites.

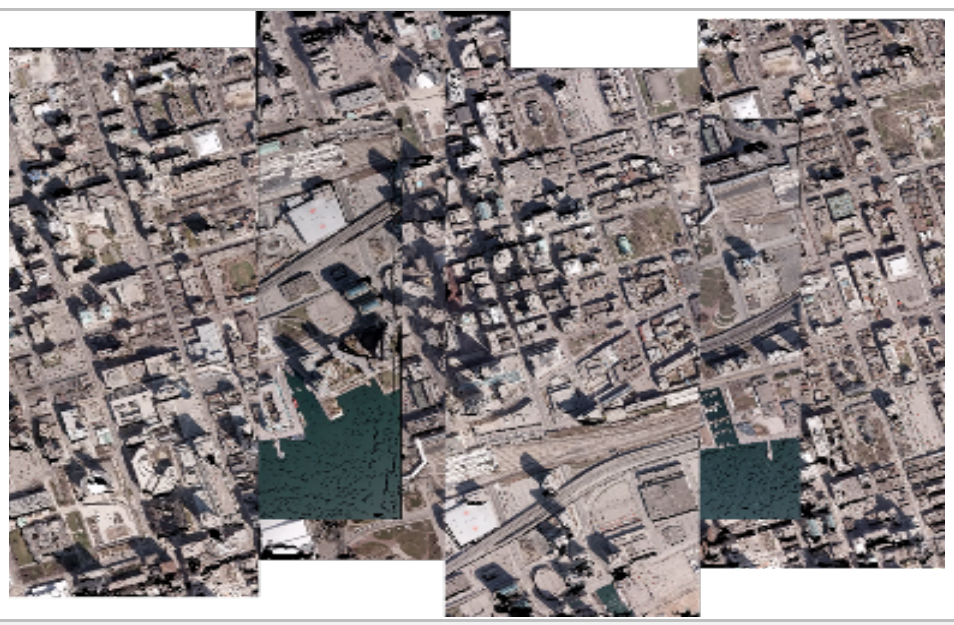

Number of stereo pairs: 8
Image dimensions: 11500 x 7500
Intersection angle range: 15°-30°
Ground truth data: LiDAR

(b) Airborne dataset stereo pair images information for Toronto, Canada test-site.

Figure 7.1. Information about the stereo pair images in the (a) satellite and (b) airborne datasets.

### 7.4.2. Ground Truth Data Derivation and Preprocessing

The ground-truth disparities for the satellite and airborne datasets are derived from LiDAR point clouds following three steps (Patil et al., 2019), first, the point clouds are aligned to the original unrectified images using correlating methods to register the images, and the LiDAR intensity data, such as using the mutual information (MI). The offsets between these two sources are further used to adjust their corresponding RPC parameters of the satellite images; second, the stereo pair images are rectified using their adjusted RPC models to the epipolar image space; third, the 3D LiDAR point clouds are projected to the satellite images and mapped to the rectified left and right images; finally, the disparity map is computed for both the left and right rectified image based on the projected 3D points. Obtaining disparity maps with sub-pixel accuracy from the LiDAR on the satellite images is very challenging. According to Wu et al. (2021), the transformation between LiDAR and ground truth disparity may produce additional errors. Whereas, Patil et al. (2019) report that the ground truth disparity in the 2019 DFC benchmark (Bosch et al., 2019) lacks adequate validation, thus may include errors that have not been acknowledged. A common source of error is the change of scene content between LiDAR and satellite images due to different acquisition times. As can be noticed in the example in Figure 7.2(a), the yellow rectangles show some trees in the ground truth disparity that are not recorded in the image. Other sources of errors include missing points in the ground truth disparity due to occlusions or shadows (see a blue rectangle in Figure 7.2(b)), random noise and artifacts (see red and green rectangles in Figure 7.2(a-b)), or systematic errors from the image rectification process. These errors will inevitably impact the training of the DL models.

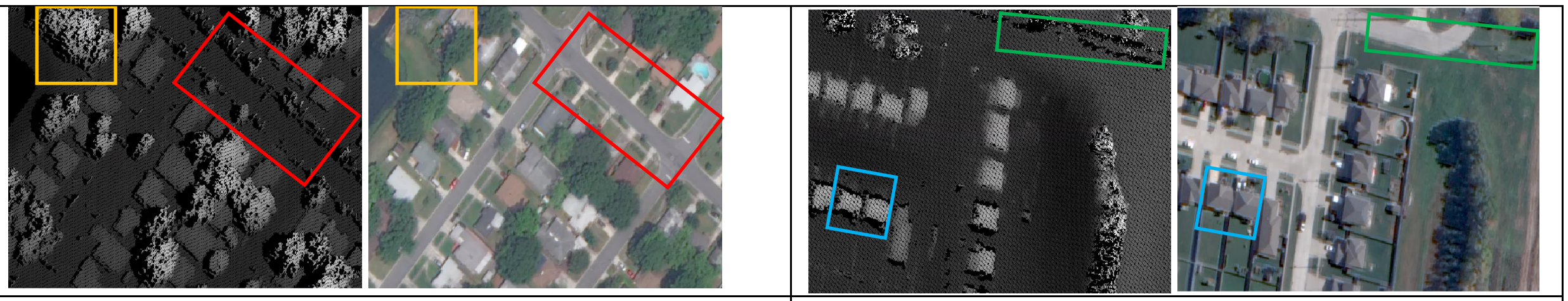

(a) Example I (b) Example II

Figure 7.2. Sample for the errors in the ground-truth disparity derived from LiDAR shown in (a) example I and (b) example II.

### 7.4.3. Stereo DIM Algorithms and Evaluation

As mentioned, we evaluate two types of DL approaches, 1) DIM methods following a classic multi-stage paradigm including feature extraction, cost matching, and cost aggregation for disparity computation, where DL component can be one of these stages, for example, the learning-based cost metric in our evaluation learns the feature extraction, or 2) DL methods with E2E learning which use trainable networks to learn from stereo data and ground-truth disparities, and directly predict disparity maps. All approaches require rectified epipolar images as inputs, which we generate using RPC Stereo Processor (RSP) software (Qin, 2016). As mentioned in section I, we use SGM for cost aggregation following Census and MC-CNN cost metrics as proposed by (Hirschmuller, 2005a, 2008b). In this subsection, we briefly introduce these stereo DIM algorithms used for our evaluations.

- *Cost Metrics for Classic Multi-Stage Stereo DIM Methods*

***Census*** maps a window of pixels across left and right images to find similar features with minimum cost (Hirschmuller, 2005a; Zabih & Woodfill, 1994). First, a fixed window (*w*) is set around every pixel in the left image and a sliding window scans the same epipolar line in the right image. The pixels in *w* are transformed into a vector of binary strings. Depending on their values relative to the central pixel, a value of 1 is assigned if a pixel is larger than the central pixel and 0 otherwise. The binary vectors are processed by a hamming distance and summed to compute the cost and indicate the number of variant pixels. A match is selected based on the minimum cost.

***MC-CNN*** uses a Siamese convolutional neural network to compute the matching cost. The training is performed using small patches of size 9×9 extracted from grayscale left and right images and their ground truth disparity. The testing, on the other hand, takes the entire image as input. During training, the network learns from each pair of patches by classifying them into similar or dissimilar. The patches are processed separately in two sub-networks that include four convolutional layers with a rectified linear activation function (ReLU), a convolutional layer, and a normalization layer. The sub-networks result in two normalized feature vectors that are fed to a cosine dot product to compute their similarity and indicate a good or bad match. MC-CNN has a small model capacity with a total number of parameters of 0.3 million (M).

- *E2E learning stereo DIM methods*

Three recently published E2E networks (with available codes) are described in this subsection. These methods vary in their model architecture, backbone building blocks, and model complexity (model capacity). Some of the architectures are designed to readily model the conceptual nature of depth sensing and recognition, such as the pyramid architectures (Chang & Chen, 2018) that process the data in a coarse-to-fine fashion. Understanding the model capacity is useful to determine the degree of generalization and applicability in practice. Generally, a network with a large number of parameters reduces the training error, increases the computational cost, and most likely has a high testing error, thus, leading to overfitting and poor generalization problems (Z. Huang et al., 2021).

***GC-Net*** begins by extracting unary feature maps from the left and right images. Then, it applies feature correlation at all disparity levels by concatenating both feature maps to build a 4D cost volume. The cost volume is then regularized via cost aggregation using 3D convolutional layers (CNNs). Finally, it uses regression to predict the final disparity map from the regularized cost volume (Kendall et al., 2017). The total number of model parameters for GCNet is 3.5M, which is the highest among the three selected E2E algorithms.

***PSMNet*** first introduces Spatial Pyramid Pooling (SPP) module to extract feature maps hieratically on multiple scales. The SPP module enhances visibility and feature matching of deformed objects (He et al., 2014). Then, it applies a stacked 3D hourglass module for cost aggregation and regularization, finally, it uses regression to estimate the disparity map (Chang & Chen, 2018). PSMNet has a total number of model parameters of 2.8M.

***LEAStereo*** uses a hierarchical neural architecture search (NAS) pipeline to find the optimal stereo matching network parameters (X. Cheng et al., 2020). The optimal parameters determined by NAS include filter size of the convolutional layer, strides, etc. The stereo matching network consists of two sub-networks the feature net to generate the feature maps and its 4D feature volume and the matching net to generate the 3D cost volume. Finally, regression is applied to compute the final estimates of the disparity map. LEAStereo has the least number of model parameters of 1.81M.

#### 7.4.4. Evaluation of the Stereo DIM

For evaluation and comparison of the stereo DIM algorithms, we triangulate the predicted disparity to DSMs. We assess each DSM *I(x,y)* against the ground-truth DSM from LiDAR *GT(x,y)* using the root mean squared error (RMSE) as follows:

$$RMSE = \sqrt{\frac{\sum_{i,j=1}^{N,M}(I(x,y) - GT(x,y))^2}{N \times M}} \quad (7.1)$$

Where *(x,y)* are the pixel positions and *(N, M)* are the DSM dimensions.

## 7.5. Results and Discussion

Based on the DSMs generated using different stereo DIM algorithms for the entire nine test-sites, we assess each class of the algorithms in the following three aspects:

1) Overall performance: we evaluate the RMSE of the resulting DSMs of each algorithm on each of the few hundred pairs based on the average RMSE, error distribution, completeness rate, and visual analysis.

2) The robustness with respect to varying acquisition configurations: we assess for each algorithm, the impact of two critical geometric/radiometric parameters, i.e., sun angle difference and intersection angle, on the resulting DSM through correlation analysis using hundreds of stereo pairs.

3) Generalization capability of DL algorithms: we evaluate for each of the DL methods, their RMSE against the ground truth by training the model on one dataset and applying to another dataset (of a different location and/or different sensor).

### 7.5.1. Training Data and Setup

- *Training Data:*

The satellite training dataset is from the 2019 DFC (track 2) benchmark. It has 4,293 rectified stereo images of size 1024 × 1024 with their ground truth disparity maps. Because of limited GPU memory, we crop the images to small patches with W=1248 and H=384, yielding a total number of 25,705 training patches. We then split the input data into 80% training and 20% testing. The airborne training dataset includes Toronto dataset from the ISPRS benchmark. There are 8 stereo images with a sufficient overlap each with a size of 11500 × 7500. We rectify these images and crop them to small patches of size W=1248 and H=384. The total number of patches is around 3000.

There are two types of DL methods based on our taxonomy (i.e., learning-based and E2E learning), and both are trained using two different datasets. MC-CNN is supposed to be trained using both the satellite and the airborne dataset. However, considering that the ground-level KITTI dataset should be more distinctively different, and it has been well-trained already as shared in its original work (Žbontar & LeCun, 2016), we therefore alternatively use the KITTI dataset and satellite dataset as more challenging testing. The same was attempted for E2E, while this overwhelmed the E2E algorithms to yield any meaningful results. Therefore, we only train the E2E networks using satellite and airborne datasets.

- *Training Setup*

The training parameters and settings for each DL network are shown in Table 7.1. The settings include preprocessing steps such as color normalization and random crop of the input images to smaller patch size, in addition to the parameters selected prior to training such as the type of optimizer, learning rate, batch sizes, etc.

Table 7.1. The training parameters and setup for the DL models

| Input parameters | MC-CNN | GCNet* | PSMNet* | LEAStereo* |
|---|---|---|---|---|
| **Color normalization** | Grayscale patches | RGB normalization | RGB normalization | RGB normalization |
| **Patch size** | 9 × 9 | 512 × 256 | 512 × 256 | 192 × 384 |
| **Optimizer** | Gradient Descent | Adam | Adam | Stochastic Gradient Descent |
| **Optimization parameters** | β = 0.9 | β1 = 0.9, β2 =0.999 | β1 = 0.9, β2 = 0.999 | β = 0.9 |
| **Learning rate (λ)** | λ=0.002 decay: 0.9 (after 10 epochs) | λ=0.001 | λ=0.001 | Cosine λ= 0.025 - 0.001 decay: 0.0003 |
| **Maximum disparity** | - | 192 | 192 | 192 |
| **Batch size** | 1 | 1 | 1 | 1 |
| **# Epochs for training using satellite dataset** | 10 | 35 | 10 | 57 |
| **# Epochs for training using airborne dataset** | - | 500 | 20 | 300 |

*E2E learning algorithms are implemented in Pytorch and trained on Nvidia GeForce RTX 2080 Ti.

### 7.5.2. The Overall Performance Analysis

Performance analysis is the key factor to assess the reliability of stereo DIM algorithms. We use standard statistical measures such as the average RMSE and distribution of errors to analyze the DSMs from the 2,681 stereo pairs over the nine test-sites. We are particularly interested in the achievable upper and lower bounds of the accuracy for all stereo pairs. We also provide an analysis of the completeness rate and visual results. Based on knowledge from prior literature, we expect DL algorithms to outperform traditional methods (i.e., Census+SGM) since they are more complex and trained on examples of ground-truth data.

Table 7.2(a) presents the average errors (RMSE) of the DSMs from all stereo pairs and test-sites. We can notice that E2E algorithms can achieve the minimum errors. This can be obvious from several aspects. First, they have minimum overall average errors as shown from PSMNet and LEAStereo where their values are 4.73 and 4.83 meters, respectively (see the last row in Table 7.2 (a)). Second, eight of nine test-sites have the lowest errors in one of GCNet, PSMNet, and LEAStereo (indicated by bold in Table 7.2 (a)). LEAStereo in particular has the most frequent lowest average RMSE among all testing sites. We can also notice that E2E algorithms can achieve the highest errors in some test-sites. Their performance varies drastically across different test-sites. Their average errors can range from 2 to 18 meters. For instance, the average RMSE for GCNet, PSMNet, and LEAStereo in JAX II is 2.73, 2.76, and 4.39 meters, respectively, while for OMA II it is high as 17.56, 8.99, and 8.84 meters, respectively. This implies that E2E methods do not predict disparity maps well from unseen data. Census+SGM and MC-CNN+SGM show opposite performance to E2E methods. They have more consistent average errors across different test-sites close to the overall average errors (See Table 7.2(a)). In addition, Census+SGM outperforms MC-CNN+SGM in terms of having a lower overall average error of 5.50 meters and five of nine test-sites with lower average RMSE (see Table 7.2 (a)). This negates the author's conclusion of MC-CNN+SGM outperforming Census+SGM. This may be due to the sensitivity of DL algorithms towards noise in the ground-truth disparity maps used for training the networks.

While E2E methods can achieve the lower and upper bounds of the average errors, their average standard deviation in Table 7.2(b) indicates higher consistency for the error distribution of stereo pairs in the same test-site. GCNet, PSMNet, LEAStereo have the lowest average standard deviations (less than 2.73) as can be seen in the last row in Table 7.2(b). This is because of their ability to learn similar context and geometry, which makes them invariant to changing radiometric properties of the stereo pairs. In contrast, Cenus+SGM and MC-CNN+SGM have higher standard deviations with averages of 4.08 and 3.40, respectively. This implies that similarity-learning algorithms are sensitive to the radiometric properties of the stereo pair images.

In general, E2E methods are able to achieve the absolute minimum errors. Their ability to learn from context makes them robust to varying radiometric properties of the stereo pairs. However, they can also have the absolute highest errors when learning from new unseen test-sites. In contrast, traditional and learning-based algorithms may have higher errors but can provide more consistent performance across different test-sites.

Table 7.2. The overall performance of the stereo DIM algorithms across different test-sites presented by the (a) average RMSE, (b) standard deviation of the RMSE, and (c) completeness rates of the DSMs.

| *Test-site ID* | *(a) Average RMSE (meters)* | | | | | *(b) Standard deviation of the RMSE* | | | | |
|---|---|---|---|---|---|---|---|---|---|---|
| | Census +SGM | MC-CNN +SGM | GC-Net | PSM-Net | LEA-Stereo | Census +SGM | MC-CNN +SGM | GC-Net | PSM-Net | LEA-Stereo |
| *OMA I* | 6.55 | 5.94 | 18.11 | 4.35 | 5.20 | 6.32 | 4.16 | 9.63 | 3.19 | 3.49 |
| *OMA II* | 6.52 | 7.46 | 17.56 | 8.99 | 8.84 | 4.71 | 4.31 | 7.61 | 5.22 | 4.77 |
| *OMAIII* | 3.32 | 3.96 | 3.30 | 2.55 | 2.14 | 2.20 | 2.04 | 1.47 | 0.75 | 0.50 |
| *JAX I* | 5.75 | 10.60 | 6.79 | 5.46 | 5.97 | 1.35 | 4.52 | 2.31 | 1.48 | 1.47 |
| *JAX II* | 5.19 | 7.32 | 3.31 | 3.30 | 3.23 | 4.72 | 3.03 | 0.19 | 0.21 | 0.19 |
| *JAX III* | 4.29 | 7.07 | 2.73 | 2.76 | 4.39 | 3.92 | 5.08 | 0.31 | 0.33 | 0.46 |
| *ARG I* | 5.31 | 5.07 | 4.25 | 5.03 | 3.88 | 4.14 | 3.39 | 1.16 | 4.08 | 2.21 |
| *ARG II* | 5.84 | 5.01 | 5.13 | 5.00 | 4.84 | 4.63 | 1.27 | 0.80 | 3.99 | 3.44 |
| *ARGIII* | 6.69 | 5.42 | 4.78 | 5.08 | 4.98 | 4.72 | 2.76 | 1.00 | 1.73 | 1.94 |
| *Overall Average errors* | 5.50 | 6.43 | 7.33 | 4.73 | 4.83 | 4.08 | 3.40 | 2.72 | 2.33 | 2.05 |

| *Test-site ID* | *(c) Completeness rate* | | | | |
|---|---|---|---|---|---|
| | Census +SGM | MC-CNN +SGM | GCNet | PSMNet | LEAStereo |
| *OMA I* | 89.29 | 87.94 | 92.53 | 92.42 | 97.42 |
| *OMA II* | 73.31 | 70.71 | 87.57 | 87.45 | 95.41 |
| *OMAIII* | 74.13 | 86.94 | 90.11 | 87.28 | 90.33 |
| *JAX I* | 90.29 | 86.37 | 93.6 | 93.54 | 93.54 |
| *JAX II* | 86.22 | 90.99 | 94.4 | 94.47 | 94.38 |
| *JAX III* | 87.08 | 81.71 | 95.64 | 95.28 | 99.29 |
| *ARG I* | 75.13 | 74.84 | 94.78 | 93.79 | 94.86 |
| *ARG II* | 63.1 | 79.22 | 89.99 | 89.23 | 95.74 |
| *ARGIII* | 81.53 | 87.66 | 94.32 | 87.08 | 93.81 |
| *Overall Average errors* | 80.01 | 82.93 | 92.55 | 91.17 | 94.98 |

Note: Bold indicates the optimal values among the DIM algorithms i.e., minimum average RMSE, minimum standard deviation, and maximum completeness rate. All DL models are trained using the satellite dataset from the 2019 DFC benchmark.

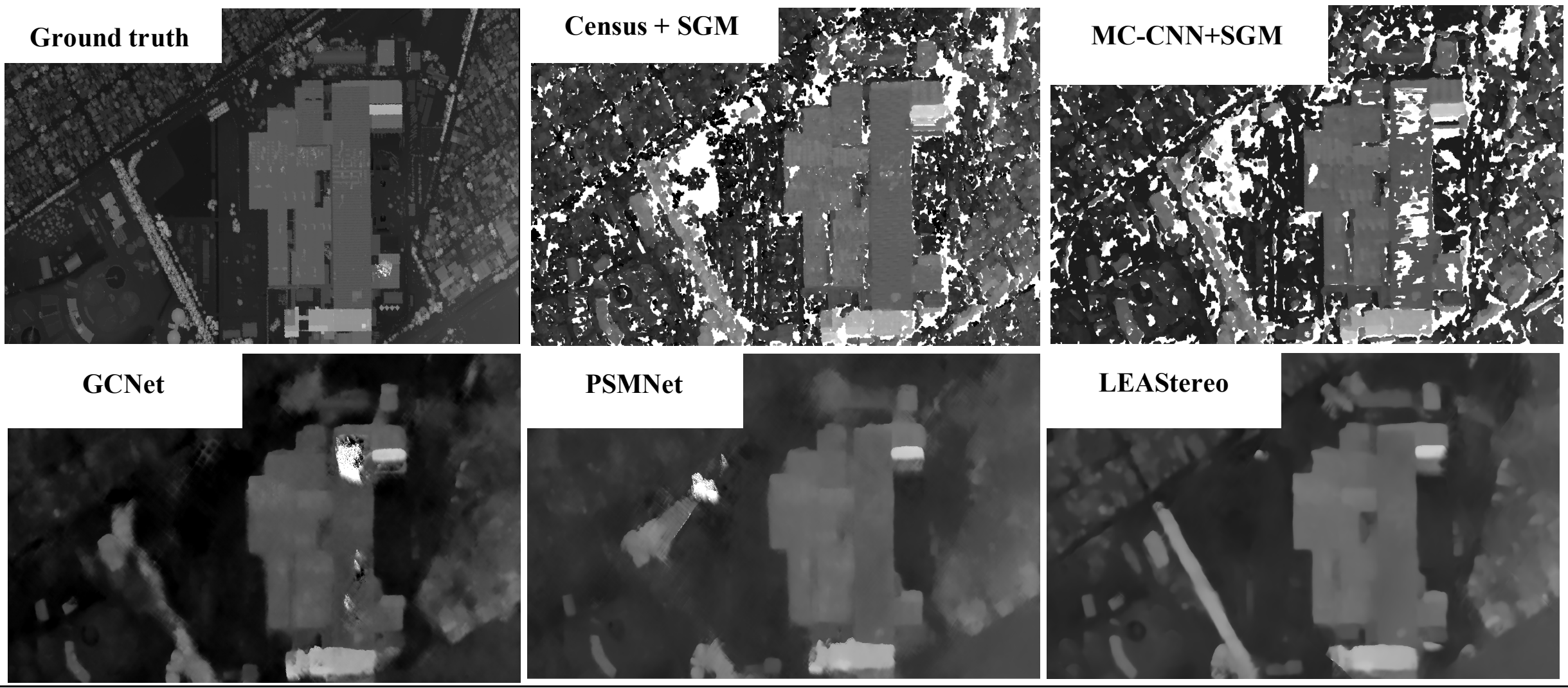

Figure 7.1. Visual illustration for sample DSMs from different stereo DIM algorithms for ARG I test-site. Note: the white pixels indicate missing values.

### 7.5.3. Robustness Analysis Using Varying Geometric Configurations of Stereo Images Pairs

The robustness measures the ability of the stereo DIM algorithm to maintain stable performance across stereo pairs with varying acquisition configurations. As reported by (Qin, 2019), geometric configuration parameters like the sun angle difference and intersection angle have a critical impact on the accuracy of DSMs generated using Census+SGM. Similarly, we aim to study their impact on the results of the DL stereo DIM algorithms. We use robust regression to fit a model between the RMSE of the DSMs and the geometric parameters (for all stereo pairs) to measure the correlation and eliminate potential outliers. The value of $R^2$ ($\in [0,1]$) represents the goodness of fit and the degree of impact. Thus, we use $R^2$ to indicate the amount of impact the geometric parameters have on the accuracy of DSMs as follows:

$$R^2 = \frac{S_0^2 - S_1^2}{S_0^2} \tag{7.2}$$

With $S_0^2 = \sum_{i=1}^{n}(y_i - \bar{y})^2$ and $S_1^2 = \sum_{i=1}^{n} {r_i}^2$, where $S_0^2$ is the sum of squares between the inputs $y_i$ and mean $\bar{y}$, and $S_1^2$ is the sum of residuals $r_i$. Given previous observations in (Qin, 2019), we expect DL algorithms to be impacted by varying acquisition configurations.

Table III presents $R^2$ showing the relationship between the RMSE of the DSMs and the geometric parameters for all stereo pair images in the test-sites. Census+SGM shows the highest $R^2$ scores between the RMSE of DSMs and each of the sun angle difference and the intersection angle with an average of 0.52 and 0.53, respectively. In addition, five to six test-sites have $R^2$ values between 0.53 and 0.82. This implies that traditional DIM methods are impacted by the geometric parameters. As for all DL algorithms, $R^2$ scores with respect to the sun angle difference are very small and their average does not exceed 0.37. While for $R^2$ with respect to the intersection angle, the learning-based MC-CNN is the least impacted by the intersection angle where it has the minimum average $R^2$ score of 0.36. On the other hand, the $R^2$ for E2E algorithms slightly increase to an average of 0.38, 0.45, and 0.48 for GCNet, PSMNet, and LEAStereo, respectively. Despite this minor increase, it is still insignificant in comparison to Census+SGM. This indicates that DL stereo DIM algorithms can be more robust to varying geometric configuration parameters that are associated with the stereo pair images

Table 7.3. Robust regression analysis denoted by $R^2$ to indicate the relationship between the DSM accuracy and geometric configuration parameters.

| ***Test-site ID*** | ***$R^2$ with respect to the sun angle difference*** | | | | | ***$R^2$ with respect to the Intersection angle*** | | | | |
|---|---|---|---|---|---|---|---|---|---|---|
| | **Census +SGM** | **MC-CNN +SGM** | **GCNet** | **PSM Net** | **LEAStereo** | **Census +SGM** | **MC-CNN +SGM** | **GCNet** | **PSM Net** | **LEAStereo** |
| ***OMA I*** | 0.80 | 0.50 | 0.02 | 0.53 | 0.39 | 0.82 | 0.50 | 0.23 | 0.55 | 0.44 |
| ***OMA II*** | 0.27 | 0.26 | 0.03 | 0.13 | 0.08 | 0.30 | 0.27 | 0.53 | 0.25 | 0.28 |
| ***OMA III*** | 0.31 | 0.20 | 0.13 | 0.17 | 0.11 | 0.53 | 0.29 | 0.40 | 0.10 | 0.06 |
| ***JAX I*** | 0.35 | 0.26 | 0.01 | 0.00 | 0.00 | 0.33 | 0.18 | 0.74 | 0.72 | 0.73 |
| ***JAX II*** | 0.53 | 0.03 | 0.21 | 0.08 | 0.09 | 0.48 | 0.12 | 0.21 | 0.19 | 0.43 |
| ***JAX III*** | 0.63 | 0.41 | 0.07 | 0.07 | 0.03 | 0.66 | 0.42 | 0.11 | 0.27 | 0.32 |
| ***ARG I*** | 0.62 | 0.67 | 0.11 | 0.50 | 0.58 | 0.60 | 0.57 | 0.14 | 0.49 | 0.61 |
| ***ARG II*** | 0.41 | 0.23 | 0.52 | 0.68 | 0.67 | 0.42 | 0.17 | 0.43 | 0.67 | 0.66 |
| ***ARG III*** | 0.76 | 0.71 | 0.70 | 0.80 | 0.79 | 0.62 | 0.69 | 0.67 | 0.81 | 0.77 |
| ***Average*** | 0.52 | 0.37 | 0.20 | 0.33 | 0.30 | 0.53 | 0.36 | 0.38 | 0.45 | 0.48 |

Note: Bold indicates the highest $R^2$ values. All DL models are trained using the satellite dataset from the 2019 DFC benchmark.

### 7.5.4. The Model Generalization Analysis for the DL Stereo DIM Algorithms

DL algorithms have difficulties generalizing to new unseen scenes. This is because of the significant differences in the characteristics of images such as intensities, illuminations, scene content, textures, noise level, etc. DL algorithms may have different generalization capabilities. To evaluate their generalization, we use similar and different training and testing datasets i.e., from different sensors and regions.

Due to the different architectures of the DL algorithms, we train and test the learning-based and E2E methods on similar and different datasets. Figure 7.4. illustrates the training and testing pipelines. For simplicity, we are going to refer to the 2019 DFC (track 2) dataset as the "satellite-training" dataset and for the OMA, JAX, and ARG datasets as the "satellite-testing" dataset. We train MC-CNN using KITTI and satellite-training datasets separately and test it on the satellite-testing dataset from the nine test-sites. While for E2E algorithms, we train them separately on airborne and satellite-training datasets and test them on the airborne dataset. Given that MC-CNN is a binary cost metric, we expect it will have good generalization. As for E2E algorithms, since we use similar training and testing datasets (i.e., similar scene content as roads, buildings, etc.), we expect them to generalize well.

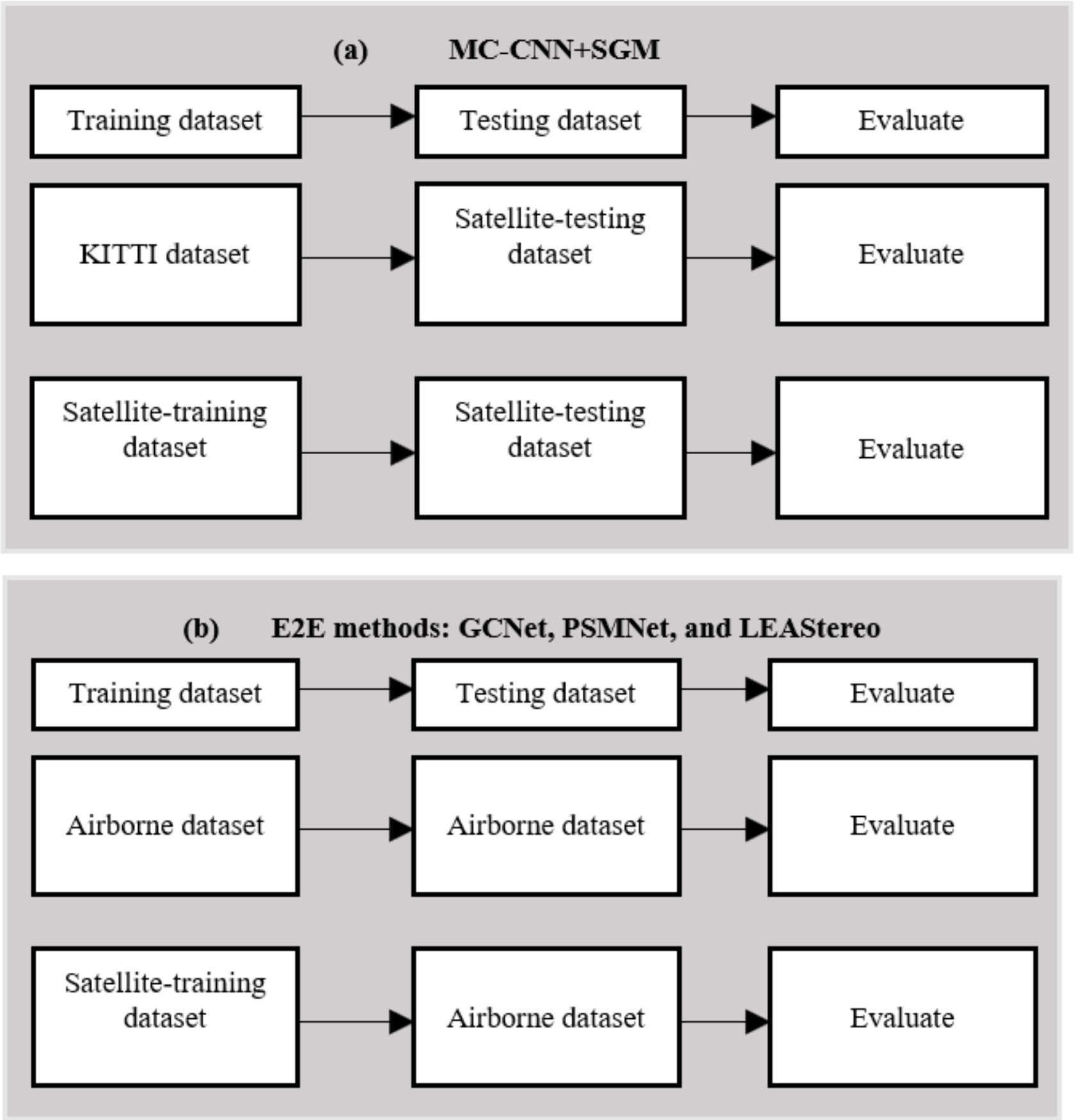

Figure 7.2. The training and testing pipelines for the generalization analysis of (a) learning-based and (b) E2E methods. Note: The satellite-training dataset refers to the DFC (track 2) benchmark, and the satellite-testing dataset refers to OMA, JAX, and ARG datasets.

In Table 7. 4(a), we compare the DSMs from MC-CNN+SGM trained independently on KITTI and satellite-training datasets and applied to the satellite-testing datasets from the nine test-sites. We observe a slight difference between the overall average RMSE of the two training models of 0.29 meters, with MC-CNN+SGM trained using the KITTI dataset performing slightly better. In addition, five test-sites have a lower average RMSE for the model trained using KITTI dataset than the satellite-training dataset. This may be due to some imperfections in the ground-truth disparity derived from LiDAR. Census+SGM seems to outperform MC-CNN+SGM (trained using KITTI and satellite-training dataset) in terms of the number of test-sites with the minimum average RMSE. This implies that Census-SGM is more stable than MC-CNN across different test-sites. Figure 7.5 shows an example for DSMs from both training models for OMA I and JAX II test-sites. It is obvious there is a large resemblance in the DSMs, however in OMA I, we can notice some errors randomly scattered over the DSM generated from the KITTI trained model. While the opposite in JAX II, more blobs and errors spread over the DSM generated from the satellite trained model. The results show a good generalization of MC-CNN+SGM and that it can perform well on different datasets. Furthermore, it indicates the importance of the quality of the training data which can directly impact the results and must be given greater attention.

In Table 7.4(b), we compare DSMs from E2E algorithms trained on each airborne and satellite-training dataset separately and applied to the airborne dataset. It is obvious that there is a significant jump in the average RMSE for GCNet and LEAStereo of about 20 to 45 meters difference between the results of airborne and satellite training models, where PSMNet shows about 6 meters difference. This indicates that the generalization degree varies based on the model architectures, and PSMNet shows significantly better results. Part of the reason is that its pyramid architecture can better adapt data with different resolutions and scales than the other two. Additionally, E2E algorithms trained on and applied to airborne dataset outperform traditional and learning-based algorithms, as they can achieve the lowest average errors (RMSE) and the most frequent minimum errors. The results from Table 7. 4(b) indicate that the source of the training data plays an important role. Example results are shown in Figure 6 for GCNet, PSMNet, and LEAStereo trained using airborne and satellite-training datasets and applied to the airborne dataset. It shows LEAStereo and GCNet perform poorly when trained using the satellite-training datasets. The PSMNet reaches better results but presents many noises and outliers (see red circles in Figure 7.6). This implies that E2E algorithms have a poor generalization capability and require domain-specific training datasets to achieve acceptable performance.

Table 7.4. The generalization analysis for the DL stereo DIM algorithms represented by (a) MC-CNN+SGM and (b) E2E algorithms.

(a) MC-CNN+SGM trained using satellite-training or KITTI datasets and tested on satellite datasets.

| ***Test-site ID*** | ***(a) Average RMSE (meters)*** | | | ***(b) Standard deviation for the RMSE*** | | |
|---|---|---|---|---|---|---|
| | ***Census +SGM*** | ***MC-CNN + SGM*** | | ***Census +SGM*** | ***MC-CNN + SGM*** | |
| | | ***Trained using satellite-training dataset*** | ***Trained using KITTI dataset*** | | ***Trained using satellite-training dataset*** | ***Trained using KITTI dataset*** |
| ***OMA I*** | 6.55 | **5.94** | 7.55 | 6.32 | **4.16** | 6.77 |
| ***OMA II*** | **6.52** | 7.46 | 7.28 | 4.71 | **4.31** | 5.02 |
| ***OMAIII*** | **3.32** | 3.96 | 3.81 | 2.20 | **2.04** | 2.53 |
| ***JAX I*** | **5.75** | 10.60 | 6.44 | **1.35** | 4.52 | 1.70 |
| ***JAX II*** | **5.19** | 7.32 | 5.38 | 4.72 | **3.03** | 3.68 |
| ***JAX III*** | **4.29** | 7.07 | 5.41 | **3.92** | 5.08 | 4.26 |
| ***ARG I*** | 5.31 | **5.07** | 6.84 | 4.14 | **3.39** | 5.90 |
| ***ARG II*** | 5.84 | **5.01** | 6.32 | 4.63 | **1.27** | 3.57 |
| ***ARGIII*** | 6.69 | **5.42** | 6.21 | 4.72 | **2.76** | 3.96 |
| ***Overall Average*** | 5.50 | 6.43 | **6.14** | 4.08 | **3.40** | 4.15 |

Note: Bold font indicates the minimum values. The satellite-training dataset refers to the 2019 DFC satellite benchmark.

(b) E2E algorithms trained using satellite-training or airborne datasets and tested on airborne datasets.

| ***Stereo pair ID*** | ***Census +SGM*** | ***MC-CNN+SGM*** | ***Trained using airborne dataset*** | | | ***Trained using satellite-training dataset*** | | |
|---|---|---|---|---|---|---|---|---|
| | | | **GCNet** | **PSMNet** | **LEAStereo** | **GCNet** | **PSMNet** | **LEAStereo** |
| *1* | 15.34 | 18.84 | 15.89 | 17.39 | 15.63 | 34.52 | 19.01 | 65.72 |
| *2* | 14.99 | 17.82 | 14.66 | 14.86 | 14.42 | 35.20 | 18.00 | 67.65 |
| *3* | 8.13 | 11.44 | 9.08 | 9.87 | 11.49 | 36.48 | 12.04 | 74.71 |
| *4* | 11.37 | 11.32 | 12.40 | 12.40 | 19.72 | 28.47 | 19.72 | 56.49 |
| *5* | 21.49 | 29.00 | 12.96 | 16.52 | 12.50 | 29.52 | 13.85 | 56.21 |
| *6* | 6.17 | 7.78 | 6.51 | 6.98 | 7.65 | 34.52 | 41.24 | 73.56 |
| *7* | 30.41 | 29.58 | 24.32 | 28.42 | 23.34 | 29.46 | 24.62 | 29.46 |
| *8* | 23.98 | 22.62 | 20.35 | 20.72 | 20.15 | 55.41 | 22.02 | 55.41 |
| ***Average RMSE (meters)*** | 16.49 | 18.55 | 14.52 | 15.90 | 15.61 | 35.45 | 21.31 | 59.90 |

Note: Bold indicates the minimum values.
MC-CNN+SGM is trained using the 2019 DFC satellite benchmark.
The satellite-training and airborne datasets used to train E2E models refer to the 2019 DFC and Toronto ISPRS benchmarks.

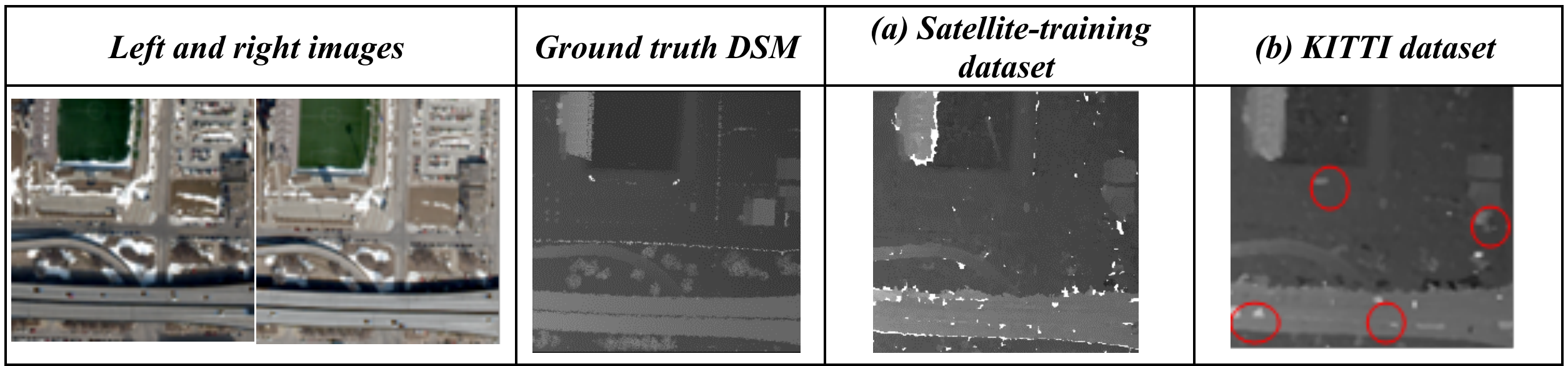

Figure 7.3. Sample DSMs for OMA I generated from MC-CNN+SGM trained using (a) satellite-training (the 2019 DFC benchmark) and (b) KITTI datasets.

| *E2E algorithms trained using* | *Left and right images* | *Ground truth DSM* | *GCNet* | *PSMNet* | *LEAStereo* |
|---|---|---|---|---|---|
| ***(a) Satellite-training dataset*** | 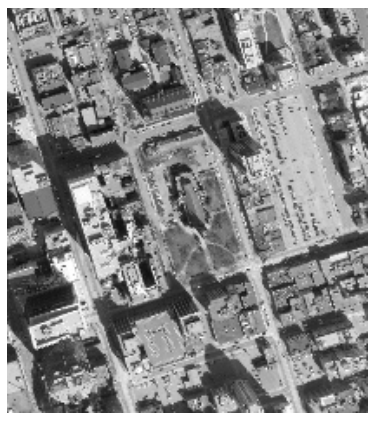 | 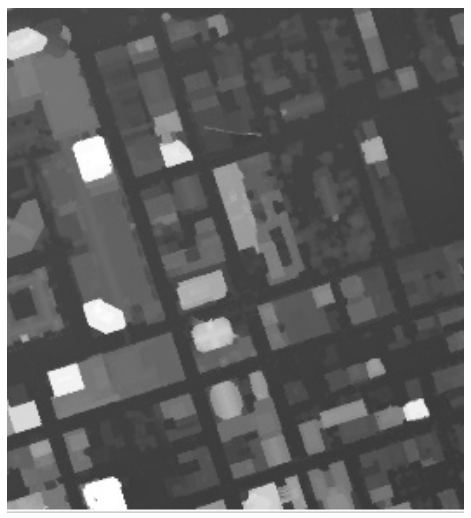 | 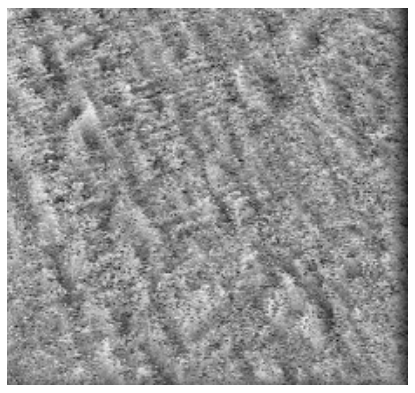 | 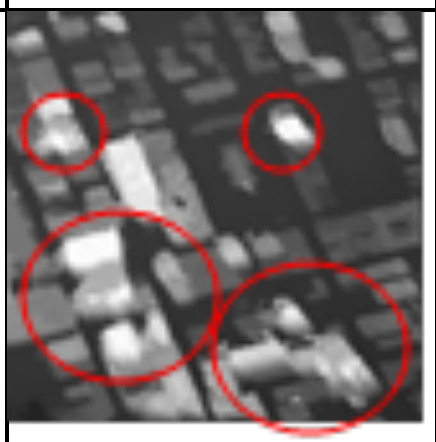 | 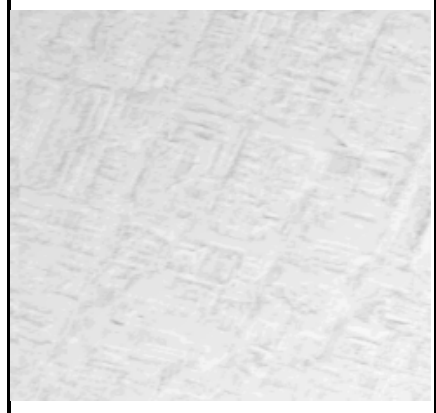 |
| ***(b) Airborne dataset*** | 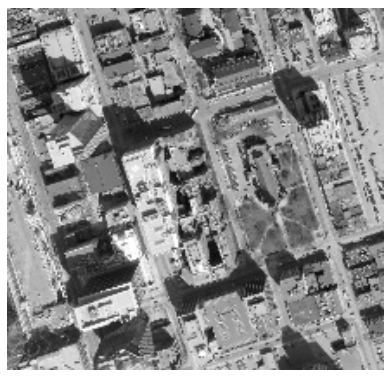 | | 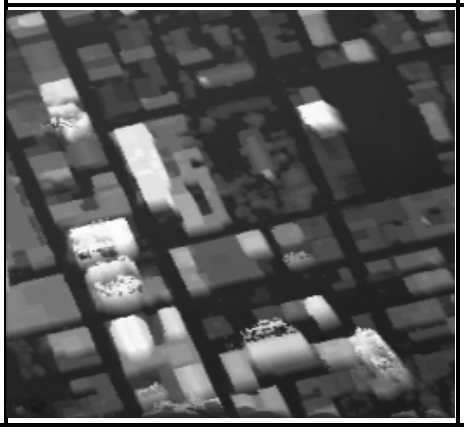 | 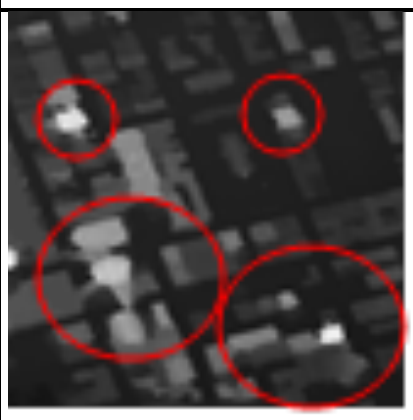 | 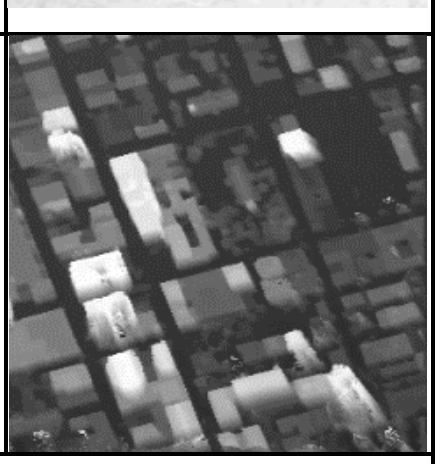 |

Figure 7.4. Sample DSMs for airborne stereo pair generated from E2E algorithms trained using (a) satellite-training (the 2019 DFC benchmark) and (b) airborne (ISPRS benchmark) datasets.

## 7.6. Conclusion

This work presents a comparative study of the traditional and DL stereo DIM algorithms, particularly for satellite datasets. We perform a comprehensive evaluation over a large volume of satellite data of 2,681 stereo pairs covering nine different regions around the world. We evaluate the stereo DIM algorithms on three bases, the overall performance, robustness to varying acquisition configurations, and generalization of the DL models. We study three classes of algorithms including: 1) traditional algorithms (e.g., Census+SGM), 2) DL learning-based algorithms (MC-CNN+SGM), and 3) E2E algorithms (e.g., GCNet, PSMNet, and LEAStereo). The results of the experiments suggest the following conclusions:

1) Overall, E2E methods outperform other types of stereo methods as they can achieve the lowest and most frequent minimum errors. They have also proven to have the highest completeness rates of DSMs and the most consistent performance within individual test-sites.

However, they can also achieve the highest errors in some test-sites, possibly due to the lack of generalization. We found LEAStereo as the top-performing algorithm, as it achieved the lowest RMSE for half of the tested pairs and has the highest completeness rates. In addition, E2E methods proved to be robust with respect to the acquisition configurations, as compared with traditional methods, and this may be largely due to their ability to learn contextual correlations. However, their poor performance on some of the pairs shows that these types of methods lack generalization. Although they achieve the best results in some of the pairs, their average RMSE over all pairs being lower than Census+SGM, suggests that E2E methods are still unpredictable when processing a large number of datasets.

2) Census+SGM as a traditional method, to our surprise, outperforms MC-CNN in terms of the overall average error and the number of test-sites having the minimum average RMSE. This shows an opposite conclusion to the original work of MC-CNN, the reason of which might be due to the noisy training data projected from the LiDAR point clouds.

3) MC-CNN+SGM (similarity-based training) proved to be robust towards acquisition configurations of the stereo pair images. In addition, due to its nature of learning similarity, has indicated experimentally in our work that it has superior generalization capability, as long as the training datasets are noise-free.

Although a well-generalized stereo matcher with superior performance is ideal, our experiments show that in practice, the choice of stereo matcher can be case-specific. It is likely a well-trained E2E network produce superior results, provided that the training datasets are available, and the testing datasets are of the same type as the training datasets. The similarity learning-based method has good generalization capability, and if well-trained by datasets, may outperform traditional stereo matcher, although may not be as competitive as the E2E methods. Future research may improve the weaknesses of DL stereo DIM algorithms, and potentially consider domain adaptation methods to address some of the challenges. As for the quality of the training data, since it is imperative to the performance of DL DIM algorithms, we recommend further investigation on its impact to understand their implications and provide potential solutions.

# Chapter 8. Finetuning Fine-tuning Strategy to Enhance the Accuracy of Digital Surface Models (DSMs)

This chapter is based on the paper “Fine-Tuning Deep Learning Models for Stereo Matching Using Results from Semi-Global Matching” that is published in the “The International Archives of the Photogrammetry, Remote Sensing and Spatial Information Sciences, Volume XLIII-B2-2022 XXIV ISPRS Congress (2022 edition” by (Albanwan & Qin, 2022).

## 8.1. Abstract:

Deep learning (DL) methods are widely investigated for stereo image matching tasks due to their reported high accuracies. However, their transferability/generalization capabilities are limited by the instances seen in the training data. With satellite images covering large-scale areas with variances in locations, content, land covers, and spatial patterns, we expect their performances to be impacted. Increasing the number and diversity of training data is always an option, but with the ground-truth disparity being limited in remote sensing due to its high cost, it is almost impossible to obtain the ground truth for all locations. Knowing that classical stereo matching methods such as Census-based semi-global-matching (SGM) are widely adopted to process different types of stereo data, we, therefore, propose a finetuning method that takes advantage of disparity maps derived from SGM on target stereo data. Our proposed method adopts a simple scheme that uses the energy map derived from the SGM algorithm to select high confidence disparity measurements, at the same utilizing the images to limit these selected disparity measurements on texture-rich regions. Our approach aims to investigate the possibility of improving the transferability of current DL methods to unseen target data without having their ground truth as a requirement. To perform a comprehensive study, we select 20 study-sites around the world to cover a variety of complexities and densities. We choose well-established DL methods like geometric and context network (GCNet), pyramid stereo matching network (PSMNet), and LEAStereo for evaluation. Our results indicate an improvement in the transferability of the DL methods across different regions visually and numerically.

## 8.2. Introduction

Recently, end-to-end deep learning (DL) networks have been the highlight for stereo matching tasks due to their ability to generate disparity maps directly using a pair of rectified images (Laga, 2019). They have been proven to provide high performance and accuracy due to learning contextual information (Chang & Chen, 2018; X. Cheng et al., 2020; Kendall et al., 2017). One of the essential criteria to evaluate DL methods in stereo matching is transferability (or generalizability), which refers to the ability to learn and predict from new environments that have not been seen previously by the model in the training dataset. Current studies show that DL stereo matching methods have poor transferability and generalization capabilities if there is a significant domain difference between training and testing datasets (Pang et al., 2018; Song et al., 2021), for example, satellite and ground view sensor images. Typically, we would assume that transferability will enhance if the datasets are from the same source as the sensor and have similar characteristics. However, for satellite images, the variation goes beyond color and geometry, the complexity and diversity of urban patterns vary greatly across cities around the world, where each region has its unique spatial distribution of buildings, landscapes, roads, etc., in addition, to the building's heights, shapes, etc. We strongly believe that these spatial variations limit the usage of DL methods in remote sensing tasks and may lead to poor performance and worse transferability.

Typically, DL methods require a large amount of training data to enhance transferability and provide good accuracies. For remote sensing data, even with satellite images becoming more accessible, the ground truth data (such as Light detection and ranging (LiDAR)) remain scarce. Some of the existing benchmarks provide the ground truth LiDAR for several regions around the world, but they are unevenly distributed around the world covering mostly major cities. With ground truth data being limited, current studies struggle to find alternative solutions for this problem, leaving this question open to be addressed.

As opposed to DL stereo matching algorithms, the Census cost metric has been proved to provide a good trade-off between the accuracy and robustness of extremely varying scenes and sensors (Chen et al., 2019; Loghman & Kim, 2013). However, it may not produce optimal results as DL methods when they are sufficiently trained and tested on similar datasets. Additionally, unlike DL stereo matching methods, Census requires cost aggregation using methods like semi-global matching (SGM) to further enhance the results. Given these properties of Census-SGM, we believe that it can be a good candidate to replace the ground-truth disparity map during the learning process. However, Census-SGM is known to perform poorly in some regions (e.g., texture-less regions) (Humenberger et al., 2010), therefore, masking reliable pixels using complementary information like SGM's energy map and the texture or edge map is important to be able to use Census-SGM as ground truth. Therefore, we propose a method that can learn directly from the target dataset using the disparity map from Census-SGM to only compare pixels with low uncertainty.

In this paper, we aim to tackle two major challenges of DL methods used in stereo matching and remote sensing satellite images, the transferability and lack of ground truth data. We first, analyze the transferability of the DL methods across 20 study sites around the world by considering three of the most used well-established methods including geometric and context network (GCNet), pyramid stereo matching (PSMNet), and LEAStereo. We then use the pre-trained satellite models for finetuning, where we only modify the loss function to compare the predicted and Census-SGM disparities around pixels with high confidence (i.e., low uncertainty). We evaluate our results qualitatively and quantitatively based on the triangulated digital surface models (DSM). In summary, our paper provides the following contributions:

- We analyze the transferability and performance of the DL methods used for stereo matching across satellite images from different regions.
- We provide a fine-tuning solution using census-SGM to improve the transferability/generalizability of DL methods by learning directly from the target dataset without requiring ground truth data.

The remainder of the paper is organized as follows, Section 7.2 discusses the related works and rationale of this paper, Section 7.3, provides an overview of the datasets and stereo methods used in this work, Section 7.4, the analysis, discussion, and results, and finally, we provide the conclusion and potential future works in Section 7.5.

## 8.3. Related Works

### 8.3.1. Deep Learning Methods for Stereo Matching

State-of-the-art DL methods have been reported to have remarkable progress in stereo matching. Unlike classical methods where disparity maps have to be generated in four different stages (i.e., cost matching, aggregation, disparity computation, and refinement) (Scharstein et al., 2001), deep learning networks have provided an alternative for a direct solution that integrates all steps into a single network to predict the disparity map using a pair of stereo images. (Mayer et al., 2016) are the first to propose a deep learning network for stereo matching, their method consists of 2D convolutional neural networks (CNNs) embedded in an encoder-decoder Siamese network; they applied their method on the KITTI dataset and proved to be able to get high accuracy than Census-SGM and DL networks as MC-CNN (Žbontar & LeCun, 2015). Later on, geometry was realized as an added value to the stereo matching problems, Kendall et al., (2017) in their geometry and context network (GCNet) have found that learning from geometry can greatly enhance the disparity map. The development of DL methods is still in progress and many current algorithms are inspired by previous advances in stereo matching (Chang & Chen, 2018; X. Cheng et al., 2020), however, due to their novelty, there still exist some limitations that need to be addressed. One of the major challenges for DL methods is the limited transferability and generalization capabilities (Pang et al., 2018; Song et al., 2021). As supervised methods, they cannot predict beyond the training data leading to poor performance around unseen instances. Additionally, they are often computationally expensive demanding a substantial amount of training time and data to assure high performance (Knöbelreiter et al., 2019) and good transferability. However, a large amount of training data is time-consuming and often unavailable for remote sensing data.

### 8.3.2. The Challenges of Remote Sensing Data Used in Stereo Image Matching

The outstanding performance of DL methods on standard benchmarks such as KITTI, Middlebury, and Sceneflow have inspired many other fields like remote sensing. Some studies have shown that they can have high accuracy than classic stereo matching algorithms when applied to satellite images (M. Chen et al., 2021), however, their performance is limited by the number of training datasets and computational time. Compared with ground-view, synthetic, or camera images, remote sensing images are more complex, and even with the continuous development of high-resolution sensors, remote sensing images are rich with information and dense in content leading to a more difficult stereo matching. The complexity in remote sensing images comes from various aspects including, 1) varying image content based on numerous regions around the world, which leads to highly varying spatial patterns, landscapes, textures, etc., 2) characteristics of objects in the image, for example, size, height, and shape of buildings (Qin, 2014), and 3) existence of occlusions from shadows or view angles (Qin, 2019). These variabilities in remote sensing images may highly compromise the performance and transferability of DL methods.

Another major challenge in remote sensing is the lack of ground truth data (Cournet et al., 2020). Typically, the ground truth disparity map for training the networks for stereo matching is often transformed and generated from Light Detection and Ranging sensor (LiDAR), however, due to its expensive price, it is impossible to collect LiDAR data for all locations around the globe. With this being an issue, this limits the capabilities of deep learning networks in many remote sensing tasks.

### 8.3.3. Transferability Solutions for DL Networks

It is a general practice that a model may be re-trained or fine-tuned when applied to unseen data. In the absence of the ground truth data re-training from scratch may not be a viable option. Instead of having ground truth data for training, studies have indicated that finetuning can be performed on alternatives to the ground truth data, for example, (Zhong et al., 2017) proposed a self-supervised DL method that used image warping errors for the loss function in the training stage. Similarly, (C. Zhou et al., 2017) proposed an unsupervised learning method that takes the loss from the left-right consistency check. Many other studies suggest unsupervised or self-supervised learning algorithms for finetuning the pre-trained model without requiring the ground truth data (Knöbelreiter et al., 2019; W. Yuan et al., 2021). Hence, in this work, we propose a training approach that allows our network to adapt to new datasets without demanding the ground truth. This can be applied by training the target domain with Census-SGM as an alternative for the ground truth. Census cost metric can work well in most environments and SGM can provide the energy map to indicate reliable pixels with low uncertainty.

## 8.4. Methodology and Analysis

In this section, we describe the data used in the experiment, present the relevant background information about the current stereo matching DL methods, and discuss the proposed finetuning method.

### 8.4.1. Dataset Description

***Source dataset*** is the training dataset that has available ground-truth disparity maps. We use this dataset for the pre-trained model. The source data is from the 2019 Institute of Electrical and Electronics Engineers (IEEE) Geoscience and Remote Sensing Society (GRSS) Data Fusion Contest (**DFC**) track 2 benchmark (Le Saux, 2019); it includes 4320 rectified satellite images from the Worldview-3 sensor with their ground truth disparity maps and dimensions of 1024 x 1024. We pre-process by normalizing and cropping the images to smaller patches of size 1248 x 384 to reduce memory consumption, thus, in total, we have 25,000 training patches, we divide the patches into 80% for training and 20% for testing.

***Target dataset*** is used for the proposed finetuning method. It is composed of stereo pair satellite images from different sensors and regions around the world. We select 20 study sites from around the world to cover a wide range of land covers, building types, shapes, heights (e.g., residential areas, industrial areas, etc.), densities, and complexities. Our study sites cover a few sub-areas in 1) Omaha, Nebraska, USA, 2) Jacksonville, Florida, USA, 3) near San Fernando Argentina, 4) London, England, 5) Haiti, and 6) Rochor and Punggol, Singapore. The stereo pair images are very high resolution (VHR) satellite images from different sensors including Worldview-3, Worldview-2, IKONOS, and GeoEye-1. We pre-process the stereo pair images and unify the spatial resolution by up-sampling to 0.3 meters, we then rectify each pair of images to be processed by the stereo matching algorithms. For more information about the datasets refer to Figure 8.1.

For finetuning in the target domain, we use the stereo pairs from target datasets with their Census-SGM disparity and energy maps. We finetune using 571 training and 393 testing patches of size (1248 x 384). Since we are using the pre-trained model from the source data for finetuning we only need a small sample from the target data to finetune.

Some benchmarks provide the ground-truth DSM which we use for evaluation of the proposed finetuning methods. The study sites with the ground truth DSM include Omaha and Jacksonville datasets from the 2019 DFC benchmark, Argentina dataset provided by the **IARPA** (The Intelligence Advanced Research Projects Activity) Multiview stereo 3D mapping challenge, and London dataset.

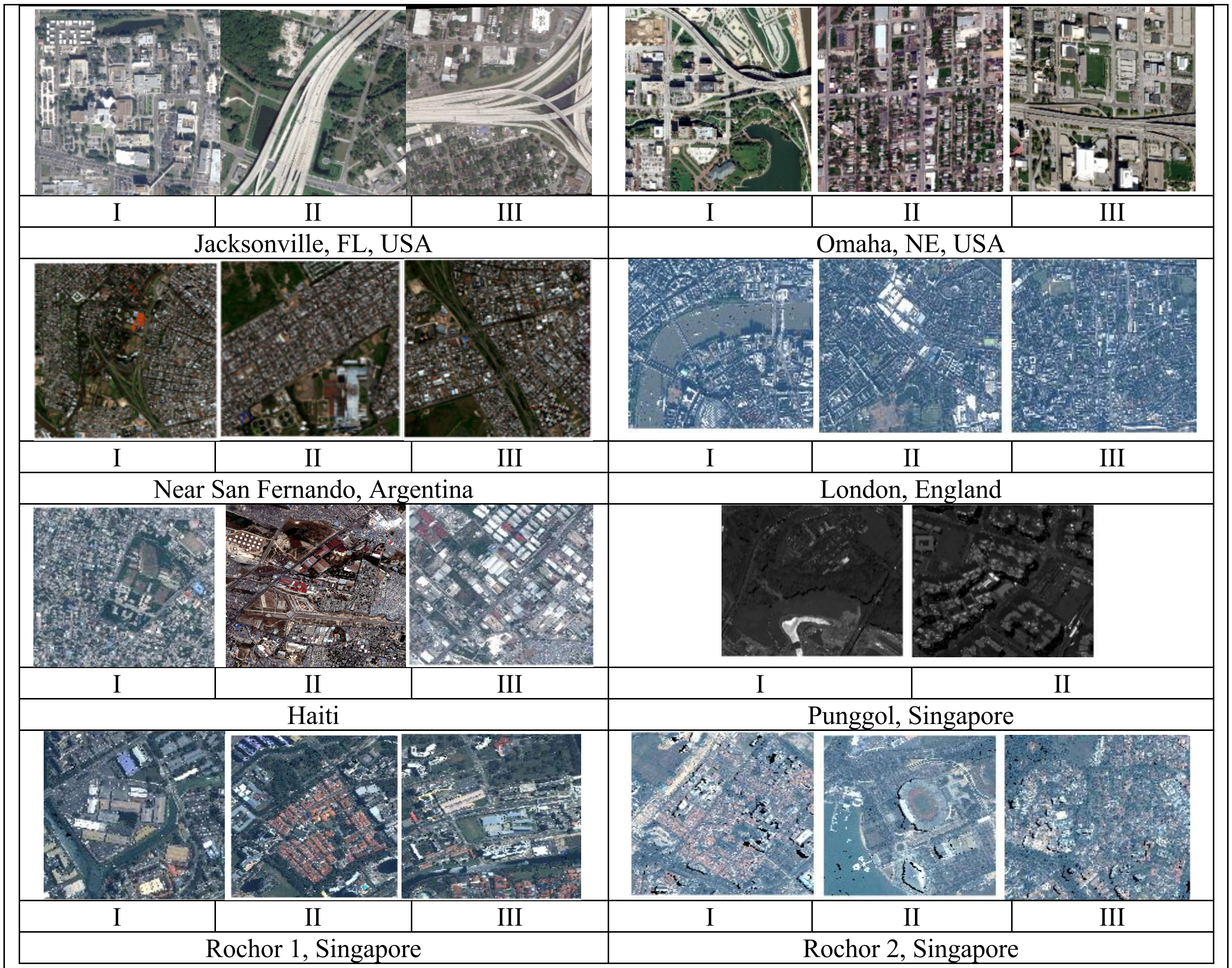

| Study site | Sensor | Ground truth DSM |
|---|---|---|
| Jacksonville, FL, USA | Worldview-3, spatial resolution 0.3 m | Available |
| Omaha, NE, USA | Worldview-3, spatial resolution 0.3 m | Available |
| Near San Fernando, Argentina | Worldview-3, spatial resolution 0.3 m | Available |
| Haiti | Worldview-2, Geo-EYE, and IKONOS, spatial resolutions 0.46, 0.51, and 0.82 m | - |
| London, England | Worldview-2, spatial resolution 0.46 m | Available |
| Punggol, Singapore | Worldview-2, spatial resolution 0.46 m | - |
| Rochor I, Singapore | IKONOS, spatial resolution 0.82 m | - |
| Rochor II, Singapore | IKONOS, spatial resolution 0.82 m | - |

Figure 8.1. Data description

### 8.4.2. The Stereo Matching Deep Learning Methods

DL methods provide direct solutions to stereo matching problems, where it takes a pair of rectified images and outputs the disparity map without further optimization as cost aggregation. We use three common DL methods that have been reported to have an outstanding performance in stereo matching computer vision tasks including CGNet, PSMNet, and LEAStereo. In this section, we briefly explain the mechanism of these methods.

- *Geometry and Context Network (GCNet)*

GCNet is developed by Kendall et al. (2017); they develop a Siamese network for their 2D convolution with shared weights to extract unary features from pair of images and generate the 4D cost volume (height, width, disparity, and feature size). The high dimension cost volume enables geometry to be preserved during cost matching. Additionally, instead of using subtraction or distance metrics to compute the cost, they rely on feature concatenation from both images, allowing better performance. They also implement a regularization step using 3D convolution to produce better feature representation using image dimensions and the disparity as an additional dimension.

- *Pyramid Stereo Matching (PSMNet)*

PSMNet learns the cost volume by integrating contextual feature cues from different scales; they construct spatial pyramid pooling (SPP) module to expand learning from pixel-level features to regional-level features on various scales. They build a stack hourglass module to learn and regularize the 4D cost volume using 3D convolutional layers.

- *LEAStereo*

As an advancement to previous methods, LEAStereo builds an automated neural architectural search (NAS) module to find optimal parameters (e.g., filter size, strides, etc.) for the convolutional layers in the networks. Their stereo matching algorithm consists of a feature extraction network and cost matching network.

### 8.4.3. Proposed Finetuning Method

- *Semi-Global Matching (SGM)*

SGM is originally proposed to impose a smoothness constraint on the matched cost. Its energy function can be expressed as follows:

$$E(D) = \sum_x (C(x, d) + \sum_{q \in N_p} P_1 T[|d_x - d_y| = 1] + \sum_{q \in N_p} P_2 T[|d_x - d_y| > 1]) \quad (8.1)$$

Where the $C(x, d)$ is the computed cost for the pixel at position x and disparity d. The $P_1$ and $P_2$ are the penalty parameters defined by users and are applied to regularize the cost function for all pixels $q$ in the neighborhood $N_p$. $P_1$ is a constant value for pixels, it accounts for pixels on slanted regions, whereas $P_2$ is proposed to impose large values on pixels discontinuities.

The key idea for SGM (Hirschmuller, 2005b, 2008a) is to aggregate the matching cost iteratively from multiple directions. The cost aggregation can be applied in eight or sixteen directions depending on the desire of users. The eight directions can provide fair results and a fast process, while the sixteen directions can produce more accurate results but at the expense of speed. The accumulated cost from SGM into a single pixel can be expressed as follows:

$$S(x, d) = \sum_r L_r(x, d) \quad (8.2)$$

Where $S(x, d)$ is the total accumulated cost from all costs $L_r(x, d)$ at different directions *r* for pixel at position x and disparity d. Each cost is computed as follows:

$$L_r(x,d) = C(x,d) + \min\Big\{ L_r(x-r,d),\ L_r(x-r,d-1) + P_1,\ L_r(x-r,d+1) + P_1,\ \min_i L_r(x-r,i) + P_2 \Big\} - \min_k L_r(x-r,k) \tag{8.3}$$

Where the $C(x,d)$ is the computed cost for the pixel x, $P_1$ and $P_2$ are the penalty parameters that are user-defined and are applied to regularize the cost function. We refer to the minimal aggregated/accumulated cost $S(x,d)$ as the energy map in which we use to guide the training of the DL algorithms. A sample for the energy map is shown in Figure 8.2.

- *Proposed Finetuning with SGM*

We suggest a generic finetuning approach that can be applied to any DL stereo matching algorithm. We introduce a method that can learn from the target data without having their ground truth disparity maps for training. Instead, we use Census-SGM as the ground-truth disparity. Census-SGM provides a good trade-off for robustness and accuracy, as it has stable performance across datasets of different sensors and regions. However, as a prediction method, it does not work well in some areas like flat and texture-less regions. Therefore, we propose a weighted loss function that assigns weights based on the importance of the information. We extract confidence maps which we regard as the most important features we want to learn from. We have two confidence maps, 1) the edge map which we extract from the left image using the Canny edge detector to get the edges and textured regions, and 2) the energy map from Census-SGM to indicate pixels with low uncertainty. For information about the confidence, maps refer to Figure 8.2.

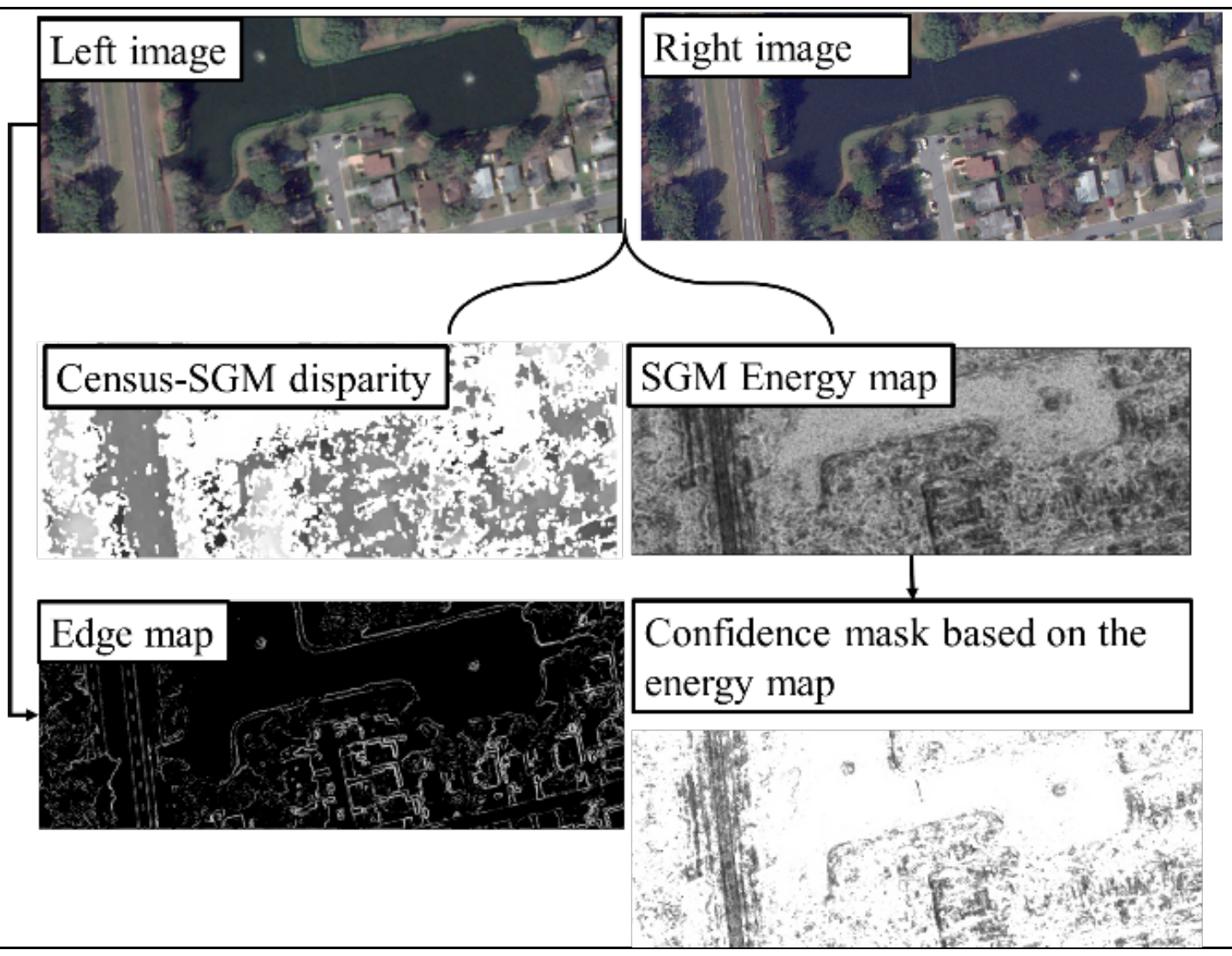

Figure 8.2. Confidence maps extracted from Census-SGM's energy map and the edge map from the left image

In addition to training using the target dataset, we have also applied some data augmentation on some of the source datasets for regularization and smoothness of the results. The architecture of the networks remains intact, the only modification we apply is to the loss function. The modified loss function is expressed as follows:

$$Loss = Loss_{augumneted\ source} + Loss_{target} \quad (8.4)$$

Where the $Loss_{augumented_source}$ computes the L1 loss function between the augmented source dataset and their provided ground truth disparity. The $Loss_{target}$ computes the loss for the target data by comparing the predicted and Census-SGM disparities. We use Huber loss for the target loss because of its robustness to outliers. To optimize the loss function, we divide our target loss into three parts: first, the loss for all pixels in the predicted disparity against the Census-SGM disparity, which can be expressed as follows:

$$Loss_1 = huber_loss \begin{bmatrix} Census\ disparity, \\ predicted\ disparity \end{bmatrix} \quad (8.5)$$

Second, the loss of pixels with high confidence based on the energy map. We apply a threshold on the energy map to mask the pixels with low uncertainty. We then use the Huber loss function as follows:

$$Loss_2 = huber_loss \begin{bmatrix} Census\ disparity(mask_{energyemap}), \\ predicted\ disparity\ (mask_{energyemap}) \end{bmatrix} \quad (8.6)$$

Where $mask_{energymap} = ENERGY_MAP < threshold$. The threshold is chosen empirically and set to 2500. Third, the loss for pixels with high confidence is based on the texture and edge map which we extract using canny edge detection. The Huber loss is then computed based on the valid pixels masked from the edge map as follows:

$$Loss_3 = huber_loss \begin{bmatrix} Census\ disparity(mask_{edgemap}), \\ predicted\ disparity\ (mask_{edgemap}) \end{bmatrix} \quad (8.7)$$

We combine these three losses into the target loss. Since Census-SGM does not work well for all areas (e.g., flat, texture-less, etc.), we further optimize the loss function by imposing weights based on the importance of information and the likelihood that it will have high confidence. We assign low weights for $Loss_1$ and high weights for $Loss_2$ and $Loss_3$ weights. The final target function can be expressed as follows:

$$Loss_{target} = w_1 * Loss_1 + w_2 * loss_2 + w_3 * loss_3 \quad (8.9)$$

The weights are determined empirically and set to $w_1 = 0.1, w_2 = 0.45, w_3 = 0.45$.

### 8.5. Experiment and Results

To evaluate the transferability of DL stereo matching algorithms using satellite images, we triangulate the disparity maps to DSMs. First, we visually analyze the transferability of DL algorithms across different study sites before finetuning. Then, we present visual results and numerical evaluation for before and after finetuning by computing the root mean square error (RMSE) against the ground truth DSM.

### 8.5.1. DL Training Setup

GCNet and PSMNet are implemented in Pytorch, they use Adam optimizer with β1=0.9 and β2=0.999, the initial learning rate of 0.001, and training and testing batch size of 1. For training, the images are randomly cropped to 256 x 512. As for LEAStereo, it is also implemented in Pytorch, the training images cropped into 288 x 576, we use SGD optimizer with β1=0.9, a cosine learning rate ranging from 0.025 to 0.001, and a batch size of 1.

### 8.5.2. Visual Analysis Before Fine-Tuning

The visual analysis allows viewing and inspecting the results based on the visible characteristics to make informative decisions and determinations about behavioral changes, consistency, variability, and overall accuracy of the data. In this section, we present the DSMs generated by GCNet, PSMNet, and LEAStereo before finetuning across different study sites, we expect their performance to vary based on region.

We present a visual comparison of the DSMs generated by Census-SGM, GCNet, PSMNet, and LEAStereo from different areas in the world in Figure 7.3. Our results show that DL methods work well for areas with sparse spatially distributed objects that have large buildings, as can be seen from Omaha, Jacksonville, and Rochor, Singapore study sites in Figure 7.3, where we can see clear outlines for different classes. While for dense areas with many buildings as in London, Argentina, and Haiti study sites, we can see that the predictions are not as meaningful and distinct as in the other study sites. For example, in the London study site, we can barely distinguish between different classes in the DSMs of the three DL algorithms (see last raw Figure 8.3). In some cases as in the Argentina study site, we can see that the DSM is very smooth, especially around dense residential areas, this can be apparent in PSMNet and LEAStereo, while for the Haiti dataset with even more dense areas having small-sized residential houses, the prediction of DL algorithms is worst. In contrast, we can see that Census-SGM has stable performance across all presented study sites, and even with dense areas, we can still distinguish between different classes as in London and Haiti study sites we can clearly differentiate between roads, ground, and buildings, while in the DL algorithms they are barely recognizable. Therefore, we can conclude that the performance of DL stereo matching algorithms is highly sensitive to the variances in sensors and regions.

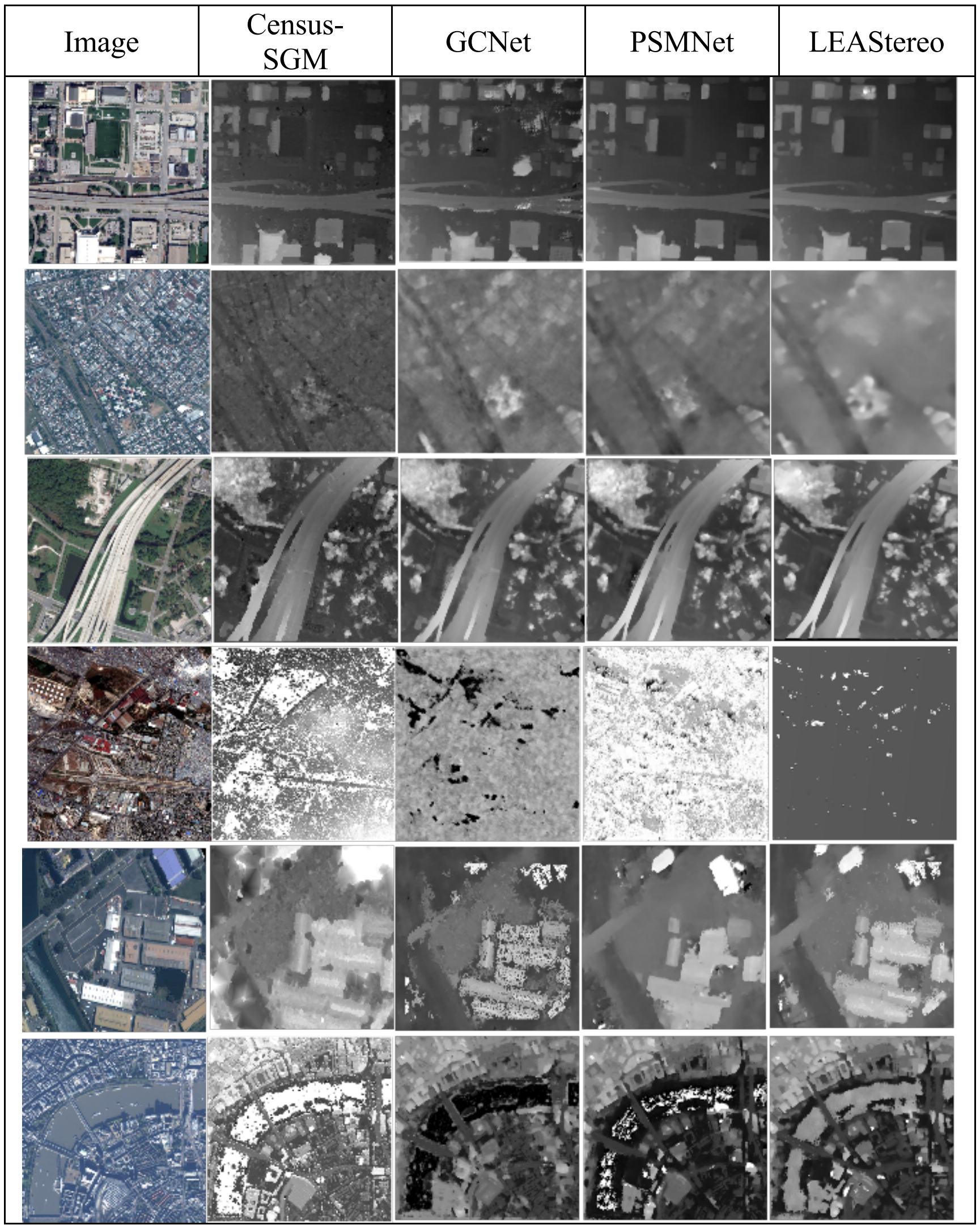

Figure 8.3. Visual analysis for the transferability of the DL algorithms across different study sites. Note: the rows from 1-6 show Omaha I, Argentina III, Jacksonville II, Haiti II, Rochor I Singapore, and London study sites, respectively.

### 8.5.3. Finetuning Results

We evaluate the DSMs generated by the GCNet, PSMNet, and LEAStereo before and after the proposed finetuning. We analyze the results visually and numerically by evaluating the DSMs and their RMSE for the study sites that have the ground-truth DSM available including Omaha, Jacksonville, Argentina, and London. We anticipate that the level of details will enhance as we add more information from the Census-SGM energy map and disparity map.

We present sample DSMs for the before and after finetuning the DL algorithms for London and Haiti study sites in Figure 8.4. and Figure 8.5, respectively. We can notice that the level of details in the London study site enhanced significantly which is visible from the enlarged highlighted region in red. Before finetuning the roads and small buildings are very blurry and almost impossible to distinguish, whilst after finetuning with Census-SGM, we can see that the different classes outlined clearly like the ground, roads, and small buildings. For the Haiti study site, before finetuning the DSMs are inaccurate and invisible (see Figure 7.5), however, after finetuning, we can see the details of the city more obvious in all DSMs from GCNet, PSMNet, and LEAStereo (see red rectangle in Figure 8.5). This indicates that finetuning with Census-SGM and target data can indeed enhance the results.

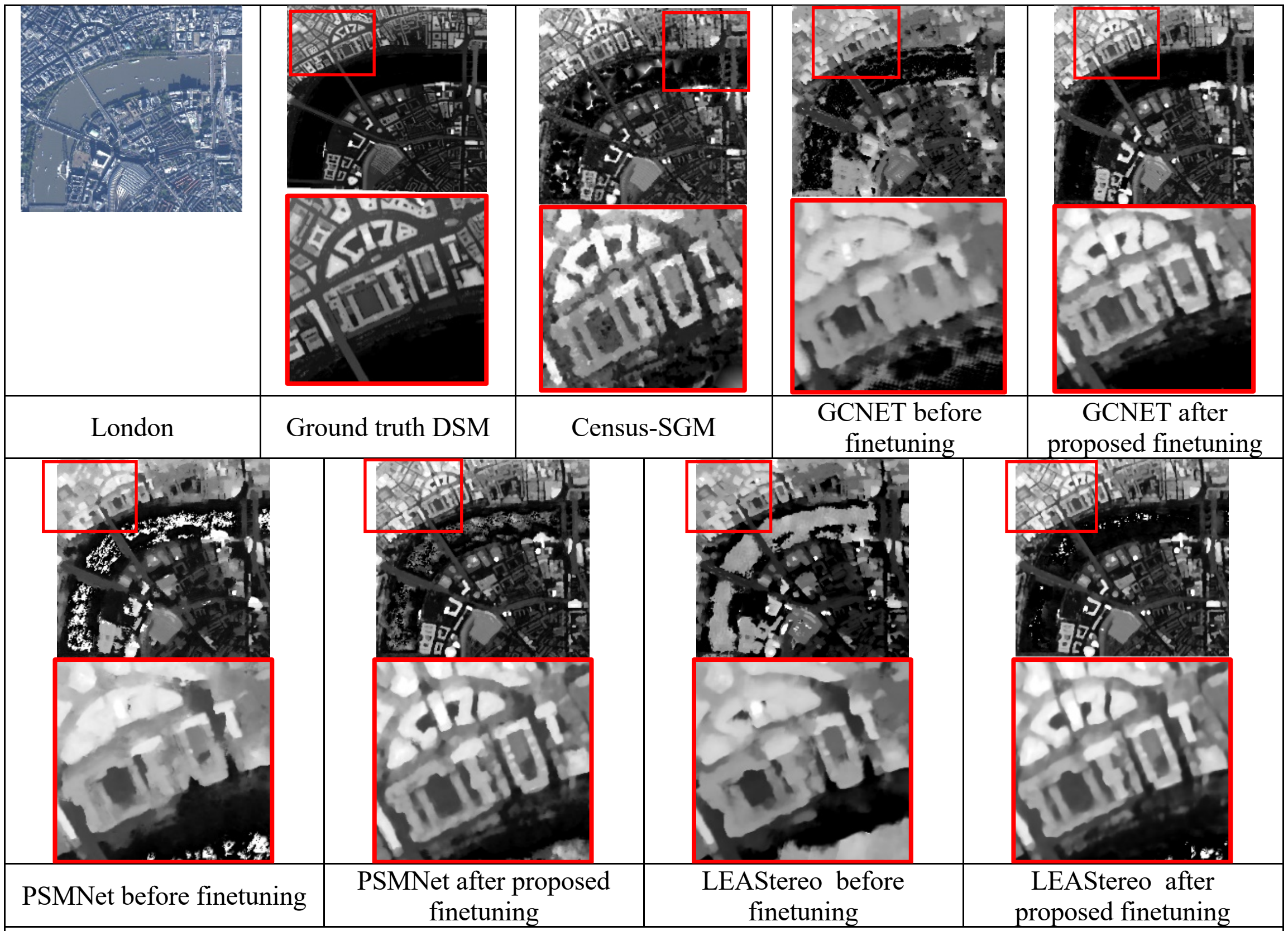

Figure 8.4. The DSMs from London I study sites before and after the proposed finetuning method. The enlarged image in the red rectangles indicates areas with a major improvement.

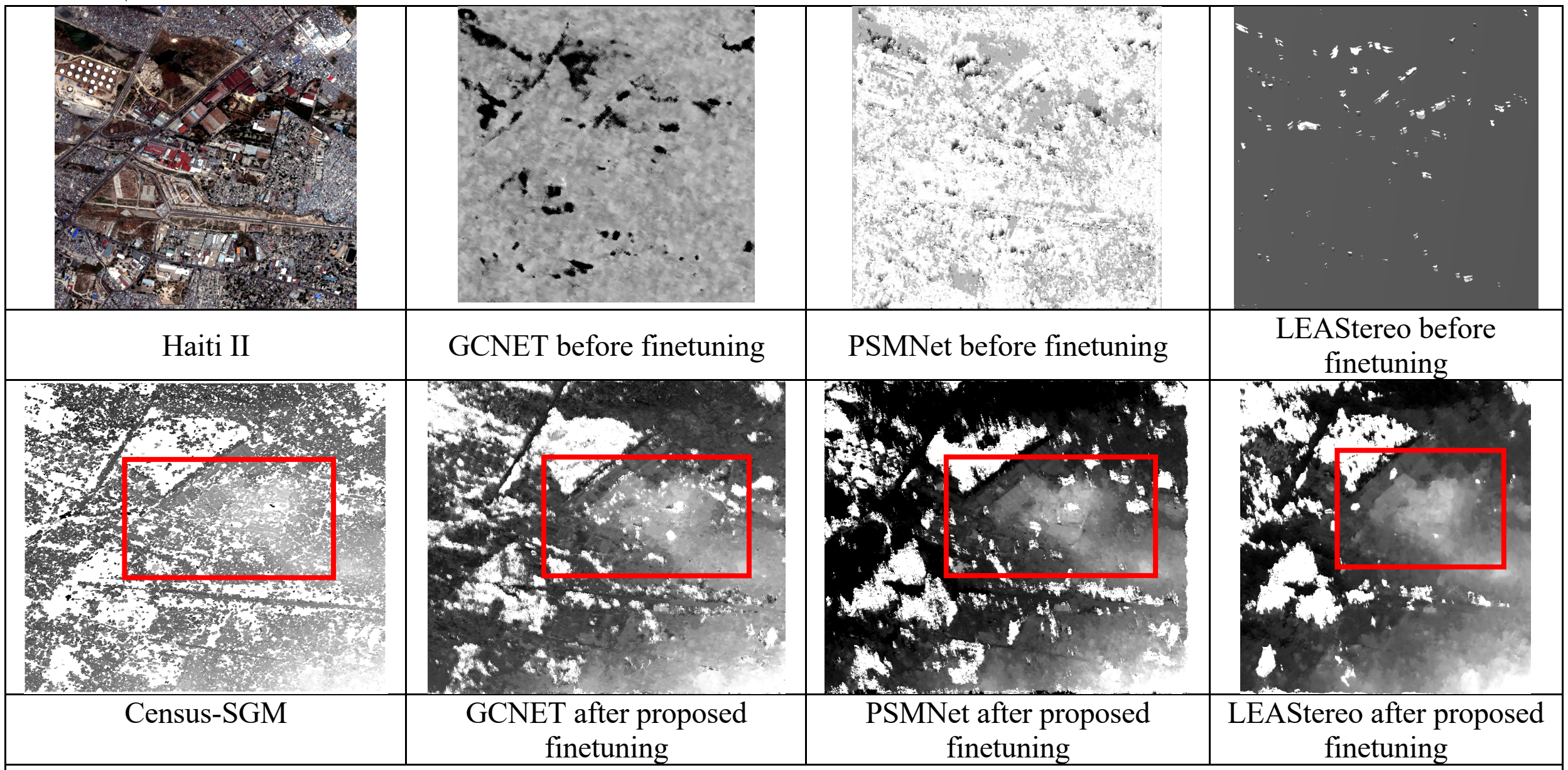

Figure 8.5. The DSMs from the Haiti II study site before and after our proposed finetuning method. The red rectangles indicate areas with a major improvement.

We present the numerical evaluation of the RMSE for the DSMs generated by GCNet, PSMNet, and LEAStereo before and after the proposed finetuning approach in two parts: first, the profile that shows the alignment between the ground-truth DSM and the DL algorithms before and after finetuning, second, Table 8.1 to show the error (i.e., RMSE) for the DSM in each study sites with the overall average RMSE.

Figure 8.6. shows the DSM profile for a line segment from London I study site comparing the ground-truth DSM and the DSMs from GCNet, PSMNet, and LEAStereo before and after finetuning. From the profile of GCNet and PSMNet, we can notice that the DSMs after finetuning become close to the shape and pattern of the ground truth DSM (See red and green lines), whereas before finetuning the blue line is almost straight and does not reflect much of the true values or patterns form the ground-truth DSM. LEAStereo profile looks much better than before finetuning, and it shows more matching patterns to the ground-truth DSM than GCNet and PSMNet, this is obvious in the middle section from curves showing the drops between the edges and boundaries and flat regions. We can see a clear separation of DSM values.

Table 8.1 presents the errors (RMSE) for the DSMs from GCNet, PSMNet, and LEAStereo before and after finetuning against the Census-SGM. We evaluate the study sites that have the ground-truth DSM available like Argentina, Omaha, Jacksonville, and London. In general, we can see a notable improvement in the results after the proposed finetuning method, this can be observed from a few aspects. First, the RMSE for the DSMs finetuned are lower than before finetuning. This can be seen by the difference ($\Delta$) column in Table 8.1 that shows the RMSE of the after finetuning minus raw results (before finetuning); the negative values indicate a drop in the RMSE. The overall average drop in the RMSE for GCNet, PSMNet, and LEAStereo ranges between 0.07 and 5.3 meters. Second, we can see that the minimum RMSE for all study sites always falls in one of the finetuned DL methods (See bold numbers in Table 8.1). Third, the average RMSE shows that the lowest errors are achieved by the finetuned DL algorithms of less than 4.80 meters, which is also less than Census-SGM whose average is 5.33 meters. In general, the RMSE values of the finetuned DSMs are better than Census-SGM, where they are lower by 0.02 to 3.24 meters. These results imply that the proposed finetuning method using target data and Census-SGM can improve the performance of stereo matching numerically and visually.

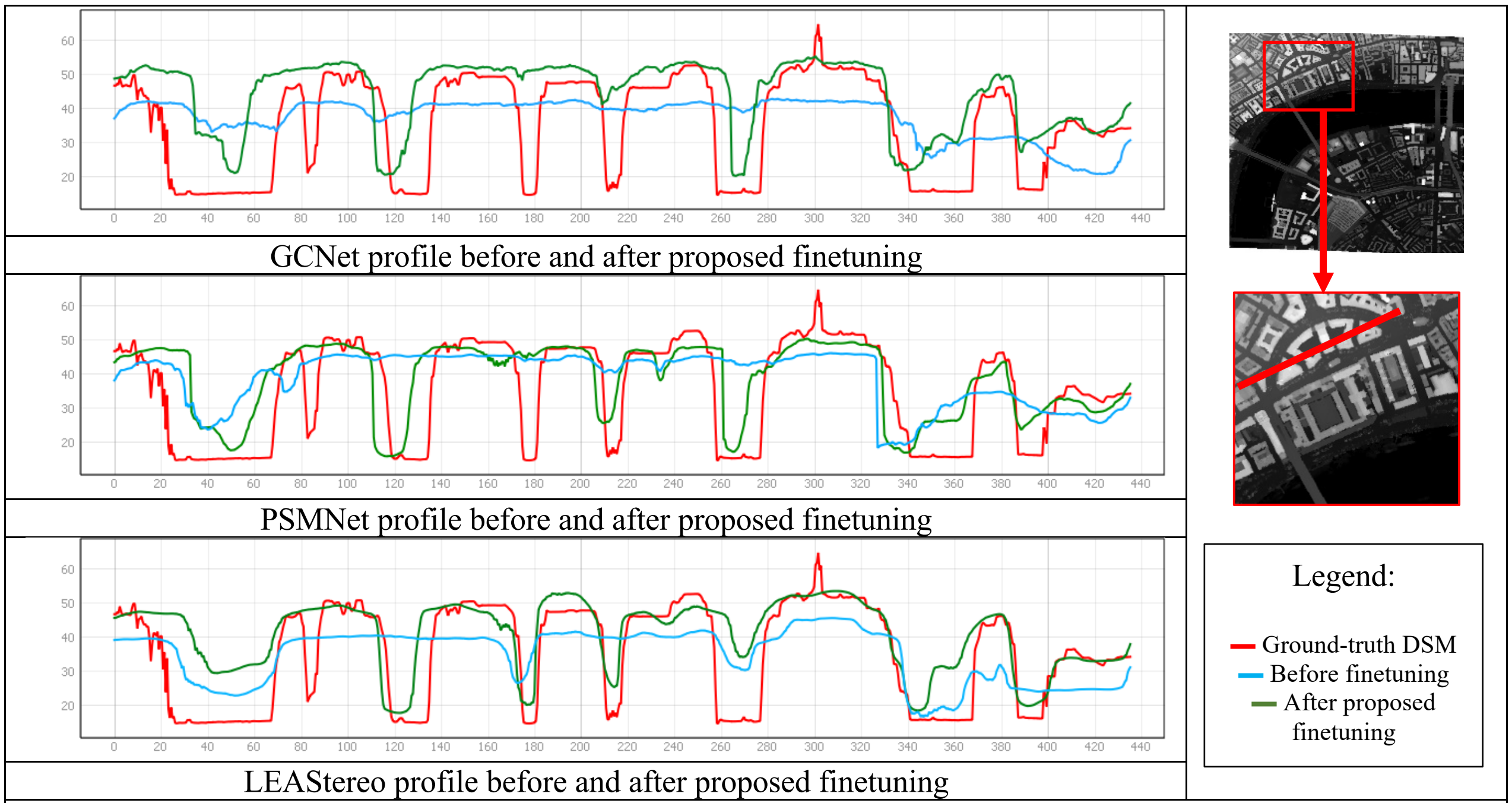

Figure 8.6. The profile for the DSMs generated using DL algorithms before and after finetuning against the ground truth DSM for London I study site. Note: the x-axis is the pixel position in the line segment and the y-axis is the DSM value.

Table 8.1. Evaluation for the RMSE of DSMs generated using different DL algorithms for all study sites. Note: Δ refers to the difference between the RMSE before and after finetuning.

| Study site | Census-SGM | GC-Net | GCNet-finetuned | Δ [GCNet finetuned-GCNet] | PSM-Net | PSMNet finetuned | Δ [PSMNet finetuned-PSMNet] | LEA-Stereo | LEAStereo finetuned | Δ [LEAStereo finetuned-LEAStereo] |
|---|---|---|---|---|---|---|---|---|---|---|
| **Argentina I** | 5.86 | 5.40 | 5.22 | -0.18 | 5.17 | 4.99 | -0.18 | 4.70 | 4.37 | -0.33 |
| **Argentina II** | 6.53 | 3.87 | 3.44 | -0.43 | 4.99 | 4.46 | -0.53 | 4.57 | 3.24 | -1.33 |
| **Argentina III** | 4.23 | 3.78 | 3.51 | -0.27 | 4.02 | 3.66 | -0.36 | 3.64 | 3.53 | -0.11 |
| **Omaha I** | 3.86 | 4.11 | 3.60 | -0.51 | 2.63 | 3.29 | 0.66 | 3.05 | 3.84 | 0.79 |
| **Omaha II** | 5.16 | 6.69 | 6.74 | 0.05 | 4.90 | 4.15 | -0.75 | 4.49 | 4.75 | 0.26 |
| **Omaha III** | 3.23 | 3.18 | 3.94 | 0.76 | 2.84 | 2.24 | -0.60 | 2.69 | 2.80 | 0.11 |
| **Jacksonville I** | 5.02 | 5.03 | 4.84 | -0.19 | 4.34 | 4.95 | 0.61 | 5.46 | 6.06 | 0.60 |
| **Jacksonville II** | 4.38 | 3.38 | 3.16 | -0.22 | 3.08 | 2.91 | -0.17 | 3.19 | 3.60 | 0.41 |
| **Jacksonville III** | 3.79 | 2.82 | 2.60 | -0.22 | 3.00 | 2.93 | -0.07 | 2.68 | 3.15 | 0.47 |
| **London I** | 9.03 | 13.77 | 9.74 | -4.03 | 14.13 | 8.82 | -5.30 | 9.61 | 8.91 | -0.70 |
| **London II** | 6.93 | 8.48 | 6.34 | -2.14 | 5.91 | 5.67 | -0.25 | 6.14 | 5.28 | -0.87 |
| **London III** | 5.93 | 5.74 | 4.75 | -0.99 | 5.02 | 4.67 | -0.36 | 5.06 | 4.78 | -0.28 |
| **Average RMSE (m)** | 5.33 | 5.52 | 4.82 | -0.70 | 5.00 | 4.40 | -0.61 | 4.61 | 4.53 | -0.08 |

## 8.6. Conclusion and Future Works

In this work, we have examined the possibility of integrating the high-performance DL stereo matching methods with the stable Census-SGM as an unsupervised learning approach to enhance transferability across different study sites. We have analyzed the performance of DL methods across different regions and found that they are highly sensitive to unseen regions and environments, which influence the accuracy of the results. We have proposed a finetuning approach that focuses on solving the lack of ground-truth data for satellite images used for stereo matching. It, therefore, aims to adapt and train using information directly from the target data and Census-SGM. The finetuning has improved the results visually and numerically. The overall average drop in the RMSE is from 0.07 to 5.3 meters after finetuning. While the drop between the finetuned and Census-SGM results is between 0.02 to 3.24 meters. Our approach shows high potential and needs to be thoroughly investigated in the future to take advantage of the complementary benefits of both methods.

# Chapter 9. Summary and Conclusion

## 9.1. Summary:

The continuous development of sensors provides unlimited access to numerous multidate, multisite, and multisource RS images. However, obtaining noise-free and accurate images in RS is extremely challenging due to the various sources of errors from the acquisition conditions, configuration parameters, and preprocessing algorithms. Fusion is one of the most popular solutions extensively used to combine several sources of information to provide better interpretability and analyzability of images. Traditional spatial and temporal fusion methods are mostly designed to handle spectral images and specific applications such as pansharpened. Today, the fusion methods extended to cover a variety of image types (e.g., elevation maps, classification maps, and spectral images) and applications (e.g., thematic mapping, 3D reconstruction, and image enhancement). Existing spatial and temporal fusion methods may still be sensitive to some factors such as the spectral heterogeneity, spatiotemporal inconsistencies, and noise in images, which can limit the practical usage of fusion algorithms.

This work has investigated the various ways to improve the performance of fusion algorithms by adapting to multimodal gridded data and their unique uncertainties. We have proposed some fusion solutions to enhance different types of images and applications, and these include the following:

- A spatiotemporal filter to achieve spectral homogeneity between multitemporal images,
- The 3D iterative spatiotemporal filtering to provide consistent classification maps,

- Adaptive and semantic-guided spatiotemporal filtering to enhance the DSM accuracy, and
- Because of the importance of the input's accuracy, we further provide a comprehensive analysis of the accuracy of the DSMs at the stereo matching level, we then propose a finetuning solution to enhance its accuracy.

We demonstrated that the key solution to most fusion approaches is the underlying information in the images, which consists of spatial, temporal, and geometric knowledge. This can be useful to build connections, understand relationships, identify and exclude outliers, and predict using the most relevant and reliable information.

### 9.2. Discussion and Future Directions:

Spatial, temporal, and geometric fusion can be implemented in many innovative ways to enhance RS images. The choice of fusion method depends on several factors like the number of available images, their types, source, and application. The number of available images defines the fusion strategy, for example, if a single image is provided, fusion is performed locally on a group of pixels, whereas fusion is performed using spatial and temporal information for multi-date images. The source of images can involve combining images from similar or different sensors. Fusing images of the same type and from the same sensor requires modeling for minimum biases, whereas combining multisource images is more complicated and requires modeling multimodal data and their unique uncertainties. The type and complexity of the application define the main strategy in which images should be fused. Therefore, no standard fusion approach/strategy works perfectly for all image types and applications. The fusion solution is often considered a case-by-case situation, where the success of fusion depends on multistage processing. Applying fusion should include a comprehensive consideration and understanding of the image type, scene complexity, and uncertainties. This section summarizes a few of the most important considerations and challenges of the fusion problem.

First, a thorough understanding of the images is the first key step toward solving any image fusion problem. Images in RS are remarkably diverse in terms of type (i.e., satellite images, elevation maps, classification maps, etc.), characteristic (e.g., spatial, spectral, and temporal resolutions), content (i.e., pixels can include color, heights, class, etc.), distortion, and uncertainty. This diversity requires specific consideration of the individual differences to extract relevant features and determine a suitable fusion method.

Second, spectral heterogeneity and spatiotemporal inconsistency are critical factors commonly existing in RS images impacting the resultant fused image. They can lead to a major decrease in accuracy if not resolved or addressed in advance. This problem requires first recognizing and differentiating between types of variations, whether due to different radiometric properties or time changes. The spectral variability can be reduced using preprocessing steps, such as RRN methods. In contrast, spatiotemporal variations require background information and image inspection to determine the cause and type of change, for example, whether the change is due to natural causes, such as disasters, or human activities, such as new construction.

Third, single-source or multisource image fusion requires different techniques and approaches. Single-source image fusion has minimum biases and uncertainties, whereas multisource image fusion is more complex because it requires compensation for multimodal data by accounting for their unique uncertainties.

Fourth, the 3D geometric information plays a vital role in improving the spatiotemporal fusion algorithm as it is considered to be more robust towards varying acquisition conditions such as sun angle and weather.

Fifth, adaptive methods have proven their efficiency in increasing the spatiotemporal coherency between images. However, some challenges, such as choosing the parameters and approximate methods to segment and divide the image into classes or categories, still hinder its efficiency.

It is common practice in RS to fuse images by incorporating spatial, spectral, temporal, and geometric information. Fusion must be implemented in an elaborate mechanism to avoid artifacts or loss of information. This work demonstrates that a few key points must be satisfied to provide an efficient fusion algorithm. First, input data should be preprocessed to minimize distortion and achieve spatiotemporal coherence. Second, the fusion algorithm should be able to recognize and suppress undesirable distortion, such as noise and outliers. Finally, the fusion algorithm should have an efficient prediction model that can produce accurate results.